%% file: ArXiv_v1.tex
\documentclass[
  10pt,
  aps,
  prx,
  superscriptaddress,
  twocolumn,
  notitlepage,
  nofootinbib,
  longbibliography,
  floatfix
]{revtex4-2}

\ifdefined\pdfminorversion\pdfminorversion=7\fi
\usepackage[T1]{fontenc}
\usepackage[utf8]{inputenc}
\usepackage{lmodern}

\usepackage{amsmath}
\usepackage{amssymb}
\usepackage{amsfonts}
\usepackage{mathtools}
\usepackage{bm}
\usepackage{amsthm}

\usepackage{graphicx}
\usepackage{ragged2e}
\usepackage{placeins}
\usepackage{booktabs}
\usepackage{array}
\usepackage{multirow}

\usepackage{xcolor}
\usepackage{tikz}
\usetikzlibrary{
  arrows.meta,
  calc,
  positioning,
  decorations.pathmorphing,
  shapes.geometric,
  shapes.misc
}

\usepackage{comment}
\usepackage[normalem]{ulem}

\definecolor{darkblue}{RGB}{0, 50, 130}
\definecolor{darkred}{RGB}{160, 20, 20}
\definecolor{darkgreen}{RGB}{0, 120, 60}
\definecolor{purplecomment}{RGB}{110, 40, 160}

\definecolor{figureBlue}{HTML}{ACDFF3}
\definecolor{figureOrange}{HTML}{FCD5A9}
\definecolor{figureGreen}{HTML}{B2E2B8}
\definecolor{gtnblue}{RGB}{116,184,216}
\definecolor{phaseRule}{HTML}{B4B8BA}

\colorlet{phaseBlue}{figureBlue!65!white}
\colorlet{phaseCream}{figureOrange!65!white}
\colorlet{aregionbase}{figureOrange}
\colorlet{bregionbase}{figureBlue}
\colorlet{cregionbase}{figureGreen}
\colorlet{aregion}{aregionbase!65!white}
\colorlet{bregion}{bregionbase!65!white}
\colorlet{cregion}{cregionbase!65!white}

\newcommand{\papertitle}{Long-range entangled criticality from coherent errors}

\usepackage[
  colorlinks=true,
  linkcolor=darkblue,
  citecolor=darkblue,
  urlcolor=darkblue
]{hyperref}

\hypersetup{
  pdftitle={\papertitle},
  pdfauthor={Amir-Reza Negari; Subhayan Sahu; Jan Behrends; Benjamin B\'eri; Timothy H. Hsieh},
  pdfsubject={Coherent errors, long-range entanglement, and conditional mutual information in surface codes},
  pdfkeywords={surface code, coherent errors, conditional mutual information, Majorana transport}
}

\usepackage[capitalise,nameinlink]{cleveref}

\crefname{equation}{Eq.}{Eqs.}
\Crefname{equation}{Equation}{Equations}
\crefname{section}{Sec.}{Secs.}
\Crefname{section}{Section}{Sections}
\crefname{appendix}{App.}{Apps.}
\Crefname{appendix}{Appendix}{Appendices}
\crefname{figure}{Fig.}{Figs.}
\Crefname{figure}{Figure}{Figures}
\crefname{table}{Table}{Tables}
\Crefname{table}{Table}{Tables}
\crefname{theorem}{Theorem}{Theorems}
\Crefname{theorem}{Theorem}{Theorems}
\crefname{lemma}{Lemma}{Lemmas}
\Crefname{lemma}{Lemma}{Lemmas}

\makeatletter
\def\th@plainpunct{\th@plain
  \thm@headpunct{{\normalfont.}}}
\def\th@definitionpunct{\th@definition
  \thm@headpunct{{\normalfont.}}}
\makeatother

\theoremstyle{plainpunct}
\newtheorem{theorem}{Theorem}
\newtheorem{lemma}{Lemma}
\newtheorem{proposition}{Proposition}
\newtheorem{corollary}{Corollary}
\newtheorem{plainpuncttheorem}[theorem]{Theorem}

\theoremstyle{definitionpunct}
\newtheorem{definition}{Definition}

\newcommand{\tcm}{T.C.M. Group, Cavendish Laboratory, University of Cambridge, J.J. Thomson Avenue, Cambridge, CB3 0US, UK}
\newcommand{\quantinuum}{Quantinuum, Terrington House, 13-15 Hills Rd, Cambridge, CB2 1NL, UK}

\DeclareMathOperator{\Tr}{Tr}

\newcommand{\ket}[1]{\left|#1\right\rangle}
\newcommand{\bra}[1]{\left\langle#1\right|}

\tikzset{
  gaussian transport/.style={
  x=0.66cm,y=0.66cm,
  tensor/.style={circle,draw=black!96,fill=gtnblue,minimum size=3.5mm,
                 inner sep=0pt,line width=0.8pt},
  scatter/.style={circle,draw=black!96,fill=gtnblue,minimum size=6.4mm,
                  inner sep=0pt,line width=0.8pt,font=\scriptsize},
  bond/.style={black!94,line width=0.9pt,line cap=round},
  fline/.style={draw=blue!72!black,line width=0.82pt},
  bline/.style={draw=orange!92!black,line width=0.82pt},
  forward/.style={-{Latex[length=1.35mm,width=1.05mm]},
                  draw=blue!72!black,line width=0.82pt},
  backward/.style={-{Latex[length=1.35mm,width=1.05mm]},
                   draw=orange!92!black,line width=0.82pt}
  }
}

\newcommand{\phaseTableSetup}{%
  \small
  \setlength{\tabcolsep}{0pt}%
  \renewcommand{\arraystretch}{1.0}%
}

\newcommand{\phasecell}[1]{\begin{tabular}[c]{@{}c@{}}#1\end{tabular}}

\newsavebox{\phaseHeader}
\newsavebox{\phaseMemory}
\newsavebox{\phaseEntanglement}
\newsavebox{\phaseCMI}
\newsavebox{\phaseSymmetry}
\newsavebox{\phaseCircuit}
\newsavebox{\phaseTable}

\newcommand{\phaserow}[2]{\sbox{#1}{\begin{minipage}{\textwidth}
      \vspace*{3pt}
      \begin{tabular*}{\textwidth}{@{\extracolsep{\fill}}
        >{\centering\arraybackslash}m{0.28\textwidth}
        >{\centering\arraybackslash}m{0.34\textwidth}
        >{\centering\arraybackslash}m{0.34\textwidth}@{}}
        #2
      \end{tabular*}
      \vspace*{3pt}
    \end{minipage}}}

\newcommand{\phaseheight}[1]{\the\dimexpr\ht#1+\dp#1\relax}

\begin{document}

\title{\papertitle}

\author{Amir-Reza~Negari}
\email{anegari@uwaterloo.ca}
\affiliation{Perimeter Institute for Theoretical Physics, Waterloo, Ontario N2L 2Y5, Canada}
\affiliation{Department of Physics and Astronomy, University of Waterloo, Ontario N2L 3G1, Canada}

\author{Subhayan~Sahu}
\affiliation{Perimeter Institute for Theoretical Physics, Waterloo, Ontario N2L 2Y5, Canada}

\author{Jan~Behrends}
\altaffiliation[Present address: ]{\quantinuum}
\affiliation{\tcm}

\author{Benjamin~B\'eri}
\affiliation{\tcm}
\affiliation{DAMTP, University of Cambridge, Wilberforce Road, Cambridge CB3 0WA, United Kingdom}

\author{Timothy~H.~Hsieh}
\affiliation{Perimeter Institute for Theoretical Physics, Waterloo, Ontario N2L 2Y5, Canada}
\affiliation{Department of Physics and Astronomy, University of Waterloo, Ontario N2L 3G1, Canada}

\begin{abstract}

We find that coherent error can induce a phase transition from the clean surface code phase into an extended long-range entangled critical regime with diverging Markov length.  To demonstrate this, we establish two general results of independent interest.  The first relates topology and quantum information to mixed-state entanglement: we show that when a noise channel acting on a topological code preserves a higher-form symmetry and nonzero coherent information, the noisy state for generic logical input is long-range entangled.  The second generalizes the local recovery construction based on finite Markov length to any locally parameterized family, thus encompassing local Lindbladian evolution, coherent errors followed by syndrome dephasing, and thermal states. We apply these results to the honeycomb surface code, in which we find an extended long-range entangled critical regime that is a quasi-long-ranged counterpart of 1-form strong-to-weak spontaneous symmetry breaking.  As a byproduct of these studies, we relate the conditional mutual information to conductance in a Majorana network model.        

\end{abstract}

\maketitle

\section{Introduction}
\label{sec:introduction}

\begin{figure}[t]
 \centering
 \includegraphics[width=\columnwidth]{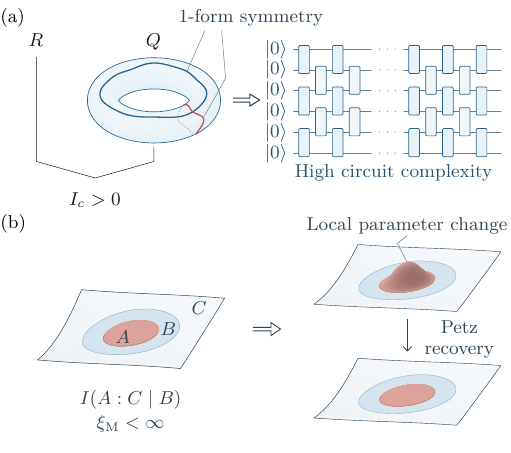}
 \caption{Two general results for noisy topological codes.
(a) When a strong 1-form symmetry is preserved, surviving quantum
information implies long-range entanglement: the noisy state cannot be
approximated by a convex mixture of pure states
prepared by shallow local circuits
(Theorem~\ref{thm:ci-lre-criterion}).
(b) In a locally parameterized family, finite Markov length
allows states related by a local parameter change to be connected by local Petz recovery,
keeping the states in the same phase
(Theorem~\ref{thm:marginal-preserving-local-reverse}).
}
 \label{fig:summary}
 \label{fig:ci-lre-schematic}
 \label{fig:locally-parameterized-family}
\end{figure}

There is a fundamental connection between error-correcting codes and phases of matter, extending back to the $d\geq 2$ classical Ising model, whose low temperature phase retains a classical bit of information encoded into a repetition code.  In recent years, this relation between error correction and phases has been extended to the non-equilibrium setting.  A canonical example is the toric code (a quantum memory) subject to bit-flip error, with a noise threshold above which the quantum memory is lost \cite{Dennis_2002}.

Defining phases in this setting requires going beyond the equilibrium paradigm of ground and thermal states.  As a coarse definition, two mixed states belong to the same phase if they can be transformed into each other using local Lindbladian evolution ~\cite{Coser_2019,SangZouHsieh2024MixedStatePhases, SangEtAl2025MixedStatePhasesLocalReversibility}. The spatial range of conditional correlations provides a way to verify when a local Lindbladian evolution admits a reverse local evolution. For two regions separated by a buffer, conditional mutual information (CMI) measures the correlations that
are not mediated by the buffer. If
these correlations decay exponentially with the buffer width, the characteristic scale
defines the Markov length $\xi_{\mathrm M}$. When $\xi_{\mathrm M}$ remains finite
along a local Lindbladian evolution, the evolution can be reversed
locally \cite{Sang_2025}, keeping the evolving state
in the same phase. Under this definition, the threshold of the toric code against bit-flip error coincides with a transition between the toric code phase and a short-range entangled phase, as the Markov length of the noisy state is finite everywhere except at the threshold \cite{Sang_2025}.  

This phase transition also coincides with a transition in the entanglement structure of the noisy mixed-state \cite{Fan_2024, ChenGrover2024SeparabilityTransitions}. In particular, the topological entanglement negativity \cite{ten} exhibits a sharp jump at the threshold \cite{Fan_2024}, and indeed there is an important structural manifestation. 
A mixed state is called short-range entangled (SRE) if it can be
approximated by a convex mixture of SRE pure states \cite{hastingstop}, and long-range entangled (LRE) otherwise.  Increasing the noise rate turns the noisy toric code from LRE to SRE right at the code threshold  ~\cite{ChenGrover2024SeparabilityTransitions,
lessa2025higherformanomalylongrangeentanglement,
li2025entanglementneededtopologicalcodes}.  Thus, it is satisfying that all three transitions (in quantum memory, phase, and entanglement) coincide in this case.

In contrast, the relations between error correction threshold and phase/entanglement structure has not been understood for coherent errors.  Coherent errors are systematic rotations of constituent qubits. These are experimentally relevant~\cite{Hashim2021RandomizedCompiling,Google2025BelowThreshold,Ishii2025CoherentSuppression,Deshmukh2026CoherentErrors}, yet they pose challenges to existing theoretical frameworks.  For example, it is only recently that the existence of a threshold for coherent error has been analytically established~\cite{HanEtAl2026CoherentThreshold, tamiya2026proofpositivecoherenterrorthreshold} on general grounds.  Regarding the question of phase connectivity, coherent errors are challenging because the noisy states after syndrome measurements are not explicitly connected by even a one-way Lindbladian evolution.  Likewise, the entanglement structure of coherently noisy states and its relation to the code threshold and quantum memory has not been characterized.  Finally, given the increasingly prevalent modern perspective of topological order as spontaneous breaking of a {\it higher-form} symmetry-- one acting on extended as opposed to point-like objects, one would also like to understand the effect of coherent noise from a symmetry point of view.     

Motivated by these challenges, in this work we first establish two general results. (1) If noise preserves one of the code's anomalous higher-form symmetries, then positive coherent information implies that the noisy state is long-range entangled for every generic encoded input. (2) We generalize the finite-Markov-length local-recovery construction beyond Lindbladian evolution to any locally parameterized family of states.  We apply these results to the honeycomb surface code subject to coherent errors and thereby establish a transition from the surface code phase (with finite Markov length) to a critical mixed-state regime (with infinite Markov length) that is nonetheless long-range entangled.  The latter is a quasi-long-ranged version of 1-form strong-to-weak symmetry breaking \cite{PRXQuantum.4.030317, lessa_strong--weak_2024, PhysRevB.110.155150, ZhangXuZhangXuBiLuo2025}. As a byproduct, we also relate conditional correlations to transport in the scattering network model description of the noisy toric code, finding that conditional mutual information and conductance exhibit the same scaling behavior.

\subsection{Main results}
\label{subsec:main-results}

\input{figures/Table}

Figure~\ref{fig:summary} illustrates our two general results. The first connects the logical information that survives noise to long-range entanglement in noisy topological codes. The codes we consider are defined by a pair of mutually anomalous higher-form symmetries. These symmetries act on extended objects such as strings, and in two dimensions, for example, one-form symmetries are represented by loop operators. Fixing the eigenvalues of the contractible loops defines the code space, while noncontractible loops act on the encoded logical information. As reviewed in Sec.~\ref{subsec:strong_strong_memory}, the noncommutation of these logical loops reflects the mutual anomaly. A noise channel preserves one of the symmetries as a strong symmetry when each of its Kraus operators commutes with the corresponding symmetry operators. We show in Sec.~\ref{subsec:noisy-one-strong-symmetry} that if the noise channel $\mathcal N$ preserves either strong one-form symmetry, then
\begin{equation}
 I_c(\mathcal N)>0
 \quad\Longrightarrow\quad
 \mathcal N(\psi)\in\mathrm{LRE}.
 \label{eq:positive-ci-implies-lre}
\end{equation}
Here $\psi$ is a generic pure encoded input. The coherent information $I_c$ measures how much quantum information survives the noise when the logical degrees of freedom are initially maximally entangled with a reference. We require $I_c$ to remain bounded away from zero in the thermodynamic limit. Appendix~\ref{app:lre-proof} gives the detailed proof. The criterion also extends to $d$-dimensional codes with mutually anomalous $k$-form and $(d-k)$-form symmetries, provided the noise preserves either strong symmetry.

Our second result, developed in Sec.~\ref{subsec:markov-recovery}, establishes phase equivalence between mixed states using locally parameterized families of states. We write the density matrix as $\rho[\boldsymbol\lambda]$, where $\boldsymbol\lambda=\{\lambda_x\}$ is a field of parameters associated with lattice sites. Varying these parameters one at a time defines a path through the family. The family is locally parameterized if changing only $\lambda_x$ leaves the reduced density matrix outside a bounded region around $x$ unchanged. Using this locality property, we generalize the finite-Markov-length recovery construction of Ref.~\cite{Sang_2025} beyond local Lindbladian evolution. When the Markov length remains finite along such a locally parametrized  path, all states on the path belong to the same phase.

In Appendix~\ref{app:unified-phase-equivalence}, we extend the construction to approximately locally parameterized families and apply it to thermal states. In these families, a local parameter change can affect distant regions, but its influence decays exponentially with distance. When both the decay length of this response and the Markov length remain uniformly finite throughout the local updates, a local channel can approximately implement each update. Composing these channels along the path in either direction gives the local circuits required for phase equivalence.

Both results are motivated by the problem of coherent rotations followed by stabilizer measurements in topological codes, which has attracted extensive attention~\cite{Bravyi_2018,Beale2018QECdecoheres,HuangDohertyFlammia2019CoherentErrors,Iverson_2020,Venn_PhysRevResearch.2.043412,MartonAsboth2023CoherentReadout,Venn_2023,ZhuTrebst2023MajoranaLiquid,Behrends_PhysRevResearch.6.013137,ChenGrover2024UnconventionalMixedStateTransition,Bao_Anand_Phases,PatoStaplesBrown2025LogicalCoherence,Cheng_2025,LeeMoon2025CoherentNoise,Behrends_generic1qb_gskb-t5ql,Darmawan2025DecoderAdaptation,Behrends_SBPC,liu2025coherenterrorinducedphase}. Because the rotations precede syndrome dephasing, varying their angles does not directly specify a local Lindbladian evolution between the resulting mixed states. These states nevertheless form a locally parameterized family, so we can apply our recovery construction wherever the uniform finite-Markov-length condition holds.

In particular, we study the surface code on a two-dimensional honeycomb lattice under uniform coherent $Z$ rotations followed by stabilizer measurements, as described in Sec.~\ref{subsec:surface-code-geometry}. For a rotation $U_\theta=e^{i\theta\sum_i Z_i}$ and an encoded input $\rho_Q$, the syndrome-dephased state is
\begin{equation}\label{eq:post-syndrome-state}
 \rho_\theta=\mathcal D_S(U_\theta\rho_Q U_\theta^\dagger).
\end{equation}
Here $\mathcal D_S(\rho)=\sum_s P_s\rho P_s$ dephases between the syndrome sectors of the code's stabilizer group $S$, and $P_s$ projects onto syndrome $s$. In Sec.~\ref{subsec:coherent-information}, we find that the coherent information approaches one bit below the maximum-likelihood Pauli-decoding threshold $\theta_c\simeq0.09\pi$~\cite{yan2026nonlinearsigmamodelsurface,yang2026decodingcoherenterrorstoric,bejan2026projectedlogicalensemblessurface} and a smaller positive plateau, $I_c=1/(2\ln2)$, above it (but below $\theta=\pi/4$), cf.~Fig.~\ref{fig:coherent-information_honeycomb_surface}. The latter result is consistent with the projected logical ensemble~\cite{bejan2026projectedlogicalensemblessurface}, the distribution of logical states over all syndrome outcomes, approaching a Haar distribution above threshold. Thus, a partial quantum memory survives the loss of reliable Pauli decoding. Since the combined rotation and dephasing channel preserves the strong $Z$-type one-form symmetry, our first result establishes LRE for generic encoded inputs on both sides of the threshold. Appendix~\ref{app:pi4-endpoint} discusses exceptional logical inputs and the solvable $\theta=\pi/4$ endpoint.

We next ask whether the decoding threshold separates distinct mixed-state phases. If the Markov length remains uniformly finite as the rotation angles vary, our recovery construction for locally parameterized families gives local circuits connecting the noisy states to the clean code in both directions. We therefore examine how the long-distance behavior of the conditional mutual information  changes across the threshold.

In Sec.~\ref{sec:cmi-markov-length}, we evaluate the CMI ($I(A:C|B)$) across an $L\times L$ honeycomb surface-code. As shown in Fig.~\ref{fig:cmi-geometry}, $A$ and $C$ are one-row regions on opposite physical boundaries, and $B$ is the bulk region separating them. Our numerical results support the scaling form
\begin{equation}
 I(A:C|B)\sim
 \begin{cases}
  \exp[-L/\xi_{\mathrm M}(\theta)], & \theta<\theta_c,\\
  O(1), & \theta=\theta_c,\\
  \ln L, & \theta_c<\theta<\pi/4.
 \end{cases}
 \label{eq:main-cmi-scaling}
\end{equation}
Below threshold, the CMI decays exponentially, defining a finite Markov length $\xi_{\mathrm M}(\theta)$ in this geometry. At the threshold, it becomes scale invariant. Above threshold, it grows logarithmically, indicating critical conditional correlations and a divergent Markov length. The states below threshold thus belong to the clean surface-code phase, while the regime above threshold combines critical conditional correlations with long-range entanglement.

We next examine the critical LRE regime more closely (in Sec.~\ref{sec:recoverability}) and show that its surviving partial quantum memory cannot be decoded locally. Following Ref.~\cite{negari2026criticalnonequilibrium}, we compare the coherent information $I_c$ with the punctured coherent information $I_c^{(p)}$ that remains after erasing a bulk region. In the clean code, the logical operators can be deformed around this region, so erasing it leaves the encoded information intact. A local decoder can spread the effect of the erasure only over its range, and we choose the puncture so that the enlarged region remains correctable in the clean code. Any information recoverable by a local decoder must therefore survive the puncture. In Sec.~\ref{subsec:punctured-coherent-information}, we find that $I_c^{(p)}$ approaches $I_c$ below threshold. Above threshold, $I_c^{(p)}$ approaches zero while the unpunctured code retains positive coherent information, showing that recovery of the surviving memory requires a global decoder.

The emergent one-form symmetries explain why the punctured coherent information approaches the full coherent information below threshold. Applying the symmetry pullback of Ref.~\cite{lessa2025higherformanomalylongrangeentanglement} to the local recovery channel of Sec.~\ref{subsec:coherent-local-recovery} gives dressed $X$ and $Z$ strings that generate an emergent pair of one-form symmetries. Both strings can be routed around the puncture while preserving their crossing algebra, so the encoded qubit remains accessible outside the erased region. Above threshold, the microscopic $Z$-type symmetry remains exact, but the loss of coherent information upon puncturing suggests that a deformable $X$-type partner no longer protects the memory.

In Sec.~\ref{subsec:one-form-fidelity}, we test for this partner by inserting an open $Z$ string $Z_\gamma$, which creates electric charges at its endpoints $x$ and $y$. A complementary $X$ loop enclosing just one endpoint anticommutes with the string, making the states with and without the charge pair perfectly distinguishable. We quantify their distinguishability using the open-string fidelity $\mathcal F_Z(x,y)=F(\rho_\theta,Z_\gamma\rho_\theta Z_\gamma)$. An exponentially localized strong $X$ partner with this crossing algebra forces the fidelity to decay exponentially with endpoint separation, as we observed below threshold (Fig.~\ref{fig:swssb-scaling}). In the critical regime above threshold, the fidelity data support algebraic decay with possible logarithmic corrections (Fig.~\ref{fig:swssb-scaling}). This slow decay obstructs an exponentially localized complementary $X$ partner. 

The fidelity correlator also serves as a disorder parameter for one-form strong-to-weak spontaneous symmetry breaking (SW-SSB)~\cite{lessa_strong--weak_2024,ZhangXuZhangXuBiLuo2025}. A nonzero limiting fidelity at large endpoint separation would signal conventional one-form SW-SSB. The algebraic decay observed here gives a quasi-long-range counterpart of this order. We use this quasi-long-range behavior of one-form SW-SSB to establish that the critical regime is distinct from the surface-code phase below threshold. In Theorem~\ref{thm:fidelity-obstruction-local-recovery}, we prove that a circuit of sublinear range that returns the noisy state to the clean code with superpolynomially small error requires the open-string fidelity to decay superpolynomially with $L$. This bound applies to arbitrary local recovery channels, including those that break the code's symmetries. The algebraic decay supported by our data above threshold violates this necessary condition, ruling out such a local return to the clean surface code.

The statistical mechanics of maximum-likelihood Pauli decoding relates the threshold of the honeycomb surface code to an insulator--metal transition in a class-DIII Majorana network~\cite{Venn_2023,yan2026nonlinearsigmamodelsurface,yang2026decodingcoherenterrorstoric,bejan2026projectedlogicalensemblessurface}. As the network becomes metallic, conditional correlations in the noisy code become long-ranged. In Sec.~\ref{sec:network-localization}, we explore this connection by relating the boundary-to-boundary CMI of the code to the conductance of the network. In the localized regime, this relation gives the rigorous bound $\xi_{\mathrm M}\leq\xi_{\mathrm{loc}}$. We further argue that the two lengths coincide, $\xi_{\mathrm M}=\xi_{\mathrm{loc}}$, in agreement with the numerical results in Fig.~\ref{fig:length-scale-comparison}.

Taken together, these information-theoretic probes distinguish two regimes of the honeycomb surface code under coherent errors. Below threshold, an emergent pair of one-form symmetries protects the full logical qubit, which can be recovered by a local decoder. Above threshold, the surviving partial memory requires a global operation, while the slow decay of the open-string fidelity obstructs a localized $X$-type partner for the exact $Z$-type symmetry. The preserved $Z$-type symmetry and positive coherent information still certify long-range entanglement for generic encoded inputs. Coherent errors thus give rise to an extended mixed-state regime in which long-range entanglement coexists with critical conditional and non linear correlations~\cite{kitptalk}.

\tableofcontents

\section{Quantum Memory, Long-Range Entanglement, and Local Recovery}
\label{sec:lre}

This section develops two
general results for noisy quantum topological error correcting codes: first, we will show that when a noise channel
preserves one of two anomalous one-form symmetries, positive coherent
information implies that generic encoded outputs are long-range entangled
(LRE).
  Second, finite Markov length along a locally parameterized family
allows us to construct local recovery channels and establish phase
equivalence. Before presenting the main results, we briefly recap the construction of quantum memory from anomalous higher form symmetries in Sec.~\ref{subsec:strong_strong_memory}, highlighting the fact that the anomalous higher form symmetries imply LRE for all noisy output states.

\subsection{Quantum memory from anomalous one-form symmetries}
\label{subsec:strong_strong_memory}

Topological codes store quantum information in encoded degrees of freedom
associated with anomalous higher-form symmetries. We consider a general two-dimensional code with two finite Abelian one-form symmetries, represented
by loop operators $W_a(\gamma)$ and $\widetilde W_b(\tau)$.
  Their
contractible loops commute, and their common $+1$ eigenspace defines the
code. We refer to these $+1$-eigenvalue conditions as the code constraints.
  Logical strings wrap around noncontractible cycles on a torus or
connect the appropriate boundaries of a planar patch.  These strings act
on the encoded degrees of freedom of the quantum memory.

The two symmetries have a mixed anomaly expressed by the crossing algebra
\begin{equation}
 W_a(\gamma)\widetilde W_b(\tau)
 =M_{ab}^{\,I([\gamma],[\tau])}
  \widetilde W_b(\tau)W_a(\gamma),
 \label{eq:ss_mixed_anomaly}
\end{equation}
where $a,b$ label symmetry elements, $[\gamma]$ and $[\tau]$ denote the
homology classes of the curves (relative to the appropriate boundaries for
open strings), $I([\gamma],[\tau])$ is their oriented intersection number,
and $M_{ab}$ is the mutual braiding phase.
We take the pairing to be nondegenerate: every nontrivial element has a
nontrivial braiding phase with some element of the other
family~\cite{BarkeshliBondersonChengWang2019}.

We consider geometries in which each logical string has equivalent
representatives separated by a distance proportional to the code distance $D$.
The dimension of the encoded Hilbert space is fixed as $D$ grows, and the joint eigenvalues of either commuting logical-string family uniquely label an encoded basis.

Deformability and this noncommuting algebra together enforce LRE, regardless of the microscopic details of the code.

\begin{definition}[Short- and long-range entanglement]
\label{def:mixed-state-lre}
We call a family of mixed states $\{\rho_L\}$ defined on a tensor product Hilbert space of linear size $L$ short-range entangled (SRE)
if each $\rho_L$ can be approximated by a convex mixture
\begin{equation}
 \rho_L\simeq\sum_\alpha p_{\alpha,L}
       \ket{\psi_{\alpha,L}}\!\bra{\psi_{\alpha,L}}.
 \label{eq:mixed-sre-decomposition}
\end{equation}
The trace-norm approximation error must vanish as $L\to\infty$, with no prescribed rate.
Every state in the mixture must be preparable from a product state by a
geometrically local unitary circuit, such that a common light-cone radius
$r_L=o(L)$\footnote{Throughout, we use Landau notation for scaling:
a quantity is $O(L)$ if it grows at most linearly with $L$,
$\Theta(L)$ if it scales linearly up to constant factors,
and $o(L)$ if it grows sublinearly, becoming negligible
compared with $L$ as $L\to\infty$.} applies to all components.

  If no such approximation exists,
the family is long-range entangled (LRE).
\end{definition}

Our notion of a local circuit is deliberately broad: allowing any sublinear
light-cone radius $r_L=o(L)$ includes both strictly finite-range circuits and
circuits whose range grows polylogarithmically with $L$.  Consequently, LRE
established with respect to this broad class also implies LRE under either
of these more restrictive notions.  Related mixed-state notions are used in
Refs.~\cite{ChenGrover2024SeparabilityTransitions,
lessa2025higherformanomalylongrangeentanglement,
li2025entanglementneededtopologicalcodes,
hsin2025higherformanomaliesimplyintrinsic,
ThorngrenGioiaZhang2026TranslationLRE,
LessaLu2026DimensionalConstraints}.
The role of approximation error and circuit complexity is discussed in
Ref.~\cite{yuhsueh,yhgrover,NegariZouHsieh2026Instability}, suggesting that such a broad
definition may have suitable stability properties.

\begin{proposition}
\label{prop:lre}
Every pure state in a code defined by these symmetries is LRE.
\end{proposition}

\begin{proof}[Proof sketch]
Suppose a pure encoded state $\ket\psi$ were SRE.  Choose two distant
representatives $W^L,W^R$ of the same logical string.  Their product is a
product of contractible loop operators, so $W^L(W^R)^\dagger = 1$ on the
logical subspace and its expectation value is one.
  The circuit preparing the SRE state from a product state has disjoint
backward light cones for the two strings, giving
$\langle W^L(W^R)^\dagger\rangle
=\langle W^L\rangle\langle W^R\rangle^*=1$.

Both factors must have unit magnitude, making $\ket\psi$ an eigenstate
of the logical string.  Repeating the argument for the other family
would give a simultaneous eigenstate of two crossing strings with a
nontrivial braiding phase, which their algebra forbids.
Appendix~\ref{app:lre-proof} gives the factorization argument and its
approximate form.
\end{proof}

\subsection{Long-range entanglement from coherent information}
\label{subsec:noisy-one-strong-symmetry}

We now turn to noisy encoded states and ask when their LRE can still be
established.  We say that a noise channel
$\mathcal N(\rho)=\sum_\mu K_\mu\rho K_\mu^\dagger$ has $W$ as a strong
one-form symmetry if every Kraus operator commutes with every $W$ operator,
including the logical representatives:
\begin{equation}
 [K_\mu,W_a(\gamma)]=0
 \qquad\text{for every }\mu,a,\gamma.
 \label{eq:strong-channel-symmetry-main}
\end{equation}
Here $a$ labels elements of the preserved $W$ one-form symmetry;
preservation of the complementary $\widetilde W$ symmetry is not required.

The symmetry is therefore preserved in every noise branch.  In particular,
two distant representatives remain perfectly correlated,
$W_a^L(W_a^R)^\dagger\mathcal N(\rho_Q)=\mathcal N(\rho_Q)$,
for every encoded input $\rho_Q$.

This constraint alone does not establish LRE.  A mixed state can satisfy
it through classical uncertainty over definite string eigenvalues,
as in classical loop ensembles~\cite{negari2026criticalnonequilibrium}.
Coherent information distinguishes a surviving quantum memory from
such classical information.

We quantify the surviving logical information by entangling the encoded
degrees of freedom with a reference $R$ that is left untouched by the noise.
Let $\{\ket{\overline{i}}_Q\}_{i=0}^{d_{\rm code}-1}$ be the
logical-charge basis of the encoded space of dimension $d_{\rm code}$.
This is the simultaneous eigenbasis of the commuting logical string
operators of the preserved $W$ symmetry. We take the reference $R$ to have
the same dimension.
Starting from the encoded Bell state
\begin{equation}
 \ket{\Phi}_{RQ}
 =\frac{1}{\sqrt{d_{\rm code}}}
   \sum_{i=0}^{d_{\rm code}-1}\ket{i}_R\ket{\overline{i}}_Q,
 \label{eq:encoded-bell-input}
\end{equation}
we write $\Phi_{RQ}=\ket\Phi\!\bra\Phi$, apply $\mathcal N$
to $Q$, and denote the resulting joint state by $\omega_{RQ}$.
  Its coherent information is
\begin{equation}
 I_c(\mathcal N)=S(Q)_\omega-S(RQ)_\omega.
 \label{eq:coherent-information}
\end{equation}
Throughout the paper, entropies and information quantities are in bits:
$S(\rho)=-\Tr(\rho\log_2\rho)$.  We reserve $\ln$ for natural logarithms.
For one encoded qubit, $I_c=1$ means that it is perfectly recoverable, while
$I_c>0$ certifies surviving quantum correlations with the reference.

The criterion is summarized schematically in Fig.~\ref{fig:summary}(a).

\begin{theorem}
\label{thm:ci-lre-criterion}
Consider a family of codes defined by an anomalous pair of Abelian one-form
symmetries with the crossing
algebra in Eq.~\eqref{eq:ss_mixed_anomaly} and the code assumptions stated
in Sec.~\ref{subsec:strong_strong_memory}.
 Suppose the noise channel $\mathcal N$ is strongly symmetric with respect to one of the one-form symmetries.
 For every fixed generic pure encoded
input $\ket\psi$,
\begin{equation}
 I_c(\mathcal N)>0
 \quad\Longrightarrow\quad
 \mathcal N(\psi)\in\mathrm{LRE}.
 \label{eq:ci-implies-lre-theorem}
\end{equation}
\end{theorem}

The coherent information is evaluated using the maximally entangled
reference--code input in Eq.~\eqref{eq:encoded-bell-input}, whereas
$\ket\psi$ is a pure encoded state of $Q$ alone.

Here $\psi=\ket\psi\!\bra\psi$, and the implication concerns the
thermodynamic family: $I_c(\mathcal N)>0$ means
$\liminf_{L\to\infty}I_c(\mathcal N_L)>0$, while LRE refers to the
family $\{\mathcal N_L(\psi_L)\}_L$, with $\psi_L$ encoding the same
logical input at each size $L$.
A pure input $\ket\psi=\sum_\eta c_\eta\ket\eta$ is generic if
$c_\eta\ne0$ for every logical-charge basis state $\ket\eta$.
Here $\eta$ labels the joint eigenvalues of the preserved
commuting logical strings.
 For one encoded
qubit, the only excluded pure inputs are the two eigenstates of the
preserved logical string.  The input is fixed as $L$ grows; no uniform
bound is claimed when one of its charge amplitudes tends to zero.
These exceptional inputs form a set of measure zero and are discussed in
Appendix~\ref{app:pi4-endpoint}.

\begin{proof}[Proof sketch]
Suppressing the size index, assume, for contradiction, that the output
$\sigma=\mathcal N(\psi)$, with $\psi=\ket\psi\!\bra\psi$, admits an
exact SRE decomposition,
$\sigma=\sum_\alpha p_\alpha
\ket{\phi_\alpha}\!\bra{\phi_\alpha}$.
Choose two representatives $W^L$ and $W^R$ of the same preserved
logical string, separated by a distance proportional to the code distance.
Their product is a product of contractible $W$ loops, and strong
symmetry preserves the corresponding code constraints in every noise branch.
  Consequently,
$W^L(W^R)^\dagger\sigma=\sigma$: the output is supported entirely in the
$+1$ eigenspace of $W^L(W^R)^\dagger$.  Because the SRE decomposition is a
positive sum, every component with $p_\alpha>0$ must lie in that same
eigenspace.  Hence
$1=\bra{\phi_\alpha}W^L(W^R)^\dagger\ket{\phi_\alpha}$.

Now write $\ket{\phi_\alpha}=U_\alpha\ket{0}^{\otimes n}$, where, by the SRE assumption in Definition~\ref{def:mixed-state-lre}, the
preparation circuit has sublinear light-cone radius $r_L=o(L)$,
common to all components.
  For sufficiently
large code distance, the backward light cones of $W^L$ and $W^R$ are
disjoint, so their expectation value on the product input factorizes:
$1=\langle W^L\rangle_\alpha\langle W^R\rangle_\alpha^*$.
Each factor has magnitude at most one, and their product has magnitude
one.  Both must therefore have unit magnitude, which for a unitary string
operator means that $\ket{\phi_\alpha}$ is its eigenstate.  Applying the
same argument to every preserved logical string shows that each SRE
component has a definite logical charge.  Thus $[\sigma,W_a^L]=0$ for all
$a$ in the preserved $W$ symmetry family, and $\sigma$ is
block-diagonal in their joint logical-charge sectors.

Such a diagonal output retains only a classical charge label, not coherent
quantum information.  For a generic encoded input, strong symmetry then
forces the channel to erase every off-diagonal logical coherence, so its
action on the encoded degree of freedom is complete dephasing.  The
joint reference--output state
$\omega_{RQ}=(\operatorname{id}_R\otimes\mathcal N)(\Phi_{RQ})$
obtained from the encoded Bell input in Eq.~\eqref{eq:encoded-bell-input}
therefore has $I_c=0$, contradicting the hypothesis.
  Appendix~
\ref{app:lre-proof} gives the channel-level calculation and the approximate
argument excluding SRE mixtures with vanishing trace-norm error.
\end{proof}

The criterion applies to both stochastic and coherent errors.  A first
application is the stochastic decoding transition, where surviving coherent
information establishes LRE at criticality.

Consider the toric code on a square lattice, defined by the Hamiltonian
$H_{\mathrm{TC}}=-\sum_{p} B_p-\sum_{s} A_s$, where $B_p=\prod_{e\in\partial p} Z_e$ is the $Z$-plaquette term and $A_s=\prod_{e\ni s} X_e$ is the $X$-star term.
Consider independent stochastic $Z$ errors in the toric code,
$\mathcal E_i^p(\rho)=(1-p)\rho+pZ_i\rho Z_i$.
The channel $\mathcal N_p=\prod_i\mathcal E_i^p$ preserves the
$Z$-type one-form symmetry.  We consider one encoded qubit, on a planar
patch or with the other logical degree of freedom on the torus fixed.

Below the threshold $p^*$, local recovery gives emergent anomalous
one-form symmetries: the clean code's loop operators can be pulled
back through a local dilation of the recovery channel while preserving their mutual
anomaly~\cite{lessa2025higherformanomalylongrangeentanglement}.
This establishes LRE below threshold.  At criticality, however, the
Markov length diverges, so this construction no longer establishes
local recovery or the emergent symmetries.  The SRE decompositions
above threshold~\cite{ChenGrover2024SeparabilityTransitions} also leave
the critical point unresolved.

This unresolved critical point provides the first nontrivial consequence
of Theorem~\ref{thm:ci-lre-criterion}:
\begin{corollary}\label{thm:Ic_toric_code_dephasing}
For toric code under $Z$ dephasing noise at the transition threshold, generic logical inputs give LRE outputs.
\end{corollary}

\begin{proof}
Theorem~\ref{thm:ci-lre-criterion} requires only a nonzero amount of
surviving coherent information.  For stochastic errors, $I_c$ is
determined by logical error probabilities, which map to domain-wall
free energies in the random-bond Ising model on the Nishimori line
\cite{Dennis_2002,Lee_2025}.  Assume the finite-size scaling form
$I_c(\mathcal N_p)=f((p-p^*)D^{1/\nu})$ where D is the code distance, up to vanishing corrections,
with a smooth, strictly decreasing crossover from $f(-\infty)=1$ to
$f(+\infty)=0$.  Then $0<f(0)<1$: some quantum information survives
at criticality, although the full qubit cannot be recovered.  Theorem~
\ref{thm:ci-lre-criterion} therefore implies that generic outputs at the
transition are LRE.
\end{proof}

This criterion is particularly useful for coherent errors followed by
syndrome measurement, as we show in Sec.~\ref{sec:surface_code_coherent_error_coherent_info}.

\begin{figure}
    \centering
    \includegraphics[width=0.8\linewidth]{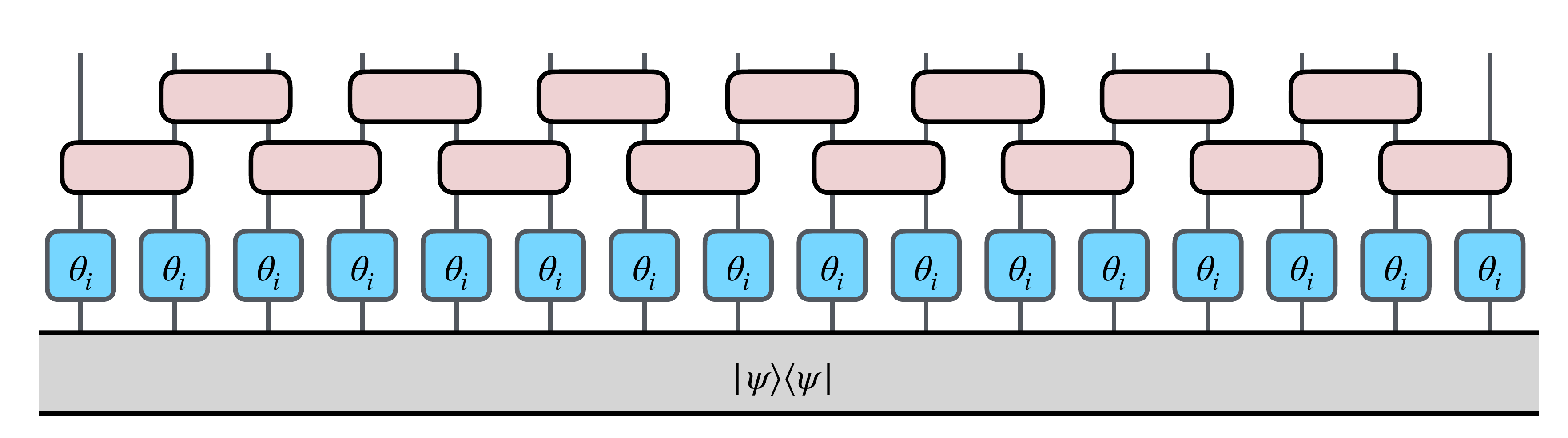}
    \caption{Example of a locally parameterized state family. The gates are all quantum channels, and the local parameter family is the set of local parameters $\{\theta_i\}$.}
    \label{fig:locally_parameterized}
\end{figure}

\subsection{Locally parameterized families and local reversibility}
\label{subsec:markov-recovery}

The coherent-information criterion establishes LRE, but does not determine
whether two noisy states belong to the same phase.  Our second result
gives a local recovery construction for locally parameterized families
with finite Markov length.

Throughout, a local channel is implemented by a circuit of \(d_L\) layers of quantum channel gates, each acting on a region of diameter at most \(w_L\). Information can therefore propagate through the full circuit over a distance bounded by its total light-cone radius \(r_L=O(d_Lw_L)\). We require \(r_L=o(L)\): the propagation distance must be sublinear in the system size. Measurements and classical feedforward, as well as unitary operations on the system and fresh ancillas initialized in a product state followed by tracing out the ancillas, are all allowed, with the full implementation subject to the same locality constraints.

\begin{definition}[Phase equivalence]
\label{def:phase-equivalence}
Two state families $\{\rho_L\}$ and $\{\sigma_L\}$ are phase equivalent
if there exist local channels $\mathcal C_L$ and $\mathcal R_L$ with
sublinear light-cone radius $r_L=o(L)$ such that
\begin{equation*}
 \|\mathcal C_L(\rho_L)-\sigma_L\|_1\leq\epsilon_L,\qquad
 \|\mathcal R_L(\sigma_L)-\rho_L\|_1\leq\epsilon_L,
\end{equation*}
with superpolynomially small error, written $\epsilon_L=L^{-\infty}$:
$L^n\epsilon_L\to0$ for every fixed $n>0$.\footnote{For pure states,
uniformly gapped paths admit local-circuit approximations compatible with
this superpolynomial definition.} 

\end{definition}

The basic tool for constructing the recovery is conditional
mutual information (CMI),
\begin{equation}
 I(A:C|B)=S(AB)+S(BC)-S(B)-S(ABC),
 \label{eq:quantum-cmi-definition}
\end{equation}
which measures the correlations between a local region $A$ and the
exterior $C$ that are not mediated by a surrounding buffer $B$.
Exponential decay with the buffer width defines a finite Markov length.
Ref.~\cite{Sang_2025} showed that a local Lindbladian evolution can be
reversed when this length remains finite along the evolution, including
states obtained partway through each Trotter layer.  The reverse acts gate
by gate: it erases the support of a local channel and reconstructs it from
a buffer using a twirled Petz map.

Extending this construction is especially useful for coherent errors,
because changing the rotation angle does not directly specify a Lindbladian
evolution of the syndrome-dephased state. We discuss
this family in Sec.~\ref{sec:surface_code_coherent_error_coherent_info}.
The
angle parametrizes the rotation before syndrome dephasing, rather than a channel acting on the already dephased
state.

\begin{definition}[Locally parameterized state family]
\label{def:locally-parameterized-family}
Let $\boldsymbol\lambda=\{\lambda_x\}$ be a field of local parameters,
with each $\lambda_x$ associated with a bounded region $A_x$.  A family
$\rho_L[\boldsymbol\lambda]$ is locally parameterized if any two profiles $\boldsymbol\lambda, \boldsymbol\lambda'$
that differ only at $x$ satisfy
\begin{equation}
 \Tr_{A_x}\rho_L[\boldsymbol\lambda]
 =\Tr_{A_x}\rho_L[\boldsymbol\lambda'].
 \label{eq:local-parameter-marginal}
\end{equation}
Thus the value of a local parameter becomes invisible once its associated
region is erased.
\end{definition}

Trotterized local Lindbladian paths form such families layer by layer. Each layer consists of local channels acting on mutually disjoint regions $A_x$. Fix the state entering a layer and let $\lambda_x$ be the evolution time of its channel on $A_x$. Trace preservation makes the marginal outside $A_x$ independent of $\lambda_x$.
Definition~\ref{def:locally-parameterized-family} retains this consequence without
requiring a channel between neighboring states on the path.
 An example of locally parameterized state family is shown in Fig.~\ref{fig:locally_parameterized}, which is relevant to question of coherent error channels, as we show in later sections.

One can organize the local parameter changes into rounds, playing the role
of Trotter layers.  We consider a bounded-density two-dimensional lattice
with $N=O(L^2)$ parameter regions of uniformly bounded diameter and
overlap, and use $K=\operatorname{polylog}L$ rounds, updating each
parameter at most once per round; a fixed number of rounds is allowed.
As in the Trotterized construction of Ref.~\cite{Sang_2025}, we assume the following uniform CMI bound to control the recovery error:
\begin{equation}
 I_{\rho_L[\boldsymbol\lambda]}(A_x:C|B)
 \leq aL^p e^{-b/\xi_*},
 \label{eq:local-markov-assumption}
\end{equation}
uniformly for every updated region $A_x$ and every intermediate profile
obtained by partially completing a round in any order, in either
direction along the path.  Here $B$ is a width-$b$ buffer around $A_x$,
truncated at a physical boundary, and $C$ is the remaining exterior.
The constants $a>0$, $p\geq0$, and $\xi_*>0$ are independent of $L$
and the intermediate profile.  The bound holds for
$b_0\leq b\leq L^\eta$, with fixed $b_0>0$ and $0<\eta<1$.
In the following, finite Markov length along a path includes this
uniformity over partially completed rounds.

\begin{figure}
    \centering
    \includegraphics[width=0.9\linewidth]{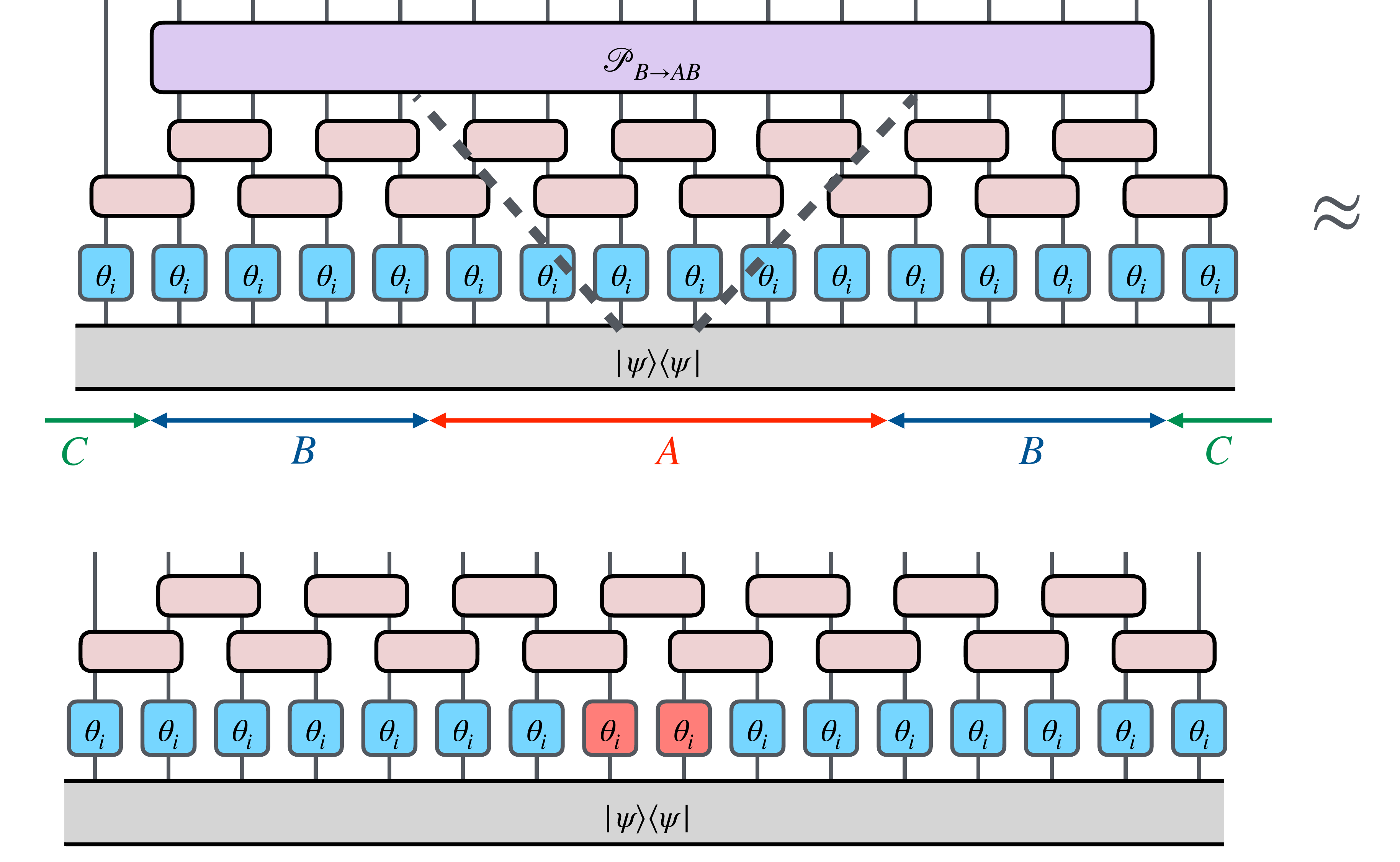}
    \caption{A state with local parameters $\{\theta_i\}$ can be converted approximately into a state with local parameters $\{\theta_i^\prime\}$ differing only in a local patch $A$, by acting with a local channel on $A$ and a buffer region $B$, with the error controlled by the CMI across the buffer. By staggering and compiling the local recovery into a polylog depth circuit, all local parameters can be updated, as long as the Markov length remains finite at all intermediate steps.}
    \label{fig:local_channel_parameterized}
\end{figure}

\begin{plainpuncttheorem}[Phase stability from finite Markov length]
\label{thm:marginal-preserving-local-reverse}

Suppose the locally parameterized path and update schedule above satisfy the uniform finite-Markov-length bound in Eq.~\eqref{eq:local-markov-assumption}. Then all states along the path belong to the same phase in the sense of Definition~\ref{def:phase-equivalence}.
\end{plainpuncttheorem}

\begin{proof}
Consider one step, with source $\rho$ and target $\sigma$ differing only
in the parameter at $x$.  For $A=A_x$, the definition gives
$\rho_{BC}=\sigma_{BC}$.  Erase $A$ and reconstruct it from the target
marginal $\sigma_{AB}$ using the twirled Petz map
$\mathcal R^\sigma_{A,b}=(\mathcal P^\sigma_{B\to AB}\otimes
\operatorname{id}_C)\circ\Tr_A$.  Approximate quantum
recovery~\cite{FawziRenner2015Recovery,Junge2018UniversalRecovery} then gives
$\|\mathcal R^\sigma_{A,b}(\rho)-\sigma\|_1\leq
2\sqrt{\ln2\,I_\sigma(A:C|B)}$.  Thus a map supported only on $AB$
recovers the full state, with an error controlled by the CMI across its
buffer. Example of this local recovery is shown in Fig.~\ref{fig:local_channel_parameterized}.

For $N=O(L^2)$ parameters and $K$ rounds, construct each map from its
ideal target marginal.

The bound above describes a single recovery step when its input
is exactly the intended state on the parameter path. Let $\rho_j$ denote
the intended state after step $j$, and $\widetilde\rho_j$ the actual state
produced by the circuit. In the full circuit, even the first recovery step
is generally approximate. Its output, which becomes the input to the next
step, may therefore differ slightly from the intended state. The same
issue arises at every later step. To bound the total error, we keep the
recovery maps fixed by the ideal path and track how these deviations
accumulate.

Consider applying the next recovery map $\mathcal R_j$ to two
inputs: the actual state $\widetilde\rho_{j-1}$ produced by the preceding
steps, and the intended state $\rho_{j-1}$ on the path. Their trace-norm
distance, $e_{j-1}=\|\widetilde\rho_{j-1}-\rho_{j-1}\|_1$, is the error
accumulated so far. Since the same quantum channel acts on both inputs,
trace-norm contractivity ensures that their outputs are no farther apart:
$\|\mathcal R_j(\widetilde\rho_{j-1})-\mathcal R_j(\rho_{j-1})\|_1
\leq e_{j-1}$. The output from the intended input is also close to the next
target $\rho_j$, by the single-step estimate already proved; denote this
recovery error by $\varepsilon_j=\|\mathcal R_j(\rho_{j-1})-\rho_j\|_1$.
We can therefore compare the actual output
$\widetilde\rho_j=\mathcal R_j(\widetilde\rho_{j-1})$ to the target in two
stages: first to the output from the intended input, and then to the
target. The triangle inequality bounds the total distance by the sum of
these two distances. The error after this step,
$e_j=\|\widetilde\rho_j-\rho_j\|_1$, thus satisfies
$e_j\leq e_{j-1}+\varepsilon_j$: it is at most the error carried over from
earlier steps plus the recovery error of the current step.

We begin with the exact initial state,
$\widetilde\rho_0=\rho_0$, so there is no error before the first update:
$e_0=0$. Repeating this argument over the $NK$ updates bounds the final
error by the sum of the individual recovery errors,
$e_{NK}\leq\sum_{j=1}^{NK}\varepsilon_j$. The CMI bound is needed only for
the intended states on the path; deviations of the actual states are
controlled by contractivity. Using the uniform bound in
Eq.~\eqref{eq:local-markov-assumption} then gives
$\epsilon_L\leq2NK\sqrt{a\ln2}\,L^{p/2}e^{-b/(2\xi_*)}$.
  Choosing
$b=c\xi_*(\ln L)^2$ with fixed $c>0$ makes $\epsilon_L=L^{-\infty}$.  In two
dimensions, color the buffered regions using $O(b^2)$ colors per round
so that same-color patches are disjoint.  Updates on these patches leave
one another's target marginals unchanged, so their Petz maps can be fixed
and applied in parallel.  This gives depth $O(Kb^2)$ and light-cone radius
$O(Kb^3)=\operatorname{polylog}L=o(L)$.  Applying the same construction
along the reversed parameter path supplies the opposite channel.
\end{proof}

The same recovery construction extends to approximately locally
parameterized families. After a local parameter change, the marginals
outside a distance $\ell$ from the changed region need not coincide
exactly: their trace-norm difference can instead decay exponentially as
$e^{-\ell/\xi_{\mathrm D}}$. This introduces a second length scale, the
\emph{deformation length} $\xi_{\mathrm D}$, which measures how far the
parameter change affects the state. (For Gibbs states, the deformation length for local Hamiltonian perturbations is the correlation length \cite{CapelMoscolariTeufelWessel2025Locality}.) The recovery error is then bounded
by this exterior mismatch together with the CMI contribution.  This combination was also discussed in the more specific contexts of effects of local Hamiltonian change in thermal states \cite{MaKhemaniSang2025ThermalPhases} and local measurement \cite{li2026unifiedframework}.  When both
the deformation and Markov lengths remain uniformly finite along the
local updates composing a path, the construction still gives two-way
local-channel phase equivalence.    

For thermal states, we can use local inverse temperatures $\{\beta_x\}$
as parameters in the Gibbs family
$\rho[\boldsymbol\beta]\propto e^{-\sum_x\beta_x h_x}$, where $h_x$ are
local energy terms. Varying them one region at a time interpolates
between states at different uniform temperatures through inhomogeneous
Gibbs states. This connects with the local-channel characterization of
thermal phases developed in Ref.~\cite{MaKhemaniSang2025ThermalPhases}.

In Appendix~\ref{app:unified-phase-equivalence}, we show that the
deformation length diverges at the thermal transition of the
two-dimensional Ising model, even though the shielded CMI vanishes. Therefore this construction establishes two-way circuit phase equivalence
within each thermal phase. Building on the Gibbs-state construction of
Ref.~\cite{MaKhemaniSang2025ThermalPhases}, our formulation applies more
generally to parameterized state families.

The construction also applies to nonequilibrium steady states of local
Lindbladians, with the local jump operators serving as parameters. For
uniformly bounded, finite-range generators with unique steady states,
uniform exponential rapid mixing and Lieb--Robinson bounds imply finite
deformation length~\cite{LuciaEtAl2015RapidMixingStability}. If the Markov
length also remains uniformly finite along the local updates, the Petz
construction connects the steady states by locally reversible channels.
CMI control is required only for the steady-state family, not during
physical relaxation. Thus, the Markov/deformation length criteria for a locally parameterized family encompasses families of Gibbs
states \cite{MaKhemaniSang2025ThermalPhases}, Lindbladian steady states, and states subject to measurement \cite{li2026unifiedframework} and coherent errors, as we will discuss.

\section{Honeycomb Surface Code under coherent errors}
\label{sec:surface-code-results}
\label{sec:setup}
\label{sec:surface_code_coherent_error_coherent_info}

We now apply the general framework to the honeycomb surface code under
coherent errors.  We first specify the code and noise channel, then
determine how much logical information survives as the error strength
increases.  Coherent information and CMI together establish a regime in
which critical conditional correlations coexist with long-range
entanglement.

\begin{figure}[tbp]
 \centering
 \includegraphics[width=0.9\columnwidth]{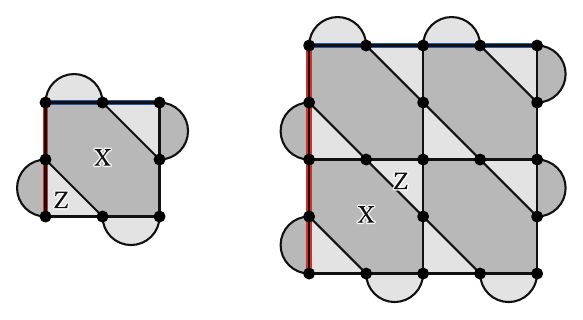}
 \caption{\justifying
Honeycomb surface-code patches with $L=1$ (left) and $L=2$ (right).
Physical qubits occupy the vertices, and dark and light plaquettes carry
$X$ and $Z$ checks, respectively. The blue and red strings represent
$\overline X$ and $\overline Z$, respectively. The strings shown are not
necessarily the shortest representatives of these logical operators.
 }
 \label{fig:lattice_surface_code}
\end{figure}

\subsection{Model and coherent-error channel}
\label{subsec:surface-code-geometry}
\label{subsec:coherent-errors-post-syndrome-state}

We use the honeycomb surface-code patch shown in
Fig.~\ref{fig:lattice_surface_code}.  Physical qubits occupy vertices, and
each plaquette supports an $X$- or $Z$-type stabilizer check.
Writing $\mathcal P_X$ and
$\mathcal P_Z$ for the two sets of plaquettes, the checks are
\begin{equation}
 S_p^X=\prod_{i\in\partial p}X_i
 \quad(p\in\mathcal P_X),
 \qquad
 S_p^Z=\prod_{i\in\partial p}Z_i
 \quad(p\in\mathcal P_Z),
 \label{eq:hex-code-stabilizers}
\end{equation}
where $\partial p$ is the set of boundary qubits of $p$.
All checks commute: checks of opposite type overlap on an even number of
qubits.  The code space is their common $+1$ eigenspace.

The truncated checks at the outer boundary leave one encoded qubit.
The boundary strings shown in blue and red represent its logical operators
$\overline X$ and $\overline Z$, which anticommute.  We use the encoded basis
$\{\ket{\overline0},\ket{\overline1}\}$, with
$\overline Z\ket{\overline z}=(-1)^z\ket{\overline z}$ for $z=0,1$.

The stabilizers realize the general symmetry structure of
Sec.~\ref{subsec:strong_strong_memory}: they can be understood as generators of
two $\mathbb Z_2$ one-form symmetries.  Each $X$- or $Z$-type plaquette
check is supported on a closed curve, and products of plaquette checks
generate larger contractible loops.  Code states have eigenvalue $+1$
under all such loops.

The blue and red strings in Fig.~\ref{fig:lattice_surface_code} are the
corresponding logical operators $\overline X$ and $\overline Z$.  They
connect the appropriate boundaries of the planar patch and cannot be
contracted to a point.  Multiplying either string by plaquette checks
deforms its path without changing its action on the code space.  The two
logical strings cross an odd number of times and anti-commute, so the two one-form symmetries
have a mixed anomaly.

The clean honeycomb code is LRE for every logical input, including arbitrary
mixtures, as established in Appendix~\ref{app:lre-proof}.

The noise is the uniform coherent $Z$ rotation, followed by syndrome dephasing defined
in Eqs.~\eqref{eq:post-syndrome-state}, now acting on this honeycomb patch,
\begin{align}\label{eq:rho_theta}
    \rho_{\theta}=\mathcal D_S(U_{\theta}\rho_Q
 U_{\theta}^{\dagger}),\quad U_{\theta} = e^{i\theta\sum_{i}Z_i}.
\end{align}

\subsection{Surviving coherent information and long-range entanglement}
\label{subsec:coherent-information}

\begin{figure}[tbp]
 \centering
 \includegraphics[width=0.95\columnwidth]{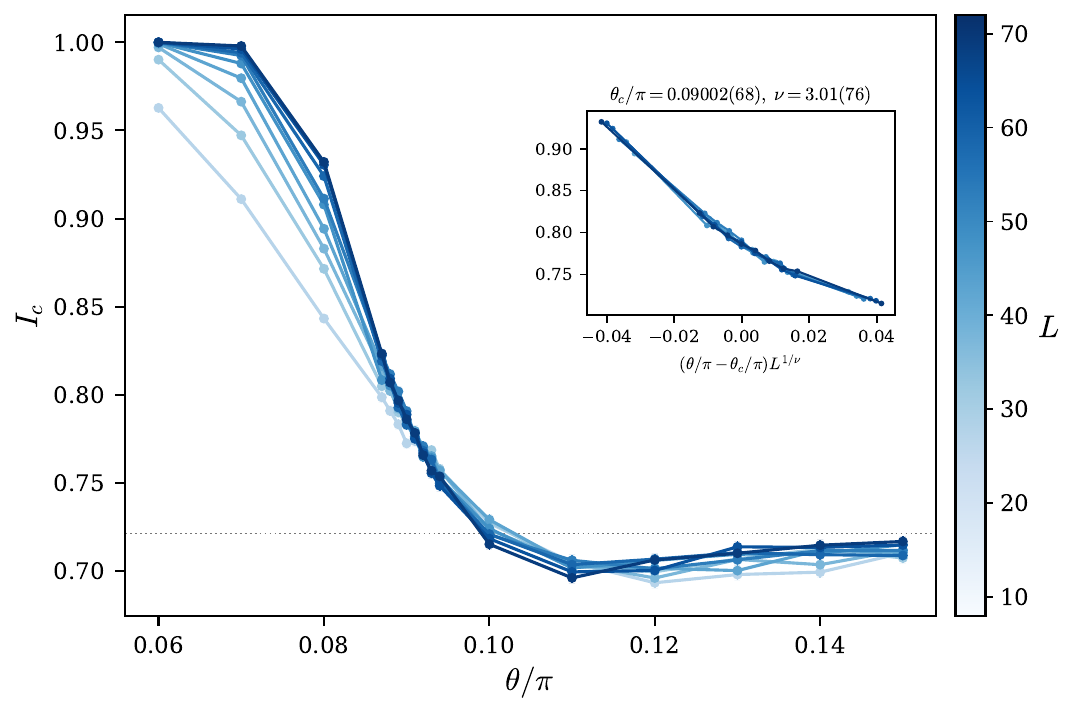}
 \caption{\justifying
 Surviving coherent information $I_c$ in bits for honeycomb patches with
 $L=8,16,\ldots,72$ unit cells along each direction, with data collapse
 at the threshold shown in the inset. The dotted line marks
 the Haar prediction $1/(2\ln2)\simeq0.721$ above threshold.}
 \label{fig:coherent-information_honeycomb_surface}
\end{figure}

We use the reference--code input of Eq.~\eqref{eq:encoded-bell-input}.
For each syndrome $s$, occurring with probability $p_s$, a $Z$-type
Pauli correction returns the physical block to the code space.
Since the errors and corrections preserve logical $Z$, the logical
qubit's reduced density matrix is diagonal
in this basis, with eigenvalues $q_s$ and $1-q_s$.  These are the
probabilities of the logical measurement outcomes $\ket{\overline0}$
and $\ket{\overline1}$.  Conditioned on $s$, the reference and logical
qubit form a pure state, so their entanglement is $h_2(q_s)$, where
$h_2(q)=-q\log_2q-(1-q)\log_2(1-q)$ is the binary entropy.

The different syndrome sectors are orthogonal, so the coherent information
is the average entanglement retained over all syndrome outcomes:
\begin{equation}
 I_c=\sum_s p_s h_2(q_s).
 \label{eq:main-coherent-information-syndrome-average}
\end{equation}
We evaluate this average by sampling the syndromes with fermionic linear
optics (FLO), using methods developed for coherent-error decoding
\cite{Bravyi_2018,Venn_PhysRevResearch.2.043412,Venn_2023,
Behrends_PhysRevResearch.6.013137,Bao_Anand_Phases}.
Appendix~\ref{app:coherent-information} gives the derivation and sampling
procedure.

Figure~\ref{fig:coherent-information_honeycomb_surface} shows how $I_c$
changes with the error strength.  Below $\theta_c\simeq0.09\pi$, it
approaches one bit as the patch grows, indicating full retention of the
encoded qubit.  Above threshold, it instead approaches a nonzero plateau
below one bit, showing that some logical information survives even after
the noise prevents full retention of the encoded qubit.  This change
occurs at a threshold consistent with that reported in
Ref.~\cite{bejan2026projectedlogicalensemblessurface}.

Using the results of Ref.~\cite{bejan2026projectedlogicalensemblessurface},
we can give a physical explanation of this plateau by considering the
logical states produced from the input
$\ket{\overline+}=(\ket{\overline0}+\ket{\overline1})/\sqrt2$.
Below threshold, maximum-likelihood Pauli correction returns these states
close to $\ket{\overline+}$ as the patch grows.  Above threshold, the
results support an approach to the Haar distribution restricted to the
hemisphere facing $\ket{\overline+}$.  This restriction comes from the
choice between corrections that differ by a logical $Z$: the decoder
selects the more likely correction, which returns the state closer to the input.

To connect this distribution to $I_c$, note that $\ket{\overline+}$ has
the same syndrome probabilities $p_s$ as the Bell input.  For the
$Z$-type channel, these probabilities depend only on the logical-$Z$
measurement probabilities, which are equal for the two inputs.
After syndrome correction, the normalized encoded state for any syndrome $s$ is 
\begin{equation}
 \ket{\overline{\psi_s}}=\sqrt{q_s}\ket{\overline0}
       +e^{i\phi_s}\sqrt{1-q_s}\ket{\overline1},
 \label{eq:conditional-logical-pure-state}
\end{equation}
where $\phi_s$ is the relative phase and $q_s$ is the same weight that
enters Eq.~\eqref{eq:main-coherent-information-syndrome-average}. \footnote{For geometries with even-weight stabilizers, the phase $\phi_s$ is restricted to specific values. For example, $\phi_s=\pm\pi/2$ for the rotated square lattice studied in Ref.~\cite{Bravyi_2018}.}
For the uniform hemisphere distribution, $q_s$ is uniform on $[0,1]$,
giving
\begin{equation}
 I_c^{\mathrm{Haar}}
 =\int_0^1 h_2(q)\,dq=\frac{1}{2\ln2},
 \label{eq:haar-coherent-information}
\end{equation}
consistent with Fig.~\ref{fig:coherent-information_honeycomb_surface}.
This Haar behavior is expected to persist as $\theta$ approaches
$\pi/4$ from below.  At the Clifford point $\theta=\pi/4$ itself,
the coherent information instead has an even--odd dependence on the
patch dimensions, as analyzed in Appendix~\ref{app:pi4-endpoint}.

Given the coherent information behavior, we can now apply the general results of Sec.~\ref{subsec:noisy-one-strong-symmetry} to this setting.  Every Kraus operator $\Pi_sU(\theta)$ of the honeycomb channel
$\mathcal N_\theta$ commutes with the $Z$-type strings, so
Theorem~\ref{thm:ci-lre-criterion} gives:

\begin{proposition}
\label{prop:honeycomb-coherent-lre}
Under coherent $Z$ errors, generic honeycomb code inputs yield LRE
outputs wherever the limiting coherent information is positive.
\end{proposition}

The numerical data in Fig.~\ref{fig:coherent-information_honeycomb_surface},
together with the Haar-randomness argument leading to
Eq.~\eqref{eq:haar-coherent-information}, indicate that above $\theta_c$
the coherent information approaches the nonzero Haar value
$I_c^{\mathrm{Haar}}=1/(2\ln 2)$.  Proposition~
\ref{prop:honeycomb-coherent-lre} therefore implies that generic noisy
honeycomb states remain LRE above the decoding threshold~\footnote{Below threshold, the local recovery
constructed in Sec.~\ref{sec:cmi-markov-length} establishes LRE
for \emph{every} logical encoding.}. The dependence on the logical input, including the exceptional Clifford
endpoint $\theta = \pi/4$, is summarized in Fig.~\ref{fig:pi4-logical-bloch-ball} and analyzed
in Appendix~\ref{app:pi4-endpoint}.  

This shows that both the above and below threshold regimes are nontrivial in the preparation sense: these mixed
states cannot be written as mixtures of states obtained from product inputs
by the allowed sublinear-range circuits. Next, we aim to determine whether the maximum-likelihood decoding threshold corresponds to a genuine mixed-state phase transition.

\subsection{Critical conditional correlations and phase equivalence}
\label{sec:cmi-markov-length}
\label{subsec:coherent-local-recovery}

\input{figures/boundary_cmi_geometry}

We now compute the CMI of the noisy code to characterize its long-distance
conditional correlations.  Away from the Clifford endpoint, it is natural
to expect the leading long-distance behavior to be independent of the
logical input.
For simplicity, we choose a fixed logical-$Z$ eigenstate
$\rho_Q=|\overline z\rangle\langle\overline z|$ and denote the resulting
post-syndrome state by $\rho_\theta$.  We use the geometry of
Fig.~\ref{fig:cmi-geometry}, where the one-row boundary regions $A,C$
are separated by a bulk strip $B$, with transverse extents and separation
proportional to $L$.

Dephasing every physical qubit in the $X$ basis produces a classical
distribution whose CMI we denote by $I_X(A:C|B)$.  This agrees with
the CMI of $\rho_\theta$ up to a constant set by the fixed logical input
and geometry.  For our choice, with entropies measured in bits,
\begin{equation}\label{eq:cmi_reduction}
 I(A:C|B)_{\rho_\theta}=I_X(A:C|B)+1.
\end{equation}
The offset is independent of $L$ and $\theta$, so the decay of $I_X$
determines the Markov length in this geometry.
Appendix~\ref{app:cmi-invariance} derives this relation.

We evaluate $I_X$ using the hybrid qubit--Majorana tensor network in
Fig.~\ref{fig:cmi-geometry}(b).  Fixing a physical outcome leaves a Gaussian
tensor on the virtual Majorana legs.  We therefore use fermionic linear
optics (FLO), tracking the Majorana covariance matrix to compute its Born
probability and the marginals entering the CMI.  We sample the
outcomes with their Born weights and average the resulting CMI estimator.
Appendices~\ref{app:gaussian-information-transport} and \ref{app:numerics}
give the contraction and numerical details.

\begin{figure}[tbp]
    \centering
    \includegraphics[width=\columnwidth]{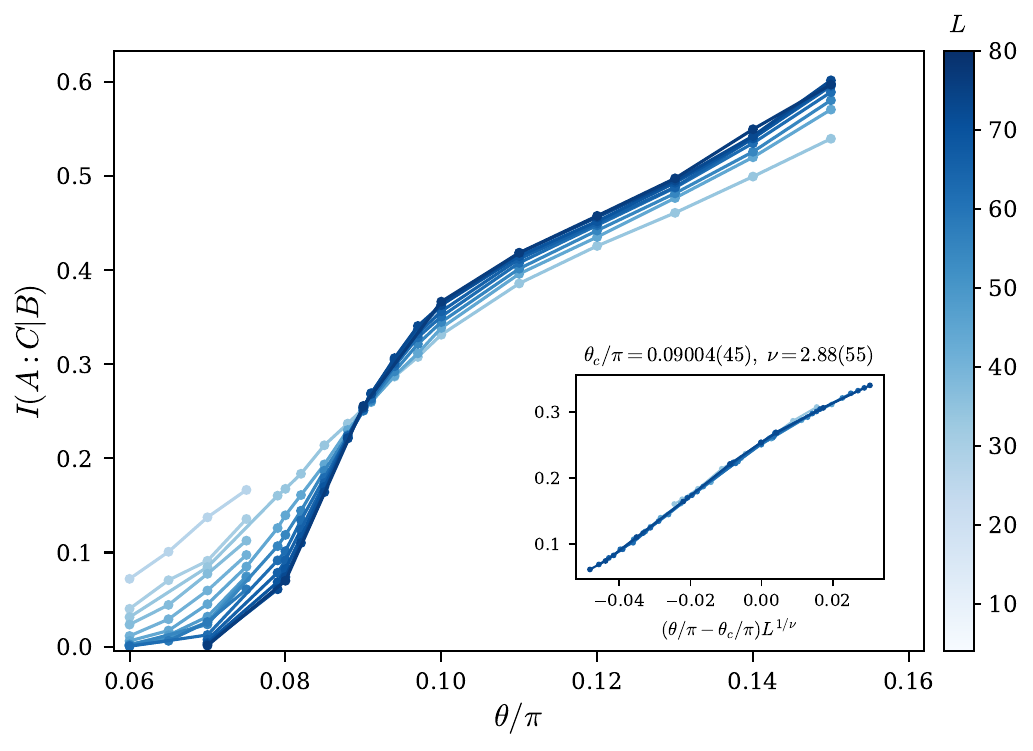}\\[-1mm]
    \includegraphics[width=\columnwidth]{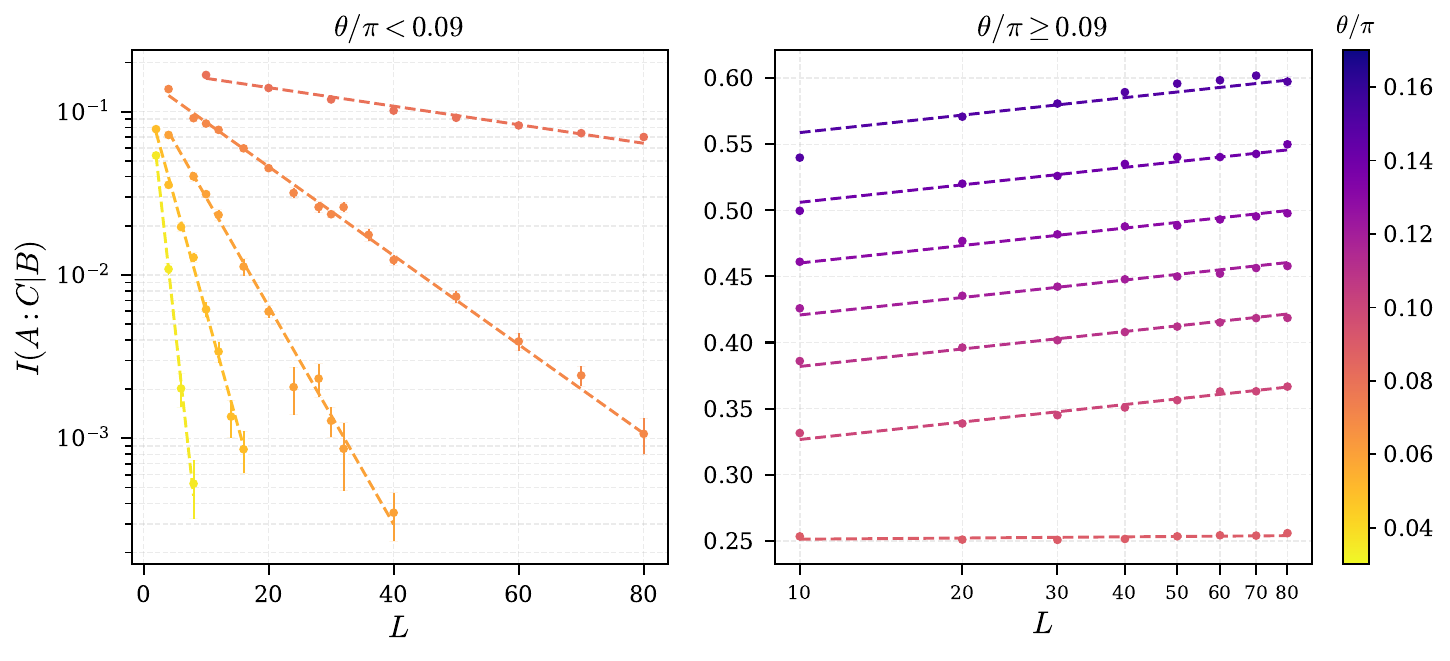}
    \caption{\justifying
 Excess CMI $I_X=I(A:C|B)-1$ in bits.
 Top: angle scan and data collapse. Bottom: size dependence below (left)
 and above (right) threshold. Dashed curves are fits; error bars are
 standard errors.}
    \label{fig:cmi-scaling}
\end{figure}

Figure~\ref{fig:cmi-scaling} shows a clear change near
$\theta_c/\pi\simeq0.09$, consistent with the insulator--metal transition of
the decoder network~\cite{yan2026nonlinearsigmamodelsurface,
yang2026decodingcoherenterrorstoric,
bejan2026projectedlogicalensemblessurface}.
Below threshold, $I_X$ decays
exponentially and defines a finite Markov length $\xi_{\mathrm M}$.  At the
transition it remains scale invariant, while above threshold it grows
logarithmically over the accessible sizes.  Thus the decoding transition is
also a transition from short-ranged to scale-dependent conditional
correlations.  The observed size dependence is summarized by
\begin{equation*}
  I_X(A:C|B)
  \sim
  \begin{cases}
    \exp[-L/\xi_{\mathrm M}(\theta)], & \theta<\theta_c,\\
    O(1), & \theta=\theta_c,\\
    a_X\ln[L/\ell_X(\theta)], & \theta_c<\theta<\pi/4,
  \end{cases}.
\end{equation*}
The above-threshold curves have approximately the same logarithmic slope,
suggesting a common coefficient $a_X>0$, with the angle dependence entering
primarily through the nonuniversal length $\ell_X(\theta)$ in lattice units.
The above-threshold form
describes the sampled intermediate angles $0.10\leq\theta/\pi\leq0.15$.
We expect this logarithmic behavior to persist throughout
$\theta_c<\theta<\pi/4$.  At the Clifford endpoint itself, the even-size
logical-$Z$ endpoint states analyzed in Appendix~\ref{app:pi4-endpoint}
have $I_X=0$.
Together with the surviving coherent information, these conditional
correlations establish an extended critical regime with LRE. 

Furthermore, now we can apply Theorem~\ref{thm:marginal-preserving-local-reverse} to connect the CMI behavior to the question of phase equivalence under two-way local channel connectivity. For a fixed encoded input $\rho_Q$, the noise itself supplies the forward
channel from the clean state to
$\rho_\theta=\mathcal N_\theta(\rho_Q)$; phase equivalence therefore hinges
on a local reverse channel from $\rho_\theta$ to the clean state. For the coherent-error family, we assume that the local-update path from
$\theta_i=\theta$ to $\theta_i=0$ has uniformly finite Markov length
below threshold.

  Theorem~\ref{thm:marginal-preserving-local-reverse} then supplies a local reverse channel with
$\|\mathcal R_{\theta,L}(\rho_\theta)-\rho_Q\|_1=L^{-\infty}$.  Hence:

\begin{proposition}
\label{prop:honeycomb-coherent-phase-equivalence}
Under coherent $Z$ errors, the honeycomb surface code has a mixed state phase transition at $\theta_c/\pi\simeq0.09$. For every angle below threshold, the corresponding noisy state
belongs to the clean surface-code phase under two-way connectivity by local quantum channels.
\end{proposition}

Its local dilation pulls the two clean one-form symmetries or loops back to dressed
loops of polylogarithmic width~\cite{SangEtAl2025MixedStatePhasesLocalReversibility}, preserving their crossing algebra on
the system and dilation ancillas.  The below-threshold states thus
inherit an anomalous pair of emergent one-form symmetries, and thus exhibit LRE in all output states. 

Note that the reverse channel is in general
distinct from local recovery of the encoded quantum information (in the sense of quantum error correction).  Recovery
must work on the entire encoded subspace and preserve its correlations with
an external reference, and not just a specific logical encoded state. Recovery of the encoded subspace further requires the same uniform decay
of $I(i:CR|B)$ for an encoded Bell input with untouched reference $R$, which we assume holds. Now, we can construct the recovery channel using the reverse local channel of Theorem~\ref{thm:marginal-preserving-local-reverse}, constructed using the Petz maps corresponding to the maximally mixed
code state.

\section{Criticality above threshold}
\label{sec:recoverability}

In the previous section, we established that the honeycomb surface code under $Z$ coherent error channel preserves LRE at all angles, and exhibits a mixed-state phase transition, separating a topologically ordered phase below threshold, and a critical LRE regime above threshold. The critical regime above threshold also has non-vanishing coherent information, which suggests the survival of partial quantum memory. Next, we characterize this regime from two further information-theoretic angles: the locality of recovery of this partial quantum information and quasi-long-range strong-to-weak one-form symmetry breaking.

First, we compute the punctured coherent information, and show that it rules out local recovery to the clean code above
threshold when recovery is required to preserve the surviving quantum
correlations with the logical reference.  This is incompatible with a
localized emergent anomalous one-form symmetry pair, whose deformable logical
strings would route the memory around the puncture.  Finally, we ask whether
a local reverse channel can recover any chosen clean state even without
preserving the reference correlations necessary for quantum error correction.  We introduce the open-string
fidelity correlator and prove the stronger result that its slow decay
obstructs such a reverse channel as well.
  The above-threshold data exhibit
this critical behavior, providing a state-level phase-separation test that does
not rely on logical degrees of freedom.

\subsection{Punctured coherent information}
\label{subsec:global-quantum-memory}
\label{subsec:punctured-coherent-information}

\begin{figure}[tbp]
 \centering
 \includegraphics[width=0.95\columnwidth]
 {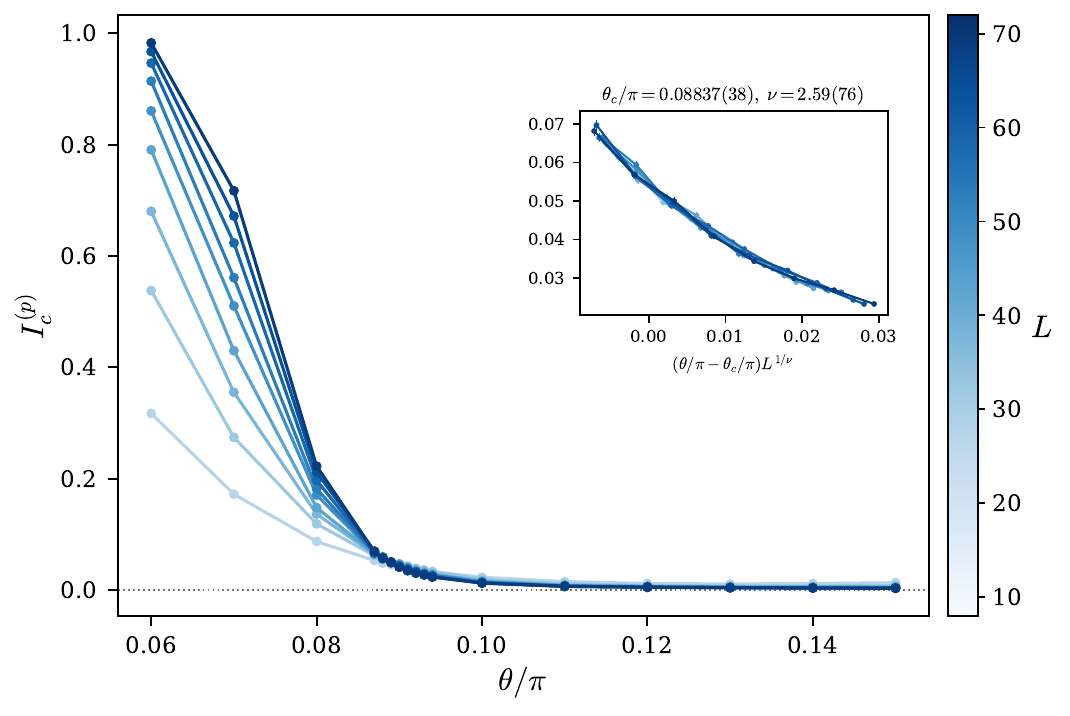}
 \caption{\justifying
 Punctured coherent information $I_c^{(p)}$ in bits for a centered puncture
 with $r_p/L=1/2$ and $L=8,16,\ldots,72$, with data collapse at the
 threshold shown in the inset.}
\label{fig:punctured-coherent-information}
\end{figure}

Topological memory can avoid a correctable region of the bulk.  If both
emergent logical strings remain deformable, they can be routed around
a puncture with a buffer wider than their dressing.  The complement
then carries the same logical algebra.  Punctured coherent
information~\cite{negari2026criticalnonequilibrium} tests this spatial
property of the surviving memory.

Prepare the encoded Bell state with reference $R$, apply the noise,
and erase a simply connected bulk region $Q_p$ of characteristic scale
$r_p=\alpha_p L$, with fixed puncture fraction $0<\alpha_p<1$.  For the remaining qubits
$\overline Q_p$, define
\begin{equation}
 I_c^{(p)}\equiv I_c(R\rangle\overline Q_p)
 =S(\overline Q_p)-S(R\overline Q_p).
 \label{eq:physical-punctured-coherent-information}
\end{equation}
The loss $\Delta_p=I_c-I_c^{(p)}\geq0$ measures how much coherent
information depends on the erased region.  Appendix~\ref{app:punctured-coherent-information}
gives the erasure convention and
its FLO evaluation.

Figure~\ref{fig:punctured-coherent-information} reveals a change in
the punctured coherent information near $\theta_c$.  For a centered puncture with
$r_p/L=1/2$, the finite-size flows suggest
\begin{equation}
 \lim_{L\to\infty}I_c^{(p)}(\theta)=
 \begin{cases}
  1, & \theta<\theta_c,\\
  c_p, & \theta=\theta_c,\\
  0, & \theta_c<\theta<\pi/4.
 \end{cases}
 \label{eq:punctured-memory-transition}
\end{equation}
Here $c_p$ is the critical value for the chosen puncture geometry and
fraction.  The above-threshold flow
refers to the angles shown in the figure; the Clifford
endpoint is discussed separately in Appendix~\ref{app:pi4-endpoint}.
Below threshold, puncturing leaves the full coherent information
$I_c\simeq1$ intact, as expected for deformable logical strings.
Above threshold, $I_c^{(p)}$ decreases toward zero even though the full
coherent information remains positive.  The surviving quantum information
is therefore no longer carried by an anomalous logical pair that remains
deformable around a macroscopic puncture.

This diagnostic also directly constrains local recovery.
If a local channel maps the noisy reference--code state to a state
supported on the clean code while preserving its surviving coherent
information, and the puncture enlarged by its light cone is correctable,
the same target can be recovered using $\overline Q_p$ alone.  The puncture
criterion of Ref.~\cite{negari2026criticalnonequilibrium}, recalled in
Appendix~\ref{app:puncture-stability-proof}, therefore requires
\begin{equation}
 I_c-I_c^{(p)}=o(1),
 \label{eq:puncture-necessary-condition}
\end{equation}
as the recovery error vanishes.  The nonzero limiting loss above threshold
therefore excludes local recovery that preserves all the surviving coherent
information, without requiring recovery of a perfect Bell pair.
This code-level requirement is stronger than the local reverse channel
needed for phase equivalence: it must preserve the reference correlations,
whereas phase equivalence concerns only a chosen state.  To rule out phase equivalence
itself, we therefore need a state-level diagnostic that does not invoke a
logical reference.  The fidelity correlator provides this second test.
\subsection{Fidelity correlations and phase inequivalence}
\label{subsec:one-form-fidelity}

Punctured coherent information constrains local recovery that preserves
the correlations with a logical reference. Phase equivalence in the sense
of Definition~\ref{def:phase-equivalence} asks for less: a local channel that returns
one chosen noisy state to its clean counterpart. We now give a state-level
diagnostic of this weaker property without requiring a logical reference
or any symmetry assumption on the reverse channel.

While we expect the main results to apply for any 2d topological codes with abelian anyons, we focus on the surface code for concreteness. Consider a surface code in any two-dimensional geometry (a planar patch, a
torus, or a sphere) of linear size $L$, with local checks of range $r_0$. We fix a
\emph{pure} clean code state $\ket{\psi}$ and write
$\psi=\ket{\psi}\!\bra{\psi}$. When the geometry encodes logical
qubits, this means fixing a logical state. Let $\mathsf{B}$ be a simply connected region of the
bulk, of linear size $\Theta(L)$, that is correctable: it contains no
noncontractible cycle and no path joining distinct physical boundaries, so
it supports no logical operator. Inside $\mathsf{B}$, products of checks of
either type generate contractible loops, as in Sec.~\ref{subsec:surface-code-geometry}.

Let $\gamma=\gamma_{xy}$ be a path in $\mathsf{B}$ with endpoints $x,y$, and
let $Z_\gamma$ be the open $Z$ string along it. On the clean state,
$Z_\gamma$ creates a pair of electric charges at $x$ and $y$, which an $X$
loop enclosing exactly one endpoint detects. We take the endpoints to be a
distance $\Theta(L)$ apart and a distance $\Theta(L)$ from the boundary of $\mathsf{B}$.

Now, the state is subjected to local noise channel $\mathcal{N}$. For the noisy state $\rho=\mathcal N(\psi)$ we define
\begin{equation}
  \mathcal F_Z(x,y)
  = F\!\left(\rho,\, Z_\gamma \rho Z_\gamma\right),
  \qquad
  F(\rho,\sigma)=\big\|\sqrt\rho\sqrt\sigma\big\|_1 ,
  \label{eq:FZ}
\end{equation}
with $F$ the Uhlmann fidelity. This is the fidelity version of the
R\'enyi-2 correlator introduced in Ref.~\cite{bao2023mixedstatetopologicalordererrorfield}, and it detects strong-to-weak spontaneous
symmetry breaking of one-form symmetries~\cite{lessa2025higherformanomalylongrangeentanglement, ZhangXuZhangXuBiLuo2025}.
Related learnability transitions involve global strong-to-weak symmetry
breaking~\cite{SinghEtAl2026MixedStateLearnability}.
If the noise
preserves the $Z$-type one-form symmetry strongly, deforming $\gamma$
within $\mathsf B$ by a contractible $Z$ loop leaves
$Z_\gamma\rho_L Z_\gamma$ unchanged. The correlator then depends only on
the endpoints. For the clean state, an $X$ loop around $x$ has eigenvalue
$+1$ on $\psi$ and $-1$ on $Z_\gamma\psi Z_\gamma$, so
$\mathcal F_Z=0$.

For noise which is strongly symmetric under open $Z$ strings,
\begin{equation}
  \mathcal N(Z_\gamma\, \omega\, Z_\gamma)
  = Z_\gamma\, \mathcal N(\omega)\, Z_\gamma
  \quad\text{for every state } \omega.
  \label{eq:covariance}
\end{equation}
For strongly symmetric noise, the fidelity
correlator compares two noisy states that differ \emph{only in their
input}:
\begin{equation}
  \mathcal F_Z(x,y)
  = F\!\left(\mathcal N(\psi),\,
             \mathcal N(\psi_\gamma)\right),
  \qquad
  \psi_\gamma = Z_\gamma \psi Z_\gamma .
  \label{eq:FZ-cov}
\end{equation}
The right-hand side is defined for any noise channel, which asks how well the
noise hides an anyon pair inserted into the clean input. The following
theorem bounds it directly, so strong symmetry is needed only to identify it
with the fidelity correlator in Eq.~\eqref{eq:FZ}. The theorem states that the existence of a local recovery channel implies that the fidelity correlator must be super-polynomially small (in $L$):  if not, the local recovery channel would map the noisy
image of $\psi_\gamma$ close to $\psi$, erasing a pair of charges
separated by $\Theta(L)$. The contrapositive of this theorem is useful in establishing the lack of a local recovery channel.

\begin{theorem}[Fidelity obstruction to local recovery]
\label{thm:fidelity-obstruction-local-recovery}
Let $Z_\gamma$ be an open
string whose endpoints lie in a correctable bulk region, $\Theta(L)$ apart
and $\Theta(L)$ from its boundary. Let $\psi$ be a pure clean code state, and $\psi_\gamma$ be the corresponding state conjugated with $Z_\gamma$. Suppose a local channel
$\mathcal R$ returns the noisy state $\mathcal N(\psi)$ to
$\psi$ with superpolynomially small error, as required for phase
equivalence (Definition~\ref{def:phase-equivalence}). Then
\begin{equation}
  F\!\left(\mathcal N(\psi),\,\mathcal N(\psi_\gamma)\right)
  = L^{-\infty}.
  \label{eq:fid-bound}
\end{equation}
For noise that is strongly symmetric with $Z_\gamma$, this means $\mathcal F_Z(x,y)=L^{-\infty}$.
Conversely, a fidelity correlator that decays only polynomially or slower
rules out such a recovery channel.
\end{theorem}

No symmetry restriction is placed on $\mathcal R$, and $\mathcal R$ is
not required to recover $\psi_\gamma$. The geometry enters only through
the correctable bulk region $\mathsf B$ and the purity of $\psi$. On a
sphere, any disk of linear size $\Theta(L)$ can serve as $\mathsf B$. We sketch the proof below, with the technical details given in Appendix~\ref{app:phase-equivalence}.

\begin{proof}[Proof Sketch]
The theorem makes precise a simple physical statement: \emph{a local
channel cannot remove a distant pair of anyons.} The theorem
says that if a local channel $\mathcal R$ brings $\mathcal N(\psi)$ back to
$\psi$, then $\mathcal N(\psi)$ and $\mathcal N(\psi_\gamma)$ must be
almost perfectly distinguishable. Fidelity can only increase under a
channel, so for $S=\mathcal R\circ\mathcal N$
\begin{align}
  F\big(\mathcal N(\psi),\mathcal N(\psi_\gamma)\big)
  \le
  F\big(\mathcal S(\psi),\mathcal S(\psi_\gamma)\big).
\end{align}
It therefore suffices to exhibit one measurement on the output system that
tells $\mathcal S(\psi)\approx\psi$ apart from $\mathcal S(\psi_\gamma)$.

The input $\psi_\gamma$ has two anyons a distance $\Theta(L)$ apart. A
channel of light-cone radius $s\ll L$ can move, split, or dress each anyon
locally, and can deposit junk in ancillas, but cannot change the
total topological charge in a disk around one endpoint, because that charge
is measured by a loop far from anything the channel can correlate across
the system. So $\mathcal S(\psi_\gamma)$ should still contain a nontrivial
charge near $x$. That charge is detected by a loop on which
$\mathcal S(\psi)\approx\psi$ has eigenvalue $+1$. If the recovery $\mathcal{S}(\psi) \approx \psi$ is only approximate with recovery error $\epsilon_L$, each step holds up to errors controlled
by $\epsilon_L$ and the number of local constraints, which gives
the quantitative superpolynomial bound, as proven in Appendix~\ref{app:phase-equivalence}.

\end{proof}

The correlator also tests for an emergent $X$-type strong symmetry
complementary to the microscopic $Z$-type one. Suppose such a symmetry is
generated by dressed $X$ loops whose charge uncertainty, and whose failure
to anticommute with bare open $Z$ strings, decrease exponentially with the
loop width. Then a loop around one endpoint distinguishes the two states in
Eq.~\eqref{eq:FZ} with exponentially small error, and
\begin{equation}
  \mathcal F_Z(x,y)\le C e^{-|x-y|/\xi_F},
  \qquad \xi_F = O(\xi_X),
  \label{eq:loop-bound}
\end{equation}
where $\xi_X$ is the localization length of the dressed loops
(Appendix~\ref{app:emergent-symmetry-fidelity-bound}). Persistently slower decay therefore rules
out such a partner. 

\begin{figure}[tbp]
 \centering
 \includegraphics[width=\linewidth]{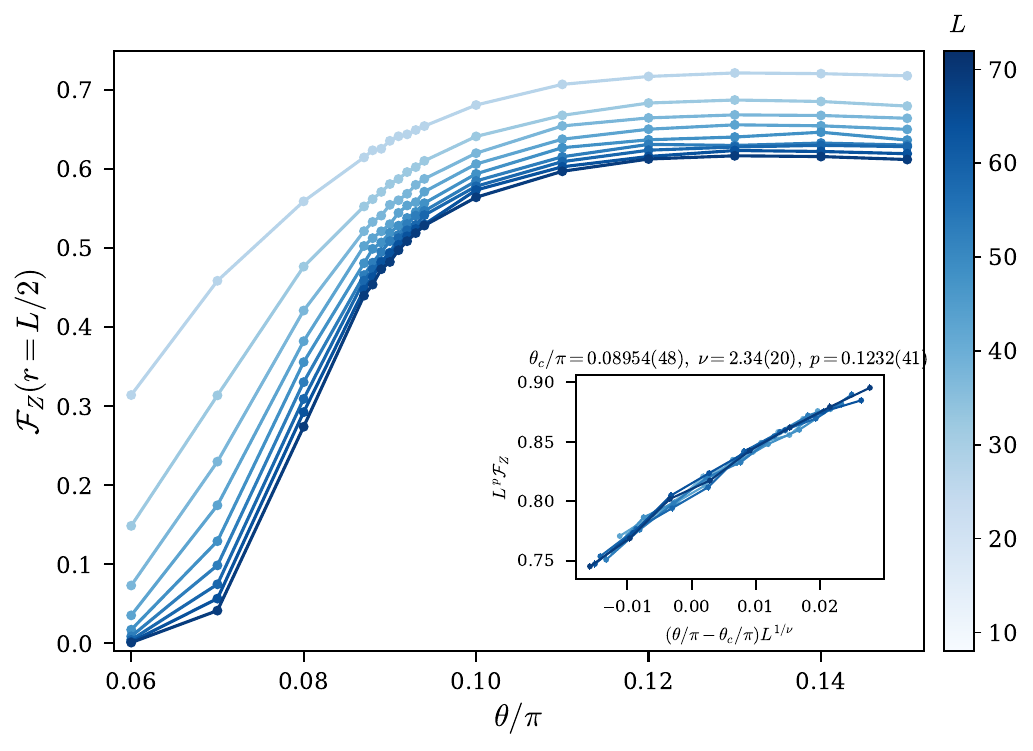}\\[-1mm]
 \includegraphics[width=\linewidth]{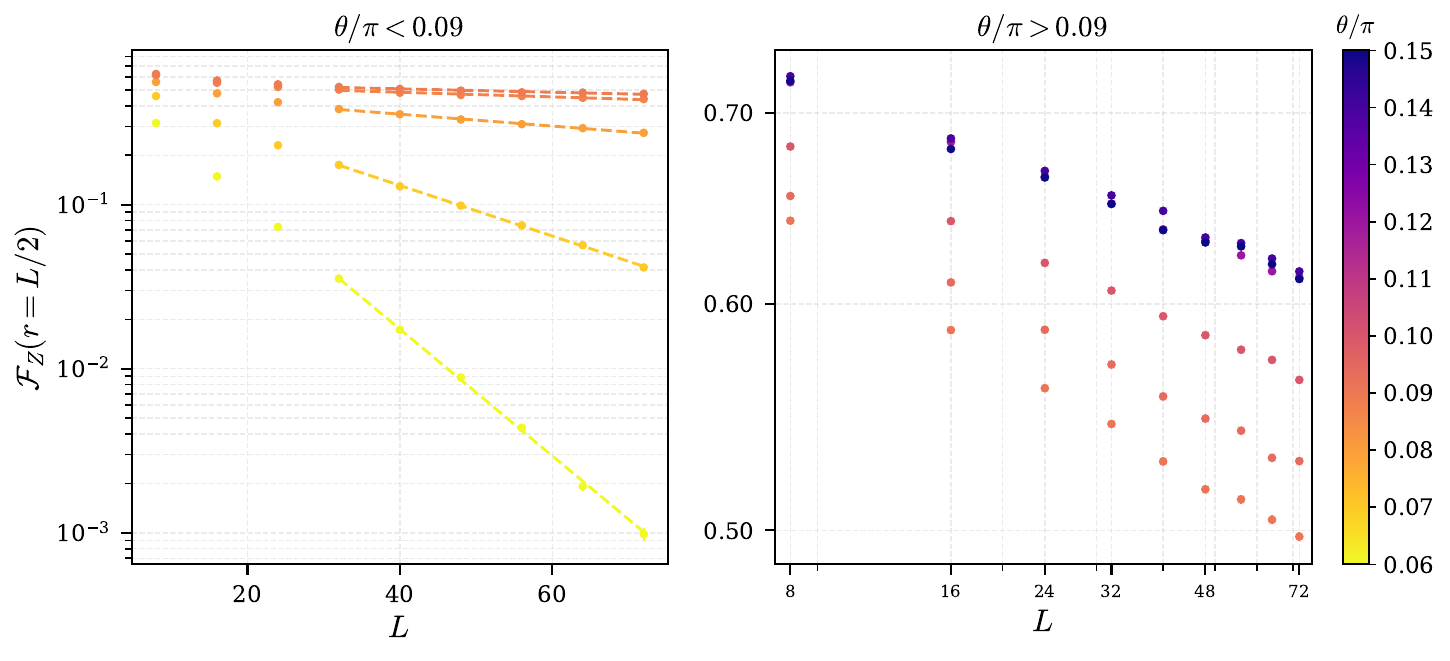}
 \caption{\justifying
 Open-string fidelity at separation $L/2$ for the logical
 $\overline Z=+1$ input and $L=8,16,\ldots,72$.
 Top: angle scan and data collapse. Bottom: size dependence below (left)
 and above (right) threshold. Dashed lines are exponential fits;
 error bars are standard errors.}
 \label{fig:swssb-scaling}
\end{figure}

\subsection{Application to coherent errors}
We evaluate $\mathcal F_Z$ on the planar honeycomb patches for the fixed
input $\ket{\overline{0}}$. We use contractible bulk strings with endpoints
$\Theta(L)$ apart and $\Theta(L)$ from the physical boundary. The
microscopic $Z$-type symmetry is strong and the covariance
Eq.~\eqref{eq:covariance} holds at every angle.

In Fig.~\ref{fig:swssb-scaling}, the fidelity decays exponentially below the
threshold and much more slowly at and above it. The data are consistent
with
\begin{equation}
  \mathcal F_Z(L/2,\theta)\sim
  \begin{cases}
    e^{-L/\xi_F(\theta)}, & \theta<\theta_c,\\[2pt]
    L^{-p}, & \theta=\theta_c,\\[2pt]
    \,L^{-\alpha}[\ln L]^{-\beta}, & \theta_c<\theta<\pi/4,
  \end{cases}
  \label{eq:FZscaling}
\end{equation}
where $\mathcal F_Z(L/2)$ denotes $|x-y|=L/2$, $p$ is the decay exponent
at threshold, and $\alpha$ and $\beta$ govern the power-law and logarithmic
factors above threshold. The above-threshold form
describes the intermediate angles probed numerically, not the Clifford
endpoint. At $\theta=\pi/4$, on the even-size patches of
Appendix~\ref{app:pi4-endpoint}, the output is diagonal in the physical $Z$
basis and $\mathcal F_Z=1$ for every $Z$ string. Exponential decay below threshold is consistent with the anomalous pair of
one-form symmetries of the clean code. Above threshold, the slow decay
rules out a localized complementary $X$-type symmetry, while the
microscopic $Z$-type symmetry remains exact. Together with the vanishing
ordinary string expectation $\mathrm{Tr}(\rho_L Z_\gamma)=0$, this is a
quasi-long-range counterpart of one-form strong-to-weak spontaneous
symmetry breaking~\cite{ lessa2025higherformanomalylongrangeentanglement, ZhangXuZhangXuBiLuo2025}: the fidelity vanishes
algebraically or logarithmically, rather than approaching the nonzero
limit of conventional SW-SSB.

By Theorem~\ref{thm:fidelity-obstruction-local-recovery}, the slow decay is also a phase obstruction.
No reverse channel with superpolynomially small error returns an
above-threshold state to the clean code. The below-threshold states are
connected to the clean code by the recovery construction of
Sec.~\ref{sec:cmi-markov-length}, under its finite-Markov-length assumption.
Suppose a local channel mapped an above-threshold state to a
below-threshold one. Composing it with that recovery would give a local
reverse channel to the clean code, contradicting Eq.~\eqref{eq:fid-bound}.
The above-threshold states are therefore inequivalent to the clean surface-code phase under two-way local channel connectivity.

\section{Critical Correlations and Majorana Transport}
\label{sec:network-localization}

In this largely self-contained section, we now relate the boundary-to-boundary CMI in the honeycomb surface code to localization and transport properties of a dual Majorana network model.  This
connects the change in conditional correlations across the decoding
threshold to the transition to diffusive metallic transport. This metallic behavior, in turn, explains the approach of the projected logical ensemble to Haar~\cite{bejan2026projectedlogicalensemblessurface}, underlying 
$I_c^{\mathrm{Haar}} =1/(2\ln2)$ in Eq.~\eqref{eq:haar-coherent-information}.  Related entanglement transitions in monitored fermions admit loop
models~\cite{KlockeBuchhold2023MajoranaLoops,KlockeEtAl2025KitaevCircuits} and
replica field theories~\cite{BuchholdEtAl2021DiracFermions,LadewigDiehlBuchhold2022MonitoredOpenFermions,ChahineBuchhold2024MonitoredFermions2D,PoboikoGornyiMirlin2024AboveOneDimension}.

\subsection{Gaussian representation and transport bounds}
\label{subsec:mie}
\label{subsec:born-gaussian-ensemble}
\label{subsec:gaussian-network}
\label{subsec:scattering-conductance}
\label{subsec:gaussian-label-ensembles}

To connect CMI to transport, we first bound it by the entanglement left
between the boundaries after measuring the bulk. Recall in Eq.~\eqref{eq:cmi_reduction} we reduced the CMI of the syndrome dephased state $I_{\rho_\theta}$ to that of the state with all qubits dephased along Pauli $X$ operator, $I_X$. We now bound the latter by the measurement-induced entanglement of the pure surface code state. Consider a pure state on
$ABC$ and measure $B$ in a local basis. 
A record $b$, obtained with probability
$p_b$, prepares a pure boundary state $|\psi_b\rangle_{AC}$.  The Holevo bound
then limits the correlations accessible by subsequently measuring $A$ and
$C$~\cite{Holevo1973InformationBound}:
\begin{equation*}
 I(X_A:X_C|X_B)\leq\sum_b p_b S(A)_{\psi_b}.
\end{equation*}
The right-hand side is the measurement-induced entanglement between the two
boundaries~\cite{Negari_2024}.  The bound is general and does not require a
Gaussian description; Appendix~\ref{app:classical-cmi-mie} gives the
argument.  In our setting, syndrome dephasing leaves the
$X$-outcome distribution unchanged, so it applies directly to $I_X$.

The Gaussian representation connects this boundary entanglement to
transport.  Each outcome $x_i$ selects a pure Gaussian tensor $T_i(x_i)$
on the virtual Majorana legs [Fig.~\ref{fig:random-gaussian-ensemble}(a)].
Contracting the network gives the amplitude
$\Psi(x)=\operatorname{tTr}[\bigotimes_iT_i(x_i)]$ of a complete record,
where $\operatorname{tTr}$ denotes contraction of all virtual tensor legs.
Its normalized Born weight is
$p(x)=|\Psi(x)|^2/\sum_{x'}|\Psi(x')|^2$.
The physical measurements thus determine both the Gaussian ensemble and
its correlated weights~\cite{Negari_2024,
jian2022criticalityentanglementnonunitaryquantum,jian2023Born}.

\input{figures/gaussian_transport}

For a bulk record $b$, contracting $B$ while leaving the virtual boundaries
$L,R$ open produces a pure Gaussian state $|\phi_b\rangle_{LR}$.  In the
following transport formulas, $L,R$ label the virtual left and right
boundaries, not the system size or reference qubit.
In the honeycomb geometry, its Schmidt spectrum agrees with that of the physical
boundary state $|\psi_b\rangle_{AC}$.  Define the Born-averaged virtual
mutual information as $E_{LR}=\sum_b p_b I(L:R)_{\phi_b}$.
Purity gives $E_{LR}=2\sum_b p_b S(L)_{\phi_b}$, so the measurement bound
becomes $I_X(A:C|B)\leq E_{LR}/2$.
This identification uses the state preparation and boundary convention
established in Appendix~\ref{app:honeycomb-boundary-entanglement}.

The same covariance matrix defines a single-particle scattering problem:
$\mathsf S_b=i\Gamma^{(b)}$, with transmission block
$\mathsf t_b=i\Gamma^{(b)}_{RL}$~\cite{jian2022criticalityentanglementnonunitaryquantum}.
The dimensionless conductance of a record is
$g(b)=\operatorname{tr}(\mathsf t_b\mathsf t_b^\dagger)$, with Born average
$g=\sum_b p_b g(b)$.  Thus $I_X$, $E_{LR}$, and $g$ probe the same
Gaussian ensemble.
Throughout this comparison, $\xi_{\mathrm{loc}}$ is extracted from this
Born-averaged conductance; its identification with a typical transfer-matrix
length is discussed below.

The statistical mechanics of optimal Pauli decoding relates
coherent-error correction to
localization in a Majorana network~\cite{Venn_2023}.  For the honeycomb
geometry studied here, the decoding threshold corresponds to an
insulator--metal transition~\cite{yan2026nonlinearsigmamodelsurface,
yang2026decodingcoherenterrorstoric}. The square-lattice problem belongs to class D.
With the physical Born weights, class DIII permits a stable thermal metal,
whereas the apparent class-D metal can exhibit a slow crossover to an insulator
\cite{yan2026nonlinearsigmamodelsurface,yang2026decodingcoherenterrorstoric}.

Entanglement and transmission are functions of the same Gaussian mode
spectrum.  For any ensemble of pure Gaussian boundary states, the spectral
bounds derived in Appendix~\ref{app:conductance-mie-bound} give
\begin{equation*}
 \frac g2\leq E_{LR}\leq
 \frac g2\log_2\frac{4eN_\partial}{g},
\end{equation*}
where $N_\partial$ is the number of boundary-mode pairs.  If
$g=e^{-L/\xi_{\mathrm{loc}}+o(L)}$ and $N_\partial$ grows at most
polynomially, the logarithmic factor is subexponential, so $E_{LR}$ and $g$
have the same decay length.  The CMI bound then implies
\begin{equation*}
 \xi_{\mathrm M}\leq\xi_{\mathrm{loc}},
\end{equation*}
for the honeycomb ensemble.  Localization of the virtual Majoranas therefore
forces short-ranged conditional correlations in the physical record.
Equality further requires the CMI to couple to the network's slowest
decaying mode.

\subsection{Markov and localization lengths}
\label{subsec:mie-conductance-numerics}

\label{subsec:boundary-hamiltonian}

To understand when the bound is saturated, consider a long strip of width
$W$.  The product of its Gaussian transfer matrices defines an effective
boundary Hamiltonian~\cite{Oseledets1968,MacKinnonKramer1983}.  If
$\Delta(W)$ is its slowest positive decay rate, then a typical record $x$
has $g_x(L,W)\sim e^{-2L\Delta(W)}$.  The same mode can also control the entropy of
the measured labels.  Writing $H_L$ for the Shannon entropy of $L$
consecutive slices, the CMI for one-slice boundary regions is the negative
discrete second derivative
\begin{equation*}
 I(X_A:X_C|X_B)=2H_{L+1}-H_L-H_{L+2}.
\end{equation*}
The extensive bulk entropy and constant boundary term cancel in this
combination.  If the leading correction overlaps with the slowest boundary mode,
the CMI inherits the same exponential decay rate as the conductance.  Under this
physical assumption, together with the Born average and large-width limit,
$\Delta=\lim_{W\to\infty}\Delta(W)$ gives
\begin{equation*}
 \xi_{\mathrm M}=\xi_{\mathrm{loc}}=\frac{1}{2\Delta}.
\end{equation*}
Appendix~\ref{app:boundary-hamiltonian} states the assumptions needed for
this equality beyond Gaussianity.  We test it by evaluating the CMI,
boundary entanglement, and conductance in the same Born-weighted ensemble;
Appendices~\ref{app:hybrid-tn} and \ref{app:numerics} give the contraction
and fitting details.

\begin{figure}[tbp]
    \centering
    \includegraphics[width=\linewidth]{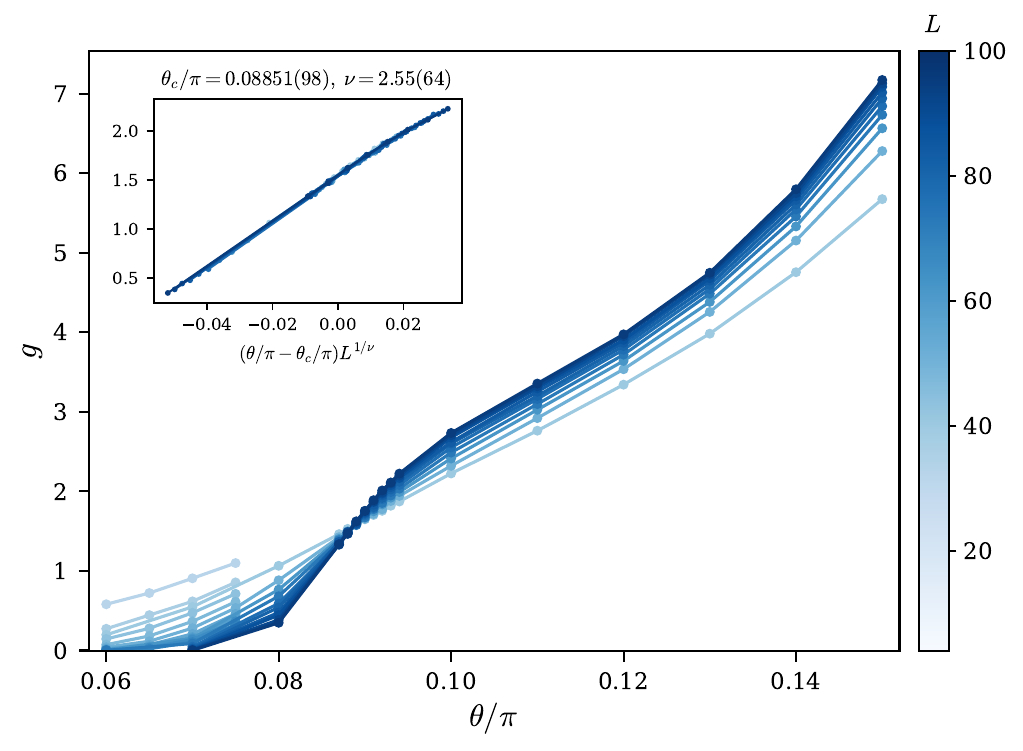}\\[-1mm]
    \includegraphics[width=\linewidth]{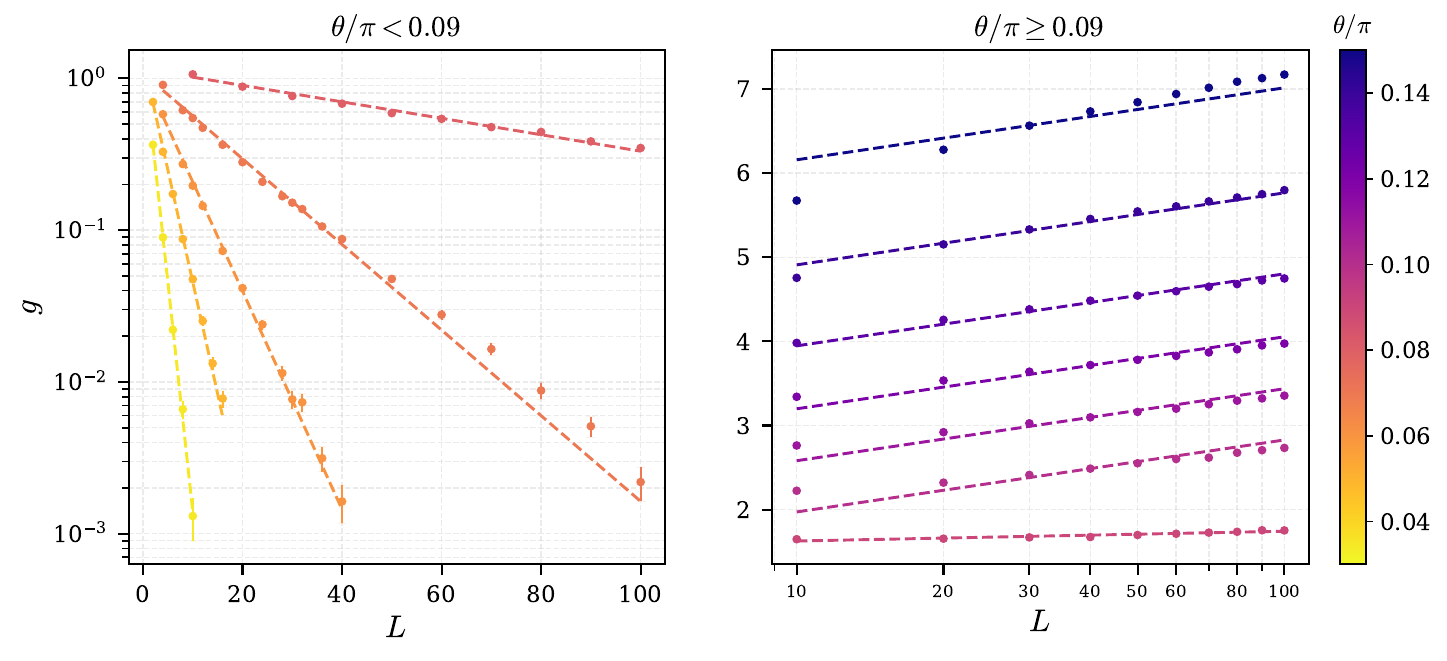}
    \caption{\justifying
 Born-averaged two-terminal conductance $g(\theta,L)$ of the Majorana network.
 Top: angle scan and data collapse. Bottom: size dependence below (left)
 and above (right) threshold. Dashed curves are fits; error bars are
 standard errors.}
    \label{fig:conductance-scaling}
\end{figure}

Figure~\ref{fig:conductance-scaling} is the transport counterpart of
Fig.~\ref{fig:cmi-scaling}.  The conductance decays exponentially below
threshold, crosses near the same angle, and grows logarithmically over the
accessible metallic sizes.  The virtual entanglement shows the same regimes
(Fig.~\ref{fig:mie-scaling-appendix} in Appendix~\ref{app:gaussian-information-transport}).  All three observables
therefore locate the same transition.  In the localized regime, where both
$I_X$ and $g$ decay exponentially, we can compare their lengths directly.

\begin{figure}[tbp]
    \centering
    \includegraphics[width=0.9\columnwidth]{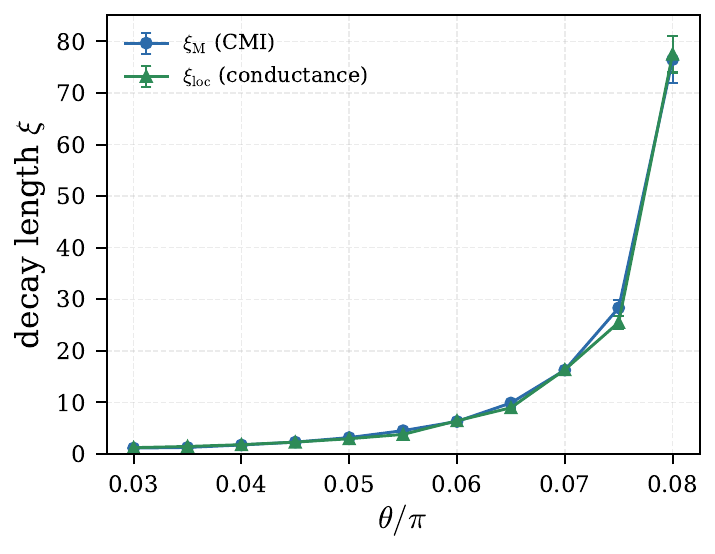}
    \caption{\justifying
 Markov length $\xi_{\mathrm M}$ and localization length
 $\xi_{\mathrm{loc}}$ below threshold. Error bars are statistical fit
 uncertainties.}
    \label{fig:length-scale-comparison}
\end{figure}

Figure~\ref{fig:length-scale-comparison} shows that the two lengths track one
another throughout the resolved localized regime.  Near $0.09\pi$ they
become comparable to the largest simulated systems, so finite-size effects
limit the comparison.  Within this range, the data support
$\xi_{\mathrm M}\simeq\xi_{\mathrm{loc}}$.
The CMI is therefore an entropic probe of Majorana localization: its Markov
length grows with the transport localization length and becomes long-ranged
as the network turns metallic.  The square-lattice construction shows the
same finite-size correspondence, although there the transition may be a long
crossover~\cite{Venn_2023,yan2026nonlinearsigmamodelsurface}; see Appendix
~\ref{app:square-lattice-results}.

For the honeycomb code, the finite Markov length below threshold supplies the
numerical input for the reverse-channel construction under the local-buffer
and parameter-path assumptions stated in Secs.~\ref{subsec:markov-recovery}
and \ref{subsec:coherent-local-recovery}.
Above threshold, the diffusive metallic regime below $\theta=\pi/4$ provides the mechanism for the projected logical ensemble---related to the syndrome-heralded ensemble underlying the mixed state $\rho_{\boldsymbol\theta}$ [Eq.~\eqref{eq:rho_theta}] through Pauli recovery---to approach the Haar ensemble~\cite{bejan2026projectedlogicalensemblessurface}, which we invoke in Sec.~\ref{subsec:coherent-information}.

\section{Discussion}
\label{sec:discussion}

Crossing the decoding threshold can change how quantum information is stored, and we found that it need not be completely lost even above it. In the honeycomb surface code under coherent error, below threshold, local recovery connects the noisy states
to the clean code and establishes an emergent anomalous one-form symmetry
pair. Above threshold, the coherent
information remains positive above threshold and certifies LRE for
generic encoded inputs, while CMI reveals critical conditional
correlations. In this critical regime, the surviving memory loses the spatial
protection of the clean code: punctured coherent information and fidelity
correlations diagnose obstructions to local recovery and to a localized
anomalous one-form symmetry pair. The slow fidelity decay is a
quasi-long-range counterpart of strong-to-weak symmetry breaking and
obstructs phase equivalence with the clean code.
The relation between CMI
and Majorana transport connects this change to delocalization in the
underlying network.

This loss of spatial protection raises the question of what kind of
entanglement survives above threshold. The clean code can route information around a
correctable region, whereas the critical memory is sensitive to
puncturing, with vanishing punctured coherent information. This recalls the fragility of LRE in GHZ states under puncturing even a single qubit. A description in terms of GHZ ensembles would need
to explain the logarithmic CMI and slow fidelity decay; furthermore, the GHZ analogy alone does not imply that the
punctured state is SRE. Whether the critical regime is characterized by a distinct form of topological entanglement remains open.

The dependence on logical input gives this question a concrete starting
point. For a maximally mixed logical input, our criterion leaves the
output's entanglement unresolved
(Fig.~\ref{fig:pi4-logical-bloch-ball}). Under the local-recovery condition
this output is LRE below threshold, but at the special endpoint it is
fully separable for even system sizes, although the channel retains one
bit of coherent information (Appendix~\ref{app:pi4-endpoint}). Following
this state across the transition would reveal where its LRE is lost
while the channel's quantum memory survives.

All states in the critical regime need not belong to one phase. The fidelity
obstruction separates the studied critical states from the clean phase, but
leaves their connectivity to one another unresolved under the coarse two-way local channel connectivity Definition~\ref{def:phase-equivalence}. Constructing local
channels between different angles, or finding an invariant that forbids
them, would determine whether the critical regime contains further phase
distinctions. The refined classification based on locally reversible
circuits~\cite{SangEtAl2025MixedStatePhasesLocalReversibility} offers
another way to examine this structure. In Appendix~\ref{app:logical-fidelity-conversion}, we show that the slow decay of fidelity correlator in the critical regime and drifting mean free path in the critical regime obstructs two-way locally reversible connectivity under super-polynomial error, under a finer phase equivalence definition than Definition~\ref{def:phase-equivalence}. Whether an equivalence under the coarse two-way local channel connectivity definition and the renormalization group flow towards the metallic fixed point in the critical regime coincide will be interesting to further investigate.

The transport mapping offers a complementary route to understanding the
critical regime. A field theory of CMI and fidelity could determine their
universal scaling and explain when the Markov and localization lengths
coincide. The square-lattice crossover provides a further test of this
connection at larger scales (Appendix~\ref{app:square-lattice-results}).

\textbf{Note added:} During the completion of this draft, we noticed a related preprint~\cite{yan2026algebraicstrongtoweaksymmetrybreaking}, which also studied the quasi-long range strong-weak spontaneous symmetry-breaking in the honeycomb surface code.

\begin{acknowledgments}
This research was supported in part by grant NSF PHY-2309135 to the Kavli Institute for Theoretical Physics.  Research at Perimeter Institute is supported in part by the Government of Canada through Innovation, Science and Economic Development Canada and by the Province of Ontario through the Ministry of Colleges and Universities.  J.~B. and B.~B. acknowledge support from EPSRC grant EP/V062654/1.  We also acknowledge the use of AI in assisting with
writing, simulations, and analytical derivations.
\end{acknowledgments}

\appendix
\section{One-Form-Symmetry Criterion}
\label{app:lre-proof}

We prove Theorem~\ref{thm:ci-lre-criterion} directly from the one-form
symmetry algebra.  The physical argument has two steps.  First, a strong one-form
constraint relating two distant representatives becomes a weak symmetry in
every SRE state.  Second, for a pure input with nonzero amplitudes in
every logical-charge basis state, this weak output symmetry would force
complete logical dephasing by the channel and hence zero coherent
information.  We also give a short robust argument for clean code states,
and then extend the noisy-state obstruction to approximate SRE mixtures.

\subsection{One-form code}

Let \(G\) be a finite Abelian group and \(\widehat G\) its character group.
The two one-form symmetries are represented by loop operators
\(W_g(\gamma)\), \(g\in G\), and
\(\widetilde W_\chi(\tau)\), \(\chi\in\widehat G\).  Their mixed anomaly is
expressed by
\begin{equation}
 W_g(\gamma)\widetilde W_\chi(\tau)
 =
 \chi(g)^{I([\gamma],[\tau])}
 \widetilde W_\chi(\tau)W_g(\gamma),
 \label{eq:app-one-form-anomaly}
\end{equation}
where \(I([\gamma],[\tau])\) is the oriented intersection number and the
pairing \(\chi(g)\) is nondegenerate.  The contractible members of the two
families commute and can be fixed simultaneously.  Their common
eigenvalue-one sector is
\begin{equation}
 \mathcal C_0=\left\{\ket\psi:\;
 \begin{gathered}
 \begin{aligned}
 W_g(c)\ket\psi&=\ket\psi,\\
 \widetilde W_\chi(\widetilde c)\ket\psi&=\ket\psi,
 \end{aligned}\\
 c,\widetilde c\ \text{contractible}
 \end{gathered}
 \right\}.
 \label{eq:app-one-form-code-space}
\end{equation}
Here the conditions hold for every \(g\in G\) and
\(\chi\in\widehat G\).  As specified in
Sec.~\ref{subsec:strong_strong_memory}, we choose an encoded subspace
\(\mathcal C\subseteq\mathcal C_0\) in which the eigenvalues of the
commuting logical \(W_g\) uniquely distinguish the basis states.
In representation-theoretic language, the crossing logical pair acts
irreducibly on this space.
This is the full code space for the planar patch with one encoded qubit.
If the geometry encodes additional logical degrees of freedom, their
commuting logical charges are fixed when choosing \(\mathcal C\).
We write \(d_{\rm code}=\dim\mathcal C=|G|\).  This restriction matters:
dephasing a logical charge with a degenerate eigenspace need not destroy
quantum information stored within that eigenspace.

Let \(A\simeq\mathbb C^{d_{\rm code}}\) be the abstract encoded input and
\(V:A\to\mathcal C\) the encoding isometry.  For an input state \(\rho_A\),
we distinguish the encoded state from the state after noise by writing
\begin{equation}
 \rho_{\mathrm{enc}}=V\rho_AV^\dagger,
 \qquad
 \rho_{\mathrm{noisy}}=\mathcal N(\rho_{\mathrm{enc}}).
 \label{eq:app-encoded-and-noisy-states}
\end{equation}
The corresponding channel from the abstract input to the noisy physical
state is
\(\mathcal E(\rho_A)=\mathcal N(V\rho_AV^\dagger)\).
The proof uses locality of the preparation circuits defining SRE; it does
not require a local recovery map or a locality assumption on \(\mathcal N\).

Suppose that every Kraus operator of \(\mathcal N\) commutes with every
\(W\) operator, including the logical representatives and, on a planar
patch, the boundary-anchored strings, as in
Eq.~\eqref{eq:app-strong-channel-symmetry}.  Then every
\(\rho_{\mathrm{noisy}}\) remains in the eigenvalue-one sector of every
contractible \(W\) loop.  For each \(g\), take two homologous representatives
\(W_g^L\) and \(W_g^R\), separated by a distance of order the code size.
Their product \(W_g^L(W_g^R)^\dagger\) is a product of contractible
constraints, and hence
\begin{equation}
 W_g^L(W_g^R)^\dagger\rho_{\mathrm{noisy}}
 =\rho_{\mathrm{noisy}},
 \label{eq:app-one-form-strong-output}
\end{equation}
for every input \(\rho_A\).  On a planar patch, the same statement holds
for strings ending on the appropriate boundaries: two representatives
differ by a product of stabilizers.

Let \(R\simeq A\) be a reference.  For
\begin{equation}
 \ket{\Phi}_{RA}
 =
 \frac{1}{\sqrt{d_{\rm code}}}
 \sum_{x=1}^{d_{\rm code}}\ket{x}_R\ket{x}_A ,
 \label{eq:app-one-form-bell-state}
\end{equation}
write
\(\omega_{RB}=(\operatorname{id}_R\otimes\mathcal E)(\Phi_{RA})\), with
\(B\) denoting the noisy physical output.  The
coherent information of the encoded channel is
\begin{equation}
 I_c(\mathcal E)
 =
 S(B)_\omega-S(RB)_\omega
 =
 -S(R|B)_\omega .
 \label{eq:app-one-form-coherent-information}
\end{equation}
We abbreviate \(I_c(\mathcal E)\) as \(I_c(\mathcal N)\) in the main text.

\subsection{From strong to weak symmetry}

A pure state is \(r\)-trivial if it can be prepared from a product state by a
geometrically local circuit with light-cone radius \(r\).  An SRE mixed state
is a convex mixture of such states with a common \(r\) that is small compared
with the separation between \(W_g^L\) and \(W_g^R\).

\begin{lemma}[Strong symmetry implies weak symmetry for SRE states]
\label{lem:app-strong-to-weak}
Let \(\sigma\) be an SRE state satisfying the strong constraint
\[
 W_g^L(W_g^R)^\dagger\sigma=\sigma .
\]
If the pulled-back supports of \(W_g^L\) and \(W_g^R\) are disjoint, then
\begin{equation}
 [W_g^L,\sigma]=[W_g^R,\sigma]=0 .
 \label{eq:app-one-form-weak-symmetry}
\end{equation}
\end{lemma}

\begin{proof}
Choose an SRE decomposition
\begin{equation}
 \begin{aligned}
 \sigma&=\sum_\alpha p_\alpha
 \ket{\phi_\alpha}\!\bra{\phi_\alpha},\\
 \ket{\phi_\alpha}&=U_\alpha\ket{0}^{\otimes n}.
 \end{aligned}
 \label{eq:app-one-form-sre-decomposition}
\end{equation}
The strong constraint and positivity of this decomposition imply
\[
 W_g^L(W_g^R)^\dagger\ket{\phi_\alpha}
 =
 \ket{\phi_\alpha}
\]
for every component of nonzero weight.  Pulling the two loop operators
through \(U_\alpha\) leaves them on disjoint regions of the product input.
Their expectation value therefore factorizes:
\begin{equation}
 \begin{aligned}
 1&=\bra{\phi_\alpha}W_g^L(W_g^R)^\dagger\ket{\phi_\alpha}\\
  &=\bra{\phi_\alpha}W_g^L\ket{\phi_\alpha}
    \bra{\phi_\alpha}(W_g^R)^\dagger\ket{\phi_\alpha}.
 \end{aligned}
 \label{eq:app-one-form-factorization}
\end{equation}
Each factor has modulus at most one, so both must have unit modulus.  Hence
\(\ket{\phi_\alpha}\) is an eigenstate of \(W_g^L\) and of \(W_g^R\).
The eigenvalues may depend on \(\alpha\), but each projector
\(\ket{\phi_\alpha}\!\bra{\phi_\alpha}\) commutes with the two loops.
Equation~\eqref{eq:app-one-form-weak-symmetry} follows after summing over
\(\alpha\).
\end{proof}

\subsection{Robust obstruction for clean code states}

The two anomalous symmetries give a direct obstruction even for arbitrary
mixed encoded inputs.  This closes the gap between exact decompositions
and the approximate definition of SRE used in the main text.

Choose two distant representatives from each family,
\(W^L,W^R\) and \(\widetilde W^L,\widetilde W^R\), such that
\(W^L\widetilde W^L=\zeta\widetilde W^LW^L\) with \(\zeta\ne1\).
The phase \(\zeta\) is fixed as the code grows.  Set
\[
 C_W=W^L(W^R)^\dagger,
 \qquad C_{\widetilde W}
 =\widetilde W^L(\widetilde W^R)^\dagger.
\]
Every code-supported state \(\rho\) has expectation one for both
constraints.  In contrast, every mixture \(\tau\) of local-circuit states
whose light cones do not join either separated pair obeys
\begin{equation}
 \|\rho-\tau\|_1\geq\frac{|1-\zeta|^2}{128}.
 \label{eq:app-clean-code-robust-lre}
\end{equation}
Only the existence of a positive, size-independent bound is important;
the constant is not optimized.

\begin{proof}
\emph{Step 1: crossing strings cannot both have a sharp charge.}
For any normalized pure state, let
\(a=\langle W^L\rangle\), \(b=\langle\widetilde W^L\rangle\), and
\(s=2-|a|-|b|\).  The residual vectors obtained by subtracting these
expectation values have norms \(\sqrt{1-|a|^2}\) and
\(\sqrt{1-|b|^2}\).  Applying the crossing algebra to the state and
using the triangle inequality gives
\begin{equation}
 \begin{aligned}
 |1-\zeta|\,|ab|
 &\leq2\sqrt{1-|a|^2}+2\sqrt{1-|b|^2}\\
 &\leq4\sqrt{s}.
 \end{aligned}
 \label{eq:app-crossing-charge-uncertainty}
\end{equation}
If \(s\leq1/2\), then \(|ab|\geq1-s\geq1/2\), so
\(s\geq|1-\zeta|^2/64\).  If \(s>1/2\), the same lower bound is
automatic.  Thus \(|a|+|b|\leq2-\kappa\), where
\(\kappa=|1-\zeta|^2/64>0\).

\emph{Step 2: local preparation turns perfect pair correlations into
sharp individual charges.}
In every pure component of \(\tau\), locality factorizes each separated
pair as in Eq.~\eqref{eq:app-one-form-factorization}.  Consequently,
\[
 \operatorname{Re}\langle C_W+C_{\widetilde W}\rangle
 \leq |a|+|b|\leq2-\kappa.
\]
The same bound holds after mixing.  Since the expectation for \(\rho\)
is two and both constraints have operator norm one,
\(2\|\rho-\tau\|_1\geq\kappa\), proving
Eq.~\eqref{eq:app-clean-code-robust-lre}.  Separated representatives at
distance proportional to \(D\) make the bound applicable to every
\(o(D)\)-radius preparation family for sufficiently large \(D\).
\end{proof}

\subsection{Coherent-information obstruction}

We first isolate the channel-theoretic statement needed below.

\begin{lemma}[Weak output symmetry implies measure-and-prepare]
\label{lem:app-weak-output-measure-prepare}
Suppose that \(\mathcal N\) admits a Kraus representation
\(\mathcal N(X)=\sum_\mu K_\mu X K_\mu^\dagger\) satisfying, for all
contractible and logical \(W\) representatives,
\begin{equation}
 [K_\mu,W_g(\gamma)]=0
 \qquad
 \text{for every }\mu,g,\gamma .
 \label{eq:app-strong-channel-symmetry}
\end{equation}
If every state $\rho$ supported on $\mathcal C$ satisfies
\begin{equation}
 [W_g^L,\mathcal N(\rho)]=0
 \qquad
 \text{for every }g,
 \label{eq:app-one-form-weak-all-outputs}
\end{equation}
then, on the logical sector \(\mathcal C\) specified above,
\(\mathcal E\) is a measure-and-prepare channel in the logical \(W\) basis.
Consequently \(I_c(\mathcal E)=0\).
\end{lemma}

\begin{proof}
Equation~\eqref{eq:app-strong-channel-symmetry} says that every noise branch
is neutral under the strong one-form symmetry.  Moving \(W_g^L\) through
each Kraus operator gives
\begin{equation}
 \begin{aligned}
 \mathcal N\!\left(W_g^L\rho(W_g^L)^\dagger\right)
 &=\sum_\mu K_\mu W_g^L\rho(W_g^L)^\dagger K_\mu^\dagger\\
 &=W_g^L\mathcal N(\rho)(W_g^L)^\dagger .
 \end{aligned}
 \label{eq:app-channel-covariance}
\end{equation}
This is covariance under the preserved logical symmetry.  Combining
Eq.~\eqref{eq:app-channel-covariance} with the weak symmetry
\eqref{eq:app-one-form-weak-all-outputs}, we find

\begin{equation}
 \mathcal N\!\left(W_g^L\rho(W_g^L)^\dagger\right)
 =\mathcal N(\rho)
 \qquad
 \text{for every }g .
 \label{eq:app-channel-forgets-conjugation}
\end{equation}
Thus conjugating the input by the preserved logical loop has no effect on the
output.

For any operator \(X\) supported on \(\mathcal C\), average over the
logical symmetry:
\begin{equation}
 \Delta_W(X)
 =
 \frac{1}{|G|}\sum_{g\in G}
 W_g^LX(W_g^L)^\dagger.
 \label{eq:app-logical-twirl}
\end{equation}
Linearity and Eq.~\eqref{eq:app-channel-forgets-conjugation} give
\begin{equation}
 \left.\mathcal N\right|_{\mathcal C}
 =
 \left.\mathcal N\circ\Delta_W\right|_{\mathcal C}.
 \label{eq:app-channel-factors-through-twirl}
\end{equation}
Let \(\{\ket\eta:\eta\in\widehat G\}\) be the joint eigenbasis of the
commuting logical \(W_g^L\) on \(\mathcal C\).  Nondegeneracy of the anomaly and the
single irreducible encoded degree of freedom considered here imply
\begin{equation}
 \Delta_W(\rho)
 =
 \sum_{\eta}
 \bra\eta\rho\ket\eta\,
 \ket\eta\!\bra\eta .
 \label{eq:app-logical-complete-dephasing}
\end{equation}
The twirl is therefore complete dephasing of the logical input.  Defining
\(\sigma_\eta=\mathcal N(\ket\eta\!\bra\eta)\), we obtain
\begin{equation}
 \mathcal N(\rho)
 =
 \sum_\eta
 \bra\eta\rho\ket\eta\,\sigma_\eta,
 \qquad \operatorname{supp}\rho\subseteq\mathcal C .
 \label{eq:app-one-form-measure-prepare}
\end{equation}
This is a measure-and-prepare channel: it first measures the logical \(W\)
charge and then prepares the physical state \(\sigma_\eta\).
Kraus commutation also preserves that charge, so the supports of
\(\sigma_\eta\) and \(\sigma_{\eta'}\) are orthogonal for
\(\eta\ne\eta'\).

Equation~\eqref{eq:app-one-form-measure-prepare} shows that \(\mathcal E\) is
entanglement breaking.  In particular, sending half of the Bell state in
Eq.~\eqref{eq:app-one-form-bell-state} through \(\mathcal E\) gives the
separable state
\begin{equation}
 \omega_{RB}
 =
 \frac{1}{d_{\rm code}}
 \sum_\eta
 \ket{\eta^*}\!\bra{\eta^*}_R\otimes\sigma_\eta,
 \label{eq:app-one-form-separable-choi}
\end{equation}
where complex conjugation is taken in the basis used to define the Bell
state.  The reference charge is classical and can be read perfectly from
the output, so \(S(R|B)_\omega=0\).  In particular,
\begin{equation}
 I_c(\mathcal E)=-S(R|B)_\omega=0 .
 \label{eq:app-one-form-nonpositive-ic}
\end{equation}
\end{proof}

To obtain the statement for generic inputs, consider
\(\ket\psi=\sum_\eta c_\eta\ket\eta\), with every
\(c_\eta\ne0\), in the logical-charge basis defined above.
Let \(P_\eta\) project onto the physical sector with charge \(\eta\),
and define
\begin{equation}
 T_{\eta\eta'}=\mathcal N(\ket\eta\!\bra{\eta'}).
 \label{eq:app-one-form-charge-block}
\end{equation}
Kraus commutation preserves each charge separately, so
\(T_{\eta\eta'}=P_\eta T_{\eta\eta'}P_{\eta'}\).
Consequently, for \(\sigma=\mathcal N(\ket\psi\!\bra\psi)\),
\begin{equation}
 P_\eta\sigma P_{\eta'}=c_\eta c_{\eta'}^*T_{\eta\eta'}.
 \label{eq:app-one-form-generic-output-block}
\end{equation}
Different input coherences thus occupy distinct physical charge blocks
and cannot cancel one another.

If \(\sigma\) were SRE, its preserved constraint
\eqref{eq:app-one-form-strong-output} and
Lemma~\ref{lem:app-strong-to-weak} would imply that it commutes with
every \(W_g^L\).  Hence \(P_\eta\sigma P_{\eta'}=0\) for
\(\eta\ne\eta'\).  Since all input amplitudes are nonzero,
Eq.~\eqref{eq:app-one-form-generic-output-block} forces
\(T_{\eta\eta'}=0\) for every distinct pair.  The channel then
discards all logical coherence and takes the measure-and-prepare form
\eqref{eq:app-one-form-measure-prepare}, giving \(I_c=0\).
Therefore \(I_c>0\) excludes an exact SRE decomposition for every
such input.  These inputs have full Haar measure; for one encoded qubit,
the only possible pure-state exceptions are the two eigenstates of the
preserved logical string.  We establish the thermodynamic statement,
including approximate SRE states, below.

For one encoded qubit, the same argument also covers mixed inputs.
Let \(\rho^{\rm L}_{01}=\bra{\overline0}\rho_Q\ket{\overline1}\).
Equation~\eqref{eq:app-one-form-generic-output-block} then reads
\(P_0\mathcal N(\rho_Q)P_1=\rho^{\rm L}_{01}T_{01}\).
If \(\rho^{\rm L}_{01}\ne0\), an SRE output would force \(T_{01}=0\), and
hence \(I_c=0\).  In the approximate argument below, the denominator
\(|c_0c_1^*|\) is simply replaced by \(|\rho^{\rm L}_{01}|\).
Thus a positive limiting coherent information certifies LRE for every
fixed pure or mixed input with nonzero logical coherence, equivalently
\(r_X^2+r_Y^2>0\).  The criterion alone does not classify inputs on
the logical \(Z\) diameter.

For a code with several logical degrees of freedom, the same exact
argument applies to their joint charge basis if every independent charge
has two separated representatives and is preserved by every Kraus
operator.  Apply Lemma~\ref{lem:app-strong-to-weak} to each charge
separately, then twirl over all of them.  When their joint eigenspaces are
one-dimensional, this again gives complete logical dephasing.  For example,
on the torus both independent logical \(Z\) loops are needed to treat the
full two-qubit input; fixing one of their eigenvalues instead gives the
single-qubit sector used above.

\subsection{Approximate SRE states}

For \(0\leq\epsilon\leq1\), we call a state \((r,\epsilon)\)-SRE if it is
within trace-norm error \(\epsilon\) of a convex mixture of pure states prepared
by local circuits with light-cone radius at most \(r\).  Thus there is a state
\begin{equation}
 \tau=\sum_\alpha p_\alpha
 \ket{\phi_\alpha}\!\bra{\phi_\alpha},
 \qquad
 \|\sigma-\tau\|_1\leq\epsilon ,
 \label{eq:app-one-form-approx-sre}
\end{equation}
where \(\ket{\phi_\alpha}=U_\alpha\ket{0}^{\otimes n}\) and every
\(U_\alpha\) has light-cone radius at most \(r\).

\begin{proposition}
\label{prop:app-approx-lre}
Suppose that \(\mathcal N\) has \(W\) as a strong symmetry in the sense of
Eq.~\eqref{eq:app-strong-channel-symmetry}, and that every encoded output
\(\mathcal E(\rho_A)\) is
\((r,\epsilon)\)-SRE, with \(r\) small enough that the pulled-back supports
of \(W_g^L\) and \(W_g^R\) are disjoint.  Then
for sufficiently small \(\epsilon>0\),
\begin{equation}
 I_c(\mathcal E)
 \leq
 C_{d_{\rm code}}\sqrt{\epsilon}
 \log_2\!\frac{1}{\epsilon},
 \label{eq:app-one-form-robust-ic}
\end{equation}
where \(C_d\) depends only on the encoded dimension \(d\).  In particular,
the right-hand side vanishes as \(\epsilon\to0\).  Equivalently, if
\(I_c(\mathcal E)>C_{d_{\rm code}}\sqrt{\epsilon}\log_2(1/\epsilon)\), at
least one noisy encoded state is not \((r,\epsilon)\)-SRE.
\end{proposition}

\begin{proof}
\emph{Step 1: approximate strong symmetry still forces approximate weak
symmetry.}
Fix \(\rho_A\), set \(\rho=V\rho_AV^\dagger\) and
\(\sigma=\mathcal N(\rho)=\mathcal E(\rho_A)\), and choose \(\tau\) as in
Eq.~\eqref{eq:app-one-form-approx-sre}.  For a fixed \(g\),
Eq.~\eqref{eq:app-one-form-strong-output} gives
\(W_g^L(W_g^R)^\dagger\sigma=\sigma\).  Since
\(\|\sigma-\tau\|_1\leq\epsilon\),
\begin{equation}
 \operatorname{Re}\operatorname{Tr}
 \!\left[W_g^L(W_g^R)^\dagger\tau\right]
 \geq1-\epsilon .
 \label{eq:app-one-form-approx-strong}
\end{equation}
Define
\[
 a_{\alpha,g}
 =
 \bra{\phi_\alpha}W_g^L\ket{\phi_\alpha},
 \qquad
 b_{\alpha,g}
 =
 \bra{\phi_\alpha}W_g^R\ket{\phi_\alpha}.
\]
The two pulled-back loops have disjoint support, so their expectation value
factorizes in each component:
\[
 \bra{\phi_\alpha}W_g^L(W_g^R)^\dagger\ket{\phi_\alpha}
 =
 a_{\alpha,g}b_{\alpha,g}^* .
\]
Equation~\eqref{eq:app-one-form-approx-strong} and
\(|b_{\alpha,g}|\leq1\) imply
\(\sum_\alpha p_\alpha|a_{\alpha,g}|\geq1-\epsilon\).

For \(P_\alpha=\ket{\phi_\alpha}\!\bra{\phi_\alpha}\),
\[
 \|[W_g^L,P_\alpha]\|_1
 =
 2\sqrt{1-|a_{\alpha,g}|^2}.
\]
Convexity and concavity of the square root give
\begin{equation}
 \begin{aligned}
 \|[W_g^L,\tau]\|_1
 &\leq2\sum_\alpha p_\alpha\sqrt{1-|a_{\alpha,g}|^2}\\
 &\leq2\sqrt{1-\bigl(\sum\nolimits_\alpha
 p_\alpha|a_{\alpha,g}|\bigr)^2}\\
 &\leq2\sqrt{2\epsilon-\epsilon^2}.
 \end{aligned}
 \label{eq:app-one-form-approx-tau-weak}
\end{equation}
Together with
\(\|[W_g^L,\sigma-\tau]\|_1\leq2\epsilon\), this yields
\begin{equation}
 \begin{aligned}
 \|[W_g^L,\sigma]\|_1
 &\leq2\sqrt{2\epsilon-\epsilon^2}+2\epsilon
 =:\delta_\epsilon\\
 &\leq5\sqrt{\epsilon} .
 \end{aligned}
 \label{eq:app-one-form-approx-weak}
\end{equation}

\emph{Step 2: the channel is close to one that retains only classical
logical charges.}
Kraus covariance, Eq.~\eqref{eq:app-channel-covariance}, now transfers this
bound to the input.  Indeed,
\begin{equation}
 \begin{aligned}
 &\bigl\|\mathcal N(\rho)
 -\mathcal N\!\bigl(W_g^L\rho(W_g^L)^\dagger\bigr)\bigr\|_1\\
 &\qquad=\|[W_g^L,\sigma]\|_1\leq\delta_\epsilon .
 \end{aligned}
 \label{eq:app-one-form-approx-channel-invariance}
\end{equation}
Averaging over \(g\) and using the logical twirl
\eqref{eq:app-logical-twirl}, we obtain
\begin{equation}
 \left\|
 \mathcal N(\rho)-\mathcal N[\Delta_W(\rho)]
 \right\|_1
 \leq\delta_\epsilon
 \leq5\sqrt{\epsilon} .
 \label{eq:app-one-form-approx-logical-twirl}
\end{equation}

Define a channel on the abstract encoded input by
\begin{equation}
 \mathcal M(\rho_A)
 =
 \mathcal N\!\left[\Delta_W(V\rho_AV^\dagger)\right].
 \label{eq:app-one-form-nearby-classical-channel}
\end{equation}
Equation~\eqref{eq:app-logical-complete-dephasing} shows that \(\mathcal M\)
is measure-and-prepare and hence entanglement breaking.  We only need to
compare the encoded Bell-pair outputs \(\omega_{RB}\) and \(\nu_{RB}\) of
\(\mathcal E\) and \(\mathcal M\).  Write
\(\mathcal T=\mathcal E-\mathcal M\) and \(d=d_{\rm code}\).
Equation~\eqref{eq:app-one-form-approx-logical-twirl} bounds
\(\|\mathcal T(P)\|_1\leq\delta_\epsilon\) for every pure-state
projector \(P\).  For distinct charge-basis states \(\ket j,\ket k\),
expressing \(\ket j\!\bra k\) in terms of the four projectors onto
\((\ket j\pm\ket k)/\sqrt2\) and
\((\ket j\pm i\ket k)/\sqrt2\) gives
\(\|\mathcal T(\ket j\!\bra k)\|_1\leq2\delta_\epsilon\).
The diagonal terms vanish because \(\mathcal T(\ket j\!\bra j)=0\).
Expanding the Bell state in this basis therefore gives
\begin{equation}
 \begin{aligned}
 \|\omega-\nu\|_1
 &\leq\frac1d\sum_{j\ne k}
   \|\mathcal T(\ket j\!\bra k)\|_1\\
 &\leq2(d-1)\delta_\epsilon .
 \end{aligned}
 \label{eq:app-one-form-bell-output-bound}
\end{equation}

\emph{Step 3: a nearly classical Bell output has small coherent
information.}
Set \(\varepsilon_B=\tfrac12\|\omega-\nu\|_1\), so
\(\varepsilon_B\leq\min\{1,5(d-1)\sqrt\epsilon\}\).
Since \(\mathcal M\) is entanglement breaking,
\(S(R|B)_\nu\geq0\).  The conditional-entropy continuity
bound in Ref.~\cite{Winter_2016_continuity} then gives
\begin{equation}
 \begin{aligned}
 I_c(\mathcal E)
 &\leq2\varepsilon_B\log_2 d_{\rm code}\\
 &\quad +(1+\varepsilon_B)h_2\!\left(
 \frac{\varepsilon_B}{1+\varepsilon_B}\right),
 \end{aligned}
 \label{eq:app-one-form-continuity-bound}
\end{equation}
where \(h_2(x)=-x\log_2x-(1-x)\log_2(1-x)\).  For fixed
\(d_{\rm code}\) and sufficiently small \(\epsilon\), the right-hand side
is bounded by
\(C_{d_{\rm code}}\sqrt\epsilon\log_2(1/\epsilon)\), proving
Eq.~\eqref{eq:app-one-form-robust-ic}.
\end{proof}

For a family of codes with fixed encoded dimension, suppose
\(\liminf_{D\to\infty}I_c(\mathcal E_D)>0\).  Choose a fixed
\(\epsilon_0>0\) small enough that the bound in
Eq.~\eqref{eq:app-one-form-robust-ic} lies below this limit.
If the loop separation is proportional to the linear size \(D\), there
is a constant \(c>0\) for which the proposition applies with \(r=cD\).
At every sufficiently large size, at least one noisy encoded output then
lies more than \(\epsilon_0\) in trace norm from every \(cD\)-trivial
mixture.

The same estimates give the stronger statement for each fixed generic
input \(\ket\psi=\sum_\eta c_\eta\ket\eta\).
Suppose its output \(\sigma\) alone is \((r,\epsilon)\)-SRE.
The argument through Eq.~\eqref{eq:app-one-form-approx-weak} applies
to this state and gives
\(\|[W_g^L,\sigma]\|_1\leq\delta_\epsilon\) for every \(g\).
Averaging over the symmetry therefore changes \(\sigma\) by at most
\(\delta_\epsilon\) in trace norm and removes its off-diagonal
charge blocks.  Projection onto any such block, together with
Eq.~\eqref{eq:app-one-form-generic-output-block}, gives
\begin{equation}
 \|T_{\eta\eta'}\|_1
 \leq\frac{\delta_\epsilon}{|c_\eta c_{\eta'}|}
 \qquad(\eta\ne\eta').
 \label{eq:app-one-form-generic-block-bound}
\end{equation}
Let \(m_\psi=\min_{\eta\ne\eta'}|c_\eta c_{\eta'}|>0\).
Expanding the Bell outputs of \(\mathcal E\) and the completely
logically dephased channel \(\mathcal M\), as above, now gives
\(\|\omega-\nu\|_1\leq(d_{\rm code}-1)\delta_\epsilon/m_\psi\).
The same conditional-entropy continuity bound yields
\begin{equation}
 I_c(\mathcal E)
 \leq C_{d_{\rm code},\psi}\sqrt\epsilon
          \log_2\!\frac1\epsilon,
 \label{eq:app-one-form-generic-robust-ic}
\end{equation}
for sufficiently small \(\epsilon\).  The constant depends only on
the encoded dimension and the fixed input amplitudes.  Thus, if
\(\liminf_{D\to\infty}I_c(\mathcal E_D)>0\), the output of every
fixed generic input remains a nonzero trace-norm distance from SRE
mixtures with sublinear preparation radius.  This proves
Theorem~\ref{thm:ci-lre-criterion} in the sense of
Definition~\ref{def:mixed-state-lre}.  The statement does not require a uniform
distance bound for inputs whose amplitudes approach zero as \(D\) grows.

\section{Unified phase-equivalence scheme}
\label{app:unified-phase-equivalence}

A useful way to connect two mixed states is to change their local
parameters one region at a time. The question is whether each change
can also be implemented by a channel acting near that region. For the
exactly locally parameterized families considered in
Sec.~\ref{sec:lre}, changing one parameter leaves the exterior state
unchanged. When the Markov length is finite, we can then erase the
affected region and reconstruct the target state from a surrounding
buffer. More generally, a local
parameter change may affect the state farther away, as happens when a
local Hamiltonian term is varied in a thermal state. The same recovery
construction still works if this effect becomes exponentially small
with distance.

\begin{figure*}[t]
 \centering
 \includegraphics[width=\textwidth]{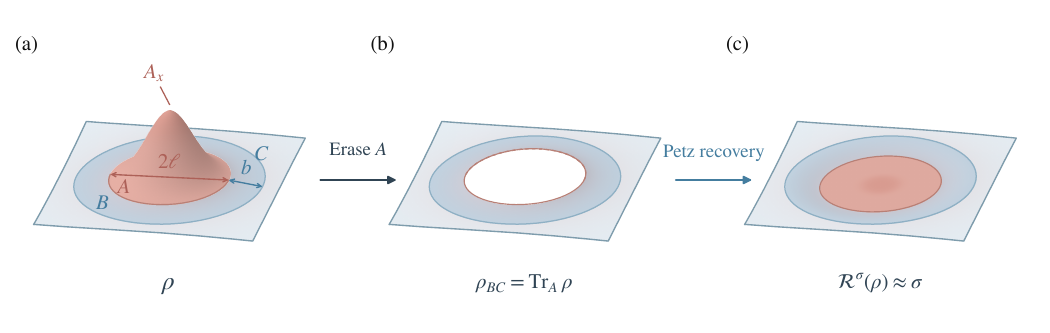}
 \caption{\justifying
 One local update in the two-length construction.
 (a) The bump schematically represents the deformation between the
 source $\rho$ and the target $\sigma$. Enlarging the parameter region
 $A_x$ to $A=A_x(\ell)$ leaves only a small exterior response,
 indicated by the faint red shading in $BC$. The deformation length
 controls how large $\ell$ must be. The diameter is drawn as $2\ell$,
 suppressing the bounded size of $A_x$.
 (b) Erase $A$, retaining $\rho_{BC}$.
 (c) Reconstruct the target patch using the Petz map. The map acts on
 $AB$, with its recovery error controlled by the buffer width $b$ and
 the Markov length. Repeating the construction in both directions
 along a parameter path gives two-way phase equivalence.}
 \label{fig:unified-phase-construction}
\end{figure*}

Two lengths now enter the construction, illustrated in
Fig.~\ref{fig:unified-phase-construction}. The \emph{deformation length}
measures how far a parameter change affects the state, while the
\emph{Markov length} controls how well an erased region can be
reconstructed from its surroundings. The first tells us how large a
region to erase; the second tells us how wide a buffer to use for
recovery. When both remain uniformly finite along the local parameter
changes, these recovery maps form circuits connecting the initial and
final states in both directions.

This gives a common phase-equivalence scheme for equilibrium and
nonequilibrium families. We first explain the construction and state
the resulting criterion. We then turn to thermal states and illustrate
why the deformation length is needed using the two-dimensional classical
Ising model: its shielded CMI vanishes even at the thermal transition,
whereas the deformation length diverges.

\subsection{Two length scales for local recovery}

Consider a parameterized family $\rho_L[\boldsymbol\lambda]$ with bounded
parameter regions $A_x$ as in
Definition~\ref{def:locally-parameterized-family}, but now allow a local
parameter change to affect exterior marginals. Take two parameter
profiles that differ only at $x$, and denote the corresponding states by
$\rho$ and $\sigma$. Enlarge the bounded parameter region $A_x$ to its
$\ell$-neighborhood $A_x(\ell)$, and define
\begin{equation}
 \Delta_x(\ell;\rho,\sigma)
 =\left\|\Tr_{A_x(\ell)}(\rho-\sigma)\right\|_1.
 \label{eq:unified-deformation-profile}
\end{equation}
This quantity measures how well an observer outside $A_x(\ell)$ can
distinguish the two states. It includes the \emph{entire exterior
marginal}, including correlations accessible to collective measurements
there.
We say that the allowed parameter changes have finite deformation length
if
\begin{equation}
 \Delta_x(\ell;\rho,\sigma)
 \leq a_{\mathrm D}L^{p_{\mathrm D}}e^{-\ell/\xi_{\mathrm D}},
 \label{eq:unified-deformation-bound}
\end{equation}
with constants $a_{\mathrm D}>0$, $p_{\mathrm D}\geq0$, and
$\xi_{\mathrm D}>0$ independent of the system size and the intermediate
parameter profile. The allowed changes, including any symmetry
restrictions, are part of the specification of the family. Exact local
parameterization is the special case
$\Delta_x(0;\rho,\sigma)=0$.

Next take $A=A_x(\ell)$, surround it by a buffer $B$ of width $b$,
and let $C$ be the remaining exterior. The second condition is a finite
Markov length for these enlarged regions:
\begin{equation}
 I_{\rho_L[\boldsymbol\lambda]}(A_x(\ell):C|B)
 \leq a_{\mathrm M}L^{p_{\mathrm M}}e^{-b/\xi_{\mathrm M}}.
 \label{eq:unified-markov-bound}
\end{equation}
As in the main text, this says that a wider buffer contains enough
information to reconstruct $A$ while preserving its correlations with
$C$. Both bounds are required uniformly for every partially updated
parameter profile used in the construction, with constants independent
of $L$. They must hold for the enlarged regions and for distances up to
a fixed positive power of $L$.

To implement one parameter change, erase $A$ and reconstruct it using
the twirled Petz map of the target state $\sigma$:
\begin{equation}
 \mathcal R^\sigma_{A,b}
 =\bigl(\mathcal P^\sigma_{B\to AB}\otimes\operatorname{id}_C\bigr)
 \circ\Tr_A.
 \label{eq:unified-recovery-map}
\end{equation}
The map acts only on $AB$. If its input were $\sigma_{BC}$, approximate
recovery would produce $\sigma$ up to an error controlled by the target
CMI. The actual input is $\rho_{BC}$, which differs from
$\sigma_{BC}$ by $\Delta_x(\ell;\rho,\sigma)$. A quantum channel cannot
increase this difference. Adding these two errors
therefore gives~\cite{FawziRenner2015Recovery,Junge2018UniversalRecovery}
\begin{equation}
 \begin{aligned}
 &\|\mathcal R^\sigma_{A,b}(\rho)-\sigma\|_1\\
 &\quad\leq\|\rho_{BC}-\sigma_{BC}\|_1
       +\|\mathcal R^\sigma_{A,b}(\sigma)-\sigma\|_1\\
 &\quad\leq\Delta_x(\ell;\rho,\sigma)
       +2\sqrt{\ln2\,I_\sigma(A:C|B)}.
 \end{aligned}
 \label{eq:unified-recovery-error}
\end{equation}
The two terms explain the two distances in the figure. Increasing
$\ell$ leaves less of the parameter change outside the erased region;
increasing $b$ improves the reconstruction inside it. Thus the errors
can be made small independently. To go in the opposite direction, use
the Petz map of $\rho$ instead. The exterior mismatch is the same, and
the recovery error is now controlled by the CMI of $\rho$.

\subsection{Two-way phase equivalence}

We now apply this local construction repeatedly along a parameter path.
Use the local-update schedule of
Theorem~\ref{thm:marginal-preserving-local-reverse}: on a
bounded-density $d$-dimensional lattice, the $N=O(L^d)$ bounded
parameter regions have bounded overlap, and each parameter is changed
at most once in each of $K=\operatorname{polylog}L$ rounds. The
intermediate profiles include the inhomogeneous states obtained when
only some parameters in a round have been updated.

\begin{samepage}
\begin{theorem}[Phase equivalence from two finite length scales]
\label{thm:unified-phase-equivalence}
Along the local updates above, assume the uniform bounds
\begin{equation*}
 \begin{aligned}
 \Delta_x(\ell;\rho,\sigma)
 &\leq a_{\mathrm D}L^{p_{\mathrm D}}e^{-\ell/\xi_{\mathrm D}},\\
 I_\sigma(A_x(\ell):C|B)
 &\leq a_{\mathrm M}L^{p_{\mathrm M}}e^{-b/\xi_{\mathrm M}},
 \end{aligned}
\end{equation*}
with finite, $L$-independent $\xi_{\mathrm D},\xi_{\mathrm M}$,
for all intermediate profiles in both directions. Then all states
along the path are phase equivalent
(Definition~\ref{def:phase-equivalence}), with light-cone radius
$\operatorname{polylog}L$ and error $L^{-\infty}$.
\end{theorem}
\end{samepage}

\begin{proof}[Proof sketch]
Choose $\ell$ and $b$ proportional to $(\ln L)^2$. Both terms in
Eq.~\eqref{eq:unified-recovery-error} are then smaller than every fixed
inverse power of $L$. For each step, construct the recovery map from
its ideal target state on the parameter path. Previous errors cannot
be amplified by the next channel, so the total error is at most the
sum of the $NK$ individual errors and remains $L^{-\infty}$.

To keep the circuit local, group updates whose recovery regions $AB$
are disjoint and apply each group in parallel. A region has diameter
$s=O(\ell+b)$, so $O(s^d)$ groups suffice in each round. Choose the
update order accordingly and construct the maps from the corresponding
ideal intermediate states. The resulting circuit has depth $O(Ks^d)$
and light-cone radius $O(Ks^{d+1})=\operatorname{polylog}L$.
Reversing the updates and using the reverse recovery maps gives the
channel in the other direction.
\end{proof}

The uniformity condition is important physically: a path must remain
locally recoverable while its parameters are being changed region by
region, not only when they all have the same value. Subject to this
condition, finite deformation and Markov lengths let us follow the
path using local operations throughout.

The same conclusion holds if the CMI bound contains an additional
factor $e^{\mu|A_x(\ell)|}$, as in available bounds for quantum Gibbs
states~\cite{chen2025quantumgibbsstateslocally}. Since
$|A_x(\ell)|=O(\ell^d)$, choosing
$b=O(\ell^d+(\ln L)^2)$ still gives a polylogarithmic circuit and
superpolynomial accuracy. For the classical example below, the shielded
CMI is exactly zero and this extra factor is unnecessary.

\subsection{Thermal phase transitions}

For a thermal state, a natural question is how much a distant region
can detect a local change of the Hamiltonian or temperature. This is
precisely the response measured by the deformation length. To describe
temperature changes within the construction, write a Gibbs family as
\begin{equation}
 \rho_L[\boldsymbol\beta]
 =\frac{e^{-\sum_x\beta_x h_x}}
 {\Tr e^{-\sum_x\beta_x h_x}},
 \label{eq:unified-gibbs-family}
\end{equation}
where $h_x$ are bounded local energy terms. A uniform
$\beta_x=\beta$ gives the ordinary thermal state. Changing the
$\beta_x$ one region at a time connects two temperatures through
inhomogeneous Gibbs states. Here the $\beta_x$ are coefficients of
local energy terms; a change in one of them is a local Hamiltonian
perturbation. Unlike exact local parameterization, this generally
changes distant marginals as well.

The locality of this response has been studied through locality of
temperature and quantum belief propagation, which relate the effect
of a local perturbation on distant regions to the decay of thermal
correlations~\cite{KlieschEtAl2014LocalityTemperature,
Hastings2007QuantumBeliefPropagation,
CapelMoscolariTeufelWessel2025Locality}.
An erase-and-recover construction for Gibbs states was developed in
Ref.~\cite{MaKhemaniSang2025ThermalPhases}, where stable clustering
controls the exterior response and the Gibbs-state Markov property
controls recovery. Here we formulate the construction for general
parameterized state families by imposing these two bounds directly,
through the deformation and Markov lengths. This applies to equilibrium
and nonequilibrium settings without requiring a Gibbs representation
or a prescribed evolution. In particular, it includes the coherent-error
families studied here, for which exact local parameterization makes
the exterior response vanish, as well as local Lindbladian
paths~\cite{Sang_2025}.

The connection to correlations also explains the relation to a
mutual-information length. MI measures correlations between two regions
in a fixed state, whereas the deformation length measures the response
to a specified parameter change. In a classical Gibbs state, a local
change reweights the probability distribution, so its distant effect
is controlled by correlations with the term being changed. Decay of MI
can therefore bound this response. The lengths need not be equal:
changing a temperature coefficient probes the corresponding local
energy term, while the full spatial MI can detect other correlations
as well. More generally, even a state with long-range MI can have
zero exterior response to a chosen deformation, such as an on-site
unitary.

For classical finite-range Gibbs states, a buffer that separates $A$
from $C$ gives $I(A:C|B)=0$, including at a thermal transition. The
Markov condition alone thus cannot distinguish the thermal phases.
The exterior response supplies the missing information: near
criticality, a local perturbation can affect the state over an
increasingly large distance. The following example makes this
mechanism explicit.

\begin{figure}[t]
 \centering
 \includegraphics[width=\columnwidth]{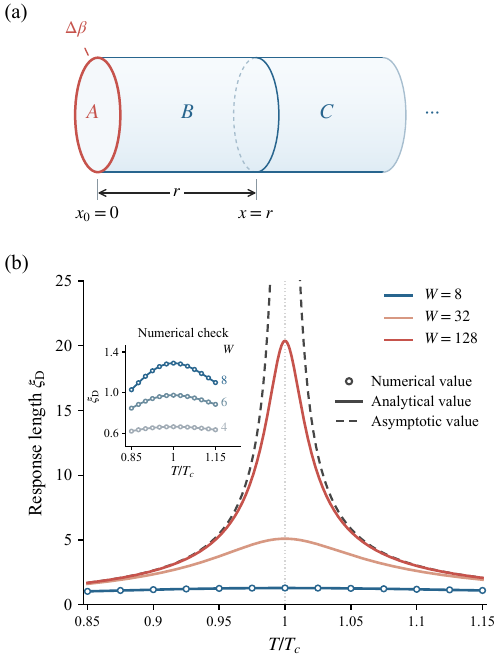}
 \caption{\justifying
 Deformation length at the Ising thermal transition.
 (a) On a cylinder of circumference $W$, change the inverse temperature
 of the bonds in $A$ by $\Delta\beta$ and measure the response of $C$,
 a distance $r$ away. Region $B$ is the intervening buffer.
 (b) Decay length of this response: circles show direct numerical fits,
 solid curves the exact finite-width results, and the dashed curve the
 infinite-width limit. The peak grows into a divergence at $T_c$ as
 $W\to\infty$, while $I(A:C|B)$ remains zero.
 The inset resolves the numerical comparison for $W=4,6,8$.
 Lengths are in lattice units.}
 \label{fig:ising-deformation-mi}
\end{figure}

\subsubsection{Two-dimensional Ising model}

The ferromagnetic square-lattice Ising model provides a simple example:
the conditional correlations are screened by a buffer at every
temperature, but the response to a local temperature change becomes
long-ranged at the transition. Its spins $s_i=\pm1$ have distribution
$p_K(s)=Z_K^{-1}\exp(K\sum_{\langle ij\rangle}s_i s_j)$, with
$K=J/T$ and $k_B=1$. The critical coupling is
$K_c=\tfrac12\ln(1+\sqrt2)$, corresponding to $T_c=J/K_c$.

We use a semi-infinite cylinder of circumference $W$, where the
transfer matrix lets us calculate the response exactly.
As shown in Fig.~\ref{fig:ising-deformation-mi}(a), $A$ is the boundary
column, $B$ is the intervening buffer, and $C$ contains all columns
starting a distance $r$ away. We change the inverse temperature of the
vertical bonds in $A$ by $\Delta\beta$, so their coupling changes
from $K$ to $K+u$, with $u=J\Delta\beta$. The resulting exterior
response is $\Delta_r=\|p_{K,u;C}-p_{K;C}\|_1$, where $p_{K,u}$
denotes the changed state. Its long-distance decay defines the
deformation length through $\Delta_r\sim\mathcal A e^{-r/\xi_{\mathrm D}}$.
For $r\geq2$, $I(A:C|B)=0$ exactly, even when this response extends
far beyond $A$.

\paragraph*{Computing the exterior response.}
Treat each column as a variable $\boldsymbol s=(s_1,\ldots,s_W)$.
The transfer matrix passes statistical weights from one column to
the next:
\begin{equation}
 \mathsf T_{\boldsymbol s\boldsymbol t}
 =\exp\!\Bigl[\tfrac K2 E(\boldsymbol s)
       +K\sum\nolimits_{y=1}^W s_y t_y+\tfrac K2 E(\boldsymbol t)\Bigr],
 \label{eq:unified-cylinder-transfer}
\end{equation}
where $E(\boldsymbol s)=\sum_y s_y s_{y+1}$ and $s_{W+1}=s_1$.
The middle term couples adjacent columns, while the two half-column
terms count each vertical bond once. Let $\lambda_0$ and $v$ be the
leading eigenvalue and positive eigenvector, with $v^Tv=1$.
The vector $v$ represents the infinite right tail, and
$\widehat{\mathsf T}=\mathsf T/\lambda_0$ propagates weights through
the buffer. At the left boundary, the weights
$h_{\boldsymbol s}=e^{K E(\boldsymbol s)/2}$ become
$h_{u,\boldsymbol s}=h_{\boldsymbol s}e^{uE(\boldsymbol s)}$.
Their normalized difference is
$\eta=h_u/(h_u^Tv)-h/(h^Tv)$.

Once the first column $\boldsymbol c$ of $C$ is fixed, the rest of
the tail has the same conditional distribution before and after the
change. The response of the entire exterior therefore equals that
of this first column. Contracting from both ends gives
\begin{equation}
 \Delta_r=\sum_{\boldsymbol c}v_{\boldsymbol c}
 \left|[\widehat{\mathsf T}^{\,r}\eta]_{\boldsymbol c}\right|.
 \label{eq:unified-cylinder-response}
\end{equation}
Thus the infinite-cylinder problem reduces to propagating a vector
with $2^W$ entries.

\paragraph*{The decay length and its divergence.}
Each transfer-matrix mode decays by its eigenvalue ratio
$\lambda_\alpha/\lambda_0$ per column. Normalization removes the
stationary mode from $\eta$, and the boundary temperature change
preserves spin-flip symmetry, so it excites only even modes. The
slowest even mode with nonzero overlap therefore sets the response
length. For the perturbation used here, this is the leading
nonstationary even mode, with eigenvalue $\lambda_{\mathrm e}$.

The free-fermion solution of the Ising transfer matrix evaluates this
ratio exactly~\cite{SchultzMattisLieb1964IsingFermions,Kastening2001IsingSpectrum}.
The lowest even excitation contains a pair of modes with momenta
$q=\pm\pi/W$, each contributing a decay exponent $\gamma(q;K)$.
Consequently,
\begin{equation}
 \begin{aligned}
 \xi_{\mathrm D}(K,W)
 &=\frac{1}{\ln(\lambda_0/\lambda_{\mathrm e})}
 =\frac{1}{2\gamma(\pi/W;K)},\\
 \cosh\gamma(q;K)
 &=\sinh(2K)+\frac{1}{\sinh(2K)}-\cos q.
 \end{aligned}
 \label{eq:unified-ising-cylinder-spectrum}
\end{equation}
At $T_c$, the length grows as $\xi_{\mathrm D}\sim W/(2\pi)$:
each finite cylinder has a rounded peak, but that peak grows without
bound as its circumference increases. In the two-dimensional limit,
\begin{equation}
 \xi_{\mathrm D}(K,\infty)
 =\frac{1}{2|2K+\ln\tanh K|}.
 \label{eq:unified-ising-bulk-length}
\end{equation}
It is finite on both sides of the transition and diverges as
$|T/T_c-1|^{-1}$ at $T_c$. The spin-flip-odd mode associated with
the ordered-spin sign does not contribute to this temperature response.

To check the result directly, we evaluate
Eq.~\eqref{eq:unified-cylinder-response} using all column configurations
for $W=4,6,8$ and $u=0.15$. Fitting the long-distance slope of
$\ln\Delta_r$ gives the circles in Fig.~\ref{fig:ising-deformation-mi}(b).
The fits use distances of approximately $9$--$15$ predicted decay
lengths and agree with the exact expression in
Eq.~\eqref{eq:unified-ising-cylinder-spectrum}. The larger-width curves
and the infinite-width limit are evaluated analytically.

We use the cylinder for analytical convenience and expect the same
bulk critical divergence for a bounded perturbation surrounded by an
annular buffer, although the decay length can depend on geometry.
For a single-bond change in the plane, exponential clustering bounds
the exterior response away from $T_c$, including in the symmetric
low-temperature state for this symmetry-preserving perturbation
\cite{DuminilCopinGoswamiRaoufi2020Truncated}. At $T_c$, critical
energy correlations give an algebraic lower bound, ruling out a finite
deformation length~\cite{IzyurovKemppainenTuisku2024Energy}.
The cylinder calculation illustrates how the deformation length detects
a transition while the shielded CMI remains zero. Establishing bulk
phase equivalence also requires the uniform bounds for bounded local
changes and intermediate profiles stated above.

\section{Logical Channel and Coherent Information}
\label{app:coherent-information}

This appendix derives the syndrome average in
Eq.~\eqref{eq:main-coherent-information-syndrome-average} and explains its
numerical evaluation in Fig.~\ref{fig:coherent-information_honeycomb_surface}.
We first describe the logical operation associated with each syndrome,
then obtain the surviving information from the conditional state of the
reference and logical qubit.  The same description gives a parity condition
for perfect logical recovery and the probability distribution used in the
sampling.  We use the encoded Bell input and notation of
Sec.~\ref{subsec:coherent-information}; all entropies are in bits.

\subsection{Logical operation at fixed syndrome}

Let \(V\) encode the logical qubit, so that
\(V\ket z_{\mathrm L}=\ket{\overline z}\) for \(z=0,1\), and write
\(Z_{\mathrm L}=V^\dagger\overline ZV\).  The barred states belong to the
physical code block, while \(\mathrm L\) denotes the decoded logical qubit.
For each syndrome \(s\), choose a \(Z\)-type Pauli correction \(C_s\) that
returns that sector to the code space.  The rotation, syndrome projection,
correction, and decoding together act on the logical qubit as
\begin{equation}
 \begin{aligned}
 K_s(\theta)=&V^\dagger C_s\Pi_sU(\theta)V,\\
 \sum_s&K_s^\dagger K_s=\mathbb I_{\mathrm L}.
 \end{aligned}
 \label{eq:coherent-logical-kraus}
\end{equation}
Here \(U(\theta)\) and \(\Pi_s\) are those of
Eq.~\eqref{eq:post-syndrome-state}.  We suppress the angle argument of
\(K_s\) below.

Both the errors and corrections are \(Z\) type, so \(K_s\) is diagonal in
the logical-\(Z\) basis:
\begin{equation}
 \begin{aligned}
 K_s&=a_s\mathbb I_{\mathrm L}+b_sZ_{\mathrm L}\\
 &=k_{s,0}\ket0_{\mathrm L}\!\bra0_{\mathrm L}
   +k_{s,1}\ket1_{\mathrm L}\!\bra1_{\mathrm L},
 \end{aligned}
 \label{eq:coherent-kraus-diagonal}
\end{equation}
where \(k_{s,z}=a_s+(-1)^zb_s\).  Thus \(k_{s,0}\) and \(k_{s,1}\) are
the amplitudes for syndrome \(s\) from the two encoded basis states.
The coefficients \(a_s,b_s\) collect errors in the two logical classes;
their explicit expressions are given with the parity argument below.

Different syndrome sectors are orthogonal in the physical output.
We can therefore represent the full output by a register \(S\) containing
\(s\) and the decoded qubit \(\mathrm L\):
\begin{equation}
 \widehat{\mathcal N}_{\theta}(\rho_{\mathrm L})
 =\sum_s |s\rangle\!\langle s|_S
 \otimes K_s\rho_{\mathrm L}K_s^\dagger.
 \label{eq:coherent-flagged-logical-channel}
\end{equation}
This is an isometric change of representation of the full block \(Q\),
so it preserves both \(S(Q)\) and \(S(RQ)\).  The register \(S\) holds
the syndrome information already present in the physical output.

\subsection{Coherent information from the conditional state}

The encoded Bell input becomes
\(\ket\Phi_{R\mathrm L}=(\ket0_R\ket0_{\mathrm L}
+\ket1_R\ket1_{\mathrm L})/\sqrt2\) after decoding.  For a fixed
syndrome \(s\), its unnormalized joint state is
\begin{equation}
 \begin{aligned}
 |\chi_s\rangle_{R\mathrm L}
 &=(\mathbb I_R\otimes K_s)|\Phi\rangle_{R\mathrm L}\\
 &=\frac{k_{s,0}\ket0_R\ket0_{\mathrm L}
        +k_{s,1}\ket1_R\ket1_{\mathrm L}}{\sqrt2}.
 \end{aligned}
 \label{eq:coherent-conditional-bell-state}
\end{equation}
Its squared norm is the syndrome probability,
\begin{equation}
 \begin{aligned}
 p_s&=\langle\chi_s|\chi_s\rangle
 =\frac12\Tr(K_s^\dagger K_s)\\
 &=\frac{|k_{s,0}|^2+|k_{s,1}|^2}{2}.
 \end{aligned}
 \label{eq:coherent-syndrome-probability}
\end{equation}
For \(p_s>0\), the normalized state is
\(\ket{\chi_s}/\sqrt{p_s}\).  This makes the conditional purity used in
Sec.~\ref{subsec:coherent-information} explicit: each syndrome selects a
single vector.  Averaging over syndromes generally gives a mixed state.

The logical qubit alone has the reduced density matrix
\begin{equation}
 \begin{aligned}
 \rho_{\mathrm L|s}&=\frac{K_sK_s^\dagger}{2p_s}\\
 &=q_s\ket0_{\mathrm L}\!\bra0_{\mathrm L}
 +(1-q_s)\ket1_{\mathrm L}\!\bra1_{\mathrm L},
 \end{aligned}
 \label{eq:coherent-conditional-logical-state}
\end{equation}
where
\begin{equation}
 q_s=\frac{|k_{s,0}|^2}{|k_{s,0}|^2+|k_{s,1}|^2}.
 \label{eq:coherent-schmidt-weight}
\end{equation}
As in the main text, \(q_s\) and \(1-q_s\) are the probabilities of the
two logical-\(Z\) measurement outcomes.  They are also the eigenvalues of
\(\rho_{\mathrm L|s}\), whose entropy \(h_2(q_s)\) is the entanglement
with the reference in this conditional pure state.  Outcomes with
\(p_s=0\) are omitted.

The orthogonal syndrome sectors contribute the same classical entropy
\(H(p)=-\sum_s p_s\log_2p_s\) to both terms of the coherent information:
\[
 \begin{aligned}
 S(Q)_\omega&=H(p)+\sum_s p_s h_2(q_s),\\
 S(RQ)_\omega&=H(p).
 \end{aligned}
\]
Subtracting gives
\begin{equation}
 I_c=\sum_s p_s h_2(q_s),
 \label{eq:coherent-information-syndrome-average}
\end{equation}
with \(h_2(q)=-q\log_2q-(1-q)\log_2(1-q)\), as in
Sec.~\ref{subsec:coherent-information}.
Thus \(0\leq I_c\leq1\).  Full retention, \(I_c=1\), requires
\(q_s=1/2\) for every nonzero-probability outcome.  Equivalently, every
nonzero \(K_s\) is proportional to a unitary that can be inverted once
the syndrome is known.  At the other extreme, \(I_c=0\) when every
conditional reference--logical state is a product state, or equivalently
every nonzero \(K_s\) has rank one.

For an arbitrary logical input \(\rho_{\mathrm L}\), the probability of
syndrome \(s\) is \(\Tr(K_s^\dagger K_s\rho_{\mathrm L})\).  Since
\(K_s\) is diagonal, this depends only on the probabilities of the two
logical-\(Z\) outcomes.  Both the Bell input and \(\ket{\overline+}\)
give probability \(1/2\) to each outcome, so both have syndrome weights
\(p_s\).  Applying \(K_s\) to \(\ket+_{\mathrm L}\) gives the conditional
state in Eq.~\eqref{eq:conditional-logical-pure-state}, with the same
\(q_s\).  This is why its distribution determines the coherent information.
In particular, a uniform Born-weighted distribution of \(q_s\) gives
\(I_c=1/(2\ln2)\), the Haar benchmark in
Eq.~\eqref{eq:haar-coherent-information}.

\subsection{Parity and perfect logical recovery}

The condition for each syndrome to preserve the full logical qubit is
\begin{equation}
 \begin{aligned}
 K_s^\dagger K_s\propto\mathbb I_{\mathrm L}
 &\Longleftrightarrow |k_{s,0}|=|k_{s,1}|\\
 &\Longleftrightarrow \operatorname{Re}(a_s^*b_s)=0.
 \end{aligned}
 \label{eq:coherent-unitary-branch-condition}
\end{equation}
The physical error amplitudes explain when this equality is enforced by
the code.  Write \(Z(E)=\prod_{i\in E}Z_i\), where \(E\) is a set of
qubits.  The rotation on \(N\) physical qubits expands as
\begin{equation}
 \begin{aligned}
 U(\theta)&=\sum_{E\subseteq Q}u_\theta(E)Z(E),\\
 u_\theta(E)&=(-i\sin\theta)^{|E|}(\cos\theta)^{N-|E|}.
 \end{aligned}
 \label{eq:coherent-physical-error-amplitude}
\end{equation}
For a given syndrome, correction leaves either a stabilizer or a
stabilizer times \(\overline Z\).  Let \(\mathcal S_Z\) be the group
generated by the \(Z\)-type checks, and choose \(C_s\) as a product of
\(Z\) operators with positive overall sign.  The two logical amplitudes are
\begin{equation}
 \begin{aligned}
 a_s&=\quad\sum_{\mathclap{E:\,C_sZ(E)\in\mathcal S_Z}}\quad u_\theta(E),\\
 b_s&=\quad\sum_{\mathclap{E:\,C_sZ(E)\in\overline Z\mathcal S_Z}}\quad u_\theta(E).
 \end{aligned}
 \label{eq:coherent-coset-amplitudes}
\end{equation}
These are coherent sums over the two logical error classes.  Changing
\(C_s\) by \(\overline Z\) exchanges \(a_s\) and \(b_s\), but leaves
\(p_s\), \(q_s\), and hence \(I_c\) unchanged.

Suppose every \(Z\)-type stabilizer has even weight and \(\overline Z\)
has odd weight.  Then all errors contributing to \(a_s\) have the same
weight parity as \(C_s\), while those contributing to \(b_s\) have the
opposite parity.  The factor \((-i)^{|E|}\) in
Eq.~\eqref{eq:coherent-physical-error-amplitude} makes one sum real and
the other imaginary, up to a common phase.  Therefore
\(\operatorname{Re}(a_s^*b_s)=0\): each nonzero logical operation is
proportional to a unitary, and \(I_c=1\).  This is the parity mechanism
for the conventional odd-distance square-lattice code
\cite{Bravyi_2018,Behrends_SBPC}.

The honeycomb patch has weight-three \(Z\) checks, so this parity argument
does not enforce unitary logical operations.\footnote{The $X$ and $Z$
labels in Refs.~\cite{yang2026decodingcoherenterrorstoric,
bejan2026projectedlogicalensemblessurface} are exchanged relative to our
convention: the odd-weight checks here are $Z$ type.}
The two logical basis states can then have different syndrome probabilities.
For the Bell input, such a difference gives \(q_s\ne1/2\) and reduces the
conditional entanglement between the reference and logical qubit.
The special Clifford endpoint \(\theta=\pi/4\) is treated
in Appendix~\ref{app:pi4-endpoint}.

\subsection{Sampling the syndrome average}

To evaluate Eq.~\eqref{eq:coherent-information-syndrome-average}, the
Gaussian calculation samples complete strings \(m\) of physical
single-qubit \(X\) outcomes.  For the encoded input \(\ket{\overline z}\),
define
\[
 P_z(m)=|\langle m_X|U(\theta)|\overline z\rangle|^2,
 \qquad z=0,1.
\]
Products of these outcomes along the \(X\) checks determine the syndrome
\(s(m)\).  If there are \(n_X\) independent \(X\) checks on \(N\)
qubits, each syndrome is consistent with \(M=2^{N-n_X}\) such strings.

Fixing the syndrome and logical \(Z\) specifies a unique stabilizer
state.  The \(Z\)-type stabilizers and logical \(Z\) map between all
\(M\) compatible \(X\)-outcome strings without changing their probabilities,
so those strings are equally likely within that sector.  Consequently,
\begin{equation}
 \begin{aligned}
 P_z(m)&=\frac{|k_{s(m),z}|^2}{M},\\
 q_{s(m)}&=\frac{P_0(m)}{P_0(m)+P_1(m)}.
 \end{aligned}
 \label{eq:coherent-refined-outcome-probability}
\end{equation}
Sampling \(m\) from \(P(m)=[P_0(m)+P_1(m)]/2\) gives each syndrome
with probability \(p_s\), because
\(\sum_{m:s(m)=s}P(m)=p_s\).  Averaging \(h_2(q_{s(m)})\) therefore
gives exactly the coherent information in
Eq.~\eqref{eq:main-coherent-information-syndrome-average}.

In practice, we choose \(z=0\) or \(1\) with equal probability and
sample \(m\) using the corresponding Gaussian network.  We then evaluate
the probability of that same \(m\) in both logical sectors.  Their ratio
in Eq.~\eqref{eq:coherent-refined-outcome-probability} gives \(q_s\), and
\(h_2(q_s)\) is the contribution to the average.  Sampling full measurement
outcomes and computing their probabilities can both be performed using
FLO by tracking the covariance matrix, as explained in
Appendix~\ref{app:gaussian-information-transport}.

\section{The \texorpdfstring{$\theta=\pi/4$}{theta = pi/4} Endpoint}
\label{app:pi4-endpoint}
\label{appsec:odd_sre}

At $\theta=\pi/4$, the coherent rotation is a Clifford unitary:
$U(\pi/4)=\prod_i e^{-i\pi Z_i/4}$ takes $X_i$ to $Y_i$ and leaves
$Z_i$ unchanged.  Syndrome dephasing then retains a Pauli operator
if it commutes with every measured check, and removes it otherwise.
The endpoint can therefore be understood directly from its surviving
stabilizers and logical operators.

We consider square honeycomb patches with $D_x=D_y=D$, where $D$
counts unit cells in each direction, as in
Fig.~\ref{fig:lattice_surface_code}.  Thus $D=L$ for the square patches in
the numerical sections; we retain $D$ here to match the endpoint figures.
The stabilizer parity pattern stated below makes even patches retain the
encoded qubit and odd patches retain only its logical-$Z$ populations.
We first summarize the dependence on logical input, then explain this
parity effect and show that odd-patch outputs are SRE, whereas even-patch
outputs with logical coherence are LRE.

\subsection{Logical input and output entanglement}
\label{app:logical-input-entanglement}

\begin{figure}[tbp]
 \centering
 \includegraphics[width=\columnwidth]{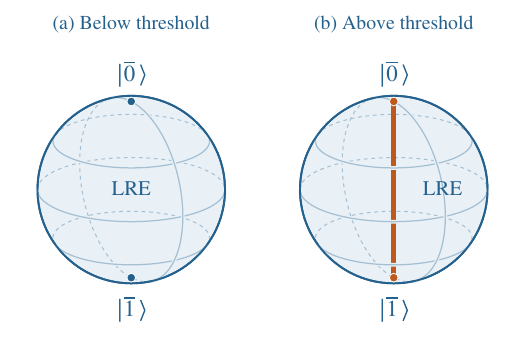}
 \caption{\justifying
 Output entanglement across the input logical Bloch ball on even square honeycomb
 patches. (a) Below threshold: LRE throughout, assuming local recovery.
 (b) Above threshold: positive limiting $I_c$ certifies LRE off the orange
 $Z$ diameter. The diameter is not determined by this criterion and is SRE
 at $\theta=\pi/4$.}
 \label{fig:pi4-logical-bloch-ball}
\end{figure}

To identify which logical inputs are covered by Proposition~
\ref{prop:honeycomb-coherent-lre}, represent the encoded state on the
logical Bloch ball shown in Fig.~\ref{fig:pi4-logical-bloch-ball}.
Write $r_X=\Tr(\rho_Q\overline X)$, and similarly for
$r_Y,r_Z$.  For one qubit, the mixed-state extension of our theorem
certifies LRE whenever $r_X^2+r_Y^2>0$
(Appendix~\ref{app:lre-proof}), which is everywhere off the $Z$ diameter.
The theorem does not say that states on this line are SRE; it simply leaves
their entanglement undetermined.

The two regimes resolve this remaining line differently.  At the Clifford
endpoint $\theta=\pi/4$, the stabilizer analysis
below shows that, on even square honeycomb patches,
every logical-$Z$-diagonal input---precisely the $Z$ diameter---produces an
output diagonal in the physical product $Z$ basis and is therefore SRE.
This occurs despite $I_c=1$, showing that a channel
can retain a full quantum memory while particular encoded inputs give SRE
outputs; on odd patches, $I_c=0$ and every output is SRE.  The all-size parity
pattern used here is stated explicitly below.  Below threshold,
by contrast, Sec.~\ref{subsec:coherent-local-recovery} constructs a local recovery
channel for every fixed $\theta<\theta_c$, arbitrarily close to the threshold.
Because this channel recovers the entire encoded subspace, it establishes LRE
for every logical encoding, including the $Z$ diameter, throughout the
below-threshold phase.

\subsection{Logical information and the parity effect}

All $Z$ checks and the logical $\overline Z$ survive at this angle.
The distinction between even and odd patches follows from the
commutation relations of the rotated $X$ checks with the measured
plaquette and boundary checks.  The required parity properties have been
verified exactly for $D=1,\ldots,24$.  We state their binary rank test below
and assume their continuation when discussing an arbitrary-size
fixed-parity sequence; the consequences for information and entanglement
then follow exactly.

\begin{figure}[htbp]
 \centering
 \includegraphics[width=\columnwidth]{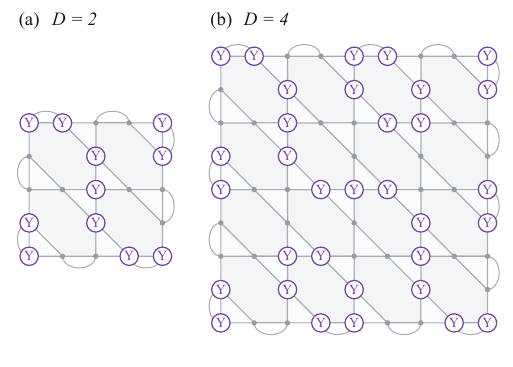}
 \caption{\justifying
 Surviving logical operator $W_X$ at $\theta=\pi/4$ for (a) $D=2$ and
 (b) $D=4$, at the same lattice spacing. Labeled vertices carry $Y$;
 unmarked vertices carry the identity. Thin lines show the lattice and
 boundary checks.}
 \label{fig:thetapi4logical}
 \label{fig:thetapi4}
\end{figure}

For even $D$, no nonidentity product of rotated $X$ checks survives.
Nevertheless, multiplying the input logical $\overline X$ by a suitable
product $S_*$ of $X$ checks gives a surviving logical operator:
\begin{equation}
 \begin{aligned}
 W_X&=U(\pi/4)(S_*\overline X)U(\pi/4)^{\dagger},\\
 W_Y&=iW_X\overline Z .
 \end{aligned}
 \label{eq:pi4-surviving-logicals}
\end{equation}
These operators commute with every measured check.  Together with
$\overline Z$, they obey the logical Pauli algebra and retain the
three components of the input logical Bloch vector.
Figure~\ref{fig:thetapi4logical} shows $W_X$ for $D=2$ and $D=4$.
Thus the full encoded qubit is preserved, and the coherent information
for an encoded Bell input is $I_c=1$.

\begin{figure}[htbp]
 \centering
 \includegraphics[width=0.96\columnwidth]{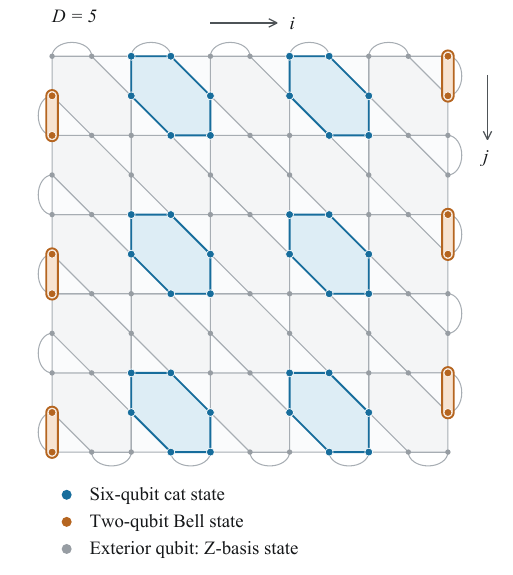}
 \caption{\justifying
 Local blocks in the SRE decomposition for $D=5$. Blue hexagons support
 six-qubit cat states and orange boundary pairs support Bell states;
 gray exterior qubits are fixed in the $Z$ basis.}
 \label{fig:odd_l_Y_stabilizers}
 \label{fig:odd_sre}
\end{figure}

For odd $D$, the boundaries instead allow one nonidentity product of
rotated $X$ checks to survive.  Let $S_0$ be the product of the
unrotated $X$ checks on the disjoint blocks in
Fig.~\ref{fig:odd_sre}.  On the input code space,
$U(\pi/4)^{\dagger}S_0U(\pi/4)$ acts as $\overline Z$.
The product of the corresponding syndrome outcomes therefore measures
the input logical $Z$ exactly.  This removes the logical-$X$ and
logical-$Y$ coherences while preserving the logical-$Z$ populations,
giving $I_c=0$.

For completeness, these properties can be checked without evolving a
density matrix.  Let $H_X$ be a binary matrix whose rows are independent
$X$-check supports, let $x_{\rm L}$ be the binary support of $\overline X$,
and set
\begin{equation}
 G_X=H_XH_X^{\mathsf T},\qquad h_X=H_Xx_{\rm L},
 \label{eq:pi4-binary-overlap-test}
\end{equation}
with arithmetic over $\mathbb F_2$ and column-vector supports.
A rotated product of $X$ checks labeled by a binary vector $a$ survives
syndrome dephasing precisely when $G_Xa=0$.  A rotated representative of
logical $X$ survives precisely when $G_Xa=h_X$ has a solution.
The even-patch pattern is therefore $\ker G_X=\{0\}$; the odd-patch
pattern is
\begin{equation}
 \ker G_X=\operatorname{span}\{a_0\},
 \qquad a_0^{\mathsf T}h_X=1.
 \label{eq:pi4-odd-rank-condition}
\end{equation}
Because $G_X$ is symmetric, the second condition makes the logical-$X$
equation inconsistent.  It also identifies $S_0$ from the support
$H_X^{\mathsf T}a_0$.  These rank and sign identities, including the local
block constraints used below, were verified by exact binary stabilizer
calculations for every square size $D=1,\ldots,24$.  An all-size parity
classification requires these same identities for every $D$; finite-size
verification alone is not an all-size rank proof.  The parity statement
here is specific to the square-patch boundary convention.

\subsection{Even patches: coherence and long-range entanglement}

Write the input logical Bloch vector as $(r_X,r_Y,r_Z)$.
To see why a retained transverse component implies LRE, choose distant
left and right representatives $Z_{\rm left},Z_{\rm right}$ of $\overline Z$.
Their product is a product of $Z$ checks, so every output obeys
$Z_{\rm left}Z_{\rm right}\rho_{\pi/4}=\rho_{\pi/4}$.
Lemma~\ref{lem:app-strong-to-weak} shows that an SRE state with this
perfect correlation must commute with $Z_{\rm left}$.
Since $W_X$ and $W_Y$ anticommute with $Z_{\rm left}$, such a state must have
$r_X=r_Y=0$.  A nonzero transverse component therefore excludes SRE.
If it remains bounded away from zero as $D\to\infty$,
the approximate bound in Eq.~\eqref{eq:app-one-form-approx-weak}
also excludes approximation by SRE mixtures with vanishing error.

Logical-$Z$-diagonal inputs behave differently.  Since no rotated
$X$-check product survives for even $D$, the output for logical-$Z$
input $\ket{\overline z}$, with $z=0,1$, is simply
\begin{equation}
 \begin{aligned}
 \rho_{\pi/4,z}&\propto P_Z(I+(-1)^z\overline Z),\\
 P_Z&=\prod_{p\in\mathcal P_Z}\frac{I+S_p^Z}{2}.
 \end{aligned}
 \label{eq:pi4-even-diagonal-output}
\end{equation}
This is diagonal in the physical product $Z$ basis and hence fully
separable.  The same holds for any mixture of the two logical-$Z$
inputs.

\subsection{Odd patches: short-range entanglement}

The blocks selected by $S_0$ are disjoint hexagons and boundary pairs.
Numbering rows and columns from zero, they comprise the hexagons in
odd columns and even rows, together with the left and right boundary
pairs in even rows.  Their rotated product is the surviving stabilizer
\begin{equation}
 \mathcal O_Y=\prod_a Y(B_a),
 \qquad Y(B_a)=\prod_{i\in B_a}Y_i .
 \label{eq:pi4-odd-surviving-stabilizer}
\end{equation}
Because logical coherence has been removed, it suffices to consider
the two logical-$Z$ inputs.  Their outputs are
\begin{equation}
 \rho_{\pi/4,z}\propto
 P_Z(I+(-1)^z\overline Z)(I+\mathcal O_Y).
 \label{eq:pi4-odd-output}
\end{equation}
Although $\mathcal O_Y$ extends across the patch, its constraint
requires only classical correlations between small entangled blocks.

An explicit SRE decomposition follows in two steps.
First resolve every exterior qubit in the $Z$ basis.
Each $Z$ check meets at most one selected block, so the remaining
constraints act separately within the blocks.  They leave two
computational-basis configurations per block, related by flipping
all its qubits.  Next resolve the $Y$ parity of each block, with
the product of these parities fixed to $+1$ by $\mathcal O_Y$.
Each block is then in a definite superposition of its two
configurations: a six-qubit cat state or a two-qubit Bell state.
Fixing $\overline Z=(-1)^z$ only restricts the allowed classical
assignments; it does not entangle different blocks.

Averaging these pure states over all allowed assignments recovers
Eq.~\eqref{eq:pi4-odd-output}.  Each component is a product of
bounded-size block states and exterior $Z$-basis states, and can
therefore be prepared by a local circuit of depth independent of $D$.
This proves SRE for every odd-patch output with the stabilizer and block
structure specified above.
The block constraints were included in the stabilizer checks above.
We also independently verified the output and its decomposition by
a direct density-matrix calculation on the eight-qubit $D=1$ patch.

These results show that boundary parity affects both logical information and physical
entanglement at this Clifford endpoint.  Endpoint size scaling must
therefore use a fixed parity sequence.  In particular, for an even-patch
logical-$Z$ input, Eq.~\eqref{eq:pi4-even-diagonal-output} gives
$\mathcal F_Z=1$, while measurement in the physical $X$ basis is uniform
and gives $I_X=0$.  The intermediate-angle scaling laws discussed in the
main text must not be extrapolated to this endpoint.

\section{Quantum CMI and \texorpdfstring{$X$}{X}-Basis Dephasing}
\label{app:cmi-invariance}

The CMI calculation in Sec.~\ref{sec:cmi-markov-length} samples
single-qubit $X$ outcomes.  For the fixed logical-$Z$ input and the
boundary tripartition in Fig.~\ref{fig:cmi-geometry}, this classical
quantity determines the quantum CMI through
\begin{equation}
 I(A:C|B)_{\rho_\theta}=1+I_X(A:C|B).
 \label{eq:excess-quantum-cmi}
\end{equation}
Here $\rho_\theta$ is the post-syndrome state and $I_X$ is the
CMI of its product-$X$ measurement outcomes.  All entropies in this
appendix are in bits.  The extra bit comes from the fixed logical
$\overline Z$ constraint.  We first determine the entropy change
under $X$ dephasing, then explain why only this logical contribution
remains in the boundary CMI.  We also show that the offset vanishes
in the local geometry needed for recovery.

\subsection{Entropy change under \texorpdfstring{$X$}{X} dephasing}

Take an encoded input $\ket{\overline z}$ with
$\overline Z\ket{\overline z}=(-1)^z\ket{\overline z}$.
As explained in Appendix~\ref{app:coherent-information}, two
physical $Z$ errors with the same syndrome differ only by
$Z$ stabilizers and, possibly, $\overline Z$.
Both act as scalars on this input.  Consequently, each syndrome
selects a fixed stabilizer state, and the rotation angle changes
only its probability:
\begin{equation}
 \begin{aligned}
 \rho_\theta
 &=\sum_s p_{s,z}(\theta)
   \ket{\phi_{s,z}}\!\bra{\phi_{s,z}},\\
 \ket{\phi_{s,z}}&=Z(E_s)\ket{\overline z}.
 \end{aligned}
 \label{eq:fixed-sector-syndrome-mixture}
\end{equation}
Here $E_s$ is any $Z$-error representative of syndrome $s$, and
$p_{s,z}(\theta)$ is the syndrome probability.  These states share all
$Z$ stabilizers, including $(-1)^z\overline Z$; only the signs of
the $X$ checks vary.

For a region $R$, let $n_X(R)$ and $n_Z(R)$ count the independent
$X$- and $Z$-type stabilizers supported entirely in $R$.
These counts include products of checks, and $n_Z(R)$ includes
the fixed logical generator whenever it has a representative in $R$.
Let $q_R$ be the probability distribution of the eigenvalues of
the independent $X$-type stabilizers in $R$.
For each such eigenvalue assignment, the reduced state is uniform
on a subspace of dimension $2^{|R|-n_X(R)-n_Z(R)}$.
Complete product-$X$ dephasing, denoted $\Delta_{X,R}$, removes the
$Z$ constraints while preserving $q_R$.
Writing $\rho_R=(\rho_\theta)_R$, the two entropies are
\begin{equation}
 \begin{aligned}
 S(\rho_R)
 &=H(q_R)+|R|-n_X(R)-n_Z(R),\\
 S(\Delta_{X,R}(\rho_R))
 &=H(q_R)+|R|-n_X(R).
 \end{aligned}
 \label{eq:regional-syndrome-state-entropy}
\end{equation}
Thus dephasing adds exactly $n_Z(R)$ bits of entropy, independently
of the syndrome probabilities and hence of $\theta$.
This is the regional version of the uniform refinement of syndrome
outcomes described in Appendix~\ref{app:coherent-information}.

\subsection{The remaining logical bit}

Substituting this entropy increase into the four terms of the CMI gives
\begin{equation}
 \begin{aligned}
 I(A:C|B)_{\rho_\theta}-I_X
 &=n_Z(B)+n_Z(ABC)\\
 &\quad-n_Z(AB)-n_Z(BC).
 \end{aligned}
 \label{eq:fixed-sector-cmi-offset}
\end{equation}
The right-hand side counts the $Z$ constraints that remain after
the regional contributions are combined.

In Fig.~\ref{fig:cmi-geometry}, $A$ and $C$ occupy opposite
boundaries and $B$ separates them.  Neither $AB$ nor $BC$ can
support a logical $\overline Z$: a logical $\overline X$ can be
placed on the omitted boundary, where it is disjoint from every
operator in the region.  An operator confined to that region
therefore cannot have the anticommutation required of $\overline Z$.

It remains to count products of local $Z$ checks.
Let $\mathsf G_Z(R)$ denote the common $Z$ stabilizers supported
in $R$.  The groups for $AB$ and $BC$ intersect precisely in
$\mathsf G_Z(B)$.  Together they generate all local $Z$-check
products, since each check lies in at least one of $AB$ and $BC$.
Their combined rank is therefore $n_Z(AB)+n_Z(BC)-n_Z(B)$.
The full region $ABC$ has just one additional independent
generator, the fixed logical $\overline Z$.
The right-hand side of Eq.~\eqref{eq:fixed-sector-cmi-offset}
is consequently one, proving
Eq.~\eqref{eq:excess-quantum-cmi}.

The logical bit is independent of angle and size.
The argument requires the opposite-boundary geometry and a buffer
wide enough that no local check meets both $A$ and $C$.
It also requires the fixed logical-$Z$ input: a generic logical
superposition or a maximally mixed logical input does not have
the common logical stabilizer used above.  For a fixed logical-$Z$ input,
the constant cancels from CMI differences along the angle evolution.

\subsection{The local recovery geometry}
\label{app:cmi-local-geometry}

For recovery, $A$ is a small correctable region, $B$ is a surrounding
buffer, and $C$ is the rest of the code.  The same entropy formulas apply
for spatially varying coherent $Z$ angles: the stabilizer states in
Eq.~\eqref{eq:fixed-sector-syndrome-mixture} stay fixed, while their
probabilities change.  Choose $B$ wide enough that no local check
meets both $A$ and $C$.  Every local $Z$ generator then lies in $AB$ or
$BC$.  Correctability of $A$ also allows the fixed logical $\overline Z$
to be represented outside $A$, hence in $BC$.  Thus
$\mathsf G_Z(AB)$ and $\mathsf G_Z(BC)$ generate the full common
$Z$-stabilizer group, with intersection $\mathsf G_Z(B)$.  The rank
combination in Eq.~\eqref{eq:fixed-sector-cmi-offset} vanishes, giving
\begin{equation}
 I_{\rho_{\boldsymbol\theta,z}}(A:C|B)=I_X(A:C|B).
 \label{eq:local-quantum-classical-cmi}
\end{equation}
There is no logical-bit subtraction in this geometry.  Products of
checks enclosing a hole must be retained when evaluating annular
marginals.  The argument also applies on a sphere, where no logical
generator is needed.  It identifies the quantum CMI relevant to recovery
of the fixed logical-$Z$ state.  The working assumptions concerning its
decay, stability under changing the logical input, and reference-inclusive
recovery are stated in Sec.~\ref{subsec:markov-recovery}.

\subsection{Numerical convention and checks}

Let $m$ be a complete set of physical $X$ outcomes, with Born
distribution $P_\theta(m)$ and marginals $P_R(m_R)$.
The Monte Carlo estimator is
\begin{equation}
 I_X
 =\mathbb E_{m\sim P_\theta}\!
 \left[
 \log_2\frac{P_{ABC}(m)P_B(m_B)}
 {P_{AB}(m_{AB})P_{BC}(m_{BC})}
 \right].
 \label{eq:classical-x-outcome-cmi-estimator}
\end{equation}
Syndrome dephasing does not change these probabilities, because
the $X$-outcome projectors commute with all $X$ checks and the
$Z$ checks remain fixed throughout the rotation.
The Gaussian contraction can therefore sample the rotated encoded
state directly.  The marginal evaluation and statistical errors
are described in Appendix~\ref{app:numerics}.

The tensor network uses the same pure logical-$Z$ input.
Its four corner Majoranas are paired along the two vertical
boundaries through the links
\begin{equation}
 \begin{aligned}
 L_{\mathrm R}
 &=i\gamma_{4,(2D_x,0)}\gamma_{3,(2D_x,2D_y)},\\
 L_{\mathrm L}
 &=i\gamma_{1,(0,2D_y)}\gamma_{2,(0,0)}.
 \end{aligned}
 \label{eq:honeycomb-logical-majorana-links}
\end{equation}
Their fixed parities select the logical-$Z$ sector; the orientation
convention determines its sign.  This boundary pairing is part of
the state preparation, as in the construction of
Appendix~\ref{app:gaussian-information-transport}.

We verified the regional entropy increase and the one-bit CMI
offset by direct density-matrix calculations on the $1\times1$,
$1\times2$, and $2\times1$ unit-cell patches, for both logical-$Z$
eigenstates and six angles spanning $0\leq\theta\leq\pi/4$.
Independent binary stabilizer counts also give the required rank
difference for every square size $D=1,\ldots,24$.

The plotted quantity in Fig.~\ref{fig:cmi-scaling} is therefore
the quantum CMI above its fixed one-bit contribution.
In the localized regime, $I_X$ decays with the Markov length
$\xi_{\mathrm M}$, while the quantum CMI approaches one bit.
The same decay length governs both statements; the constant
logical contribution does not enter the fit.
\section{Punctured Coherent Information}
\label{app:punctured-coherent-information}

Punctured coherent information tests how spatially distributed the surviving
logical information is.  The quantity and its relation to local recovery
were introduced in Ref.~\cite{negari2026criticalnonequilibrium}.  Here we
recall the recovery criterion and explain the numerical calculation for
coherent errors.  The essential step is to combine the syndrome sectors
that become indistinguishable after erasure, before evaluating their entropy.

\subsection{Physical erasure and local recovery}
\label{app:puncture-definition}
\label{app:puncture-stability-proof}

Let $\ket{\Phi}_{RQ}$ be the encoded Bell state of Eq.~\eqref{eq:encoded-bell-input},
with reference $R$.  Coherent rotation followed by syndrome dephasing gives
\begin{equation}
 \begin{aligned}
 \omega_{RQ}&=\sum_s\ket{\psi_s}\!\bra{\psi_s},\\
 \ket{\psi_s}&=(\mathbb I_R\otimes\Pi_sU(\theta))\ket{\Phi}_{RQ}.
 \end{aligned}
 \label{eq:noisy-reference-code-state}
\end{equation}
For an interior puncture $Q_p$, write $\overline Q_p=Q\setminus Q_p$ and
trace out $Q_p$.  No external copy of the erased syndrome information
is retained.  With all entropies in bits,
\begin{align}
 I_c^{(p)}&=S(\overline Q_p)_\omega-S(R\overline Q_p)_\omega,
 \label{eq:punctured-coherent-information}\\
 \Delta_p&=I_c-I_c^{(p)}
 =I(R:Q_p\mid\overline Q_p)_\omega\geq0.
 \label{eq:puncture-gap}
\end{align}
The reference in this CMI distinguishes it from the spatial label CMI
studied in Appendix~\ref{app:cmi-invariance}.

We recall the punctured-recovery criterion of
Ref.~\cite{negari2026criticalnonequilibrium}.  Its target may retain only part of the initial
logical information.  Write $T(\rho,\sigma)=\tfrac12\|\rho-\sigma\|_1$.

\begin{theorem}[Punctured recovery criterion, Ref.~\cite{negari2026criticalnonequilibrium}]
\label{thm:puncture-stability}
Let $\sigma_{RQ}$ be a target supported on the clean code, with
$\sigma_R=\omega_R$, and let a local channel on $Q$ map $\omega_{RQ}$ to
$\sigma_{RQ}$ with trace-distance error $\epsilon$.
If the enlarged puncture
$Q_{\mathrm{LC}}=\mathsf N_{r_L}(Q_p)$ is correctable in the clean code,
then $\sigma_{RQ}$ can be recovered from $\overline Q_p$ with the same
error.  For a qubit reference and $\epsilon\leq1/2$,
\begin{equation}
 \begin{aligned}
 0\leq\Delta_p&\leq\delta_I+2\epsilon
 +(1+\epsilon)h_2\!\Bigl(\frac{\epsilon}{1+\epsilon}\Bigr),\\[2pt]
 \delta_I&=\bigl|I_c(R\rangle Q)_\omega-I_c(R\rangle Q)_\sigma\bigr|.
 \end{aligned}
 \label{eq:punctured-ic-continuity}
\end{equation}
Here $h_2$ is the binary entropy in bits and $\mathsf N_r$ is the
metric $r$-neighborhood.  Thus $\epsilon,\delta_I\to0$ imply
$\Delta_p\to0$.
\end{theorem}

The local-recovery step is Appendix A, Lemma 1 of the cited work.  In
physical terms, erasure can influence the recovered output only inside
$Q_{\mathrm{LC}}$, which clean-code erasure correction can restore.
The displayed bound then follows from data processing and
conditional-entropy continuity~\cite{Winter_2016_continuity}.

\begin{figure}[t]
 \centering
 \includegraphics[width=\columnwidth]{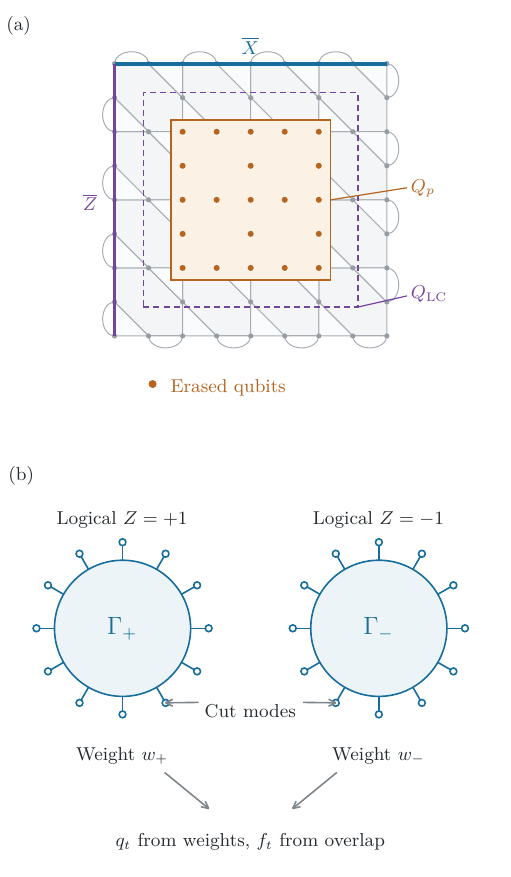}
 \caption{\justifying
 Puncture geometry and Gaussian evaluation.
 (a) Centered puncture $r_p=2$ on a $D_x=D_y=4$ patch. Orange qubits are
 erased; the dashed outline marks the enlarged puncture $Q_{\mathrm{LC}}$.
 (b) Contractions in the two logical-$Z$ sectors with puncture-boundary
 modes left open, used to obtain $q_t$ and $f_t$. Exterior outcomes are
 sampling variables, not a register available to the decoder.}
 \label{fig:puncture-parity-regions}
\end{figure}

For the centered punctures considered here, a light-cone radius $o(L)$
preserves a macroscopic margin from the physical boundary
[Fig.~\ref{fig:puncture-parity-regions}(a)].
Hence the cited lemma applies to the sublinear-radius circuits used here.
Thus a nonzero limiting $\Delta_p$ excludes vanishing-error local recovery
to a code-supported target with the same surviving coherent information.
The target need not be the original Bell state.  Conversely, $\Delta_p\to0$ alone does not
establish the existence of a local decoder.

\subsection{Syndrome information after a puncture}
\label{app:puncture-syndrome-resolved}

On the full code block, the syndrome sectors are orthogonal, giving
$I_c=\sum_s p_s h_2(q_s)$ as in
Appendix~\ref{app:coherent-information}.  A puncture removes some of the
checks that distinguish these sectors.  The surviving label $t$ consists
of the eigenvalues of all products of $X$ checks supported entirely in
$\overline Q_p$.  This includes products whose individual generators
intersect the puncture.

Choose the logical $\overline X$ along the outer boundary, away from
$Q_p$, and choose the $Z$-type Pauli corrections $C_s$ to commute with
this representative.  This fixes the relative logical phases in
$k_{s,z}$.
For a given retained label $t$ and logical label $z=0,1$, define
\begin{equation}
 \begin{aligned}
 P_{t,z}&=\sum_{s\mapsto t}|k_{s,z}|^2,\\
 C_t&=\sum_{s\mapsto t}k_{s,0}k_{s,1}^{*},\\
 p_t&=\frac{P_{t,0}+P_{t,1}}{2}.
 \end{aligned}
 \label{eq:puncture-retained-sector-weights}
\end{equation}
The sums run over the full syndromes compatible with $t$.  Stabilizer
reduction on the retained region separates a common ancillary factor
from the reference and logical qubit.  In the basis
$\{\ket0_R\ket0_{\mathrm L},\ket1_R\ket1_{\mathrm L}\}$, their normalized block is
\begin{equation}
 \sigma_t=\frac{1}{2p_t}
 \begin{pmatrix}
 P_{t,0}&C_t\\[2pt] C_t^*&P_{t,1}
 \end{pmatrix}.
 \label{eq:puncture-retained-logical-block}
\end{equation}
The ancillary entropy and the Shannon entropy of $p_t$ cancel in
$S(\overline Q_p)-S(R\overline Q_p)$.  Only the diagonal entropy of
$\sigma_t$ and its two eigenvalues are needed.  Write
$q_t=P_{t,0}/(P_{t,0}+P_{t,1})$ and
$f_t=|C_t|/\sqrt{P_{t,0}P_{t,1}}$.  Then
\begin{equation}
 \begin{aligned}
 I_c^{(p)}&=\sum_t p_t\!
 \left[h_2(q_t)-h_2\!\left(\frac{1+r_t}{2}\right)\right],\\
 r_t&=\sqrt{(2q_t-1)^2+4q_t(1-q_t)f_t^2}.
 \end{aligned}
 \label{eq:puncture-block-estimator}
\end{equation}
Zero-weight sectors are omitted.  If either diagonal weight vanishes,
the sector contributes zero.

This expression has a simple interpretation.  The weight $q_t$ describes
how well the surviving syndrome distinguishes the two logical $Z$
sectors, while $f_t$ measures their remaining coherence.  For $f_t=1$
the reference--logical block is pure and contributes $h_2(q_t)$; for
$f_t=0$ it retains only classical correlations and contributes no coherent
information.

Keeping the full syndrome in an external register would give a different
answer.  Each branch is a Pauli transform of a state in the original
code space, so a clean-code correctable puncture preserves its coherent
information.  The externally flagged quantity therefore equals $I_c$
at every angle.  Physical erasure instead requires
Eq.~\eqref{eq:puncture-block-estimator}, with the lost syndrome labels
summed before the entropy is taken.

\subsection{Gaussian evaluation on the puncture boundary}
\label{app:puncture-majorana-boundary}

Enumerating the sums in Eq.~\eqref{eq:puncture-retained-sector-weights}
would be exponential.  Instead, we use the physical $X$-outcome
sampling of Appendix~\ref{app:coherent-information} and the Majorana
representation of Appendix~\ref{app:gaussian-information-transport}.
The exterior outcomes $\mathbf m_{\overline Q_p}$ uniformly refine
the retained label $t$, just as the complete outcome string refines
the full syndrome.  Thus they can be used to sample $p_t$ without
enumerating the erased syndrome information.

For each sampled exterior string, contract its physical tensors in both
logical $Z$ sectors, leaving the modes crossing the puncture boundary
uncontracted [Fig.~\ref{fig:puncture-parity-regions}(b)].  Each contraction
gives a squared norm $w_\pm$ and a normalized pure Gaussian boundary state
$\ket{\chi_\pm}$.  The common normalization cancels, so that
$q_t=w_+/(w_++w_-)$, while
$f_t=|\langle\chi_-|\chi_+\rangle|$.  The latter is the overlap of the
unobserved degrees of freedom associated with the two logical sectors:
their distinguishability reduces the retained logical coherence.
Only exterior outcomes enter these contractions; interior outcomes
are discarded.

Let $2n_{\partial}$ be the number of cut Majorana modes, so that
$n_{\partial}$ is the number of complex fermionic boundary modes.
Use the same mode ordering in both contractions, with covariance
$\Gamma_{\pm,ab}=\tfrac{i}{2}
\langle\chi_\pm|[\gamma_a,\gamma_b]|\chi_\pm\rangle$.
The connected puncture has a single total-parity constraint,
\begin{equation}
 \begin{aligned}
 \Pi_{\partial p}&=i^{n_{\partial}}\gamma_1\cdots\gamma_{2n_{\partial}},\\
 P_\eta&=\frac{\mathbb I+\eta\Pi_{\partial p}}{2},
 \qquad\eta=\pm1.
 \end{aligned}
 \label{eq:puncture-component-parity-tensor}
\end{equation}
In this ordering and covariance convention, define the independent parity
signs $\eta_\pm=\langle\chi_\pm|\Pi_{\partial p}|\chi_\pm\rangle
=\operatorname{Pf}(\Gamma_\pm)\in\{+1,-1\}$.
The subscripts $\pm$ label the logical $Z$ sectors, not the parity signs.
Opposite parities, $\eta_+\neq\eta_-$, give $f_t=0$.  For matching parities, the standard
Gaussian overlap formula~\cite{Bravyi_2005_FLO} gives
\begin{equation}
 f_t=2^{-n_{\partial}/2}
 \det(\mathbb I-\Gamma_+\Gamma_-)^{1/4}.
 \label{eq:puncture-gaussian-overlap}
\end{equation}
Here $f_t$ is the overlap amplitude, not its square.

The implementation accumulates contraction weights and determinants in
logarithmic form, checks the common mode ordering, and retains the
Pfaffian sign.  These checks matter near zero-probability outcomes and
orthogonal boundary states.  The calculation uses covariance matrices
on the active boundary and requires polynomial resources.  We restrict
it to one connected, correctable puncture; disconnected erasures require
separate parity constraints for their components.

\subsection{Sampling, checks, and finite-size analysis}
\label{app:puncture-numerical-checks}

One realization begins by choosing either logical $Z$ sector with
probability $1/2$ and sampling a complete $X$-outcome string with its
Born probability.  Replaying that string in both sectors gives the
full-information contribution $h_2(q_s)$.  Replaying only its exterior
gives $q_t$ and $f_t$, and hence the punctured contribution in
Eq.~\eqref{eq:puncture-block-estimator}.  Averaging these paired
contributions estimates $I_c$ and $I_c^{(p)}$.  The gap is averaged
sample by sample, so its standard error includes their covariance.
Although the average gap is nonnegative, individual differences
need not be.

We checked the Gaussian estimator against exact density-matrix
calculations on small patches.  The exact calculation starts from the
encoded Bell state, applies the rotation and syndrome dephasing, and
traces the puncture before taking either entropy.  Exact enumeration
of the Gaussian estimator agrees within $2\times10^{-12}$ bits on both
13-qubit orientations, using three connected one- and two-qubit punctures at
$\theta/\pi=0,0.04,0.09,0.12,0.25$.
These checks include the even-patch Clifford endpoint, where the full
information is one bit but the tested puncture reduces $I_c^{(p)}$ to
zero.  This distinguishes physical erasure from the externally flagged
calculation.  On the 21-qubit square patch with a five-qubit central
puncture, we also compared 750-sample estimates with exact diagonalization
at the same five angles and found agreement within sampling uncertainty.

For Fig.~\ref{fig:punctured-coherent-information}, the code coordinates
span $0\leq x,y\leq2L$, and the puncture contains sites with
$|x-L|\leq r_p$ and $|y-L|\leq r_p$, with $r_p=L/2$.
Thus $r_p$ is the puncture half-width and $r_p/L$ is its fraction of
the full side length.
The scan uses $L=8,16,\ldots,72$, 17 angles in
$0.06\leq\theta/\pi\leq0.15$, and 8000--8560 independent samples per
point.  The stored natural-log estimates are divided by $\ln2$ to obtain
information quantities in bits.
The inset fits the scaling variable
$(\theta/\pi-\theta_c/\pi)L^{1/\nu}$ using
$16\leq L\leq72$ and $0.085\leq\theta/\pi\leq0.097$.
The fit minimizes the variance between interpolated size curves,
weighted by their statistical uncertainties.  The quoted uncertainties
combine, in quadrature, the spread of 300 Gaussian bootstrap fits
and the spread over 36 size and angle windows.
This finite-size collapse supports the transition discussed in the
main text; it does not establish the asymptotic critical value of
$I_c^{(p)}$.
\section{One-Form Fidelity Correlators}
\label{app:one-form-swssb}

An open \(Z\) string inserts two electric charges into the noisy code
state.  Its fidelity correlator asks how well the state with this pair
can be distinguished from the state without it.  A complementary strong
one-form symmetry supplies a loop measurement that distinguishes the two
states, forcing the fidelity to be small.  We explain this connection
first, then give the estimator used in Fig.~\ref{fig:swssb-scaling}.  For the logical input used
there, the calculation reduces to comparing two Born probabilities.

\subsection{Fidelity and one-form charges}

We use the Uhlmann fidelity throughout:
\begin{equation}
 \begin{aligned}
 F(\rho,\sigma)&=\Tr\sqrt{\sqrt\rho\,\sigma\sqrt\rho},\\
 \mathcal F_Z(x,y)
 &=F(\rho_\theta,Z_{\gamma_{xy}}\rho_\theta Z_{\gamma_{xy}})\\
 &=\bigl\|\sqrt{\rho_\theta}\,Z_{\gamma_{xy}}\sqrt{\rho_\theta}\bigr\|_1 .
 \end{aligned}
 \label{eq:swssb-fidelity-string}
\end{equation}
The path \(\gamma_{xy}\) joins distinct bulk endpoints.  Deforming it by
a contractible closed \(Z\) loop leaves the inserted state unchanged,
because that loop is an exact stabilizer of \(\rho_\theta\).  Thus the
correlator measures the endpoint pair, rather than the length of a
particular path.  On a manifold with nontrivial topology, this statement
applies within a fixed homology class.

A closed-loop operator \(W(C)\) is a strong symmetry when its charge is
fixed on the support of the density matrix; a weak symmetry only leaves
the density matrix invariant under conjugation
~\cite{lessa_strong--weak_2024,ZhangXuZhangXuBiLuo2025}:
\begin{equation}
 \begin{alignedat}{2}
 \text{strong:}&\quad &W(C)\rho &=e^{i\varphi(C)}\rho,\\
 \text{weak:}&\quad &W(C)\rho W(C)^\dagger &=\rho.
 \end{alignedat}
 \label{eq:swssb-strong-symmetry}
\end{equation}
The coherent \(Z\) rotation and syndrome dephasing preserve the strong
contractible \(Z\) loops.  The microscopic \(X\) loops are weak
symmetries after dephasing, but their charges can fluctuate.  The
low-error phase is expected to recover a strong \(X\)-type partner
through dressed loops.  Together with the \(Z\) symmetry, this partner
has the nontrivial crossing algebra of the toric code: an \(X\) loop
enclosing one electric charge changes sign.

Local recovery gives a natural origin for dressed symmetries.  If a
channel recovers the clean state exactly, let \(U\) be its local unitary
dilation with product ancillas.  A reference string \(W^{(0)}(\gamma)\)
pulls back to
\begin{equation}
 \widetilde W(\gamma)
 =U^\dagger[W^{(0)}(\gamma)\otimes\mathbb I_a]U .
 \label{eq:swssb-symmetry-pullback}
\end{equation}
The pulled-back contractible loops are strong symmetries of the state
with these ancillas included.  Their supports broaden by the recovery
light cone, and conjugation preserves the crossing algebra of the two
reference string families
~\cite{lessa2025higherformanomalylongrangeentanglement}.
This does not, by itself, fix their crossing algebra with a
\emph{bare} \(Z\) string.  We state that additional condition explicitly
in the exponential bound below.

The ordinary expectation \(\Tr(\rho_\theta Z_{\gamma_{xy}})\) vanishes:
the open string connects different syndrome sectors.  The fidelity can
nevertheless be nonzero because it compares the weights and states in
those sectors.  For a pure state it reduces to the modulus of the
ordinary expectation, so this distinction is specific to mixed states.
A nonzero limiting fidelity, together with a vanishing ordinary
correlator, gives the one-form strong-to-weak symmetry-breaking pattern
~\cite{lessa_strong--weak_2024,ZhangXuZhangXuBiLuo2025}.
The algebraic or inverse-logarithmic behavior considered here instead
has a vanishing limit.  We refer to it as a quasi-long-range counterpart
of this order, as in Sec.~\ref{subsec:one-form-fidelity}.
The diagnostic is the Uhlmann fidelity in
Eq.~\eqref{eq:swssb-fidelity-string}; replacing it by a normalized
R\'enyi-2 overlap need not give the same transition
~\cite{lessa_strong--weak_2024}.

\subsection{Why a strong complementary symmetry suppresses fidelity}
\label{app:emergent-symmetry-fidelity-bound}

We show here why an emergent $X$-type one-form symmetry, complementary to
the preserved $Z$-type symmetry, would force the open-string fidelity to
decay exponentially. The emergent symmetry must be strong and
exponentially localized, and its loops must retain the crossing algebra
with open $Z$ strings. A loop surrounding one endpoint has a definite
charge in the original state, which the open string reverses. Measuring
this charge therefore distinguishes the original state from the state
with the string inserted. Exponential localization makes the errors in
this distinction exponentially small as the endpoints separate, bounding
their fidelity accordingly. This is the basis for using slow fidelity
decay in the main text to rule out such an emergent symmetry pair.

To make this argument precise, apply any channel
\(\mathcal E\), followed by a two-outcome measurement \(Q_\pm\), to
states \(\rho,\sigma\).  If the two wrong-charge probabilities are
\(a=\Tr[Q_-\mathcal E(\rho)]\) and
\(b=\Tr[Q_+\mathcal E(\sigma)]\), monotonicity of fidelity gives
\begin{equation}
 \begin{aligned}
 F(\rho,\sigma)
 &\leq\sqrt{(1-a)b}+\sqrt{a(1-b)}\\
 &\leq\sqrt a+\sqrt b .
 \end{aligned}
 \label{eq:swssb-charge-measurement}
\end{equation}
This inequality also underlies the recovery argument in
Appendix~\ref{app:phase-equivalence}.

To obtain exponential decay, choose a dressed \(X\) loop surrounding
\(x\) but not \(y\).  Let its truncation to width \(R\) be a Hermitian
unitary \(\widetilde X_R\), with projectors
\(P_\pm=(\mathbb I\pm\widetilde X_R)/2\).  Product ancillas may be
included, so write
\(\widehat\rho=\rho_\theta\otimes|\mathbf0\rangle\langle\mathbf0|_a\).
The required localization and crossing assumptions are
\begin{equation}
 \begin{aligned}
 \Tr(P_-\widehat\rho)&\leq\epsilon_s(R),\\
 \bigl\|Z_\gamma^\dagger P_+Z_\gamma-P_-\bigr\|_\infty
 &\leq\epsilon_a(R),\\
 \epsilon_s(R)+\epsilon_a(R)&\leq c e^{-R/\xi_X}.
 \end{aligned}
 \label{eq:swssb-approximate-anomaly}
\end{equation}
Here \(Z_\gamma\) acts trivially on the ancillas.  The first error
measures uncertainty in the strong charge; the second measures failure
of the open string to reverse it.  The constants are uniform for the
loops and truncations under consideration.

\begin{proposition}
\label{prop:emergent-symmetry-fidelity-bound}
Under Eq.~\eqref{eq:swssb-approximate-anomaly}, for bulk endpoints at
separation \(r\),
\begin{equation}
 \mathcal F_Z(x,y)\leq C e^{-r/\xi_F},
 \qquad \xi_F=O(\xi_X).
 \label{eq:swssb-exponential-bound}
\end{equation}
\end{proposition}
\begin{proof}
The two wrong-charge probabilities for \(\widehat\rho\) and
\(Z_\gamma\widehat\rho Z_\gamma\) are at most \(\epsilon_s\) and
\(\epsilon_s+\epsilon_a\).  Equation~\eqref{eq:swssb-charge-measurement}
therefore bounds the fidelity by \(2\sqrt c\,e^{-R/(2\xi_X)}\).
Choose \(R\) to be a fixed fraction of \(r\), so the thickened loop
still separates the endpoints.  Adding the product ancillas leaves
fidelity unchanged.
\end{proof}

In the exact limit the two states have orthogonal loop charges and
zero fidelity.  Exponential localization allows only exponentially
small overlap.  Consequently, persistent slower decay rules out a
strong complementary symmetry satisfying the stated crossing condition.
To exclude recovery without assuming this condition, one must show
that locality itself supplies a suitable charge measurement.  That is
the purpose of Appendix~\ref{app:phase-equivalence}.

\subsection{Reduction for the logical input used in the numerics}

The simulations use a fixed logical \(\overline Z=+1\) code state
\(|\overline0\rangle\).  This choice greatly simplifies the fidelity.
After the coherent rotation and syndrome measurement, write
\(\rho_\theta=\sum_s|\phi_s\rangle\langle\phi_s|\), where
\(|\phi_s\rangle=\Pi_sU(\theta)|\overline0\rangle\) and
\(p_{s,0}=\langle\phi_s|\phi_s\rangle\).
Every \(Z\)-error string with syndrome \(s\) differs from a chosen
representative \(Z(E_s)\) by \(Z\) stabilizers and logical \(Z\)
operators.  These act as scalars on the input.  Hence
\(|\phi_s\rangle=c_sZ(E_s)|\overline0\rangle\):
coherent interference changes the amplitude \(c_s\), but not the
normalized state within that syndrome sector.

Let \(\bar s=s\oplus\delta_\gamma\), where \(\delta_\gamma\) flips the
two endpoint checks.  The string \(Z_\gamma\) maps the normalized
\(\bar s\) branch to the \(s\) branch up to a phase.  Orthogonality of
syndrome sectors then gives the exact expression
\begin{equation}
 \mathcal F_Z(\gamma)=\sum_s\sqrt{p_{s,0}p_{\bar s,0}}.
 \label{eq:swssb-syndrome-fidelity}
\end{equation}
The same simplification holds for the unique ground state on the
sphere.  It does not remove coherent interference: that interference
is contained in the probabilities.

The Gaussian sampler resolves the syndrome further into a complete
physical \(X\)-measurement record \(\mathbf m\), with probability
\(P_\theta(\mathbf m)\).  Inserting \(Z_\gamma\) flips the outcomes
on the string support; denote this involution by \(T_\gamma\).
Writing \(P_\theta(\mathbf m)=p_{s,0}R_s(\mathbf m)\), the branch
relation above implies
\(R_{\bar s}(T_\gamma\mathbf m)=R_s(\mathbf m)\).
Summing over the records within each syndrome gives
\begin{equation}
 \mathcal F_Z(\gamma)
 =\sum_{\mathbf m}
 \sqrt{P_\theta(\mathbf m)
       P_\theta(T_\gamma\mathbf m)} .
 \label{eq:swssb-classical-fidelity}
\end{equation}
Thus full \(X\) dephasing preserves this particular fidelity.  There is
no logical-sector offset and no conditional Gaussian overlap to
evaluate.

This reduction depends on the input state.  For an arbitrary pure
logical input, each syndrome block is still rank one.  Writing its
normalized state as \(|\psi_s\rangle\), the corresponding term in
Eq.~\eqref{eq:swssb-syndrome-fidelity} uses the probabilities of that input
and acquires the factor \(|\langle\psi_s|Z_\gamma|\psi_{\bar s}\rangle|\).
For a mixed input, with unnormalized blocks \(\rho_s\), the general
formula is
\begin{equation}
 \mathcal F_Z(\gamma)=
 \sum_s\bigl\|\sqrt{\rho_s}\,Z_\gamma\sqrt{\rho_{\bar s}}\bigr\|_1 .
 \label{eq:swssb-mixed-block-formula}
\end{equation}
These are fidelities of the physical output state.  If syndrome labels
are copied into an auxiliary flag as a representation of its block
structure, the insertion must shift that flag as well.

\subsection{Sampling, checks, and finite-size analysis}

We evaluate Eq.~\eqref{eq:swssb-classical-fidelity} with the sequential
fermionic linear-optics sampler.  Each independent draw produces
\(\mathbf m\sim P_\theta\) and its log probability.  Replaying the
contraction with the prescribed record \(T_\gamma\mathbf m\)
gives the second log probability.  The bounded estimator is
\begin{equation}
 \begin{aligned}
 \mathcal F_Z(\gamma)&=\mathbb E_{\mathbf m\sim P_\theta}
 [w_\gamma(\mathbf m)],\\
 w_\gamma(\mathbf m)
 &=\frac{2\sqrt{P_\theta(\mathbf m)
                    P_\theta(T_\gamma\mathbf m)}}
           {P_\theta(\mathbf m)+P_\theta(T_\gamma\mathbf m)}\\
 &=\operatorname{sech}\!\left[
 \frac{\ln P_\theta(\mathbf m)
       -\ln P_\theta(T_\gamma\mathbf m)}{2}\right].
 \end{aligned}
 \label{eq:swssb-bounded-estimator}
\end{equation}
To see why ordinary Born sampling suffices, note that
\(w_\gamma(T_\gamma\mathbf m)=w_\gamma(\mathbf m)\).
Its expectation under \(P_\theta\) therefore equals that under
\([P_\theta+P_\theta\circ T_\gamma]/2\), which immediately gives
Eq.~\eqref{eq:swssb-classical-fidelity}.
Every contribution lies in \([0,1]\); an outcome with a forbidden
partner contributes zero.

The implementation accumulates base-two log probabilities, so the
argument of \(\operatorname{sech}\) includes the factor \(\ln2/2\).
The Gaussian covariance is periodically restored to its pure-state
form to control roundoff during long contractions.  Standard errors
come from the sample variance of the bounded weights.  No sampling of
conditional logical overlaps is required.

We checked this calculation by exact enumeration on the 8-qubit patch
and both orientations of the 13-qubit patch, at
\(\theta/\pi=0,0.04,0.09,0.12,0.25\).
For two string supports per patch and angle, the singular-value
expression for the physical density-matrix fidelity, the complete
\(X\)-outcome sum, and the Gaussian estimator agree within
\(9\times10^{-14}\).  Contractible path deformations give the same
answer.  These checks include the zero-error limit and the Clifford
point, without replacing coherent noise by stochastic Pauli errors.

Figure~\ref{fig:swssb-scaling} uses \(r=L/2\),
\(L=8,16,\ldots,72\), and at least \(32000\) independent samples per
point.  The critical analysis fits
\begin{equation}
 \mathcal F_Z(L/2,\theta)
 =L^{-p}\Phi\!\left[
 (\theta/\pi-\theta_c/\pi)L^{1/\nu}\right].
 \label{eq:swssb-finite-size-scaling}
\end{equation}
The central estimate of \(p\) comes from a power-law fit at
\(\theta/\pi=0.09\), using \(L\geq16\); it is then held fixed in
the collapse over \(0.087\leq\theta/\pi\leq0.094\) and \(L\geq16\).
The reported uncertainties combine 80 parametric bootstrap fits,
which refit both stages, with variations of the angle and size
windows.  The two contributions are added in quadrature.
The below-threshold exponential fits use \(L\geq32\); the
above-threshold log--log curves are shown without imposing a fit.

\section{Phase Equivalence: Recovery and Fidelity Bounds}
\label{app:phase-equivalence}

This appendix proves the fidelity obstruction to local recovery,
Theorem~\ref{thm:fidelity-obstruction-local-recovery}. We first state its
technical form, Theorem~\ref{thm:fidelity-formal}. We consider a surface code on a two-dimensional lattice of bounded density, in any geometry, with checks acting within balls of diameter $r_0$. $\mathsf N_r(X)$ denotes the metric $r$-neighbourhood of a region $X$. Let $\ket{\psi}$ be a pure code state, with $\psi=\ket\psi\!\bra\psi$, and $\mathsf{B}$ be a correctable disk  in the bulk and let $\mathcal N$ and $\mathcal R$ be local channels in the sense of Definition~\ref{def:phase-equivalence}. Their composition has total light-cone radius $s$, and we set $\ell_0=s+r_0$.
Finally, $Z_\gamma$ is an open $Z$ string  along a path $\gamma$ with endpoints $x,y$, and $\psi_\gamma=Z_\gamma\psi Z_\gamma$.

\begin{theorem}[Fidelity obstruction to local recovery]
\label{thm:fidelity-formal}
In the setting above, suppose $\big\|\mathcal R\big(\mathcal N(\psi)\big)-\psi\big\|_1\le\epsilon$, $
|x-y|\ge 12\,\ell_0$ and $\mathsf N_{10\ell_0}(\gamma)\subseteq \mathsf{B}.$
Then
\begin{equation}
  F\!\left(\mathcal N(\psi),\,\mathcal N(\psi_\gamma)\right)
  \le\min\!\bigl\{1,\;8\sqrt{M}\,\epsilon^{1/4}\bigr\},
  \label{eq:fid-bound-formal}
\end{equation}
where $M$ is the number of checks plus the number of lattice sites. No
symmetry or covariance is assumed of $\mathcal N$ or $\mathcal R$, and
$\mathcal R$ need not recover $\psi_\gamma$. If $\mathcal N$ is covariant
under open $Z$ strings, Eq.~\eqref{eq:covariance}, the left-hand side
equals the fidelity correlator $\mathcal F_Z(x,y)$ of Eq.~\eqref{eq:FZ}, by
Eq.~\eqref{eq:FZ-cov}.
\end{theorem}

\noindent\textbf{Relation to the informal statement.}
For a family of sizes $L$ with $s_L=o(L)$ and $M_L=O(L^2)$, take endpoints
$\Theta(L)$ apart and $\Theta(L)$ from the boundary of a correctable bulk
region. The hypotheses then hold for all sufficiently large $L$. A recovery
error $\epsilon_L=L^{-\infty}$ makes the right-hand side of
Eq.~\eqref{eq:fid-bound-formal} equal to $L^{-\infty}$. This is
Eq.~\eqref{eq:fid-bound} of
Theorem~\ref{thm:fidelity-obstruction-local-recovery}.

The proof is organized as follows.
Sec.~\ref{app:phase-equivalence-fidelity-obstruction} uses fidelity
monotonicity to reduce the bound to finding one loop measurement that
distinguishes the two recovered states.
Sec.~\ref{app:sphere-exact} treats exact recovery. It pulls clean output
strings back through a unitary dilation of the channel and shows that each
one acts on the input as a canonical anyon string, dressed only near its
endpoints (Lemma~\ref{lem:sphere-exact-charge}). The induced map on anyon
labels is a permutation, and this yields the distinguishing loop.
Sec.~\ref{app:sphere-approximate} repeats each step with controlled errors
and gives Eq.~\eqref{eq:fid-bound-formal}. The argument uses the geometry
only through the correctable disk \(\mathsf B\) and the purity of
\(\psi\), so it applies on spheres, tori and planar patches alike.

\subsection{Fidelity and local recovery}
\label{app:phase-equivalence-fidelity-obstruction}

The Uhlmann fidelity
\begin{equation}
 F(\omega,\omega')=\|\sqrt\omega\sqrt{\omega'}\|_1
\end{equation}
cannot decrease under a quantum channel \(\mathcal E\):
\begin{equation}
 F(\omega,\omega')
 \leq F\bigl(\mathcal E(\omega),\mathcal E(\omega')\bigr).
 \label{eq:sphere-monotonicity}
\end{equation}
We apply this inequality to a putative recovery channel
\(\mathcal R\), followed by a loop measurement that distinguishes
the two recovered states.

Write the noisy states (under a noise channel $\mathcal{N}$ strongly symmetric under $Z_\gamma$ strings),
\begin{equation}
 \begin{aligned}
 \rho&=\mathcal N(\psi)\\
 \rho_\gamma&=Z_\gamma\rho Z_\gamma
 =\mathcal N(\psi_\gamma)
 \end{aligned}
\end{equation}

Write \(\mathcal S=\mathcal R\circ\mathcal N\), and denote its
two outputs by
\begin{equation}
 \sigma_0=\mathcal S(\psi),
 \qquad \sigma_\gamma=\mathcal S(\psi_\gamma).
\end{equation}
Monotonicity under \(\mathcal R\) gives
\begin{equation}
 \mathcal F_Z(x,y)
 =F(\rho,\rho_\gamma)
 \leq F(\sigma_0,\sigma_\gamma).
 \label{eq:sphere-recovery-comparison}
\end{equation}
Recovery fixes \(\sigma_0\) to the clean state, exactly or
approximately.  The proof establishes that a loop on the recovered
system has the opposite charge in \(\sigma_\gamma\).  This
distinguishes the outputs without requiring recovery of the
string-inserted input itself.

Both channels admit local unitary dilations with product input
ancillas.  Their composition has light-cone radius \(s=o(L)\),
including any classical communication.  The lattice has
bounded-degree, bounded-range checks and \(O(L^2)\) sites.  All
paths and endpoint disks below are chosen with separations much
larger than \(s\); we suppress the size index \(L\) until the
final bound.

\subsection{Exact recovery}
\label{app:sphere-exact}

Assume first that \(\sigma_0=\psi\).  We will find a Hermitian
unitary loop \(W_C\) on the output system such that
\begin{equation}
 \Tr(W_C\sigma_0)=+1,
 \qquad \Tr(W_C\sigma_\gamma)=-1.
 \label{eq:sphere-measurement-goal}
\end{equation}
Its two outcomes distinguish the outputs perfectly.  Equation
~\eqref{eq:sphere-recovery-comparison} will then imply
\(\mathcal F_Z(x,y)=0\).

\subsubsection{Unitary dilation}

Let \(U\) dilate the full channel \(\mathcal S\), so that
\begin{equation}
 \mathcal S(\omega)
 =\Tr_a\!\left[U(\omega\otimes|\mathbf0\rangle
 \langle\mathbf0|_a)U^\dagger\right].
\end{equation}
The unitary \(U\) represents the full channel \(\mathcal S\).
Define
\begin{equation}
 \begin{aligned}
 |\Omega\rangle&=|\psi\rangle|\mathbf0\rangle_a,\\
 |\Psi_0\rangle&=U|\Omega\rangle
              =|\psi\rangle|\phi\rangle_a,\\
 |\Psi_\gamma\rangle&=UZ_\gamma|\Omega\rangle.
 \end{aligned}
 \label{eq:sphere-dilated-states}
\end{equation}
The factorization of \(|\Psi_0\rangle\) follows from the purity
of its system marginal.  The state \(|\phi\rangle_a\) may be
entangled among ancillary sites.  Tracing out the ancillas in the
last two states gives
\(\sigma_0\) and \(\sigma_\gamma\).

Orthogonality of the joint states is insufficient because the
ancillas are discarded.  We obtain a system measurement by relating
clean output strings and their charge-measuring loops to operators
on the input.  For a clean output string \(W(\lambda)\), define
\begin{equation}
 V=U^\dagger W(\lambda)U.
 \label{eq:sphere-pullback}
\end{equation}
All bare strings act trivially on the ancillas.  The operator \(V\)
can act on both system and ancillas, but is supported in an
\(s\)-neighborhood of \(\lambda\).  The next lemma identifies
the charge it creates in the input toric code.

\subsubsection{Reduction to endpoint charges}

The pulled-back string has a definite total charge in each endpoint
region.  Its action on the input is therefore equivalent to a
canonical toric-code string followed by a unitary in the endpoint
disks.  The latter can change local excitations while leaving the
total charge measured outside each disk fixed.

\begin{lemma}[Exact endpoint-charge reduction]
\label{lem:sphere-exact-charge}
Let \(W(\lambda)\) be a clean open Pauli string along a path
\(\lambda\) with endpoints \(a,b\), creating a pair of charges of type
\(e\), \(m\), or \(em\), and let \(V\) be its pullback.  Assume
\(|a-b|\geq12\ell_0\) and \(\mathsf N_{6\ell_0}(\lambda)\subseteq\mathsf B\),
and let \(D_a,D_b\) be the disks of radius \(3\ell_0\) centred at
\(a,b\).  There are a charge \(c\in\{1,e,m,em\}\), a canonical system
Pauli string \(P_c\) along \(\lambda\) creating this charge in each
disk, and a unitary \(E\) supported in \(D_a\cup D_b\), such that
\begin{equation}
 V|\Omega\rangle=EP_c|\Omega\rangle.
 \label{eq:sphere-exact-replacement}
\end{equation}
\end{lemma}

Here \(P_c\) acts only on the input system, while \(E\) may also
act on ancillary registers in the endpoint disks.  We choose one
canonical string for each charge, with fixed endpoint locations
inside the disks, and set \(P_1=I\).  Deforming a canonical path
multiplies it by stabilizers and leaves its action on the clean
state unchanged.  The lemma concerns the action on
\(|\Omega\rangle\); it does not assert an operator identity.
The unitary \(E\) may act jointly on the two disks and is used
only to compare states.

The sizes serve three purposes.
\begin{itemize}
 \item For every check not contained in \(D_a\cup D_b\), the
       \(\ell_0\)-neighbourhood of its support avoids the endpoints of
       \(W\).
 \item The thickened disks \(\mathsf N_{2s}(D_a)\) and
       \(\mathsf N_{2s}(D_b)\) are separated by more than \(r_0\).
 \item Every support appearing in the proof lies inside \(\mathsf B\),
       away from its boundary.
\end{itemize}

 \begin{proof}
The only property of \(U\) used below is that it maps the clean input to
the clean output with decoupled ancillas,
\(U|\Omega\rangle=|\Psi_0\rangle=|\psi\rangle|\phi\rangle_a\).  Each step
turns a statement about \(V\) acting on the input into a statement about
the bare string \(W\) acting on the clean output, where strings behave in
a known way.

\emph{Step 1: nothing happens away from the endpoints.}
Let \(h\) be an input stabilizer-defect projector, except for checks
entirely contained in \(D_a\) or \(D_b\).  Also include the projectors
\(I-|\mathbf0_j\rangle\langle\mathbf0_j|\) onto nonzero input ancillas at
sites outside these disks.  Each retained \(h\) annihilates
\(|\Omega\rangle\).  Then
\begin{equation}
 \begin{aligned}
 \langle\Omega|V^\dagger hV|\Omega\rangle
 &=\langle\Psi_0|W^\dagger(UhU^\dagger)W|\Psi_0\rangle\\
 &=\langle\Psi_0|UhU^\dagger|\Psi_0\rangle\\
 &=\langle\Omega|h|\Omega\rangle=0.
 \end{aligned}
 \label{eq:sphere-exact-bulk}
\end{equation}
In the second equality we removed \(W\) using local indistinguishability.
The operator \(UhU^\dagger\) is a local probe supported within
\(\ell_0\) of \(h\), and so away from the endpoints of \(W\).  Multiplying
\(W\) by checks inside \(\mathsf B\) deforms it off the support of this
probe while keeping its endpoints fixed, and leaves its action on
\(|\psi\rangle\) unchanged.  Because \(|\Psi_0\rangle\) is a product, the
same holds when the probe also acts on ancillas. Hence \(V|\Omega\rangle\) has no defects outside the endpoint disks, and
all ancillas there remain in \(|\mathbf0\rangle\). 

\emph{Step 2: the two endpoint charges are equal.}
Measure the total charge of a disk \(D\) with its boundary loops
\(W_X(D)\) and \(W_Z(D)\).  These are the products of all \(X\)- and
\(Z\)-type checks contained in \(D\), and they measure the electric and
magnetic charge in \(D\).  Let \(\Pi_{a,c}\) and \(\Pi_{b,c}\) be the
joint charge projectors of \(D_a\) and \(D_b\).

To compare the two disks, use a loop enclosing both.  For
\(q\in\{X,Z\}\),
\begin{equation}
 W_q(\mathsf B)=W_q(D_a)\,W_q(D_b)\prod S,
\end{equation}
where the product runs over the retained \(q\)-type checks in
\(\mathsf B\).  By Step~1, these checks are \(+1\) on \(V|\Omega\rangle\).
The outer loop \(W_q(\mathsf B)\) sits near \(\partial\mathsf B\), far from
\(\operatorname{supp}V\subseteq\mathsf N_s(\lambda)\).  It therefore
commutes with \(V\), and it is \(+1\) on \(|\Omega\rangle\).  Hence
\(W_q(D_a)W_q(D_b)=+1\) on \(V|\Omega\rangle\): whatever charge sits in
\(D_a\), the same charge sits in \(D_b\).  Only a loop enclosing both
endpoints is needed here, not a closed surface.

\emph{Step 3: the charge is sharp.}
Step~2 allows the charge in \(D_a\) to be random, as long as \(D_b\)
always matches it.  We now show that the two charges are statistically
independent.  Two independent variables that always agree must both be
deterministic.

First, the pulled-back charge measurements are local.  Although \(W\)
runs from \(a\) to \(b\), it is a product of single-site Pauli operators.
For an operator \(O_A\) supported in a region \(A\), split
\(W=W_AW_{\bar A}\).  The factor outside \(A\) commutes with \(O_A\) and
cancels:
\begin{equation}
 W^\dagger O_AW=W_A^\dagger O_AW_A .
\end{equation}
Apply this to \(O'_a=U\Pi_{a,c}U^\dagger\), which is supported in
\(A=D_a^{(s)}\).  Since \(V=U^\dagger WU\),
\begin{equation}
 \begin{gathered}
 V^\dagger\Pi_{a,c}V=U^\dagger W_A^\dagger O'_aW_AU,\\
 \operatorname{supp}(V^\dagger\Pi_{a,c}V)\subseteq D_a^{(2s)} ,
 \end{gathered}
\end{equation}
and similarly at \(b\).  By the choice of sizes, \(D_a^{(2s)}\) and
\(D_b^{(2s)}\) are separated by more than \(r_0\).

Second, the input factorizes on these two separated regions.  Their
union lies in the correctable region \(\mathsf B\), so its reduced state
is the same for every code state.  We may therefore compute it from the
stabilizer expansion of the maximally mixed code state, with the
ancillas in their product state.  Consider a stabilizer supported on
the union.  It splits into a factor on each region, since no check
touches both.  Each factor therefore commutes with every check.  Nothing
supported in \(\mathsf B\) is a logical operator, so each factor is
itself a stabilizer, up to sign.  The expansion, and hence the reduced
state, factorizes.

Writing the two charge distributions as \(p_c,q_c\), we obtain
\begin{equation}
 \Pr(c_a=c,c_b=c')=p_cq_{c'},
 \qquad \sum_c p_cq_c=1,
\end{equation}
where the second equality is Step~2.  Since
\(\sum_cp_cq_c\leq\max_cp_c\), both distributions are concentrated on a
single charge \(c\).

\emph{Step 4: everything else is junk at the endpoints.}
Compare \(V|\Omega\rangle\) with \(P_c|\Omega\rangle\) outside the
disks.  Both states satisfy three conditions:
\begin{enumerate}
 \item[(i)] all retained checks are satisfied and all ancillas outside
            the disks are in \(|\mathbf0\rangle\);
 \item[(ii)] both disks carry charge \(c\);
 \item[(iii)] every logical operator \(\ell\) has the same expectation
              value as in \(|\psi\rangle\).
\end{enumerate}
For \(P_c|\Omega\rangle\), (i) and (ii) hold by construction.  For
\(V|\Omega\rangle\), they are Steps~1 to~3.  Condition (iii) holds for
both states.  By the cleaning lemma~\cite{BravyiTerhal2009Cleaning},
\(\ell\) has a representative supported outside \(\mathsf B\), and that
representative commutes with \(V\) and \(P_c\).

These three conditions fix the reduced state on the complement of
\(D_a\cup D_b\).  Expand it in Pauli operators \(P\) supported there.
\begin{itemize}
 \item If \(P\) anticommutes with a retained check, its expectation
       value vanishes.
 \item Otherwise \(P\) commutes with every check, since the omitted
       checks lie inside the disks.  By cleaning, \(P\) is then, up to a
       phase, a stabilizer times a logical operator supported outside
       \(\mathsf B\).
\end{itemize}
In the second case, the stabilizer factor acts trivially on the disks.
It therefore contains either all or none of the checks of each type
inside each disk, so it is a product of retained checks and the disks'
boundary loops.  Its eigenvalue is fixed by (i) and (ii), and the
expectation value of the logical factor is fixed by (iii).

The two pure states are therefore purifications of the same density
matrix on the complement of the disks.  By Uhlmann's theorem, a unitary
\(E\) on the purifying degrees of freedom relates them.  These are the
system and ancillas in \(D_a\cup D_b\).  This proves
Eq.~\eqref{eq:sphere-exact-replacement}.

In summary, the locality of the channel and the local
indistinguishability of the clean code confine the action of \(V\) to
its endpoints.  A loop enclosing both endpoints forces their charges to
match.  The factorization of the input across separated regions makes
the two charges independent.  Matching and independence together give
a definite charge, and Uhlmann's theorem then identifies \(V\) with the
bare string up to a unitary at the endpoints.
\end{proof}

\subsubsection{Loop measurements and charge labels}

We now place \(D_a,D_b\) around the physical endpoints \(x,y\)
of \(Z_\gamma\), and choose the canonical electric string
\(P_e=Z_\gamma\).  Choose clean output strings \(W_e,W_m,W_{em}\)
that create the corresponding pair of charges in these same disks.
The electric and magnetic strings are of \(Z\) and \(X\) type;
their product, with an overall phase if needed, gives the \(em\)
string.  We take them along \(\gamma\), or along an adjacent dual
path.  They are Pauli products and satisfy the lemma's assumptions,
because \(|x-y|\geq12\ell_0\) and
\(\mathsf N_{10\ell_0}(\gamma)\subseteq\mathsf B\).  Include the vacuum case \(W_1=I\), and write
\begin{equation}
 \begin{gathered}
 \begin{aligned}
 V_\alpha&=U^\dagger W_\alpha U,\\
 V_\alpha|\Omega\rangle
 &=E_\alpha P_{c_\alpha}|\Omega\rangle,
 \end{aligned}\\
 \alpha\in\{1,e,m,em\}.
 \end{gathered}
 \label{eq:sphere-charge-map}
\end{equation}
For the vacuum, take \(c_1=1\) and \(E_1=I\).

Take \(L_X=W_X(D')\) and \(L_Z=W_Z(D')\), the boundary loops of
the disk \(D'\) of radius \(7\ell_0\) centred at \(x\).  This disk
contains \(D_a\) but not \(D_b\).  The \(s\)-neighbourhoods of the loops
avoid both endpoint disks and lie in \(\mathsf B\).  The \(X\) loop detects
electric charge and the \(Z\) loop detects magnetic charge.  Thus
\begin{equation}
 L_qW_\alpha=\chi_q(\alpha)W_\alpha L_q,
 \qquad q\in\{X,Z\},
 \label{eq:sphere-loop-signs}
\end{equation}
For the vacuum, both signs are \(+1\).  Electric charge reverses
the \(X\)-loop sign, magnetic charge reverses the \(Z\)-loop
sign, and \(em\) reverses both.  The two loop measurements
therefore distinguish all four charges.

Pull back the loops as \(\widehat L_q=U^\dagger L_qU\).
Their supports still avoid the endpoint disks, so
\([\widehat L_q,E_\alpha]=0\).  Exact recovery and the clean
loop eigenvalues give
\begin{equation}
 \begin{aligned}
 \widehat L_q|\Omega\rangle&=|\Omega\rangle,\\
 \widehat L_qV_\alpha|\Omega\rangle
 &=\chi_q(\alpha)V_\alpha|\Omega\rangle.
 \end{aligned}
 \label{eq:sphere-pulled-loop-signs}
\end{equation}

For any nontrivial \(\alpha\), choose a loop with
\(\chi_q(\alpha)=-1\).  If its pulled-back charge were the
vacuum, the lemma would give
\(V_\alpha|\Omega\rangle=E_\alpha|\Omega\rangle\).
But then
\begin{align*}
 \widehat L_qV_\alpha|\Omega\rangle
 &=\widehat L_qE_\alpha|\Omega\rangle\\
 &=E_\alpha\widehat L_q|\Omega\rangle
 =V_\alpha|\Omega\rangle,
\end{align*}
contradicting Eq.~\eqref{eq:sphere-pulled-loop-signs}.
Hence \(c_\alpha\ne1\) for \(\alpha\ne1\).

The two loop measurements also distinguish the input charges
associated with different reference strings.  Substituting the
endpoint-charge reduction into
Eq.~\eqref{eq:sphere-pulled-loop-signs} and commuting
\(\widehat L_q\) through \(E_\alpha\) gives
\begin{equation}
 \widehat L_qP_{c_\alpha}|\Omega\rangle
 =\chi_q(\alpha)P_{c_\alpha}|\Omega\rangle.
 \label{eq:sphere-canonical-loop-signs}
\end{equation}
Suppose two different output charges \(\alpha,\beta\) gave
the same input charge \(c_\alpha=c_\beta=c\).  At least one
loop assigns opposite signs to \(\alpha\) and \(\beta\).
This would give the nonzero state \(P_c|\Omega\rangle\) both
eigenvalues \(+1\) and \(-1\) under the same pulled-back
loop.  Thus \(c_\alpha\ne c_\beta\).

The four output charges therefore give four different input charges.
Since the toric code has exactly four charge types, this is a
permutation of \(\{1,e,m,em\}\), fixing the vacuum.  In
particular, some nontrivial \(\alpha_*\) satisfies
\(c_{\alpha_*}=e\), whose canonical string is the physical
\(Z_\gamma\) chosen above.

Choose a loop with \(\chi_q(\alpha_*)=-1\).  Equation
~\eqref{eq:sphere-canonical-loop-signs} now gives
\begin{equation}
 \begin{aligned}
 \widehat L_q|\Omega\rangle&=|\Omega\rangle,\\
 \widehat L_qZ_\gamma|\Omega\rangle
 &=-Z_\gamma|\Omega\rangle.
 \end{aligned}
\end{equation}
Multiplying by \(U\), we obtain
\begin{equation}
 \begin{aligned}
 L_q|\Psi_0\rangle&=|\Psi_0\rangle,\\
 L_q|\Psi_\gamma\rangle&=-|\Psi_\gamma\rangle.
 \end{aligned}
\end{equation}
Taking \(W_C=L_q\) proves
Eq.~\eqref{eq:sphere-measurement-goal}.  Since the loop
acts only on the output system, tracing out the ancillas preserves
its eigenvalue.  The measurement distributions are \((1,0)\) and
\((0,1)\), so fidelity monotonicity gives
\begin{equation}
 \mathcal F_Z(x,y)
 \leq F(\sigma_0,\sigma_\gamma)=0.
 \label{eq:sphere-exact-conclusion}
\end{equation}

\subsection{Approximate recovery}
\label{app:sphere-approximate}

Now assume
\begin{equation}
 \|\sigma_0-\psi\|_1\leq\epsilon,
 \label{eq:sphere-approximation}
\end{equation}
where \(\epsilon\) is the full trace-norm error.  The comparison
in Eq.~\eqref{eq:sphere-recovery-comparison} remains exact.
We only need to bound the two wrong-outcome probabilities of the
same system loop measurement.

\subsubsection{Approximate endpoint-charge reduction}

Write \(|\Psi\rangle=U|\Omega\rangle\).  Its projection onto
the clean system state has probability
\(q=\langle\psi|\sigma_0|\psi\rangle\geq1-\epsilon/2\).
The normalized projection is a product
\(|\widetilde\Psi\rangle=|\psi\rangle|\phi\rangle_a\)
satisfying
\begin{equation}
 \frac12\bigl\||\Psi\rangle\langle\Psi|
 -|\widetilde\Psi\rangle\langle\widetilde\Psi|\bigr\|_1
 \leq\sqrt{\epsilon/2}.
 \label{eq:sphere-purification-error}
\end{equation}
We work below in the small-error regime; the final bound is trivial
outside it.

For each retained defect projector \(h\) from the exact proof,
compare the expectation of \(W^\dagger UhU^\dagger W\) in
\(|\Psi\rangle\) and \(|\widetilde\Psi\rangle\).  In the
product state \(|\widetilde\Psi\rangle\), local
indistinguishability removes the two \(W\)'s.  Comparing back to
\(|\Psi\rangle\), where the expectation without \(W\) is
zero, gives
\begin{equation}
 \langle\Omega|V^\dagger hV|\Omega\rangle
 \leq2\sqrt{\epsilon/2}=\sqrt{2\epsilon}.
\end{equation}
Let \(M=O(L^2)\) bound the number of retained constraints, counting
one ancillary projector per site and grouping all ancillary
registers at that site.  Since these projectors commute, their joint
no-defect projector \(G\) satisfies
\begin{equation}
 \langle\Omega|V^\dagger GV|\Omega\rangle\geq1-\delta,
 \qquad \delta=M\sqrt{2\epsilon}.
 \label{eq:sphere-bulk-error}
\end{equation}

The two endpoint-charge measurements are still exactly independent:
this used only the input state and the locality of \(V\), neither
of which has changed.  They agree on the range of \(G\), so
\begin{equation}
 \sum_c p_cq_c\geq1-\delta,
 \qquad \max_c p_c\geq1-\delta.
\end{equation}
Choose a maximizing charge \(c\).  Projecting onto \(G\) and
\(\Pi_{a,c}\) succeeds with probability at least
\(1-2\delta\).  These commuting projections enforce the bulk
constraints and the common charge \(c\) exactly.  They also preserve
the logical expectation values.  Retained projectors far from
\(\lambda\) commute with \(V\) and fix \(|\Omega\rangle\), so they can
be dropped.  The remaining pulled-back projection is then supported in
\(\mathsf B\).  Correctability of \(\mathsf B\) makes its outcome
probability independent of the logical state, so conditioning on
success leaves each \(\langle\ell\rangle\) unchanged.  The exterior
reduced-state argument in the exact lemma therefore identifies the
normalized projection with \(EP_c|\Omega\rangle\).
The projection error is bounded by
\begin{equation}
 \|(V-EP_c)|\Omega\rangle\|\leq\eta,
 \qquad \eta=\sqrt{8\delta}.
 \label{eq:sphere-approximate-replacement}
\end{equation}
Apply this bound to each of the three nontrivial reference strings,
using the same endpoint disks and canonical strings.  For the vacuum
we still take \(V_1=E_1=P_1=I\), with zero replacement error.

\subsubsection{Loop-charge measurements}

The string-loop commutation relations and the separation of the
pulled-back loops from the endpoint disks remain exact.  Recovery
error only allows a small probability of the wrong loop outcome.
For any reference charge \(\alpha\) and either loop, define
\begin{equation}
 Q_{q,\alpha}
 =\frac{I-\chi_q(\alpha)\widehat L_q}{2}.
\end{equation}
This projects onto the wrong sign for that charge.  Using
Eq.~\eqref{eq:sphere-loop-signs},
\begin{equation}
 \|Q_{q,\alpha}V_\alpha|\Omega\rangle\|^2
 =\Tr\!\left[\frac{I-L_q}{2}\sigma_0\right]
 \leq\epsilon/2.
 \label{eq:sphere-reference-loop-error}
\end{equation}
The last inequality follows because the clean state has loop
eigenvalue \(+1\).  Since \(Q_{q,\alpha}\) commutes with
\(E_\alpha\), replacing \(V_\alpha|\Omega\rangle\) by
\(E_\alpha P_{c_\alpha}|\Omega\rangle\) gives
\begin{equation}
 \begin{gathered}
 \|Q_{q,\alpha}P_{c_\alpha}|\Omega\rangle\|\leq\zeta,\\
 \zeta=\eta+\sqrt{\epsilon/2}.
 \end{gathered}
 \label{eq:sphere-canonical-loop-error}
\end{equation}
This bounds the probability of assigning the wrong loop charge to
the canonical state.

If \(c_\alpha=c_\beta=c\) for two different charges, choose
a loop that gives them opposite signs.  The two wrong-sign
projectors are then complementary, so the same normalized state
would satisfy
\begin{equation}
 1=\|Q_{q,\alpha}P_c|\Omega\rangle\|^2
   +\|Q_{q,\beta}P_c|\Omega\rangle\|^2
   \leq2\zeta^2.
\end{equation}
For \(\delta<1/32\), we have
\(\eta<1/2\) and \(\sqrt{\epsilon/2}=\delta/(2M)<1/64\),
so \(\zeta<33/64<1/\sqrt2\).  The inequality above is
impossible.  The four charge labels therefore remain distinct,
and some nontrivial \(\alpha_*\) still has
\(c_{\alpha_*}=e\).

Choose a loop with \(\chi_q(\alpha_*)=-1\) and set
\(W_C=L_q\).  As in the exact proof, take \(P_e=Z_\gamma\).
Measure this loop on the two system outputs and denote the wrong
outcome probabilities by
\begin{equation}
 \begin{aligned}
 a&=\Tr\!\left[\frac{I-W_C}{2}\sigma_0\right],\\
 b&=\Tr\!\left[\frac{I+W_C}{2}\sigma_\gamma\right].
 \end{aligned}
\end{equation}
Because the clean state has no \(-1\) outcome,
Eq.~\eqref{eq:sphere-approximation} gives
\(a\leq\epsilon/2\).  For the charged output,
\(Q_{q,\alpha_*}=(I+\widehat L_q)/2\).  Evaluating the
measurement in its purification and using
Eq.~\eqref{eq:sphere-canonical-loop-error} gives
\begin{equation}
 b=\|Q_{q,\alpha_*}Z_\gamma|\Omega\rangle\|^2
 \leq\zeta^2.
\end{equation}
The measured distributions are \((1-a,a)\) and \((b,1-b)\).
Their classical fidelity bounds the quantum fidelity from above:
\begin{equation}
 \begin{aligned}
 \mathcal F_Z(x,y)
 &\leq F(\sigma_0,\sigma_\gamma)\\
 &\leq\sqrt{(1-a)b}+\sqrt{a(1-b)}\\
 &\leq\sqrt{\epsilon/2}+\zeta\\
 &=\eta+\sqrt{2\epsilon}\\
 &\leq8\sqrt M\,\epsilon^{1/4}.
 \end{aligned}
\end{equation}
For \(\delta\geq1/32\), the last expression is already greater
than one, so the trivial fidelity bound covers the remaining regime.
Restoring the size index, we obtain
\begin{equation}
 \mathcal F_Z(x,y)
 \leq\min\bigl\{1,\,8\sqrt{M_L}\epsilon_L^{1/4}\bigr\}.
 \label{eq:spherical-fidelity-bound}
\end{equation}
Since \(M_L=O(L^2)\), recovery with full trace-norm error
\(\epsilon_L=L^{-\infty}\) implies
\(\mathcal F_Z(L/2)=L^{-\infty}\).  Algebraic or slower decay
of this fidelity is therefore incompatible with the local reverse
channel required for phase equivalence.

\section{Gaussian Tensor Networks and Transport Bounds}
\label{app:gaussian-information-transport}

This appendix supplies the two ingredients connecting the classical CMI
to transport in Sec.~\ref{sec:network-localization}.  Pure Gaussian boundary
states obey spectral bounds relating their entanglement to conductance.
Separately, the Holevo bound limits classical measurement correlations by
the entanglement of the conditional physical state.  We give the Gaussian
construction and spectral bounds first, then prove the identification of
physical and virtual entanglement for the honeycomb boundary convention.
All entropies are in bits.

\subsection{Hybrid tensors and the boundary state}
\label{app:hybrid-tn}

We use the qubit--Majorana representation of
Ref.~\cite{Negari_2024}, adapted to the honeycomb patch in
Fig.~\ref{fig:cmi-geometry}.  Each tensor contains one physical
qubit and four virtual Majoranas.  Up to orientation-dependent
signs, it is fixed by the commuting stabilizers
\begin{equation}
 \gamma_u\gamma_d\gamma_l\gamma_r,\qquad
 iX\gamma_u\gamma_d,\qquad iZ\gamma_u\gamma_l.
 \label{eq:hybrid-local-stabilizers}
\end{equation}
Neighboring virtual modes are contracted by fixing their bond
parity, with projector $(I+i\gamma_a\gamma_b)/2$ in the chosen
orientation.  The bond pattern and local choice of axes reproduce
the physical $X$ and $Z$ checks.  The corner pairing fixes the
logical input as specified in
Eq.~\eqref{eq:honeycomb-logical-majorana-links}.

The local stabilizers identify a physical Pauli operator with a
virtual Majorana bilinear.  Thus $e^{-i\theta Z}$ becomes a
Gaussian rotation through angle $2\theta$ on the corresponding
Majorana pair.  Fixing a physical $X$ outcome leaves a pure
Gaussian tensor on the virtual legs.  The full contraction can
therefore be performed using covariance matrices: rotations act
as $\Gamma\mapsto O\Gamma O^{\mathsf T}$, and parity projections
give the conditional covariance and Born probability.
The sampling conventions are given in Appendix~\ref{app:numerics}.

The symmetry-class convention is important here.  The global exchange
$X\leftrightarrow Z$ relative to
Ref.~\cite{yang2026decodingcoherenterrorstoric} maps our odd-weight $Z$
checks to its odd-weight $X$ checks.  In our convention the bulk code
respects the antiunitary $(\prod_i Z_i)\mathcal K$, where $\mathcal K$
is complex conjugation in the physical $Z$ basis, but a generic coherent
$Z$ rotation breaks it.  The associated Majorana monitored dynamics and
scattering network are in class DIII.  For the square lattice, the bulk
checks and coherent rotations instead preserve
$(\prod_i iY_i)\mathcal K$, giving class D
\cite{yang2026decodingcoherenterrorstoric}.
These class labels refer to the network associated with the monitored
evolution, not to a physical-qubit Hamiltonian.  In the scattering
construction, the odd-weight checks require doubling the propagating
channels; even-weight square-lattice checks allow two decoupled blocks
\cite{bejan2026projectedlogicalensemblessurface}.
Born weighting selects the replica limit $n\to1$, in which the
nonbipartite syndrome lattice admits the stable class-DIII metal
\cite{yan2026nonlinearsigmamodelsurface}.

For a bulk outcome $b$, contract the tensors in $B$ and leave
the virtual modes at its two ends open.  The normalized result is
a pure Gaussian state $\ket{\phi_b}_{LR}$.  Its covariance is
\begin{equation}
 \begin{aligned}
  \Gamma^{(b)}_{jk}
  &=\frac{i}{2}\bra{\phi_b}[\gamma_j,\gamma_k]\ket{\phi_b},\\
  (\Gamma^{(b)})^2&=-I .
 \end{aligned}
 \label{eq:boundary-pure-covariance}
\end{equation}
This matrix is real and antisymmetric.  In particular, the
eigenvalues used below are those of the Hermitian matrix
$i\Gamma^{(b)}_{RR}$, which occur in pairs
$\pm\lambda_i^{(b)}$ with $0\leq\lambda_i^{(b)}\leq1$.
The averages use the same bulk Born probabilities $p_b$ as the
CMI calculation.

The same covariance defines the scattering matrix
$\mathsf S=i\Gamma$, as derived in
Ref.~\cite{jian2022criticalityentanglementnonunitaryquantum}.
The tensor and scattering networks have the same connectivity,
with an incoming and an outgoing channel on each leg
(Fig.~\ref{fig:random-gaussian-ensemble}).
Suppressing $b$, the transmission block is $\mathsf t=i\Gamma_{RL}$.
The $RR$ block of $\Gamma^2=-I$ gives
\begin{equation}
 \mathsf t\mathsf t^\dagger=-\Gamma_{RL}\Gamma_{LR}
 =I_R+\Gamma_{RR}^2.
 \label{eq:covariance-transmission-identity}
\end{equation}

\subsection{Entanglement and conductance}
\label{app:conductance-mie-bound}

Write $s_2(\lambda)=h_2((1-\lambda)/2)$, where $h_2$ is the
binary entropy in bits.  For a fixed $b$, the Gaussian entropy
formula gives $S(R)_{\phi_b}=\sum_i s_2(\lambda_i^{(b)})$.
We define
\begin{equation}
 E_{LR}(b)=I(L:R)_{\phi_b}
 =2\sum_i s_2(\lambda_i^{(b)}),
 \label{eq:gaussian-mutual-information-modes}
\end{equation}
and $E_{LR}=\sum_b p_b E_{LR}(b)$.
The factor of two follows from purity of $\phi_b$.

By Eq.~\eqref{eq:covariance-transmission-identity}, each pair
$\pm\lambda_i$ contributes two transmission
eigenvalues $x_i=1-\lambda_i^2$.  The conductance convention used
in the numerics is
\begin{equation}
 g(b)=\operatorname{tr}(\mathsf t\mathsf t^\dagger)
 =2\sum_{i=1}^{N_\partial}x_i^{(b)}.
 \label{eq:paired-channel-conductance}
\end{equation}
Here $N_\partial$ counts mode pairs on the right boundary;
unused modes may be included with $x_i=0$.
This explains why weak transmission and small boundary
entanglement occur together: both are controlled by $x_i\to0$.

\paragraph{Step 1: Bound one transmitting mode.}
The relation follows from the elementary binary-entropy bound
\begin{equation}
 x\leq2s_2\bigl(\sqrt{1-x}\bigr)
 \leq x\log_2\frac{2e}{x},
 \qquad 0\leq x\leq1.
 \label{eq:correct-gaussian-mode-bound}
\end{equation}
For a short proof, set $q=(1-\sqrt{1-x})/2$.
Then $x/4\leq q\leq x/2$, and
$2q\leq h_2(q)\leq q\log_2(e/q)$.
The lower bound gives $2h_2(q)\geq4q\geq x$.
For the upper bound, use that $q\log_2(e/q)$ increases on
$0\leq q\leq1/2$ and substitute $q\leq x/2$.

\paragraph{Step 2: Sum modes, then average records.}
Summing Eq.~\eqref{eq:correct-gaussian-mode-bound} over modes and using
concavity of $x\log_2(2e/x)$ yields
\begin{equation}
 \frac{g(b)}{2}\leq E_{LR}(b)
 \leq\frac{g(b)}{2}
 \log_2\frac{4eN_\partial}{g(b)}.
 \label{eq:outcome-gaussian-transport-bound}
\end{equation}
One more application of concavity gives the Born-averaged bound,
\begin{equation}
 \begin{gathered}
  \frac{g}{2}\leq E_{LR}
  \leq\frac{g}{2}\log_2\frac{4eN_\partial}{g},\\
  g=\sum_b p_b g(b).
 \end{gathered}
 \label{eq:averaged-gaussian-transport-bound}
\end{equation}
The upper expressions are understood by continuity at $g=0$.
This bound holds for any ensemble of pure Gaussian boundary
states with at most $N_\partial$ mode pairs.

\begin{figure}[tbp]
 \centering
 \includegraphics[width=\linewidth]{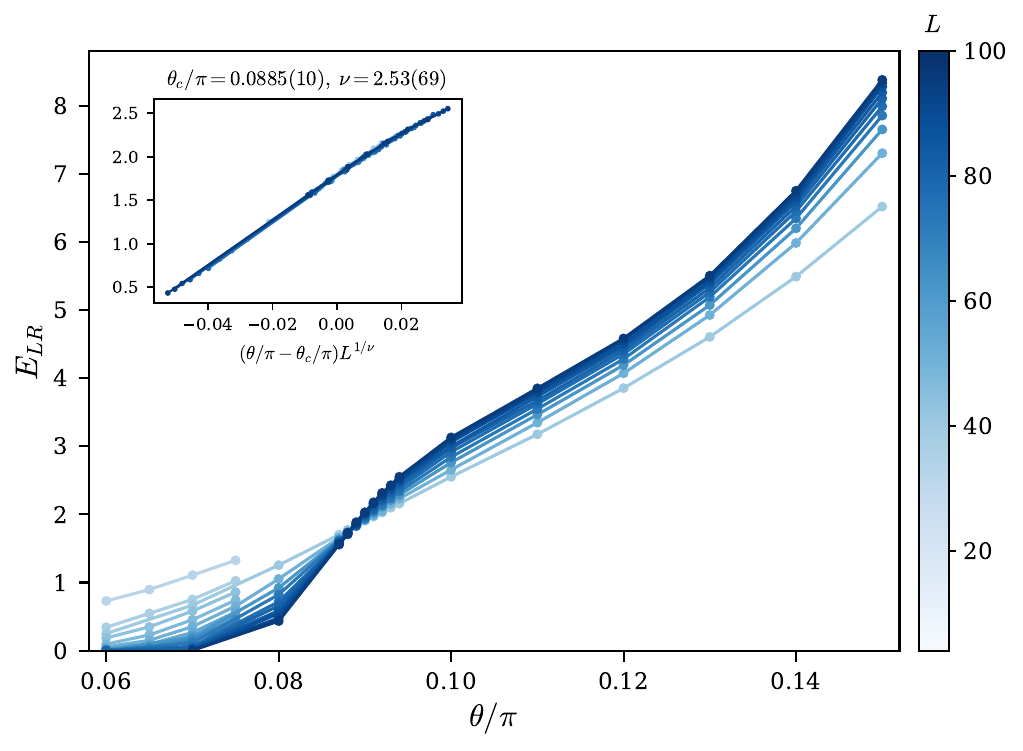}\\[-1mm]
 \includegraphics[width=\linewidth]{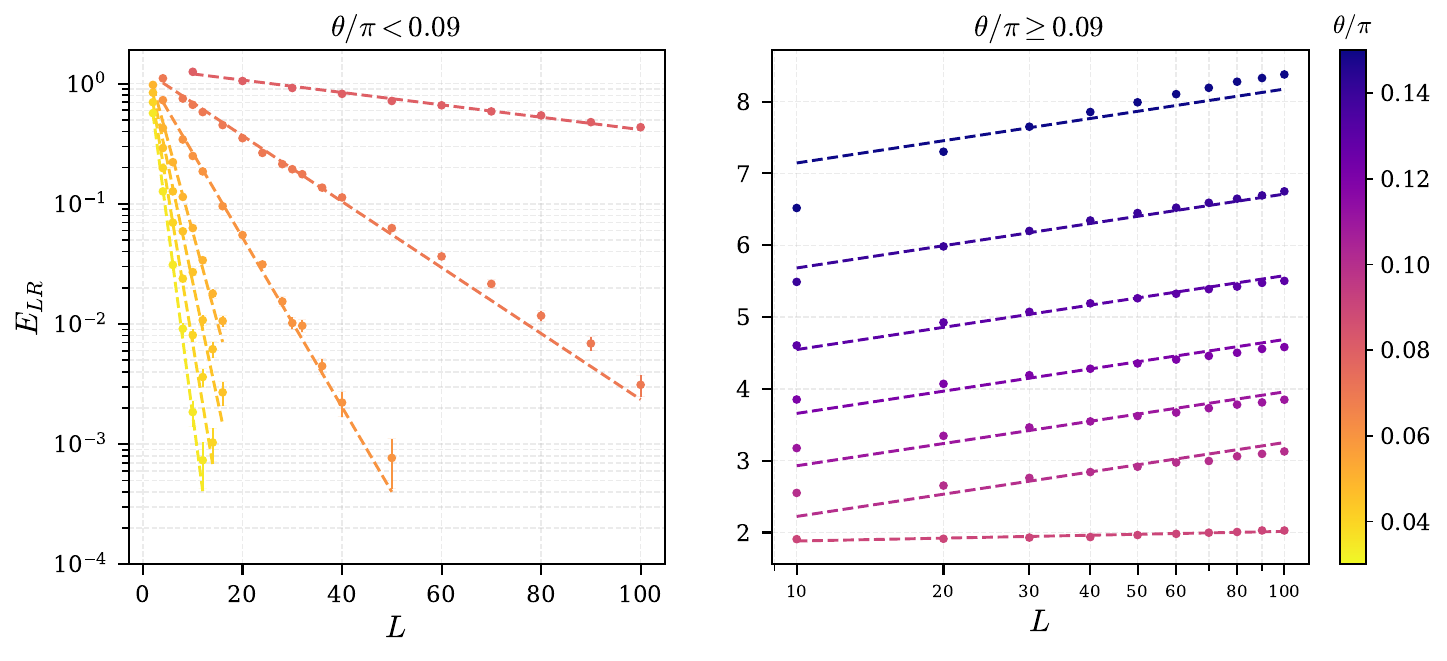}
 \caption{\justifying
 Born-averaged virtual mutual information
 $E_{LR}=2\sum_b p_b S(R)_{\phi_b}$ in bits.
 Top: angle scan and data collapse. Bottom: size dependence below (left)
 and above (right) threshold. Dashed curves are fits; error bars are
 standard errors.}
 \label{fig:mie-scaling-appendix}
\end{figure}

The points in Fig.~\ref{fig:mie-scaling-appendix} have at least $4000$
samples and signal-to-noise ratio at least two. Values below
$E_{LR}=10^{-4}$ are omitted from the lower-left panel.
The two observables are evaluated from the same covariance in
each numerical sample.  We checked the entropy formula against
direct two-mode density matrices, the transmission identity on
random pure Gaussian covariances, and both bounds on mode spectra
and finite ensembles.  These checks preserve the factor of two
in Eqs.~\eqref{eq:gaussian-mutual-information-modes}
and~\eqref{eq:paired-channel-conductance}.

\subsection{Physical measurement-induced entanglement}
\label{app:classical-cmi-mie}

Let a complete rank-one measurement of $B$ in a pure state on $ABC$
prepare $\ket{\psi_b}_{AC}$ with probability $p_b$.  A rank-one measurement
of $A$ then prepares pure states $\ket{\chi_{a,b}}_C$ with probabilities
$p(a|b)$.  Their average is the reduced state of $C$ in $\psi_b$.
The Holevo bound on any subsequent measurement of $C$ is therefore
\begin{equation}
 \begin{aligned}
  I(X_A:X_C|b)&\leq
  S\!\left(\sum_a p(a|b)\ket{\chi_{a,b}}\!\bra{\chi_{a,b}}\right)\\
  &=S(C)_{\psi_b}.
 \end{aligned}
 \label{eq:conditional-physical-holevo}
\end{equation}
The conditional states are pure, so the average of their individual
entropies vanishes.  Since $S(C)_{\psi_b}=S(A)_{\psi_b}$, averaging over $b$
proves the measurement-induced-entanglement bound in
Sec.~\ref{subsec:mie}.  This argument concerns the
physical state and holds independently of its tensor-network representation.

In our application, the pure state is the coherently rotated code state
with fixed logical $Z$.  Its physical $X$-outcome distribution is unchanged
by syndrome dephasing, as shown in Appendix~\ref{app:cmi-invariance}.
The bound thus applies to the classical CMI $I_X$ obtained in the numerics.
To evaluate its right-hand side from the Gaussian contraction, we must also
account for the projection from virtual Majoranas to physical spins.

\subsection{Physical and virtual boundary entanglement}
\label{app:honeycomb-boundary-entanglement}

We now show that, for the honeycomb boundary convention used here,
\begin{equation}
 S(A)_{\psi_b}=S(L)_{\phi_b},
 \label{eq:honeycomb-physical-virtual-entropy}
\end{equation}
for every bulk record of nonzero probability.  The argument uses the local
qubit--Majorana tensors and the boundary dimers, and is the two-boundary
version of the projection and Jordan--Wigner correspondence in
Ref.~\cite{Negari_2024}.

The covariance used for $\phi_b$ contains the modes exposed by contracting
$B$.  Dimers lying wholly within $A$ or $C$, and the two outer corner links
in Eq.~\eqref{eq:honeycomb-logical-majorana-links}, are not part of this
covariance.  To reconstruct the physical state, first restore these dimers
and then impose the local parity constraints on the boundary sites.
Boundary coherent rotations act only within $A$ or $C$ and do not change
their entanglement.

\paragraph{Step 1: Reduce each boundary projection to total parity.}
Consider $A$.  Its internal dimers form a connected graph.  Contracting
them with the local parity tensors gives an encoding $K_A$ from the exposed
modes $L$ and the two corner Majoranas on $A$ to the physical boundary spins.
The two corner Majoranas form one fermionic mode, denoted $\ell$.
If $P_L$ and $P_\ell$ are the corresponding parity operators, then
\begin{equation}
 K_A^\dagger K_A
 =c_A\frac{I+\eta_A P_LP_\ell}{2}.
 \label{eq:honeycomb-boundary-parity-encoding}
\end{equation}
Here $c_A>0$, and the sign $\eta_A=\pm1$ is fixed by the bond orientations.
A short way to obtain this
identity is to expand the product of local parity projectors.  A nonempty
proper subset of boundary sites cuts at least one internal dimer; the
corresponding term contains only one Majorana of that dimer and has zero
expectation.  Only the empty subset and the entire connected boundary
survive.  They give the identity and the total external parity, respectively.
Thus, after an overall normalization, $K_A$ preserves inner products in the
allowed parity sector.  The same statement holds for $C$, its open modes
$R$, and its corner mode $r$.

\paragraph{Step 2: Use the corner pair to accommodate both parities.}
The corner links do not impose an additional selection on the Schmidt
coefficients of $\phi_b$.  Together they form a maximally entangled state
of $\ell$ and $r$, with two components of equal magnitude in the local
parity basis.  Also, $\phi_b$ has definite total fermion parity.  Its
Schmidt vectors can therefore be chosen with definite local parities:
\begin{equation}
 \ket{\phi_b}=\sum_j\sqrt{q_j}\,
 \ket{u_j}_{L}\ket{v_j}_{R}.
 \label{eq:honeycomb-parity-schmidt}
\end{equation}
\paragraph{Step 3: Preserve the Schmidt probabilities.}
For each $j$, Eq.~\eqref{eq:honeycomb-boundary-parity-encoding} fixes the
parity of $\ell$ from that of $u_j$, and similarly fixes the parity of $r$
from $v_j$.  The fixed total parity and the chosen logical sector make
these selections compatible for every term.  Exactly one of the two
equal-magnitude corner components survives for each $j$.  Every Schmidt
coefficient is therefore multiplied by the same magnitude before
normalization.  The encodings $K_A,K_C$ preserve orthogonality of the
surviving vectors, so the physical state has the same Schmidt probabilities
$q_j$, up to zero padding.  This proves
Eq.~\eqref{eq:honeycomb-physical-virtual-entropy}.

Combining this identity with the physical measurement bound gives
\begin{equation}
 I_X(A:C|B)\leq\sum_b p_b S(L)_{\phi_b}
 =\frac{E_{LR}}2.
 \label{eq:classical-cmi-mie-bound}
\end{equation}
The fixed logical contribution to the quantum CMI is handled separately in
Appendix~\ref{app:cmi-invariance}; $I_X$ is already the excess above it.

As a check, we compared the physical Schmidt spectrum obtained from exact
code-state amplitudes with the spectrum reconstructed from the Gaussian
covariance for $676$ nonzero conditional states.  These include all nonzero
bulk records on the $1\times1$, $2\times1$, and $1\times2$ patches at five
angles, and sampled records on the $2\times2$ patch.  The largest entropy
difference was below $1.3\times10^{-14}$ bits, and all conditional Holevo
bounds were satisfied.  These checks supplement the parity argument above.

The distinction between physical and virtual entanglement matters for
generality.  Local measurement tensors in a fixed parity sector obey
$\sum_x\ket{T_x}\!\bra{T_x}\propto P_{\rm parity}$, with the appropriate
local weights included.  That identity alone is not completeness on an
arbitrary open boundary: the contracted parity projectors can select a
sector of its state.  In the present geometry, the corner components
account for this selection without changing the Schmidt spectrum.  An
arbitrary Gaussian tensor ensemble or a different boundary closure requires
its own identification before Eq.~\eqref{eq:classical-cmi-mie-bound} can be
applied.  The physical Holevo bound and the Gaussian spectral bounds do not
require that extra identification.

\subsection{Consequence for decay lengths}

Suppose $g(L)=e^{-L/\xi_{\mathrm{loc}}+o(L)}$ and the number of boundary
mode pairs grows at most polynomially with $L$.  The upper and lower bounds
in Eq.~\eqref{eq:averaged-gaussian-transport-bound} then fix the same
exponential rate for $E_{LR}$ and $g$.  Equation~\eqref{eq:classical-cmi-mie-bound}
therefore gives $\xi_{\mathrm M}\leq\xi_{\mathrm{loc}}$ for the honeycomb
measurement ensemble.  All averages here use the physical marginal $p_b$;
no replacement by a typical transfer-matrix rate is needed for this bound.
Equality of the two lengths requires additional physical input, discussed
in Appendix~\ref{app:boundary-hamiltonian}.

The square-lattice comparison in Appendix~\ref{app:square-lattice-results}
retains a maximally entangled boundary mode.  Removing its fixed entropy
and transmission contributions leaves a Gaussian spectrum to which the
same spectral bounds apply.  Relating that remaining spectrum to the
physical CMI is a separate question: subtracting constants from a bound
on the full entropy does not establish a bound on the remainder.
We therefore use the square-lattice data as an independent numerical test
of the common decay length.

\section{Boundary Hamiltonian and Decay Lengths}
\label{app:boundary-hamiltonian}

We give a physical argument for why the Markov and localization lengths
can coincide.  The idea is that the gap of an effective boundary
Hamiltonian controls both transmission through a long strip and the
leading correction to its entropy.  We obtain the entropy correction
from the partition function, stating the assumptions about the boundaries
and the Born average along the way.

\subsection{The boundary Hamiltonian and conductance}

Consider a strip of length $L$ and width $W$.  For a measurement record
$x$, contracting the Gaussian tensors along the strip gives a transfer
operator $\widehat M_x$.  Its spectrum defines a quadratic boundary
Hamiltonian~\cite{jian2022criticalityentanglementnonunitaryquantum,Venn_2023}:
\begin{equation}
 \begin{aligned}
  \widehat M_x\widehat M_x^\dagger
  &\propto e^{-L\widehat H_{L,x}},\\
  \widehat H_{L,x}
  &=2\sum_j\frac{X_j}{L}
  \left(f_j^\dagger f_j-\frac12\right).
 \end{aligned}
 \label{eq:boundary-hamiltonian-definition}
\end{equation}
Here $f_j$ are fermionic modes and $X_j\geq0$ determine how strongly each
mode is suppressed along the strip.  At fixed $W$, we assume that
$X_j/L\to\epsilon_j(W)$ for typical records as $L$ grows
\cite{Oseledets1968,MacKinnonKramer1983}.
Writing $\Delta(W)=\min_j\epsilon_j(W)>0$, the single-fermion gap is
$2\Delta(W)$.  Since $L$ plays the role of inverse temperature, the
lowest excitation has relative weight $e^{-2L\Delta(W)}$.

The same numbers determine the transmission probabilities
\cite{Beenakker1997,Venn_2023}:
\begin{equation}
 \tau_j(L,W;x)=\operatorname{sech}^2 X_j(L,W;x).
 \label{eq:boundary-conductance-spectrum}
\end{equation}
Conductance is their sum, so the smallest gap sets its exponential
decay, $g_x(L,W)=e^{-2L\Delta(W)+o(L)}$.

\subsection{Entropy and the Markov length}

Let $H_L$ be the Shannon entropy, in bits, of the measurement outcomes
on $L$ consecutive slices.  We take these distributions from the same
ensemble, whose statistics are unchanged by shifting along the strip.
To relate the entropy to the Hamiltonian, we consider boundary conditions
for which the Born probability is the squared norm of the Gaussian
transfer state.  It then has the partition-function form
\begin{equation}
 \begin{aligned}
  p_L(x)&=c_{L,x}\mathcal Z_{L,x},\\
  \mathcal Z_{L,x}
  &=\operatorname{Tr}e^{-L\widehat H_{L,x}}\\
  &=e^{-LE_{0,L,x}}\prod_j\bigl(1+e^{-2X_j}\bigr),
 \end{aligned}
 \label{eq:boundary-partition-function}
\end{equation}
where $c_{L,x}>0$ contains the scalar tensor normalizations and
$E_{0,L,x}$ is the ground-state energy.  The product includes the full
fermionic mode space; a boundary projection that removes the lowest
excitation would change the gap entering this argument.

Using $H_L=-\mathbb E_x\log_2 p_L(x)$ now gives
\begin{equation}
 \begin{aligned}
  H_L&=H_L^{(0)}
  -\frac{1}{\ln2}\mathbb E_x\sum_j\ln\bigl(1+e^{-2X_j}\bigr),\\
  H_L^{(0)}&=-\mathbb E_x\log_2
  \bigl(c_{L,x}e^{-LE_{0,L,x}}\bigr).
 \end{aligned}
 \label{eq:boundary-entropy-partition-identity}
\end{equation}
All averages use the full Born distribution $p_L$.
The second term is therefore an explicit correction from boundary
excitations.  For a long strip, $\ln(1+e^{-2X_j})\simeq e^{-2X_j}$,
so its leading exponential scale is $e^{-2L\Delta(W)}$.

The remaining assumption concerns $H_L^{(0)}$: we take its long-strip
form to be a bulk term $h(W)L$ plus a constant end contribution
$h_\partial(W)$, with any remaining contribution to CMI decaying faster.
This excludes a competing correction from the ground-state term that
could cancel the excitation contribution.  Provided the Born average
preserves the excitation's decay rate, we then obtain, for a constant
leading prefactor,
\begin{equation}
 H_L=h(W)L+h_\partial(W)-A_W e^{-2L\Delta(W)}+R_L,
 \label{eq:boundary-strip-entropy}
\end{equation}
with $A_W>0$ and $R_L$ contributing only faster-decaying terms to CMI.
Thus the entropy correction follows from the partition function under
these assumptions.

For one-slice regions $A$ and $C$ separated by $L$ slices of $B$, the CMI
is the negative discrete second derivative of the total entropy:
\begin{equation}
 I(X_A:X_C|X_B)=2H_{L+1}-H_L-H_{L+2}.
 \label{eq:boundary-cmi-discrete-curvature}
\end{equation}
The bulk and constant terms cancel, leaving
\begin{equation}
 I(X_A:X_C|X_B)
 \simeq A_W(1-e^{-2\Delta(W)})^2e^{-2L\Delta(W)}.
 \label{eq:boundary-cmi-kernel}
\end{equation}
Thus CMI and conductance decay with the same exponential rate.
Smooth power-law factors in the entropy correction do not change this
rate.  Size-independent logical contributions are removed when present,
as in Appendices~\ref{app:cmi-invariance}
and~\ref{app:square-lattice-results}.

The averaging assumption matters for both quantities: rare outcomes
with unusually high transmission can change the exponential rate of
the Born average.  To apply the argument to the numerics, we also need
the strip decay rates to remain valid as $W$ grows, without exponentially
large changes in the prefactors.  Under these assumptions, if
$\Delta(W)\to\Delta>0$, then
\begin{equation}
 \xi_{\mathrm M}=\xi_{\mathrm{loc}}=\frac{1}{2\Delta}.
 \label{eq:boundary-asymptotic-equality}
\end{equation}
A closing gap then makes both lengths diverge.  This picture motivates
the numerical comparison in Fig.~\ref{fig:length-scale-comparison} and
Appendix~\ref{app:square-lattice-results}.

\section{Numerical Comparison of CMI and Transport}
\label{app:numerical-methods}
\label{app:numerics}

The honeycomb numerics test two aspects of the proposed
Markov--localization correspondence: whether conditional correlations
and transport change behavior at the same threshold, and whether
their decay lengths agree in the localized regime.  We describe
the common ensemble and the fits used for these comparisons.

\subsection{Observables and sampling}
\label{app:honeycomb-geometry}

We use the fixed logical-\(Z\) input on a patch with
\(L\) unit cells along each direction and \(3L^2+4L+1\) qubits.
The boundary regions \(A\) and \(C\) are the rows \(y=0\) and
\(y=2L\) in the contraction coordinates; the intervening sites
form \(B\), as in Fig.~\ref{fig:cmi-geometry}.
The logical input is fixed by the outer Majorana pairings in
Eq.~\eqref{eq:honeycomb-logical-majorana-links}.

Sequential Gaussian contraction samples the physical \(X\)
outcomes with their Born probabilities.  For each full outcome,
contraction in the order \(A,B,C\) gives \(P_{ABC}\) and
\(P_{AB}\); replaying its restriction in the order \(B,C\)
gives \(P_{BC}\) and \(P_B\).  These four weights give the
CMI estimator in Eq.~\eqref{eq:classical-x-outcome-cmi-estimator}.
The plotted quantity is \(I_X\) in bits, equal to the quantum
CMI minus the fixed logical contribution of one bit
(Appendix~\ref{app:cmi-invariance}).

Contracting only \(B\), with its outcomes drawn from \(P_B\),
leaves a pure Gaussian state on the exposed virtual modes.
Its boundary entanglement \(E_{LR}\) and conductance \(g\)
are evaluated from the same covariance matrix using
Eqs.~\eqref{eq:gaussian-mutual-information-modes}
and~\eqref{eq:paired-channel-conductance}.
CMI is sampled in separate runs of the same Born ensemble.
Error bars are standard errors of the means, with independent
batches pooled at each parameter point.

\subsection{A common transition region}

Figures~\ref{fig:cmi-scaling}, \ref{fig:conductance-scaling},
and~\ref{fig:mie-scaling-appendix} compare the angle scans and
size dependence of \(I_X\), \(g\), and \(E_{LR}\).
All three cross near \(\theta/\pi=0.09\).
Below this region they decay exponentially with \(L\);
above it they grow over the accessible sizes.
The entanglement data provide a direct link between the two
descriptions: the same boundary covariance controls both
the virtual correlations and the transmission probabilities.

To compare the crossings quantitatively, we fit
\(O=I_X,E_{LR},g\) to
\begin{equation}
 O_L(\theta)=F_O\!\left[
 (\theta/\pi-\theta_c/\pi)L^{1/\nu}\right].
 \label{eq:numerical-finite-size-scaling}
\end{equation}
The central fits use \(0.079\leq\theta/\pi\leq0.097\),
with \(L=10,20,\ldots,70\) for CMI and
\(L=10,20,\ldots,100\) for entanglement and conductance.
We minimize the variance between interpolated curves on their
common scaled interval, weighted by their statistical errors.
Points require at least \(4000\) samples and signal-to-noise
ratio at least two, with at least five angles per size curve.
The uncertainties quoted in the figure captions combine
300 parametric bootstrap fits and the spread over 24 angle
windows in quadrature.  Window dependence is the larger
contribution.  The fits support a common transition region,
while their window sensitivity limits a more precise comparison
of the critical points.

The large-angle panels also test logarithmic growth.
For each observable, the curves at
\(\theta/\pi=0.10,0.11,\ldots,0.15\) are fitted with a common
slope in \(\ln L\) and separate intercepts; the \(0.09\) curve
is fitted independently.  These fits describe the leading
finite-size trend, with substantial residual corrections.

\subsection{Comparison of decay lengths}

The localized regime allows a more direct test of the
correspondence.  Figure~\ref{fig:length-scale-comparison}
compares the decay of the Born-averaged CMI and conductance:
\begin{equation}
 I_X(L)\propto e^{-L/\xi_{\mathrm M}},
 \qquad g(L)\propto e^{-L/\xi_{\mathrm{loc}}}.
 \label{eq:numerical-exponential-fit}
\end{equation}
Using the same sizes for both fits is essential: curvature at
finite \(L\) can otherwise produce an apparent difference
between the two lengths.

We fit \(\ln O_L\) against \(L\), weighted by
\(\sigma_{\ln O}=\operatorname{SEM}(O_L)/O_L\), retaining
only sizes that pass the sampling and signal-to-noise cuts for
both observables.  At least three sizes are required.
The common windows are \(10\leq L\leq40\) at
\(\theta/\pi=0.06\) and \(10\leq L\leq80\) at \(0.07\)
and \(0.08\); the other angles use all common retained sizes.
Smaller patches, down to \(L=2\), resolve the shorter decay
lengths deep in the localized regime.
Fits with either length greater than twice the largest fitted
size are excluded from the comparison.

The two lengths track one another and grow rapidly toward
the transition.  Near threshold, the available sizes no longer
separate the decay scale cleanly from finite-size corrections.
Fit uncertainties are enlarged by
\(\sqrt{\max(1,\chi^2/\mathrm{dof})}\), but do not include
window systematics.  The numerical agreement therefore supports a common
asymptotic decay scale; it does not establish an exact equality of
finite-size estimates.

\section{Square-Lattice Correspondence and Crossover}
\label{app:square-lattice-results}

The square lattice provides a second test of the correspondence between
Markov and localization lengths.  In the localized regime, the CMI and
conductance decay on the same scale, with boundary entanglement providing
an intermediate diagnostic.  At larger angles, all three show a crossover
to slow decay or growth over the accessible sizes.  We first present this
comparison, then explain its connection to the Gaussian circuits used in
earlier decoding studies.  Recent work favors a crossover interpretation
of the apparent square-lattice transition
\cite{yan2026nonlinearsigmamodelsurface,yang2026decodingcoherenterrorstoric};
this distinction motivates our focus on the honeycomb lattice in the main
text.

\subsection{CMI and the localization length}

We use rectangular patches with $L_x=5L_y$ and write $L=L_y$.
The regions $A$ and $C$ are the bottom and top boundary rows, and $B$
contains the intervening rows.  Thus the boundary separation grows with
$L$, at fixed aspect ratio.  All entropies are measured in bits.
The CMI $I_X$ is evaluated from Born-sampled physical $X$ outcomes using
Eq.~\eqref{eq:classical-x-outcome-cmi-estimator}.  Contracting the bulk
outcomes leaves a pure Gaussian state on the open virtual boundaries;
its covariance matrix gives boundary entanglement and conductance through
Eqs.~\eqref{eq:gaussian-mutual-information-modes}
and~\eqref{eq:paired-channel-conductance}.
The square-lattice plots label the virtual boundary entropy
$S_\partial=S(R)=E_{LR}/2$ as MIE; it differs by a factor of two
from the mutual information $E_{LR}$ used in the main text.
This boundary convention retains one maximally entangled mode.  In the
notation of Appendix~\ref{app:conductance-mie-bound}, it has $\lambda=0$
and hence contributes one bit to $S_\partial$ and two Majorana transmission
channels to $g$.  We therefore compare
\begin{equation}
 \delta S_\partial=S_\partial-1,
 \qquad
 \delta g=g-2.
 \label{eq:square-offsets}
\end{equation}
These subtractions remove the fixed mode and isolate the size-dependent
part of the boundary spectrum.  We checked this mode by Gaussian
contraction for $108$ sampled bulk states on
$3\times3$, $5\times3$, and $5\times5$ patches at three angles; each state
was pure and retained exactly this pair of zero eigenvalues of
$i\Gamma_{RR}$.

The square-lattice sampler averages over the corner degrees of freedom,
whereas the honeycomb construction fixes their pairings to select the
logical input.  We therefore do not assume the honeycomb identification
of physical and virtual Schmidt spectra for this convention.
The Gaussian spectral bounds apply to the remaining modes, while the
comparison with CMI below is numerical.  As a small-system check, exact
enumeration on a $3\times3$ patch gives
$I(X_A:X_C|b)\leq S_\partial(b)-1$ for every bulk record at
$\theta/\pi=0.03,0.07,0.12$; this check does not establish the subtracted
CMI bound for arbitrary sizes.

\begin{figure}[tbp]
 \centering
 \includegraphics[width=\columnwidth]{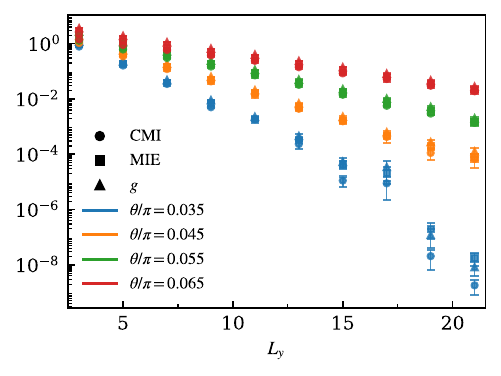}
 \caption{\justifying
 Square-lattice $I_X$ (circles), excess boundary entropy
 $\delta S_\partial$ (squares, labeled MIE), and excess conductance
 $\delta g$ (triangles) for $L_y=3,5,\ldots,21$.
 Colors denote rotation angles.}
 \label{fig:square-joint-localized}
\end{figure}

\begin{figure}[tbp]
 \centering
 \includegraphics[width=\columnwidth]{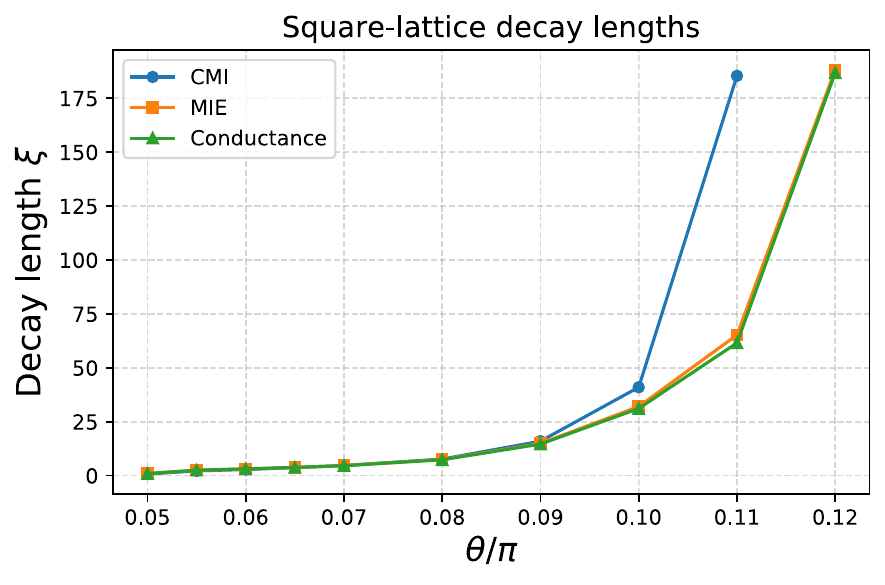}
 \caption{\justifying
 Square-lattice decay lengths for $I_X$, excess boundary entropy (MIE),
 and excess conductance, extracted using Eq.~\eqref{eq:square-cmi-fit}.}
 \label{fig:square-length-results}
\end{figure}

Figure~\ref{fig:square-joint-localized} gives the most direct comparison:
$I_X$, $\delta S_\partial$, and $\delta g$ are evaluated on the same Born
samples and decay exponentially with almost coincident slopes.  The
low-angle scan uses $L=3,5,\ldots,31$ and
$\theta/\pi=0.030,0.035,\ldots,0.070$; the displayed selected-angle
curves extend through $L=21$.  Error bars are standard errors
of the means.  For $O\in\{I_X,\delta S_\partial,\delta g\}$, we extract
decay lengths from
\begin{equation}
 \ln O(L,\theta)=a_O(\theta)-L/\xi_O(\theta).
 \label{eq:square-cmi-fit}
\end{equation}
Figure~\ref{fig:square-length-results} shows that the resulting lengths
track one another and grow rapidly with angle.  Their agreement is
clearest where exponential decay is resolved; near the crossover the
fits give effective lengths over a limited size range.

The physical interpretation is the same as on the honeycomb lattice:
CMI measures the residual dependence between the two boundaries after
conditioning on the bulk, while conductance measures transmission
between them.  The boundary entanglement relates these two descriptions,
as discussed in Appendix~\ref{app:gaussian-information-transport}.
The square-lattice data support a common decay scale without requiring
a sharp transition at larger angles.

\subsection{Connection to monitored Majorana circuits}
\label{app:gaussian-circuit}

\begin{figure}[tbp]
 \centering
 \includegraphics[width=0.85\columnwidth]{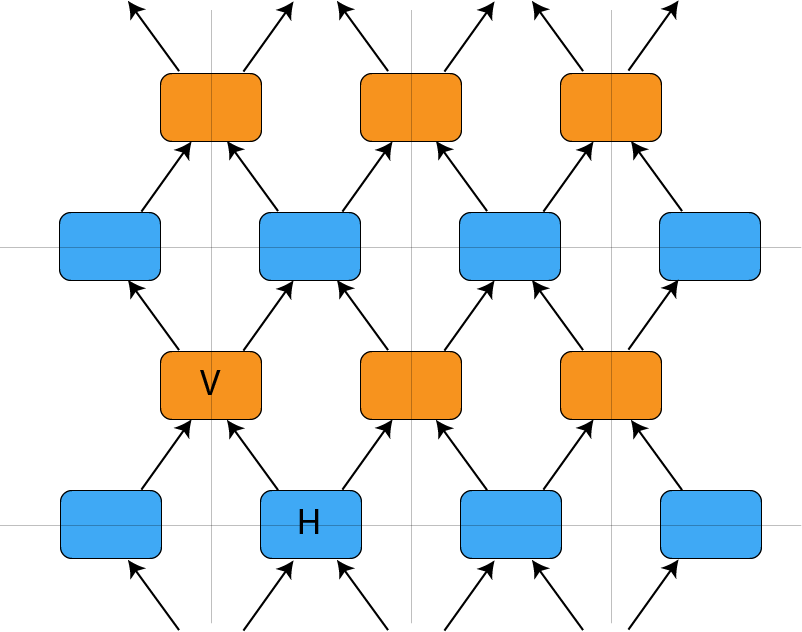}
 \caption{\justifying
 Monitored Majorana circuit from the square-lattice tensor network.
 Evolution follows the upward arrows. Blue gates $H_m$ combine parity
 measurements and Majorana exchanges; orange gates $V_m$ are unitary
 rotations. Outcomes carry Born weights.}
 \label{fig:square-gaussian-circuit}
\end{figure}

The circuit representation explains why this comparison also probes the
physics of the square-lattice decoding problem.  Slicing the tensor
network after a $45^\circ$ rotation turns its contraction into a
one-dimensional evolution of virtual Majoranas
\cite{Negari_2024}.  A physical $X$ outcome $m=\pm1$ selects a local
gate, and the two tensor sublattices give alternating layers
(Fig.~\ref{fig:square-gaussian-circuit}).  For
$0<\theta<\pi/4$, define
\begin{equation}
 \alpha_\theta=-\frac12\ln\tan\theta.
 \label{eq:alpha-theta}
\end{equation}
In the Majorana convention of Appendix~\ref{app:gaussian-information-transport},
the two gates are, up to scalar normalization,
\begin{align}
 H_m&\propto
 e^{i m\alpha_\theta\gamma_j\gamma_{j+1}}\,
 e^{m\pi\gamma_j\gamma_{j+1}/4},
 \label{eq:first-sublattice-gate}\\
 V_m&\propto
 e^{[-\theta+(1-m)\pi/4]\gamma_j\gamma_{j+1}}.
 \label{eq:second-sublattice-gate}
\end{align}
The first factor of $H_m$ weakly measures the local fermion parity
$i\gamma_j\gamma_{j+1}$; its second factor exchanges the two Majoranas.
The gate $V_m$ is unitary.  At small $\theta$ the parity measurement is
strong and suppresses propagation through the bulk.  As $\theta$ increases,
the measurement weakens, allowing much larger correlation lengths.
At $\theta=\pi/4$, $\alpha_\theta=0$ and the gates reduce to unitary
Majorana exchanges, giving the special ballistic limit.

These are the local gates underlying the complex random-bond Ising
representation of Ref.~\cite{Venn_2023} and the monitored-circuit
formulation of Ref.~\cite{yang2026decodingcoherenterrorstoric}.
For example, the Ising coupling satisfies
$e^{2J}=i\tan\theta$, so
$J=-\alpha_\theta+i\pi/4$ reproduces $H_m$; the dual coupling
$-\tfrac12\ln\tanh(mJ)$ gives $V_m$ after matching bond orientations
and outcome conventions.  Scalar normalizations cancel from a normalized
covariance matrix, but must be retained in the Born weights.
With those weights, these constructions describe the same bulk
coherent-error problem.  This common circuit is why CMI and the decoding
network are expected to follow the same bulk transition or crossover.
Different boundaries and aspect ratios can still shift their apparent
crossings at finite size.

\subsection{Finite-size crossover and the choice of lattice}

\begin{figure}[tbp]
 \centering
 \includegraphics[width=\columnwidth]{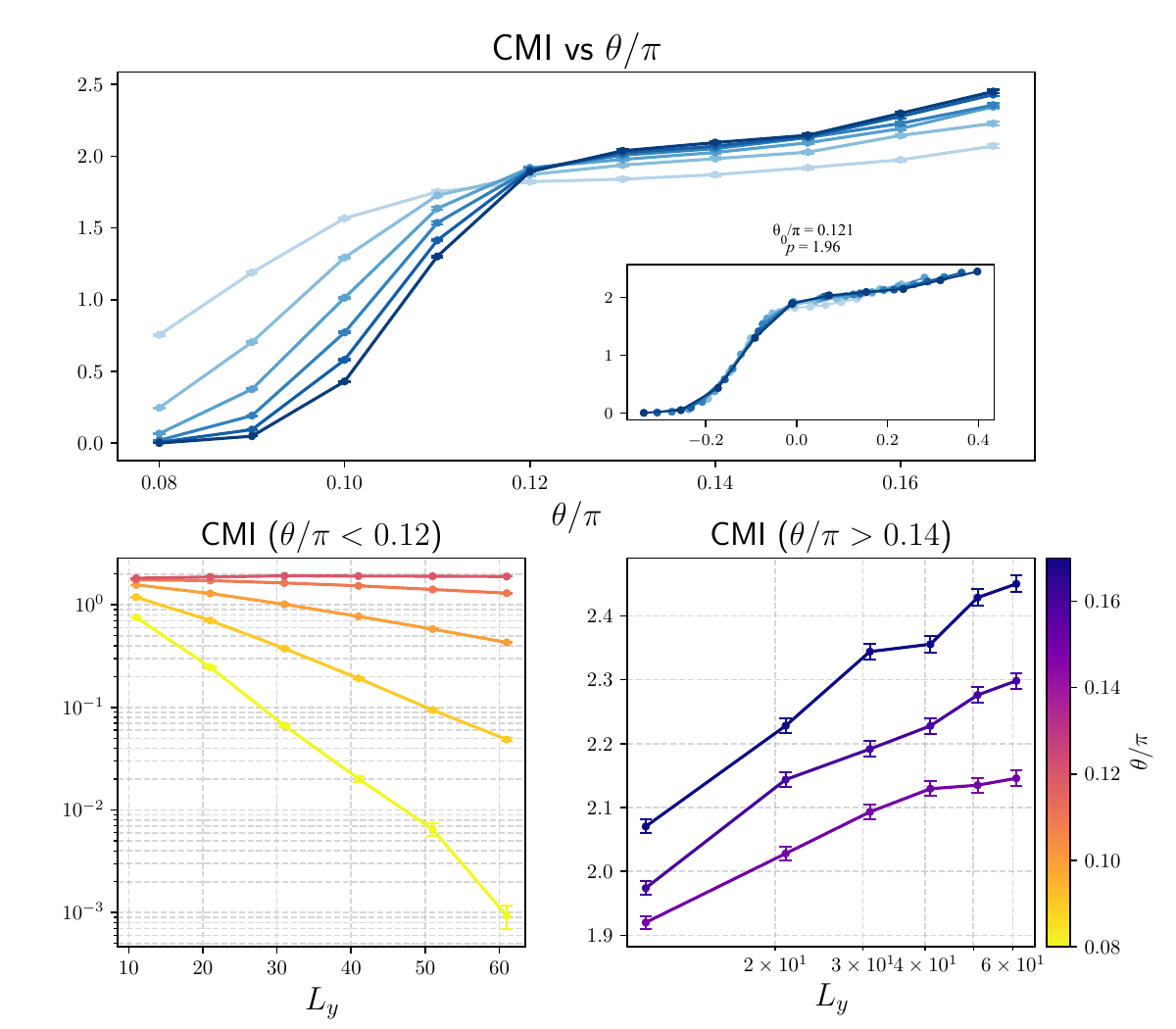}
 \caption{\justifying
 Square-lattice excess CMI $I_X$ for $L_x=5L_y$ and
 $L_y=11,21,\ldots,61$ (light to dark blue). Top: angle scan with a
 finite-size collapse in the inset. Bottom: size dependence at smaller
 (left) and larger (right) angles.}
 \label{fig:square-cmi-results}
\end{figure}

\begin{figure}[tbp]
 \centering
 \includegraphics[width=\columnwidth]{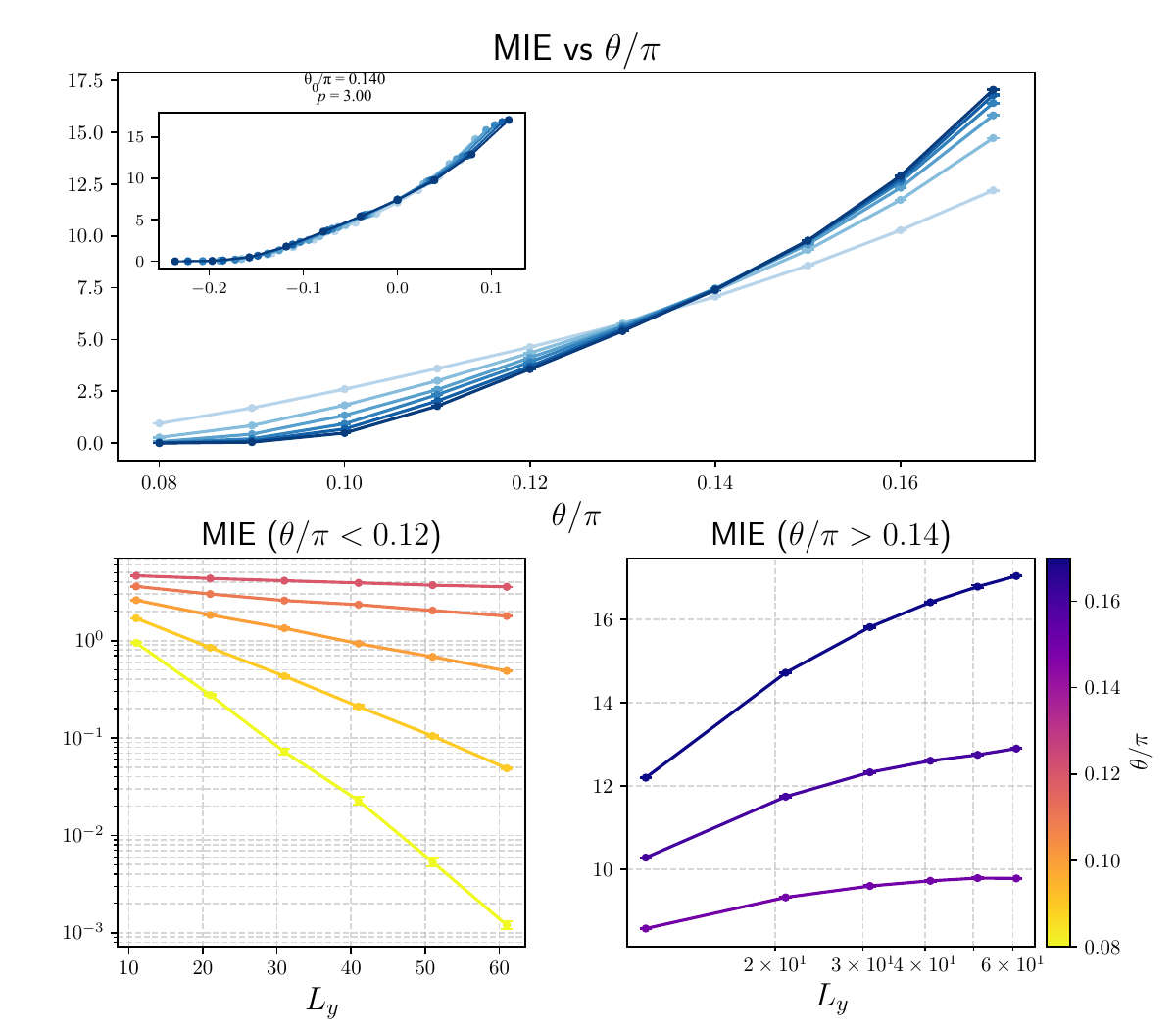}
 \caption{\justifying
 Square-lattice excess boundary entropy
 $\delta S_\partial=S_\partial-1$, labeled MIE. Layout and sizes follow
 Fig.~\ref{fig:square-cmi-results}, with a finite-size collapse in the inset.}
 \label{fig:square-mie-results}
\end{figure}

\begin{figure}[tbp]
 \centering
 \includegraphics[width=\columnwidth]{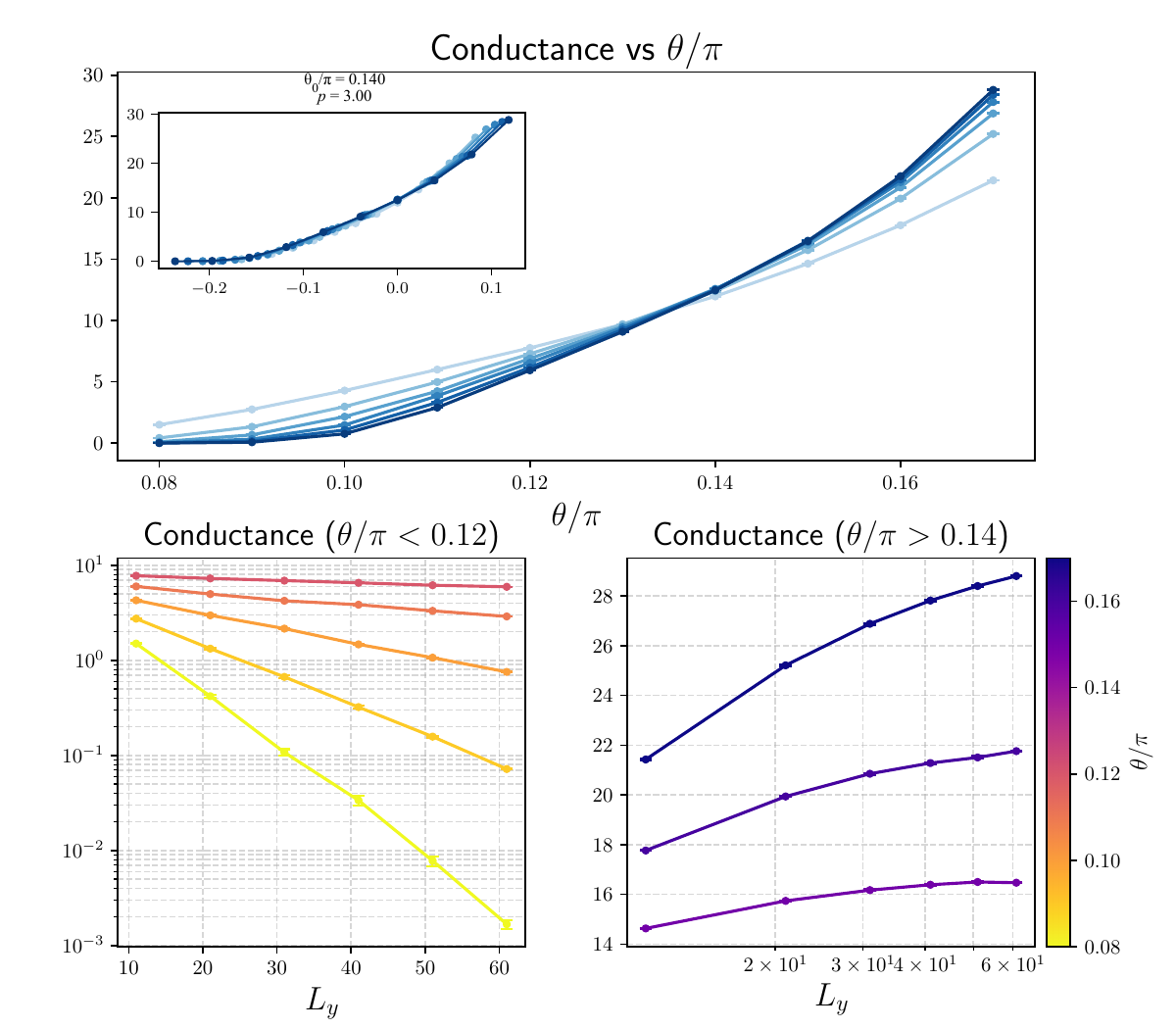}
 \caption{\justifying
 Square-lattice excess conductance $\delta g=g-2$ in the Majorana-channel
 convention. Layout and sizes follow Fig.~\ref{fig:square-cmi-results},
 with a finite-size collapse in the inset.}
 \label{fig:square-conductance-results}
\end{figure}

Figures~\ref{fig:square-cmi-results}--\ref{fig:square-conductance-results}
show the wider angle scans.  The CMI curves cross broadly near
$\theta/\pi\simeq0.12$, whereas the boundary-entanglement and conductance
curves cross near $0.14$.  On the displayed sizes, each observable changes
from decay at smaller angles to growth at larger angles.  The CMI
therefore exhibits the same qualitative crossover seen in transport,
although their finite-size crossings do not coincide precisely.

Earlier transport calculations interpreted the square-lattice behavior
as an insulator--metal transition~\cite{Venn_2023}.
Recent field-theory and circuit analyses instead find that the metallic
fixed point is unstable for the Born ensemble relevant to optimal
Pauli decoding
\cite{yan2026nonlinearsigmamodelsurface,yang2026decodingcoherenterrorstoric}.
The flow away from it can be extremely slow, producing a large crossover
length and apparent metallic behavior on accessible systems.  For uniform
rotation angles, these works favor eventual localization for
$\theta<\pi/4$.  Our CMI data are consistent with this picture: the
change in size dependence can reflect a length exceeding the available
boundary separations.  They do not resolve the eventual large-size
turnover predicted by the crossover interpretation.
Long crossovers also arise in monitored number-conserving fermion
chains~\cite{PoboikoEtAl2023ProjectiveMeasurements}.

The collapse insets provide a compact comparison of the finite-size
curves.  The rescaled coordinate is
$(\theta/\pi-\theta_0/\pi)L^{1/p}$, with reference angle $\theta_0$
and effective exponent $p$. We use $\theta_0/\pi=0.121$, $p=1.96$ for
CMI and $\theta_0/\pi=0.140$, $p=3.00$ for boundary entropy and conductance.  No statistical uncertainties
were established for these values, and the collapses alone do not
distinguish a transition from a long crossover.

This distinction explains our choice of the honeycomb lattice for the
main analysis.  Its nonbipartite syndrome lattice gives class DIII and
permits a stable thermal metal in the Born ensemble, unlike the class-D
square-lattice network
\cite{yan2026nonlinearsigmamodelsurface,yang2026decodingcoherenterrorstoric},
providing a cleaner setting for studying a genuine transition.
The square lattice remains a useful independent example: the Markov and
localization lengths agree where they can be resolved, and CMI follows
the same crossover to correlations extending beyond the sizes studied.

\section{Logical Fidelity and Local Conversion error}
\label{app:logical-fidelity-conversion}

We bound the accuracy of local conversion between two noisy
code states using the fidelity between opposite logical inputs.
The conversion is required to prepare only the logical-$+$
state. Locality then constrains its action on the logical-$-$
state, allowing us to compare the fidelities at the two angles.

Let
$|\overline{\pm}\rangle
=(|\overline0\rangle\pm|\overline1\rangle)/\sqrt2$,
and define
\begin{equation}
 \begin{aligned}
 \rho_{\theta,\pm}
 =\mathcal N_\theta
 \bigl(|\overline{\pm}\rangle
       \langle\overline{\pm}|\bigr),\quad
 \mathcal F_\pm(\theta,L)
 =F(\rho_{\theta,+},\rho_{\theta,-}).
 \end{aligned}
 \label{eq:logical-plus-minus-fidelity}
\end{equation}
Here $\mathcal N_\theta$ consists of coherent $Z$ rotations
followed by syndrome dephasing, and
$F(\rho,\sigma)=\|\sqrt\rho\sqrt\sigma\|_1$
is the unsquared fidelity. The noise preserves logical $Z$,
so every equivalent logical-$Z$ representative satisfies
\[
 \rho_{\theta,-}
 =\overline Z\rho_{\theta,+}\overline Z.
\]

We use a planar patch with three parallel logical-$Z$
representatives separated by distances proportional to $L$.
A strip of width $o(L)$ around each representative supports
no logical $X$ or $Y$ operator. Local channels have product
input ancillas, and their light-cone radius includes all
classical communication.

\begin{lemma}[Logical fidelity bounds local conversion error]
\label{lem:logical-fidelity-conversion}
Suppose a local channel $\mathcal C_L$ satisfies
\[
 \|\mathcal C_L(\rho_{\theta,+})-\rho_{\phi,+}\|_1
 \leq\epsilon_L<\frac12,
\] 
and $\mathcal C_L\circ\mathcal N_\theta$ has light-cone
radius $o(L)$. Then, for sufficiently large $L$,
\begin{equation}
 \mathcal F_\pm(\theta,L)-\mathcal F_\pm(\phi,L)
 \leq(2+\sqrt2)\sqrt{\epsilon_L}.
 \label{eq:logical-fidelity-conversion-bound}
\end{equation}
\end{lemma}

\begin{proof}
Suppress the size index and write
$\tau_\pm=\mathcal C_L(\rho_{\theta,\pm})$.
We establish three bounds and then combine them.

\emph{Step 1: Apply fidelity monotonicity.}
Applying the same channel to two states cannot decrease
their fidelity. Therefore,
\begin{equation}
 \mathcal F_\pm(\theta,L)
 \leq F(\tau_+,\tau_-).
 \label{eq:logical-fidelity-data-processing}
\end{equation}
This step holds for any channel. Locality is needed in the
next step, which relates $\tau_-$ to a logical $Z$ applied
to $\tau_+$.

\emph{Step 2: Control the logical partner using locality.}
Let $V$ be a Stinespring isometry for the map from the clean
logical input qubit to the output of
$\mathcal C_L\circ\mathcal N_\theta$.
We write $Z$ for the Pauli operator on this input qubit
and $E$ for the discarded environment.

Choose two distant output logical-$Z$ representatives,
$Z_1,Z_2$. Their backward light cones lie in disjoint strips,
with a third logical-$Z$ representative outside their union.
The effective measurements on the input qubit have the form
\[
 T_i:=V^\dagger(Z_i\otimes I_E)V=a_iI+b_iZ.
\]
Moreover,
\begin{equation}
 V^\dagger(Z_1Z_2\otimes I_E)V=T_1T_2.
 \label{eq:logical-strip-factorization}
\end{equation}

These statements follow from the structure of the clean
code on the two strips. Conditional on a fixed logical-$Z$
eigenvalue, their reduced state factorizes. Indeed, a
stabilizer supported on their union splits into two Pauli
factors. Since no local check meets both strips, each factor
commutes with every check. Each strip supports only logical
$I$ or $Z$, so both factors have definite eigenvalues in
the chosen logical-$Z$ state. The stabilizer expansion of
the reduced state therefore factorizes. The third
logical-$Z$ representative removes all off-diagonal
logical matrix elements on the strip union. Together with
the product input ancillas and disjoint backward light
cones, these facts give
Eq.~\eqref{eq:logical-strip-factorization}.

Each $T_i$ is a Hermitian contraction, so its eigenvalues
$a_i\pm b_i$ lie in $[-1,1]$. Equivalently,
\[
 |a_i|+|b_i|\leq1.
\]
In the target $\rho_{\phi,+}$, the two logical charges
agree perfectly and each is unbiased:
\[
 \Tr(Z_1\rho_{\phi,+})=0,
 \qquad
 \Tr(Z_1Z_2\rho_{\phi,+})=1.
\]
The preparation error consequently gives
\[
 |a_1|\leq\epsilon,
 \qquad
 a_1a_2+b_1b_2\geq1-\epsilon.
\]
Combining these estimates,
\[
 \begin{aligned}
 1-\epsilon
 &\leq a_1a_2+b_1b_2\\
 &\leq
 \max\{|a_1|,|b_1|\}(|a_2|+|b_2|)\\
 &\leq\max\{|a_1|,|b_1|\}.
 \end{aligned}
\]
Since $|a_1|\leq\epsilon<1-\epsilon$, we obtain
$|b_1|\geq1-\epsilon$.
Thus the output logical charge almost perfectly follows
the input logical charge, possibly with its sign reversed.

Set $s=\operatorname{sign}(b_1)$. Since
$\langle -|T_1|+\rangle=b_1$,
\begin{equation}
 \begin{aligned}
 \bigl\|V|-\rangle
 -s(Z_1\otimes I_E)V|+\rangle\bigr\|^2=2-2|b_1|\leq2\epsilon.
 \end{aligned}
 \label{eq:logical-twist-purification-bound}
\end{equation}
The two vectors in this comparison purify $\tau_-$ and
$Z_1\tau_+Z_1$, respectively.

We now use a standard consequence of Uhlmann's theorem:
replacing a purification by another normalized vector
changes its fidelity with a fixed state by at most the
distance between the two vectors. Applying this property
to Eq.~\eqref{eq:logical-twist-purification-bound} gives
\begin{equation}
 F(\tau_+,\tau_-)
 \leq F(\tau_+,Z_1\tau_+Z_1)+\sqrt{2\epsilon}.
 \label{eq:logical-fidelity-partner-bound}
\end{equation}
This is the required control of the second input. It does
not require covariance of the conversion channel.

\emph{Step 3: Compare with the target using continuity.}
Fidelity obeys the one-argument continuity bound
\[
 |F(\rho,\omega)-F(\sigma,\omega)|
 \leq\sqrt{\|\rho-\sigma\|_1}.
\]
By assumption, $\tau_+$ is within trace-norm distance
$\epsilon$ of $\rho_{\phi,+}$. Unitary conjugation preserves
this distance, and
$Z_1\rho_{\phi,+}Z_1=\rho_{\phi,-}$.
Thus $Z_1\tau_+Z_1$ is also within distance $\epsilon$
of $\rho_{\phi,-}$. Applying continuity to each argument
yields
\begin{equation}
 F(\tau_+,Z_1\tau_+Z_1)
 \leq\mathcal F_\pm(\phi,L)+2\sqrt\epsilon.
 \label{eq:logical-fidelity-target-bound}
\end{equation}

\emph{Combining the bounds.}
Equations~\eqref{eq:logical-fidelity-data-processing},
\eqref{eq:logical-fidelity-partner-bound},
and~\eqref{eq:logical-fidelity-target-bound} give
\[
 \begin{aligned}
 \mathcal F_\pm(\theta,L)
 &\leq F(\tau_+,\tau_-)\\
 &\leq F(\tau_+,Z_1\tau_+Z_1)+\sqrt{2\epsilon}\\
 &\leq\mathcal F_\pm(\phi,L)
       +(2+\sqrt2)\sqrt\epsilon,
 \end{aligned}
\]
which proves the lemma.
\end{proof}

For conversion in both directions, let $\epsilon_L<1/2$
bound the larger preparation error. Applying
Lemma~\ref{lem:logical-fidelity-conversion} in each direction
gives
\begin{equation}
 \epsilon_L\geq
 \frac{
 |\mathcal F_\pm(\theta,L)-\mathcal F_\pm(\phi,L)|^2
 }{(2+\sqrt2)^2}.
 \label{eq:two-way-logical-fidelity-error}
\end{equation}
An independently established fidelity gap of order
$(\ln L)^{-b}$ therefore requires a preparation error
at least of order $(\ln L)^{-2b}$, excluding two-way
conversion with inverse-polynomial or superpolynomial
accuracy. A nonzero limiting fidelity gap excludes
two-way conversion with any vanishing error.
These conclusions concern the logical-$+$ state families
defined above; the required asymptotic fidelity gap remains
a separate question.

\clearpage

\end{document}

%% file: figures/Table.tex
\begin{table*}[t]
\caption{Summary of the two regimes of the honeycomb surface code
under coherent $Z$ errors, with $\theta_c\simeq0.09\pi$. }
\label{tab:phase-summary}
\centering

\begingroup
\phaseTableSetup

\phaserow{\phaseHeader}{
  \textbf{Property}
  & \phasecell{
      \textbf{Topological code phase}\\[1pt]
      $0\leq\theta<\theta_c$}
  & \phasecell{
      \textbf{Critical LRE regime}\\[1pt]
      $\theta_c<\theta<\pi/4$}
  \\
}

\phaserow{\phaseMemory}{
  Quantum memory
  & \phasecell{
      Full quantum memory\\[1pt]
      {\footnotesize Admits local decoding}}
  & \phasecell{
      Partial quantum memory\\[1pt]
      {\footnotesize Global decoding required}}
  \\
}

\phaserow{\phaseEntanglement}{
  Long-range entanglement
  & \phasecell{
      $\rho\not\approx\sum_i p_i\,|\mathrm{Trivial}_i\rangle\langle\mathrm{Trivial}_i|$\\[1pt]
      {\footnotesize Every logical input}}
  & \phasecell{
      $\rho\not\approx\sum_i p_i\,|\mathrm{Trivial}_i\rangle\langle\mathrm{Trivial}_i|$\\[1pt]
      {\footnotesize Generic logical inputs}}
  \\
}

\phaserow{\phaseCMI}{
  \phasecell{
    Conditional mutual information\\[1pt]
    and conductance}
  & \phasecell{
      $\mathrm{CMI},\,g\sim e^{-L/\xi_{\mathrm M}}$\\[1pt]
      {\footnotesize Finite Markov and localization lengths}}
  & \phasecell{
      $\mathrm{CMI},\,g\sim\ln L$\\[1pt]
      {\footnotesize Divergent Markov and localization lengths}}
  \\
}

\phaserow{\phaseSymmetry}{
  \phasecell{
    One-form strong-to-weak\\[1pt]
    symmetry breaking}
  & \phasecell{
      Exponential fidelity correlator decay\\[1pt]
      $\mathcal F_Z(L)\sim e^{-L/\xi_F}$\\[1pt]
      {\footnotesize Exact $Z$; emergent $X$}}
  & \phasecell{
      Power-law fidelity correlator decay\\[1pt]
      $\mathcal F_Z(L)\sim L^{-\alpha}\ln(L)^{-\beta}$\\[1pt]
      {\footnotesize Exact $Z$; no localized strong $X$ partner}}
  \\
}

\phaserow{\phaseCircuit}{
  \phasecell{
    Circuit complexity\\[1pt]
    {\footnotesize $r_L$-range circuits}}
  & \phasecell{
      All points two-way connected\\[1pt]
      to the clean surface code\\[1pt]
      and each other with complexity\\[1pt]
      $r_L=\operatorname{polylog}L$}
  & \phasecell{
      No two-way connection\\[1pt]
      to the clean surface code\\[1pt]
      with complexity $r_L=o(L)$}
  \\
}

\sbox{\phaseTable}{%
  \vbox{\offinterlineskip
    \hbox{\usebox{\phaseHeader}}
    \hbox{\usebox{\phaseMemory}}
    \hbox{\usebox{\phaseEntanglement}}
    \hbox{\usebox{\phaseCMI}}
    \hbox{\usebox{\phaseSymmetry}}
    \hbox{\usebox{\phaseCircuit}}
  }}

\begin{tikzpicture}[x=1pt,y=1pt]
  \pgfmathsetmacro{\W}{\the\textwidth}
  \pgfmathsetmacro{\H}{\phaseheight{\phaseTable}}
  \pgfmathsetmacro{\ya}{-\phaseheight{\phaseHeader}}
  \pgfmathsetmacro{\yb}{\ya-\phaseheight{\phaseMemory}}
  \pgfmathsetmacro{\yc}{\yb-\phaseheight{\phaseEntanglement}}
  \pgfmathsetmacro{\yd}{\yc-\phaseheight{\phaseCMI}}
  \pgfmathsetmacro{\ye}{\yd-\phaseheight{\phaseSymmetry}}

  \path[use as bounding box]
    (0,0) rectangle (\W,-\H);

  \shade[top color=phaseBlue!68!white,bottom color=phaseBlue]
    ({0.29*\W},0) rectangle ({0.65*\W},-\H);

  \shade[top color=phaseCream!68!white,bottom color=phaseCream]
    ({0.65*\W},0) rectangle (\W,-\H);

  \foreach \a/\b in {0.29/0.65,0.65/1}
    \draw[white,opacity=0.7,line width=0.55pt]
      ({\a*\W+0.7},{-\H+0.7})
      --({\a*\W+0.7},-0.7)
      --({\b*\W-0.7},-0.7);

  \node[anchor=north west,inner sep=0pt,outer sep=0pt]
    at (0,0) {\usebox{\phaseTable}};

  \draw[black!75,line width=0.65pt]
    (0,0)--(\W,0)
    (0,-\H)--(\W,-\H);

  \draw[black!75,line width=0.65pt]
    (0,0)--(0,-\H)
    (\W,0)--(\W,-\H);

  \draw[black!65,line width=0.45pt]
    (0,\ya)--(\W,\ya);

  \foreach \y in {\yb,\yc,\yd,\ye}
    \draw[phaseRule,line width=0.3pt]
      (0,\y)--(\W,\y);

  \foreach \x in {0.29,0.65}
    \draw[phaseRule,line width=0.35pt]
      ({\x*\W},0)--({\x*\W},-\H);
\end{tikzpicture}

\endgroup
\end{table*}

%% file: figures/boundary_cmi_geometry.tex
\begin{figure}[tbp]
    \centering
    \begin{minipage}[b]{0.47\columnwidth}
        \centering
        \includegraphics[height=1.34in]{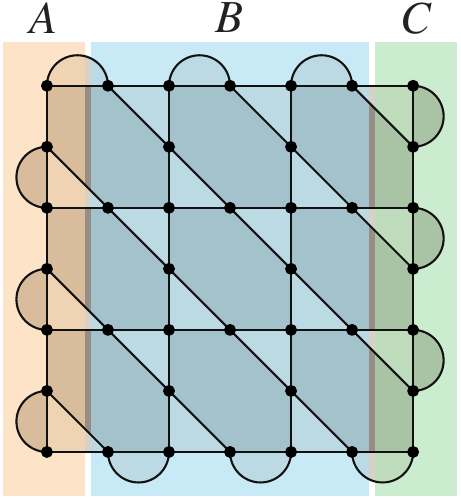}\\[-1mm]
        {\small (a)}
    \end{minipage}
    \hspace{0.015\columnwidth}
    \begin{minipage}[b]{0.47\columnwidth}
        \centering
        \begin{tikzpicture}
            \node[anchor=south west,inner sep=0] (tn) at (0,0)
            {\includegraphics[height=1.34in,trim=0 36 374 0,clip]
            {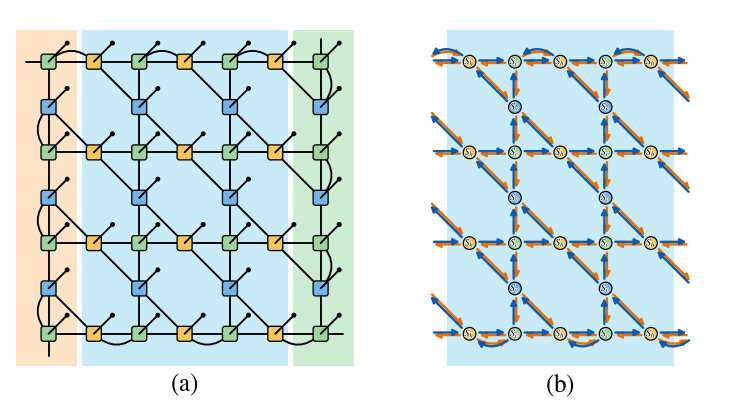}};
            \begin{scope}[x={(tn.south east)},y={(tn.north west)}]
                \node[font=\small] at (0.125,0.97) {$A$};
                \node[font=\small] at (0.50,0.97) {$B$};
                \node[font=\small] at (0.875,0.97) {$C$};
            \end{scope}
        \end{tikzpicture}\\[-1mm]
        {\small (b)}
    \end{minipage}
    \caption{\justifying
 Boundary CMI geometry. (a) One-row regions $A,C$ separated by the bulk
 strip $B$, with transverse extents and separation proportional to $L$.
 (b) Corresponding tensor network, with one physical qubit and four
 virtual Majorana modes per local tensor.}
    \label{fig:cmi-geometry}
\end{figure}

%% file: figures/gaussian_transport.tex
\begin{figure}[tbp]
\centering
\begin{tikzpicture}[gaussian transport]
  \coordinate (q) at (2.00,7.00);
  \draw[bond] ($(q)+(-1.20,0)$)--($(q)+(1.20,0)$);
  \draw[bond] ($(q)+(0,-1.20)$)--($(q)+(0,1.20)$);
  \draw[draw=violet!80!black,line width=0.95pt]
    ($(q)+(0.20,0.20)$)--($(q)+(0.88,0.88)$);
  \node[font=\scriptsize,anchor=south west,text=violet!80!black]
    at ($(q)+(0.80,0.82)$) {$\bra{x_i}$};
  \node[tensor,minimum size=6.4mm] at (q) {$T_i$};
  \node[font=\scriptsize,anchor=south west] at (-0.72,8.05) {(a)};

  \coordinate (s) at (7.60,7.00);
  \draw[fline] ($(s)+(-1.20,0.10)$)--($(s)+(0,0.10)$);
  \draw[forward] ($(s)+(-1.08,0.10)$)--($(s)+(-0.72,0.10)$);
  \draw[bline] ($(s)+(-1.20,-0.10)$)--($(s)+(0,-0.10)$);
  \draw[backward] ($(s)+(-0.72,-0.10)$)--($(s)+(-1.08,-0.10)$);
  \draw[fline] ($(s)+(0,0.10)$)--($(s)+(1.20,0.10)$);
  \draw[forward] ($(s)+(0.72,0.10)$)--($(s)+(1.08,0.10)$);
  \draw[bline] ($(s)+(0,-0.10)$)--($(s)+(1.20,-0.10)$);
  \draw[backward] ($(s)+(1.08,-0.10)$)--($(s)+(0.72,-0.10)$);
  \draw[fline] ($(s)+(0.10,1.20)$)--($(s)+(0.10,0)$);
  \draw[forward] ($(s)+(0.10,1.08)$)--($(s)+(0.10,0.72)$);
  \draw[bline] ($(s)+(-0.10,1.20)$)--($(s)+(-0.10,0)$);
  \draw[backward] ($(s)+(-0.10,0.72)$)--($(s)+(-0.10,1.08)$);
  \draw[fline] ($(s)+(0.10,0)$)--($(s)+(0.10,-1.20)$);
  \draw[forward] ($(s)+(0.10,-0.72)$)--($(s)+(0.10,-1.08)$);
  \draw[bline] ($(s)+(-0.10,0)$)--($(s)+(-0.10,-1.20)$);
  \draw[backward] ($(s)+(-0.10,-1.08)$)--($(s)+(-0.10,-0.72)$);
  \node[scatter] at (s) {$\mathsf S_i$};
  \node[font=\scriptsize,anchor=south west] at (5.48,8.05) {(b)};

  \fill[aregion] (-0.46,-0.46) rectangle (0.46,4.46);
  \fill[bregion] (0.54,-0.46) rectangle (3.46,4.46);
  \fill[cregion] (3.54,-0.46) rectangle (4.46,4.46);
  \node[font=\normalsize] at (0,4.65) {$A$};
  \node[font=\normalsize] at (2,4.65) {$B$};
  \node[font=\normalsize] at (4,4.65) {$C$};
  \foreach \yy in {0,...,4}{
    \foreach \xx in {0,...,4}{
      \ifnum\xx<4 \draw[bond] (\xx,\yy)--(\xx+1,\yy); \fi
      \ifnum\yy<4 \draw[bond] (\xx,\yy)--(\xx,\yy+1); \fi
    }
  }
  \foreach \yy in {0,...,4}{
    \foreach \xx in {0,...,4}{
      \ifnum\xx=0
        \draw[draw=violet!80!black,line width=0.8pt] (\xx,\yy)--(-0.32,{\yy+0.30});
        \node[tensor,fill=aregionbase] at (\xx,\yy) {};
      \else\ifnum\xx=4
        \draw[draw=violet!80!black,line width=0.8pt] (\xx,\yy)--(4.32,{\yy+0.30});
        \node[tensor,fill=cregionbase] at (\xx,\yy) {};
      \else
        \node[tensor] at (\xx,\yy) {};
      \fi\fi
    }
  }
  \node[font=\scriptsize,anchor=south west] at (-0.72,4.78) {(c)};

  \fill[bregion] (6.14,-0.46) rectangle (9.06,4.46);
  \node[font=\scriptsize,anchor=south west] at (5.48,4.78) {(d)};
  \node[font=\normalsize] at (7.6,4.65) {$B$};
  \foreach \yy in {0,...,4}{
    \foreach \xa/\xb in {5.78/6.6,6.6/7.6,7.6/8.6,8.6/9.42}{
      \pgfmathsetmacro{\xone}{0.66*\xa+0.34*\xb}
      \pgfmathsetmacro{\xtwo}{0.34*\xa+0.66*\xb}
      \draw[fline] (\xa,{\yy+0.10})--(\xb,{\yy+0.10});
      \draw[forward] (\xone,{\yy+0.10})--(\xtwo,{\yy+0.10});
      \draw[bline] (\xa,{\yy-0.10})--(\xb,{\yy-0.10});
      \draw[backward] (\xtwo,{\yy-0.10})--(\xone,{\yy-0.10});
    }
  }
  \foreach \xx in {6.6,7.6,8.6}{
    \foreach \yy in {0,...,3}{
      \pgfmathsetmacro{\yone}{\yy+0.66}
      \pgfmathsetmacro{\ytwo}{\yy+0.34}
      \draw[fline] ({\xx+0.10},{\yy+1})--({\xx+0.10},\yy);
      \draw[forward] ({\xx+0.10},\yone)--({\xx+0.10},\ytwo);
      \draw[bline] ({\xx-0.10},\yy)--({\xx-0.10},{\yy+1});
      \draw[backward] ({\xx-0.10},\ytwo)--({\xx-0.10},\yone);
    }
  }
  \foreach \xx in {6.6,7.6,8.6}{
    \foreach \yy in {0,...,4}{
      \node[tensor,minimum size=3.2mm] at (\xx,\yy) {};
    }
  }
  \node[font=\scriptsize,left] at (5.72,2) {$L$};
  \node[font=\scriptsize,right] at (9.48,2) {$R$};
\end{tikzpicture}
\caption{\justifying
 Tensor-network mapping to Gaussian transport.
 (a) Fixing the physical outcome $x_i$ selects $T_i(x_i)$.
 (b) Corresponding scattering matrix.
 (c) Physical boundary state on $A,C$ after measuring $B$.
 (d) Virtual boundary state on $L,R$ and its two-terminal scattering
 representation. Arrows denote scattering channels.}
\label{fig:random-gaussian-ensemble}
\end{figure}

%% file: ArXiv_v1.bbl
\begin{thebibliography}{85}%
\makeatletter
\providecommand \@ifxundefined [1]{%
 \@ifx{#1\undefined}
}%
\providecommand \@ifnum [1]{%
 \ifnum #1\expandafter \@firstoftwo
 \else \expandafter \@secondoftwo
 \fi
}%
\providecommand \@ifx [1]{%
 \ifx #1\expandafter \@firstoftwo
 \else \expandafter \@secondoftwo
 \fi
}%
\providecommand \natexlab [1]{#1}%
\providecommand \enquote  [1]{``#1''}%
\providecommand \bibnamefont  [1]{#1}%
\providecommand \bibfnamefont [1]{#1}%
\providecommand \citenamefont [1]{#1}%
\providecommand\href[2]{#2}%
\providecommand\url[1]{#1}%
\providecommand\Eprint[2][]{#2}%
\providecommand \doibase [0]{https://doi.org/}%
\providecommand \selectlanguage [0]{\@gobble}%
\providecommand \bibinfo  [0]{\@secondoftwo}%
\providecommand \bibfield  [0]{\@secondoftwo}%
\providecommand \translation [1]{[#1]}%
\providecommand \BibitemOpen [0]{}%
\providecommand \bibitemStop [0]{}%
\providecommand \bibitemNoStop [0]{.\EOS\space}%
\providecommand \EOS [0]{\spacefactor3000\relax}%
\providecommand \BibitemShut  [1]{\csname bibitem#1\endcsname}%
\let\auto@bib@innerbib\@empty
\bibitem [{\citenamefont {Dennis}\ \emph {et~al.}(2002)\citenamefont {Dennis}, \citenamefont {Kitaev}, \citenamefont {Landahl},\ and\ \citenamefont {Preskill}}]{Dennis_2002}%
  \BibitemOpen
  \bibfield  {author} {\bibinfo {author} {\bibfnamefont {E.}~\bibnamefont {Dennis}}, \bibinfo {author} {\bibfnamefont {A.}~\bibnamefont {Kitaev}}, \bibinfo {author} {\bibfnamefont {A.}~\bibnamefont {Landahl}},\ and\ \bibinfo {author} {\bibfnamefont {J.}~\bibnamefont {Preskill}},\ }\bibfield  {title} {\bibinfo {title} {Topological quantum memory},\ }\href {https://doi.org/10.1063/1.1499754} {\bibfield  {journal} {\bibinfo  {journal} {J. Math. Phys.}\ }\textbf {\bibinfo {volume} {43}},\ \bibinfo {pages} {4452} (\bibinfo {year} {2002})}\BibitemShut {NoStop}%
\bibitem [{\citenamefont {Coser}\ and\ \citenamefont {P{\'{e}}rez-Garc{\'{i}}a}(2019)}]{Coser_2019}%
  \BibitemOpen
  \bibfield  {author} {\bibinfo {author} {\bibfnamefont {A.}~\bibnamefont {Coser}}\ and\ \bibinfo {author} {\bibfnamefont {D.}~\bibnamefont {P{\'{e}}rez-Garc{\'{i}}a}},\ }\bibfield  {title} {\bibinfo {title} {Classification of phases for mixed states via fast dissipative evolution},\ }\href {https://doi.org/10.22331/q-2019-08-12-174} {\bibfield  {journal} {\bibinfo  {journal} {Quantum}\ }\textbf {\bibinfo {volume} {3}},\ \bibinfo {pages} {174} (\bibinfo {year} {2019})}\BibitemShut {NoStop}%
\bibitem [{\citenamefont {Sang}\ \emph {et~al.}(2024)\citenamefont {Sang}, \citenamefont {Zou},\ and\ \citenamefont {Hsieh}}]{SangZouHsieh2024MixedStatePhases}%
  \BibitemOpen
  \bibfield  {author} {\bibinfo {author} {\bibfnamefont {S.}~\bibnamefont {Sang}}, \bibinfo {author} {\bibfnamefont {Y.}~\bibnamefont {Zou}},\ and\ \bibinfo {author} {\bibfnamefont {T.~H.}\ \bibnamefont {Hsieh}},\ }\bibfield  {title} {\bibinfo {title} {Mixed-state quantum phases: Renormalization and quantum error correction},\ }\href {https://doi.org/10.1103/PhysRevX.14.031044} {\bibfield  {journal} {\bibinfo  {journal} {Phys. Rev. X}\ }\textbf {\bibinfo {volume} {14}},\ \bibinfo {pages} {031044} (\bibinfo {year} {2024})}\BibitemShut {NoStop}%
\bibitem [{\citenamefont {Sang}\ \emph {et~al.}(2026)\citenamefont {Sang}, \citenamefont {Lessa}, \citenamefont {Mong}, \citenamefont {Grover}, \citenamefont {Wang},\ and\ \citenamefont {Hsieh}}]{SangEtAl2025MixedStatePhasesLocalReversibility}%
  \BibitemOpen
  \bibfield  {author} {\bibinfo {author} {\bibfnamefont {S.}~\bibnamefont {Sang}}, \bibinfo {author} {\bibfnamefont {L.~A.}\ \bibnamefont {Lessa}}, \bibinfo {author} {\bibfnamefont {R.~S.~K.}\ \bibnamefont {Mong}}, \bibinfo {author} {\bibfnamefont {T.}~\bibnamefont {Grover}}, \bibinfo {author} {\bibfnamefont {C.}~\bibnamefont {Wang}},\ and\ \bibinfo {author} {\bibfnamefont {T.~H.}\ \bibnamefont {Hsieh}},\ }\bibfield  {title} {\bibinfo {title} {Mixed-state phases from local reversibility},\ }\href {https://doi.org/10.1103/1kr5-jlcb} {\bibfield  {journal} {\bibinfo  {journal} {Phys. Rev. X}\ }\textbf {\bibinfo {volume} {16}},\ \bibinfo {pages} {041014} (\bibinfo {year} {2026})},\ \Eprint {https://arxiv.org/abs/2507.02292} {arXiv:2507.02292 [quant-ph]} \BibitemShut {NoStop}%
\bibitem [{\citenamefont {Sang}\ and\ \citenamefont {Hsieh}(2025)}]{Sang_2025}%
  \BibitemOpen
  \bibfield  {author} {\bibinfo {author} {\bibfnamefont {S.}~\bibnamefont {Sang}}\ and\ \bibinfo {author} {\bibfnamefont {T.~H.}\ \bibnamefont {Hsieh}},\ }\bibfield  {title} {\bibinfo {title} {Stability of mixed-state quantum phases via finite {Markov} length},\ }\href {https://doi.org/10.1103/PhysRevLett.134.070403} {\bibfield  {journal} {\bibinfo  {journal} {Phys. Rev. Lett.}\ }\textbf {\bibinfo {volume} {134}},\ \bibinfo {pages} {070403} (\bibinfo {year} {2025})}\BibitemShut {NoStop}%
\bibitem [{\citenamefont {Fan}\ \emph {et~al.}(2024)\citenamefont {Fan}, \citenamefont {Bao}, \citenamefont {Altman},\ and\ \citenamefont {Vishwanath}}]{Fan_2024}%
  \BibitemOpen
  \bibfield  {author} {\bibinfo {author} {\bibfnamefont {R.}~\bibnamefont {Fan}}, \bibinfo {author} {\bibfnamefont {Y.}~\bibnamefont {Bao}}, \bibinfo {author} {\bibfnamefont {E.}~\bibnamefont {Altman}},\ and\ \bibinfo {author} {\bibfnamefont {A.}~\bibnamefont {Vishwanath}},\ }\bibfield  {title} {\bibinfo {title} {Diagnostics of mixed-state topological order and breakdown of quantum memory},\ }\href {https://doi.org/10.1103/prxquantum.5.020343} {\bibfield  {journal} {\bibinfo  {journal} {PRX Quantum}\ }\textbf {\bibinfo {volume} {5}},\ \bibinfo {pages} {020343} (\bibinfo {year} {2024})}\BibitemShut {NoStop}%
\bibitem [{\citenamefont {Chen}\ and\ \citenamefont {Grover}(2024{\natexlab{a}})}]{ChenGrover2024SeparabilityTransitions}%
  \BibitemOpen
  \bibfield  {author} {\bibinfo {author} {\bibfnamefont {Y.-H.}\ \bibnamefont {Chen}}\ and\ \bibinfo {author} {\bibfnamefont {T.}~\bibnamefont {Grover}},\ }\bibfield  {title} {\bibinfo {title} {Separability transitions in topological states induced by local decoherence},\ }\href {https://doi.org/10.1103/PhysRevLett.132.170602} {\bibfield  {journal} {\bibinfo  {journal} {Phys. Rev. Lett.}\ }\textbf {\bibinfo {volume} {132}},\ \bibinfo {pages} {170602} (\bibinfo {year} {2024}{\natexlab{a}})}\BibitemShut {NoStop}%
\bibitem [{\citenamefont {Lu}\ \emph {et~al.}(2020)\citenamefont {Lu}, \citenamefont {Hsieh},\ and\ \citenamefont {Grover}}]{ten}%
  \BibitemOpen
  \bibfield  {author} {\bibinfo {author} {\bibfnamefont {T.-C.}\ \bibnamefont {Lu}}, \bibinfo {author} {\bibfnamefont {T.~H.}\ \bibnamefont {Hsieh}},\ and\ \bibinfo {author} {\bibfnamefont {T.}~\bibnamefont {Grover}},\ }\bibfield  {title} {\bibinfo {title} {Detecting topological order at finite temperature using entanglement negativity},\ }\href {https://doi.org/10.1103/PhysRevLett.125.116801} {\bibfield  {journal} {\bibinfo  {journal} {Phys. Rev. Lett.}\ }\textbf {\bibinfo {volume} {125}},\ \bibinfo {pages} {116801} (\bibinfo {year} {2020})}\BibitemShut {NoStop}%
\bibitem [{\citenamefont {Hastings}(2011)}]{hastingstop}%
  \BibitemOpen
  \bibfield  {author} {\bibinfo {author} {\bibfnamefont {M.~B.}\ \bibnamefont {Hastings}},\ }\bibfield  {title} {\bibinfo {title} {Topological order at nonzero temperature},\ }\href {https://doi.org/10.1103/PhysRevLett.107.210501} {\bibfield  {journal} {\bibinfo  {journal} {Phys. Rev. Lett.}\ }\textbf {\bibinfo {volume} {107}},\ \bibinfo {pages} {210501} (\bibinfo {year} {2011})}\BibitemShut {NoStop}%
\bibitem [{\citenamefont {Lessa}\ \emph {et~al.}(2025{\natexlab{a}})\citenamefont {Lessa}, \citenamefont {Sang}, \citenamefont {Lu}, \citenamefont {Hsieh},\ and\ \citenamefont {Wang}}]{lessa2025higherformanomalylongrangeentanglement}%
  \BibitemOpen
  \bibfield  {author} {\bibinfo {author} {\bibfnamefont {L.~A.}\ \bibnamefont {Lessa}}, \bibinfo {author} {\bibfnamefont {S.}~\bibnamefont {Sang}}, \bibinfo {author} {\bibfnamefont {T.-C.}\ \bibnamefont {Lu}}, \bibinfo {author} {\bibfnamefont {T.~H.}\ \bibnamefont {Hsieh}},\ and\ \bibinfo {author} {\bibfnamefont {C.}~\bibnamefont {Wang}},\ }\href {https://arxiv.org/abs/2503.12792} {\bibinfo {title} {Higher-form anomaly and long-range entanglement of mixed states}} (\bibinfo {year} {2025}{\natexlab{a}}),\ \Eprint {https://arxiv.org/abs/2503.12792} {arXiv:2503.12792 [quant-ph]} \BibitemShut {NoStop}%
\bibitem [{\citenamefont {Li}\ \emph {et~al.}(2025)\citenamefont {Li}, \citenamefont {Lee},\ and\ \citenamefont {Yoshida}}]{li2025entanglementneededtopologicalcodes}%
  \BibitemOpen
  \bibfield  {author} {\bibinfo {author} {\bibfnamefont {Z.}~\bibnamefont {Li}}, \bibinfo {author} {\bibfnamefont {D.}~\bibnamefont {Lee}},\ and\ \bibinfo {author} {\bibfnamefont {B.}~\bibnamefont {Yoshida}},\ }\bibfield  {title} {\bibinfo {title} {How much entanglement is needed for topological codes and mixed states with anomalous symmetry?},\ }\href {https://doi.org/10.1103/pw12-kdjx} {\bibfield  {journal} {\bibinfo  {journal} {Phys. Rev. X}\ }\textbf {\bibinfo {volume} {15}},\ \bibinfo {pages} {021090} (\bibinfo {year} {2025})},\ \Eprint {https://arxiv.org/abs/2405.07970} {arXiv:2405.07970 [quant-ph]} \BibitemShut {NoStop}%
\bibitem [{\citenamefont {Hashim}\ \emph {et~al.}(2021)\citenamefont {Hashim}, \citenamefont {Naik}, \citenamefont {Morvan}, \citenamefont {Ville}, \citenamefont {Mitchell}, \citenamefont {Kreikebaum}, \citenamefont {Davis}, \citenamefont {Smith}, \citenamefont {Iancu}, \citenamefont {O'Brien}, \citenamefont {Hincks}, \citenamefont {Wallman}, \citenamefont {Emerson},\ and\ \citenamefont {Siddiqi}}]{Hashim2021RandomizedCompiling}%
  \BibitemOpen
  \bibfield  {author} {\bibinfo {author} {\bibfnamefont {A.}~\bibnamefont {Hashim}}, \bibinfo {author} {\bibfnamefont {R.~K.}\ \bibnamefont {Naik}}, \bibinfo {author} {\bibfnamefont {A.}~\bibnamefont {Morvan}}, \bibinfo {author} {\bibfnamefont {J.-L.}\ \bibnamefont {Ville}}, \bibinfo {author} {\bibfnamefont {B.}~\bibnamefont {Mitchell}}, \bibinfo {author} {\bibfnamefont {J.~M.}\ \bibnamefont {Kreikebaum}}, \bibinfo {author} {\bibfnamefont {M.}~\bibnamefont {Davis}}, \bibinfo {author} {\bibfnamefont {E.}~\bibnamefont {Smith}}, \bibinfo {author} {\bibfnamefont {C.}~\bibnamefont {Iancu}}, \bibinfo {author} {\bibfnamefont {K.~P.}\ \bibnamefont {O'Brien}}, \bibinfo {author} {\bibfnamefont {I.}~\bibnamefont {Hincks}}, \bibinfo {author} {\bibfnamefont {J.~J.}\ \bibnamefont {Wallman}}, \bibinfo {author} {\bibfnamefont {J.}~\bibnamefont {Emerson}},\ and\ \bibinfo {author} {\bibfnamefont {I.}~\bibnamefont {Siddiqi}},\ }\bibfield  {title} {\bibinfo {title} {Randomized compiling for scalable quantum computing on a noisy
  superconducting quantum processor},\ }\href {https://doi.org/10.1103/PhysRevX.11.041039} {\bibfield  {journal} {\bibinfo  {journal} {Phys. Rev. X}\ }\textbf {\bibinfo {volume} {11}},\ \bibinfo {pages} {041039} (\bibinfo {year} {2021})}\BibitemShut {NoStop}%
\bibitem [{\citenamefont {{Google Quantum AI and Collaborators}}(2025)}]{Google2025BelowThreshold}%
  \BibitemOpen
  \bibfield  {author} {\bibinfo {author} {\bibnamefont {{Google Quantum AI and Collaborators}}},\ }\bibfield  {title} {\bibinfo {title} {Quantum error correction below the surface code threshold},\ }\href {https://doi.org/10.1038/s41586-024-08449-y} {\bibfield  {journal} {\bibinfo  {journal} {Nature}\ }\textbf {\bibinfo {volume} {638}},\ \bibinfo {pages} {920} (\bibinfo {year} {2025})}\BibitemShut {NoStop}%
\bibitem [{\citenamefont {Ishii}\ \emph {et~al.}(2025)\citenamefont {Ishii}, \citenamefont {Qassim}, \citenamefont {Kurita}, \citenamefont {Emerson}, \citenamefont {Maruyama}, \citenamefont {Oshima},\ and\ \citenamefont {Sato}}]{Ishii2025CoherentSuppression}%
  \BibitemOpen
  \bibfield  {author} {\bibinfo {author} {\bibfnamefont {M.}~\bibnamefont {Ishii}}, \bibinfo {author} {\bibfnamefont {H.}~\bibnamefont {Qassim}}, \bibinfo {author} {\bibfnamefont {T.}~\bibnamefont {Kurita}}, \bibinfo {author} {\bibfnamefont {J.}~\bibnamefont {Emerson}}, \bibinfo {author} {\bibfnamefont {K.}~\bibnamefont {Maruyama}}, \bibinfo {author} {\bibfnamefont {H.}~\bibnamefont {Oshima}},\ and\ \bibinfo {author} {\bibfnamefont {S.}~\bibnamefont {Sato}},\ }\href {https://arxiv.org/abs/2503.05344} {\bibinfo {title} {Implementation and verification of coherent error suppression using randomized compiling for {Grover}'s algorithm on a trapped-ion device}} (\bibinfo {year} {2025}),\ \Eprint {https://arxiv.org/abs/2503.05344} {arXiv:2503.05344 [quant-ph]} \BibitemShut {NoStop}%
\bibitem [{\citenamefont {Deshmukh}\ \emph {et~al.}(2026)\citenamefont {Deshmukh}, \citenamefont {Das}, \citenamefont {Nambiar}, \citenamefont {Chand}, \citenamefont {Salunkhe}, \citenamefont {Umasankar}, \citenamefont {Kamat}, \citenamefont {Bhowal}, \citenamefont {Agarwal}, \citenamefont {Bharadwaj}, \citenamefont {Chawla}, \citenamefont {Jain}, \citenamefont {Patankar},\ and\ \citenamefont {Vijay}}]{Deshmukh2026CoherentErrors}%
  \BibitemOpen
  \bibfield  {author} {\bibinfo {author} {\bibfnamefont {J.}~\bibnamefont {Deshmukh}}, \bibinfo {author} {\bibfnamefont {S.}~\bibnamefont {Das}}, \bibinfo {author} {\bibfnamefont {B.}~\bibnamefont {Nambiar}}, \bibinfo {author} {\bibfnamefont {M.}~\bibnamefont {Chand}}, \bibinfo {author} {\bibfnamefont {K.~V.}\ \bibnamefont {Salunkhe}}, \bibinfo {author} {\bibfnamefont {G.}~\bibnamefont {Umasankar}}, \bibinfo {author} {\bibfnamefont {A.}~\bibnamefont {Kamat}}, \bibinfo {author} {\bibfnamefont {K.}~\bibnamefont {Bhowal}}, \bibinfo {author} {\bibfnamefont {G.}~\bibnamefont {Agarwal}}, \bibinfo {author} {\bibfnamefont {L.}~\bibnamefont {Bharadwaj}}, \bibinfo {author} {\bibfnamefont {E.}~\bibnamefont {Chawla}}, \bibinfo {author} {\bibfnamefont {P.}~\bibnamefont {Jain}}, \bibinfo {author} {\bibfnamefont {M.~P.}\ \bibnamefont {Patankar}},\ and\ \bibinfo {author} {\bibfnamefont {R.}~\bibnamefont {Vijay}},\ }\bibfield  {title} {\bibinfo {title} {Coherent errors in a superconducting six-qubit network with
  beyond-nearest-neighbor connectivity},\ }\href {https://doi.org/10.1103/6jdh-pm39} {\bibfield  {journal} {\bibinfo  {journal} {Phys. Rev. Research}\ }\textbf {\bibinfo {volume} {8}},\ \bibinfo {pages} {033049} (\bibinfo {year} {2026})}\BibitemShut {NoStop}%
\bibitem [{\citenamefont {Han}\ \emph {et~al.}(2026)\citenamefont {Han}, \citenamefont {Zhao}, \citenamefont {Xu}, \citenamefont {Wang},\ and\ \citenamefont {Liu}}]{HanEtAl2026CoherentThreshold}%
  \BibitemOpen
  \bibfield  {author} {\bibinfo {author} {\bibfnamefont {Z.}~\bibnamefont {Han}}, \bibinfo {author} {\bibfnamefont {Y.}~\bibnamefont {Zhao}}, \bibinfo {author} {\bibfnamefont {Y.}~\bibnamefont {Xu}}, \bibinfo {author} {\bibfnamefont {Y.}~\bibnamefont {Wang}},\ and\ \bibinfo {author} {\bibfnamefont {Z.-W.}\ \bibnamefont {Liu}},\ }\href {https://arxiv.org/abs/2609.20537} {\bibinfo {title} {Coherent error threshold for quantum {LDPC} codes}} (\bibinfo {year} {2026}),\ \Eprint {https://arxiv.org/abs/2609.20537} {arXiv:2609.20537 [quant-ph]} \BibitemShut {NoStop}%
\bibitem [{\citenamefont {Tamiya}\ and\ \citenamefont {Koashi}(2026)}]{tamiya2026proofpositivecoherenterrorthreshold}%
  \BibitemOpen
  \bibfield  {author} {\bibinfo {author} {\bibfnamefont {S.}~\bibnamefont {Tamiya}}\ and\ \bibinfo {author} {\bibfnamefont {M.}~\bibnamefont {Koashi}},\ }\href {https://arxiv.org/abs/2609.20708} {\bibinfo {title} {Proof of a positive coherent-error threshold for topological quantum codes}} (\bibinfo {year} {2026}),\ \Eprint {https://arxiv.org/abs/2609.20708} {arXiv:2609.20708 [quant-ph]} \BibitemShut {NoStop}%
\bibitem [{\citenamefont {Lee}\ \emph {et~al.}(2023)\citenamefont {Lee}, \citenamefont {Jian},\ and\ \citenamefont {Xu}}]{PRXQuantum.4.030317}%
  \BibitemOpen
  \bibfield  {author} {\bibinfo {author} {\bibfnamefont {J.~Y.}\ \bibnamefont {Lee}}, \bibinfo {author} {\bibfnamefont {C.-M.}\ \bibnamefont {Jian}},\ and\ \bibinfo {author} {\bibfnamefont {C.}~\bibnamefont {Xu}},\ }\bibfield  {title} {\bibinfo {title} {Quantum criticality under decoherence or weak measurement},\ }\href {https://doi.org/10.1103/PRXQuantum.4.030317} {\bibfield  {journal} {\bibinfo  {journal} {PRX Quantum}\ }\textbf {\bibinfo {volume} {4}},\ \bibinfo {pages} {030317} (\bibinfo {year} {2023})}\BibitemShut {NoStop}%
\bibitem [{\citenamefont {Lessa}\ \emph {et~al.}(2025{\natexlab{b}})\citenamefont {Lessa}, \citenamefont {Ma}, \citenamefont {Zhang}, \citenamefont {Bi}, \citenamefont {Cheng},\ and\ \citenamefont {Wang}}]{lessa_strong--weak_2024}%
  \BibitemOpen
  \bibfield  {author} {\bibinfo {author} {\bibfnamefont {L.~A.}\ \bibnamefont {Lessa}}, \bibinfo {author} {\bibfnamefont {R.}~\bibnamefont {Ma}}, \bibinfo {author} {\bibfnamefont {J.-H.}\ \bibnamefont {Zhang}}, \bibinfo {author} {\bibfnamefont {Z.}~\bibnamefont {Bi}}, \bibinfo {author} {\bibfnamefont {M.}~\bibnamefont {Cheng}},\ and\ \bibinfo {author} {\bibfnamefont {C.}~\bibnamefont {Wang}},\ }\bibfield  {title} {\bibinfo {title} {Strong-to-weak spontaneous symmetry breaking in mixed quantum states},\ }\href {https://doi.org/10.1103/PRXQuantum.6.010344} {\bibfield  {journal} {\bibinfo  {journal} {PRX Quantum}\ }\textbf {\bibinfo {volume} {6}},\ \bibinfo {pages} {010344} (\bibinfo {year} {2025}{\natexlab{b}})}\BibitemShut {NoStop}%
\bibitem [{\citenamefont {Sala}\ \emph {et~al.}(2024)\citenamefont {Sala}, \citenamefont {Gopalakrishnan}, \citenamefont {Oshikawa},\ and\ \citenamefont {You}}]{PhysRevB.110.155150}%
  \BibitemOpen
  \bibfield  {author} {\bibinfo {author} {\bibfnamefont {P.}~\bibnamefont {Sala}}, \bibinfo {author} {\bibfnamefont {S.}~\bibnamefont {Gopalakrishnan}}, \bibinfo {author} {\bibfnamefont {M.}~\bibnamefont {Oshikawa}},\ and\ \bibinfo {author} {\bibfnamefont {Y.}~\bibnamefont {You}},\ }\bibfield  {title} {\bibinfo {title} {Spontaneous strong symmetry breaking in open systems: Purification perspective},\ }\href {https://doi.org/10.1103/PhysRevB.110.155150} {\bibfield  {journal} {\bibinfo  {journal} {Phys. Rev. B}\ }\textbf {\bibinfo {volume} {110}},\ \bibinfo {pages} {155150} (\bibinfo {year} {2024})}\BibitemShut {NoStop}%
\bibitem [{\citenamefont {Zhang}\ \emph {et~al.}(2025)\citenamefont {Zhang}, \citenamefont {Xu}, \citenamefont {Zhang}, \citenamefont {Xu}, \citenamefont {Bi},\ and\ \citenamefont {Luo}}]{ZhangXuZhangXuBiLuo2025}%
  \BibitemOpen
  \bibfield  {author} {\bibinfo {author} {\bibfnamefont {C.}~\bibnamefont {Zhang}}, \bibinfo {author} {\bibfnamefont {Y.}~\bibnamefont {Xu}}, \bibinfo {author} {\bibfnamefont {J.-H.}\ \bibnamefont {Zhang}}, \bibinfo {author} {\bibfnamefont {C.}~\bibnamefont {Xu}}, \bibinfo {author} {\bibfnamefont {Z.}~\bibnamefont {Bi}},\ and\ \bibinfo {author} {\bibfnamefont {Z.-X.}\ \bibnamefont {Luo}},\ }\bibfield  {title} {\bibinfo {title} {Strong-to-weak spontaneous breaking of 1-form symmetry and intrinsically mixed topological order},\ }\href {https://doi.org/10.1103/PhysRevB.111.115137} {\bibfield  {journal} {\bibinfo  {journal} {Phys. Rev. B}\ }\textbf {\bibinfo {volume} {111}},\ \bibinfo {pages} {115137} (\bibinfo {year} {2025})}\BibitemShut {NoStop}%
\bibitem [{\citenamefont {Bravyi}\ \emph {et~al.}(2018)\citenamefont {Bravyi}, \citenamefont {Englbrecht}, \citenamefont {K{\"{o}}nig},\ and\ \citenamefont {Peard}}]{Bravyi_2018}%
  \BibitemOpen
  \bibfield  {author} {\bibinfo {author} {\bibfnamefont {S.}~\bibnamefont {Bravyi}}, \bibinfo {author} {\bibfnamefont {M.}~\bibnamefont {Englbrecht}}, \bibinfo {author} {\bibfnamefont {R.}~\bibnamefont {K{\"{o}}nig}},\ and\ \bibinfo {author} {\bibfnamefont {N.}~\bibnamefont {Peard}},\ }\bibfield  {title} {\bibinfo {title} {Correcting coherent errors with surface codes},\ }\href {https://doi.org/10.1038/s41534-018-0106-y} {\bibfield  {journal} {\bibinfo  {journal} {npj Quantum Inf.}\ }\textbf {\bibinfo {volume} {4}},\ \bibinfo {pages} {55} (\bibinfo {year} {2018})}\BibitemShut {NoStop}%
\bibitem [{\citenamefont {Beale}\ \emph {et~al.}(2018)\citenamefont {Beale}, \citenamefont {Wallman}, \citenamefont {Guti{\'{e}}rrez}, \citenamefont {Brown},\ and\ \citenamefont {Laflamme}}]{Beale2018QECdecoheres}%
  \BibitemOpen
  \bibfield  {author} {\bibinfo {author} {\bibfnamefont {S.~J.}\ \bibnamefont {Beale}}, \bibinfo {author} {\bibfnamefont {J.~J.}\ \bibnamefont {Wallman}}, \bibinfo {author} {\bibfnamefont {M.}~\bibnamefont {Guti{\'{e}}rrez}}, \bibinfo {author} {\bibfnamefont {K.~R.}\ \bibnamefont {Brown}},\ and\ \bibinfo {author} {\bibfnamefont {R.}~\bibnamefont {Laflamme}},\ }\bibfield  {title} {\bibinfo {title} {Quantum error correction decoheres noise},\ }\href {https://doi.org/10.1103/PhysRevLett.121.190501} {\bibfield  {journal} {\bibinfo  {journal} {Phys. Rev. Lett.}\ }\textbf {\bibinfo {volume} {121}},\ \bibinfo {pages} {190501} (\bibinfo {year} {2018})}\BibitemShut {NoStop}%
\bibitem [{\citenamefont {Huang}\ \emph {et~al.}(2019)\citenamefont {Huang}, \citenamefont {Doherty},\ and\ \citenamefont {Flammia}}]{HuangDohertyFlammia2019CoherentErrors}%
  \BibitemOpen
  \bibfield  {author} {\bibinfo {author} {\bibfnamefont {E.}~\bibnamefont {Huang}}, \bibinfo {author} {\bibfnamefont {A.~C.}\ \bibnamefont {Doherty}},\ and\ \bibinfo {author} {\bibfnamefont {S.}~\bibnamefont {Flammia}},\ }\bibfield  {title} {\bibinfo {title} {Performance of quantum error correction with coherent errors},\ }\href {https://doi.org/10.1103/PhysRevA.99.022313} {\bibfield  {journal} {\bibinfo  {journal} {Phys. Rev. A}\ }\textbf {\bibinfo {volume} {99}},\ \bibinfo {pages} {022313} (\bibinfo {year} {2019})}\BibitemShut {NoStop}%
\bibitem [{\citenamefont {Iverson}\ and\ \citenamefont {Preskill}(2020)}]{Iverson_2020}%
  \BibitemOpen
  \bibfield  {author} {\bibinfo {author} {\bibfnamefont {J.~K.}\ \bibnamefont {Iverson}}\ and\ \bibinfo {author} {\bibfnamefont {J.}~\bibnamefont {Preskill}},\ }\bibfield  {title} {\bibinfo {title} {Coherence in logical quantum channels},\ }\href {https://doi.org/10.1088/1367-2630/ab8e5c} {\bibfield  {journal} {\bibinfo  {journal} {New J. Phys.}\ }\textbf {\bibinfo {volume} {22}},\ \bibinfo {pages} {073066} (\bibinfo {year} {2020})}\BibitemShut {NoStop}%
\bibitem [{\citenamefont {Venn}\ and\ \citenamefont {B{\'{e}}ri}(2020)}]{Venn_PhysRevResearch.2.043412}%
  \BibitemOpen
  \bibfield  {author} {\bibinfo {author} {\bibfnamefont {F.}~\bibnamefont {Venn}}\ and\ \bibinfo {author} {\bibfnamefont {B.}~\bibnamefont {B{\'{e}}ri}},\ }\bibfield  {title} {\bibinfo {title} {Error-correction and noise-decoherence thresholds for coherent errors in planar-graph surface codes},\ }\href {https://doi.org/10.1103/PhysRevResearch.2.043412} {\bibfield  {journal} {\bibinfo  {journal} {Phys. Rev. Research}\ }\textbf {\bibinfo {volume} {2}},\ \bibinfo {pages} {043412} (\bibinfo {year} {2020})}\BibitemShut {NoStop}%
\bibitem [{\citenamefont {M{\'{a}}rton}\ and\ \citenamefont {Asb{\'{o}}th}(2023)}]{MartonAsboth2023CoherentReadout}%
  \BibitemOpen
  \bibfield  {author} {\bibinfo {author} {\bibfnamefont {{\'{A}}.}~\bibnamefont {M{\'{a}}rton}}\ and\ \bibinfo {author} {\bibfnamefont {J.~K.}\ \bibnamefont {Asb{\'{o}}th}},\ }\bibfield  {title} {\bibinfo {title} {Coherent errors and readout errors in the surface code},\ }\href {https://doi.org/10.22331/q-2023-09-21-1116} {\bibfield  {journal} {\bibinfo  {journal} {Quantum}\ }\textbf {\bibinfo {volume} {7}},\ \bibinfo {pages} {1116} (\bibinfo {year} {2023})}\BibitemShut {NoStop}%
\bibitem [{\citenamefont {Venn}\ \emph {et~al.}(2023)\citenamefont {Venn}, \citenamefont {Behrends},\ and\ \citenamefont {B{\'{e}}ri}}]{Venn_2023}%
  \BibitemOpen
  \bibfield  {author} {\bibinfo {author} {\bibfnamefont {F.}~\bibnamefont {Venn}}, \bibinfo {author} {\bibfnamefont {J.}~\bibnamefont {Behrends}},\ and\ \bibinfo {author} {\bibfnamefont {B.}~\bibnamefont {B{\'{e}}ri}},\ }\bibfield  {title} {\bibinfo {title} {Coherent-error threshold for surface codes from {Majorana} delocalization},\ }\href {https://doi.org/10.1103/PhysRevLett.131.060603} {\bibfield  {journal} {\bibinfo  {journal} {Phys. Rev. Lett.}\ }\textbf {\bibinfo {volume} {131}},\ \bibinfo {pages} {060603} (\bibinfo {year} {2023})}\BibitemShut {NoStop}%
\bibitem [{\citenamefont {Zhu}\ and\ \citenamefont {Trebst}(2023)}]{ZhuTrebst2023MajoranaLiquid}%
  \BibitemOpen
  \bibfield  {author} {\bibinfo {author} {\bibfnamefont {G.-Y.}\ \bibnamefont {Zhu}}\ and\ \bibinfo {author} {\bibfnamefont {S.}~\bibnamefont {Trebst}},\ }\href {https://arxiv.org/abs/2311.08450} {\bibinfo {title} {Qubit fractionalization and emergent {Majorana} liquid in the honeycomb {Floquet} code induced by coherent errors and weak measurements}} (\bibinfo {year} {2023}),\ \Eprint {https://arxiv.org/abs/2311.08450} {arXiv:2311.08450 [quant-ph]} \BibitemShut {NoStop}%
\bibitem [{\citenamefont {Behrends}\ \emph {et~al.}(2024)\citenamefont {Behrends}, \citenamefont {Venn},\ and\ \citenamefont {B{\'{e}}ri}}]{Behrends_PhysRevResearch.6.013137}%
  \BibitemOpen
  \bibfield  {author} {\bibinfo {author} {\bibfnamefont {J.}~\bibnamefont {Behrends}}, \bibinfo {author} {\bibfnamefont {F.}~\bibnamefont {Venn}},\ and\ \bibinfo {author} {\bibfnamefont {B.}~\bibnamefont {B{\'{e}}ri}},\ }\bibfield  {title} {\bibinfo {title} {Surface codes, quantum circuits, and entanglement phases},\ }\href {https://doi.org/10.1103/PhysRevResearch.6.013137} {\bibfield  {journal} {\bibinfo  {journal} {Phys. Rev. Research}\ }\textbf {\bibinfo {volume} {6}},\ \bibinfo {pages} {013137} (\bibinfo {year} {2024})}\BibitemShut {NoStop}%
\bibitem [{\citenamefont {Chen}\ and\ \citenamefont {Grover}(2024{\natexlab{b}})}]{ChenGrover2024UnconventionalMixedStateTransition}%
  \BibitemOpen
  \bibfield  {author} {\bibinfo {author} {\bibfnamefont {Y.-H.}\ \bibnamefont {Chen}}\ and\ \bibinfo {author} {\bibfnamefont {T.}~\bibnamefont {Grover}},\ }\bibfield  {title} {\bibinfo {title} {Unconventional topological mixed-state transition and critical phase induced by self-dual coherent errors},\ }\href {https://doi.org/10.1103/PhysRevB.110.125152} {\bibfield  {journal} {\bibinfo  {journal} {Phys. Rev. B}\ }\textbf {\bibinfo {volume} {110}},\ \bibinfo {pages} {125152} (\bibinfo {year} {2024}{\natexlab{b}})}\BibitemShut {NoStop}%
\bibitem [{\citenamefont {Bao}\ and\ \citenamefont {Anand}(2024)}]{Bao_Anand_Phases}%
  \BibitemOpen
  \bibfield  {author} {\bibinfo {author} {\bibfnamefont {Y.}~\bibnamefont {Bao}}\ and\ \bibinfo {author} {\bibfnamefont {S.}~\bibnamefont {Anand}},\ }\href {https://arxiv.org/abs/2411.05785} {\bibinfo {title} {Phases of decodability in the surface code with unitary errors}} (\bibinfo {year} {2024}),\ \Eprint {https://arxiv.org/abs/2411.05785} {arXiv:2411.05785 [quant-ph]} \BibitemShut {NoStop}%
\bibitem [{\citenamefont {Pato}\ \emph {et~al.}(2025)\citenamefont {Pato}, \citenamefont {Staples},\ and\ \citenamefont {Brown}}]{PatoStaplesBrown2025LogicalCoherence}%
  \BibitemOpen
  \bibfield  {author} {\bibinfo {author} {\bibfnamefont {B.}~\bibnamefont {Pato}}, \bibinfo {author} {\bibfnamefont {J.~W.}\ \bibnamefont {Staples}},\ and\ \bibinfo {author} {\bibfnamefont {K.~R.}\ \bibnamefont {Brown}},\ }\bibfield  {title} {\bibinfo {title} {Logical coherence in two-dimensional compass codes},\ }\href {https://doi.org/10.1103/PhysRevA.111.032424} {\bibfield  {journal} {\bibinfo  {journal} {Phys. Rev. A}\ }\textbf {\bibinfo {volume} {111}},\ \bibinfo {pages} {032424} (\bibinfo {year} {2025})}\BibitemShut {NoStop}%
\bibitem [{\citenamefont {Cheng}\ \emph {et~al.}(2025)\citenamefont {Cheng}, \citenamefont {Huang}, \citenamefont {Khemani}, \citenamefont {Gullans},\ and\ \citenamefont {Ippoliti}}]{Cheng_2025}%
  \BibitemOpen
  \bibfield  {author} {\bibinfo {author} {\bibfnamefont {Z.}~\bibnamefont {Cheng}}, \bibinfo {author} {\bibfnamefont {E.}~\bibnamefont {Huang}}, \bibinfo {author} {\bibfnamefont {V.}~\bibnamefont {Khemani}}, \bibinfo {author} {\bibfnamefont {M.~J.}\ \bibnamefont {Gullans}},\ and\ \bibinfo {author} {\bibfnamefont {M.}~\bibnamefont {Ippoliti}},\ }\bibfield  {title} {\bibinfo {title} {Emergent unitary designs for encoded qubits from coherent errors and syndrome measurements},\ }\href {https://doi.org/10.1103/bnld-2chd} {\bibfield  {journal} {\bibinfo  {journal} {PRX Quantum}\ }\textbf {\bibinfo {volume} {6}},\ \bibinfo {pages} {030333} (\bibinfo {year} {2025})}\BibitemShut {NoStop}%
\bibitem [{\citenamefont {Lee}\ and\ \citenamefont {Moon}(2025)}]{LeeMoon2025CoherentNoise}%
  \BibitemOpen
  \bibfield  {author} {\bibinfo {author} {\bibfnamefont {S.}~\bibnamefont {Lee}}\ and\ \bibinfo {author} {\bibfnamefont {E.-G.}\ \bibnamefont {Moon}},\ }\bibfield  {title} {\bibinfo {title} {Mixed-state topological order under coherent noise},\ }\href {https://doi.org/10.1103/hchr-rqq9} {\bibfield  {journal} {\bibinfo  {journal} {PRX Quantum}\ }\textbf {\bibinfo {volume} {6}},\ \bibinfo {pages} {030355} (\bibinfo {year} {2025})}\BibitemShut {NoStop}%
\bibitem [{\citenamefont {Behrends}\ and\ \citenamefont {B{\'{e}}ri}(2025{\natexlab{a}})}]{Behrends_generic1qb_gskb-t5ql}%
  \BibitemOpen
  \bibfield  {author} {\bibinfo {author} {\bibfnamefont {J.}~\bibnamefont {Behrends}}\ and\ \bibinfo {author} {\bibfnamefont {B.}~\bibnamefont {B{\'{e}}ri}},\ }\bibfield  {title} {\bibinfo {title} {Statistical mechanical mapping and maximum-likelihood thresholds for the surface code under generic single-qubit coherent errors},\ }\href {https://doi.org/10.1103/gskb-t5ql} {\bibfield  {journal} {\bibinfo  {journal} {PRX Quantum}\ }\textbf {\bibinfo {volume} {6}},\ \bibinfo {pages} {040305} (\bibinfo {year} {2025}{\natexlab{a}})}\BibitemShut {NoStop}%
\bibitem [{\citenamefont {Darmawan}(2025)}]{Darmawan2025DecoderAdaptation}%
  \BibitemOpen
  \bibfield  {author} {\bibinfo {author} {\bibfnamefont {A.~S.}\ \bibnamefont {Darmawan}},\ }\bibfield  {title} {\bibinfo {title} {Optimal adaptation of surface-code decoders to local noise},\ }\href {https://doi.org/10.1103/r2dc-qcrx} {\bibfield  {journal} {\bibinfo  {journal} {Phys. Rev. A}\ }\textbf {\bibinfo {volume} {112}},\ \bibinfo {pages} {042431} (\bibinfo {year} {2025})}\BibitemShut {NoStop}%
\bibitem [{\citenamefont {Behrends}\ and\ \citenamefont {B{\'{e}}ri}(2025{\natexlab{b}})}]{Behrends_SBPC}%
  \BibitemOpen
  \bibfield  {author} {\bibinfo {author} {\bibfnamefont {J.}~\bibnamefont {Behrends}}\ and\ \bibinfo {author} {\bibfnamefont {B.}~\bibnamefont {B{\'{e}}ri}},\ }\bibfield  {title} {\bibinfo {title} {The surface code beyond {Pauli} channels: Logical noise coherence, information-theoretic measures, and errorfield-double phenomenology},\ }\href {https://doi.org/10.1103/psf5-b6j2} {\bibfield  {journal} {\bibinfo  {journal} {PRX Quantum}\ }\textbf {\bibinfo {volume} {6}},\ \bibinfo {pages} {040350} (\bibinfo {year} {2025}{\natexlab{b}})},\ \Eprint {https://arxiv.org/abs/2412.21055} {arXiv:2412.21055 [quant-ph]} \BibitemShut {NoStop}%
\bibitem [{\citenamefont {Liu}\ and\ \citenamefont {Chen}(2026)}]{liu2025coherenterrorinducedphase}%
  \BibitemOpen
  \bibfield  {author} {\bibinfo {author} {\bibfnamefont {H.}~\bibnamefont {Liu}}\ and\ \bibinfo {author} {\bibfnamefont {X.}~\bibnamefont {Chen}},\ }\bibfield  {title} {\bibinfo {title} {Coherent error induced phase transition},\ }\href {https://doi.org/10.1103/jly5-q9mx} {\bibfield  {journal} {\bibinfo  {journal} {Phys. Rev. B}\ }\textbf {\bibinfo {volume} {114}},\ \bibinfo {pages} {214106} (\bibinfo {year} {2026})},\ \Eprint {https://arxiv.org/abs/2506.00650} {arXiv:2506.00650 [quant-ph]} \BibitemShut {NoStop}%
\bibitem [{\citenamefont {Yan}\ \emph {et~al.}(2026{\natexlab{a}})\citenamefont {Yan}, \citenamefont {Bao},\ and\ \citenamefont {Vijay}}]{yan2026nonlinearsigmamodelsurface}%
  \BibitemOpen
  \bibfield  {author} {\bibinfo {author} {\bibfnamefont {S.~W.}\ \bibnamefont {Yan}}, \bibinfo {author} {\bibfnamefont {Y.}~\bibnamefont {Bao}},\ and\ \bibinfo {author} {\bibfnamefont {S.}~\bibnamefont {Vijay}},\ }\href {https://arxiv.org/abs/2603.25665} {\bibinfo {title} {Non-linear sigma model for the surface code with coherent errors}} (\bibinfo {year} {2026}{\natexlab{a}}),\ \Eprint {https://arxiv.org/abs/2603.25665} {arXiv:2603.25665 [cond-mat.stat-mech]} \BibitemShut {NoStop}%
\bibitem [{\citenamefont {Yang}\ \emph {et~al.}(2026)\citenamefont {Yang}, \citenamefont {Ludwig},\ and\ \citenamefont {Jian}}]{yang2026decodingcoherenterrorstoric}%
  \BibitemOpen
  \bibfield  {author} {\bibinfo {author} {\bibfnamefont {Z.}~\bibnamefont {Yang}}, \bibinfo {author} {\bibfnamefont {A.~W.~W.}\ \bibnamefont {Ludwig}},\ and\ \bibinfo {author} {\bibfnamefont {C.-M.}\ \bibnamefont {Jian}},\ }\href {https://arxiv.org/abs/2604.08650} {\bibinfo {title} {Decoding coherent errors in toric codes on honeycomb and square lattices: duality to {Majorana} monitored dynamics and symmetry classes}} (\bibinfo {year} {2026}),\ \Eprint {https://arxiv.org/abs/2604.08650} {arXiv:2604.08650 [cond-mat.stat-mech]} \BibitemShut {NoStop}%
\bibitem [{\citenamefont {Bejan}\ \emph {et~al.}(2026)\citenamefont {Bejan}, \citenamefont {Behrends}, \citenamefont {McGinley},\ and\ \citenamefont {B{\'{e}}ri}}]{bejan2026projectedlogicalensemblessurface}%
  \BibitemOpen
  \bibfield  {author} {\bibinfo {author} {\bibfnamefont {M.}~\bibnamefont {Bejan}}, \bibinfo {author} {\bibfnamefont {J.}~\bibnamefont {Behrends}}, \bibinfo {author} {\bibfnamefont {M.}~\bibnamefont {McGinley}},\ and\ \bibinfo {author} {\bibfnamefont {B.}~\bibnamefont {B{\'{e}}ri}},\ }\href {https://arxiv.org/abs/2606.17140} {\bibinfo {title} {Projected logical ensembles in surface codes via the random-matrix theory of quantum dots}} (\bibinfo {year} {2026}),\ \Eprint {https://arxiv.org/abs/2606.17140} {arXiv:2606.17140 [quant-ph]} \BibitemShut {NoStop}%
\bibitem [{\citenamefont {Negari}\ \emph {et~al.}(2026{\natexlab{a}})\citenamefont {Negari}, \citenamefont {Sahu}, \citenamefont {Behrends}, \citenamefont {B{\'{e}}ri},\ and\ \citenamefont {Hsieh}}]{negari2026criticalnonequilibrium}%
  \BibitemOpen
  \bibfield  {author} {\bibinfo {author} {\bibfnamefont {A.-R.}\ \bibnamefont {Negari}}, \bibinfo {author} {\bibfnamefont {S.}~\bibnamefont {Sahu}}, \bibinfo {author} {\bibfnamefont {J.}~\bibnamefont {Behrends}}, \bibinfo {author} {\bibfnamefont {B.}~\bibnamefont {B{\'{e}}ri}},\ and\ \bibinfo {author} {\bibfnamefont {T.~H.}\ \bibnamefont {Hsieh}},\ }\href {https://arxiv.org/abs/2601.10792} {\bibinfo {title} {Critical non-equilibrium phases from noisy topological memories}} (\bibinfo {year} {2026}{\natexlab{a}}),\ \Eprint {https://arxiv.org/abs/2601.10792} {arXiv:2601.10792 [quant-ph]} \BibitemShut {NoStop}%
\bibitem [{\citenamefont {Hsieh}(2025)}]{kitptalk}%
  \BibitemOpen
  \bibfield  {author} {\bibinfo {author} {\bibfnamefont {T.~H.}\ \bibnamefont {Hsieh}},\ }\href {https://doi.org/10.26081/K6H15R} {\bibinfo {title} {Tutorial: Mixed-state phases}},\ \bibinfo {howpublished} {Kavli Institute for Theoretical Physics} (\bibinfo {year} {2025}),\ \bibinfo {note} {lecture, 21 August 2025}\BibitemShut {NoStop}%
\bibitem [{\citenamefont {Barkeshli}\ \emph {et~al.}(2019)\citenamefont {Barkeshli}, \citenamefont {Bonderson}, \citenamefont {Cheng},\ and\ \citenamefont {Wang}}]{BarkeshliBondersonChengWang2019}%
  \BibitemOpen
  \bibfield  {author} {\bibinfo {author} {\bibfnamefont {M.}~\bibnamefont {Barkeshli}}, \bibinfo {author} {\bibfnamefont {P.}~\bibnamefont {Bonderson}}, \bibinfo {author} {\bibfnamefont {M.}~\bibnamefont {Cheng}},\ and\ \bibinfo {author} {\bibfnamefont {Z.}~\bibnamefont {Wang}},\ }\bibfield  {title} {\bibinfo {title} {Symmetry fractionalization, defects, and gauging of topological phases},\ }\href {https://doi.org/10.1103/PhysRevB.100.115147} {\bibfield  {journal} {\bibinfo  {journal} {Phys. Rev. B}\ }\textbf {\bibinfo {volume} {100}},\ \bibinfo {pages} {115147} (\bibinfo {year} {2019})},\ \Eprint {https://arxiv.org/abs/1410.4540} {arXiv:1410.4540 [cond-mat.str-el]} \BibitemShut {NoStop}%
\bibitem [{\citenamefont {Hsin}\ \emph {et~al.}(2025)\citenamefont {Hsin}, \citenamefont {Kobayashi},\ and\ \citenamefont {Prem}}]{hsin2025higherformanomaliesimplyintrinsic}%
  \BibitemOpen
  \bibfield  {author} {\bibinfo {author} {\bibfnamefont {P.-S.}\ \bibnamefont {Hsin}}, \bibinfo {author} {\bibfnamefont {R.}~\bibnamefont {Kobayashi}},\ and\ \bibinfo {author} {\bibfnamefont {A.}~\bibnamefont {Prem}},\ }\href {https://arxiv.org/abs/2504.10569} {\bibinfo {title} {Higher-form anomalies imply intrinsic long-range entanglement}} (\bibinfo {year} {2025}),\ \Eprint {https://arxiv.org/abs/2504.10569} {arXiv:2504.10569 [quant-ph]} \BibitemShut {NoStop}%
\bibitem [{\citenamefont {Thorngren}\ \emph {et~al.}(2026)\citenamefont {Thorngren}, \citenamefont {Gioia},\ and\ \citenamefont {Zhang}}]{ThorngrenGioiaZhang2026TranslationLRE}%
  \BibitemOpen
  \bibfield  {author} {\bibinfo {author} {\bibfnamefont {R.}~\bibnamefont {Thorngren}}, \bibinfo {author} {\bibfnamefont {L.}~\bibnamefont {Gioia}},\ and\ \bibinfo {author} {\bibfnamefont {C.}~\bibnamefont {Zhang}},\ }\href {https://arxiv.org/abs/2605.15200} {\bibinfo {title} {Translation symmetry-enforced long-range entanglement in mixed states}} (\bibinfo {year} {2026}),\ \Eprint {https://arxiv.org/abs/2605.15200} {arXiv:2605.15200 [quant-ph]} \BibitemShut {NoStop}%
\bibitem [{\citenamefont {Lessa}\ and\ \citenamefont {Lu}(2026)}]{LessaLu2026DimensionalConstraints}%
  \BibitemOpen
  \bibfield  {author} {\bibinfo {author} {\bibfnamefont {L.~A.}\ \bibnamefont {Lessa}}\ and\ \bibinfo {author} {\bibfnamefont {T.-C.}\ \bibnamefont {Lu}},\ }\href {https://arxiv.org/abs/2605.15201} {\bibinfo {title} {Mixed-state long-range entanglement from dimensional constraints}} (\bibinfo {year} {2026}),\ \Eprint {https://arxiv.org/abs/2605.15201} {arXiv:2605.15201 [quant-ph]} \BibitemShut {NoStop}%
\bibitem [{\citenamefont {Chen}(2026)}]{yuhsueh}%
  \BibitemOpen
  \bibfield  {author} {\bibinfo {author} {\bibfnamefont {Y.-H.}\ \bibnamefont {Chen}},\ }\href {https://arxiv.org/abs/2609.02141} {\bibinfo {title} {Renormalization group and long-range conditional mutual information in hierarchical models}} (\bibinfo {year} {2026}),\ \Eprint {https://arxiv.org/abs/2609.02141} {arXiv:2609.02141 [quant-ph]} \BibitemShut {NoStop}%
\bibitem [{\citenamefont {Chen}\ and\ \citenamefont {Grover}(2026)}]{yhgrover}%
  \BibitemOpen
  \bibfield  {author} {\bibinfo {author} {\bibfnamefont {Y.-H.}\ \bibnamefont {Chen}}\ and\ \bibinfo {author} {\bibfnamefont {T.}~\bibnamefont {Grover}},\ }\href {https://arxiv.org/abs/2610.11235} {\bibinfo {title} {Markov length can diverge in systems whose universal physics is spatially markovian}} (\bibinfo {year} {2026}),\ \Eprint {https://arxiv.org/abs/2610.11235} {arXiv:2610.11235 [cond-mat.stat-mech]} \BibitemShut {NoStop}%
\bibitem [{\citenamefont {Negari}\ \emph {et~al.}(2026{\natexlab{b}})\citenamefont {Negari}, \citenamefont {Zou},\ and\ \citenamefont {Hsieh}}]{NegariZouHsieh2026Instability}%
  \BibitemOpen
  \bibfield  {author} {\bibinfo {author} {\bibfnamefont {A.-R.}\ \bibnamefont {Negari}}, \bibinfo {author} {\bibfnamefont {Y.}~\bibnamefont {Zou}},\ and\ \bibinfo {author} {\bibfnamefont {T.~H.}\ \bibnamefont {Hsieh}},\ }\href {https://arxiv.org/abs/2610.18983} {\bibinfo {title} {Instability of {Markov} chains under infinitesimal coherent dissipation}} (\bibinfo {year} {2026}{\natexlab{b}}),\ \Eprint {https://arxiv.org/abs/2610.18983} {arXiv:2610.18983 [quant-ph]} \BibitemShut {NoStop}%
\bibitem [{\citenamefont {Lee}(2025)}]{Lee_2025}%
  \BibitemOpen
  \bibfield  {author} {\bibinfo {author} {\bibfnamefont {J.~Y.}\ \bibnamefont {Lee}},\ }\bibfield  {title} {\bibinfo {title} {Exact calculations of coherent information for toric codes under decoherence: Identifying the fundamental error threshold},\ }\href {https://doi.org/10.1103/hlfh-86yz} {\bibfield  {journal} {\bibinfo  {journal} {Phys. Rev. Lett.}\ }\textbf {\bibinfo {volume} {134}},\ \bibinfo {pages} {250601} (\bibinfo {year} {2025})}\BibitemShut {NoStop}%
\bibitem [{\citenamefont {Fawzi}\ and\ \citenamefont {Renner}(2015)}]{FawziRenner2015Recovery}%
  \BibitemOpen
  \bibfield  {author} {\bibinfo {author} {\bibfnamefont {O.}~\bibnamefont {Fawzi}}\ and\ \bibinfo {author} {\bibfnamefont {R.}~\bibnamefont {Renner}},\ }\bibfield  {title} {\bibinfo {title} {Quantum conditional mutual information and approximate {Markov} chains},\ }\href {https://doi.org/10.1007/s00220-015-2466-x} {\bibfield  {journal} {\bibinfo  {journal} {Commun. Math. Phys.}\ }\textbf {\bibinfo {volume} {340}},\ \bibinfo {pages} {575} (\bibinfo {year} {2015})},\ \Eprint {https://arxiv.org/abs/1410.0664} {arXiv:1410.0664 [quant-ph]} \BibitemShut {NoStop}%
\bibitem [{\citenamefont {Junge}\ \emph {et~al.}(2018)\citenamefont {Junge}, \citenamefont {Renner}, \citenamefont {Sutter}, \citenamefont {Wilde},\ and\ \citenamefont {Winter}}]{Junge2018UniversalRecovery}%
  \BibitemOpen
  \bibfield  {author} {\bibinfo {author} {\bibfnamefont {M.}~\bibnamefont {Junge}}, \bibinfo {author} {\bibfnamefont {R.}~\bibnamefont {Renner}}, \bibinfo {author} {\bibfnamefont {D.}~\bibnamefont {Sutter}}, \bibinfo {author} {\bibfnamefont {M.~M.}\ \bibnamefont {Wilde}},\ and\ \bibinfo {author} {\bibfnamefont {A.}~\bibnamefont {Winter}},\ }\bibfield  {title} {\bibinfo {title} {Universal recovery maps and approximate sufficiency of quantum relative entropy},\ }\href {https://doi.org/10.1007/s00023-018-0716-0} {\bibfield  {journal} {\bibinfo  {journal} {Ann. Henri Poincar{\'{e}}}\ }\textbf {\bibinfo {volume} {19}},\ \bibinfo {pages} {2955} (\bibinfo {year} {2018})},\ \Eprint {https://arxiv.org/abs/1509.07127} {arXiv:1509.07127 [quant-ph]} \BibitemShut {NoStop}%
\bibitem [{\citenamefont {Capel}\ \emph {et~al.}(2025)\citenamefont {Capel}, \citenamefont {Moscolari}, \citenamefont {Teufel},\ and\ \citenamefont {Wessel}}]{CapelMoscolariTeufelWessel2025Locality}%
  \BibitemOpen
  \bibfield  {author} {\bibinfo {author} {\bibfnamefont {{\'{A}}.}~\bibnamefont {Capel}}, \bibinfo {author} {\bibfnamefont {M.}~\bibnamefont {Moscolari}}, \bibinfo {author} {\bibfnamefont {S.}~\bibnamefont {Teufel}},\ and\ \bibinfo {author} {\bibfnamefont {T.}~\bibnamefont {Wessel}},\ }\bibfield  {title} {\bibinfo {title} {From decay of correlations to locality and stability of the {Gibbs} state},\ }\href {https://doi.org/10.1007/s00220-024-05198-x} {\bibfield  {journal} {\bibinfo  {journal} {Commun. Math. Phys.}\ }\textbf {\bibinfo {volume} {406}},\ \bibinfo {pages} {43} (\bibinfo {year} {2025})},\ \Eprint {https://arxiv.org/abs/2310.09182} {arXiv:2310.09182 [math-ph]} \BibitemShut {NoStop}%
\bibitem [{\citenamefont {Ma}\ \emph {et~al.}(2026)\citenamefont {Ma}, \citenamefont {Khemani},\ and\ \citenamefont {Sang}}]{MaKhemaniSang2025ThermalPhases}%
  \BibitemOpen
  \bibfield  {author} {\bibinfo {author} {\bibfnamefont {R.}~\bibnamefont {Ma}}, \bibinfo {author} {\bibfnamefont {V.}~\bibnamefont {Khemani}},\ and\ \bibinfo {author} {\bibfnamefont {S.}~\bibnamefont {Sang}},\ }\bibfield  {title} {\bibinfo {title} {Circuit-based characterization of finite-temperature quantum phases and self-correcting quantum memory},\ }\href {https://doi.org/10.1103/sh48-wmy4} {\bibfield  {journal} {\bibinfo  {journal} {PRX Quantum}\ }\textbf {\bibinfo {volume} {7}},\ \bibinfo {pages} {033052} (\bibinfo {year} {2026})},\ \Eprint {https://arxiv.org/abs/2509.15204} {arXiv:2509.15204 [quant-ph]} \BibitemShut {NoStop}%
\bibitem [{\citenamefont {Li}\ \emph {et~al.}(2026)\citenamefont {Li}, \citenamefont {Firanko},\ and\ \citenamefont {Hsieh}}]{li2026unifiedframework}%
  \BibitemOpen
  \bibfield  {author} {\bibinfo {author} {\bibfnamefont {Z.}~\bibnamefont {Li}}, \bibinfo {author} {\bibfnamefont {R.}~\bibnamefont {Firanko}},\ and\ \bibinfo {author} {\bibfnamefont {T.~H.}\ \bibnamefont {Hsieh}},\ }\href {https://arxiv.org/abs/2605.00088} {\bibinfo {title} {A unified framework for locally stable phases}} (\bibinfo {year} {2026}),\ \Eprint {https://arxiv.org/abs/2605.00088} {arXiv:2605.00088 [quant-ph]} \BibitemShut {NoStop}%
\bibitem [{\citenamefont {Lucia}\ \emph {et~al.}(2015)\citenamefont {Lucia}, \citenamefont {Cubitt}, \citenamefont {Michalakis},\ and\ \citenamefont {P{\'{e}}rez-Garc{\'{i}}a}}]{LuciaEtAl2015RapidMixingStability}%
  \BibitemOpen
  \bibfield  {author} {\bibinfo {author} {\bibfnamefont {A.}~\bibnamefont {Lucia}}, \bibinfo {author} {\bibfnamefont {T.~S.}\ \bibnamefont {Cubitt}}, \bibinfo {author} {\bibfnamefont {S.}~\bibnamefont {Michalakis}},\ and\ \bibinfo {author} {\bibfnamefont {D.}~\bibnamefont {P{\'{e}}rez-Garc{\'{i}}a}},\ }\bibfield  {title} {\bibinfo {title} {Rapid mixing and stability of quantum dissipative systems},\ }\href {https://doi.org/10.1103/PhysRevA.91.040302} {\bibfield  {journal} {\bibinfo  {journal} {Phys. Rev. A}\ }\textbf {\bibinfo {volume} {91}},\ \bibinfo {pages} {040302} (\bibinfo {year} {2015})},\ \Eprint {https://arxiv.org/abs/1409.7809} {arXiv:1409.7809 [quant-ph]} \BibitemShut {NoStop}%
\bibitem [{\citenamefont {Bao}\ \emph {et~al.}(2026)\citenamefont {Bao}, \citenamefont {Fan}, \citenamefont {Vishwanath},\ and\ \citenamefont {Altman}}]{bao2023mixedstatetopologicalordererrorfield}%
  \BibitemOpen
  \bibfield  {author} {\bibinfo {author} {\bibfnamefont {Y.}~\bibnamefont {Bao}}, \bibinfo {author} {\bibfnamefont {R.}~\bibnamefont {Fan}}, \bibinfo {author} {\bibfnamefont {A.}~\bibnamefont {Vishwanath}},\ and\ \bibinfo {author} {\bibfnamefont {E.}~\bibnamefont {Altman}},\ }\bibfield  {title} {\bibinfo {title} {Mixed-state topological order and the errorfield double formulation of decoherence-induced transitions},\ }\href {https://doi.org/10.1103/6f98-tvb8} {\bibfield  {journal} {\bibinfo  {journal} {Phys. Rev. Lett.}\ }\textbf {\bibinfo {volume} {136}},\ \bibinfo {pages} {220402} (\bibinfo {year} {2026})},\ \Eprint {https://arxiv.org/abs/2301.05687} {arXiv:2301.05687 [quant-ph]} \BibitemShut {NoStop}%
\bibitem [{\citenamefont {Singh}\ \emph {et~al.}(2026)\citenamefont {Singh}, \citenamefont {Vasseur}, \citenamefont {Potter},\ and\ \citenamefont {Gopalakrishnan}}]{SinghEtAl2026MixedStateLearnability}%
  \BibitemOpen
  \bibfield  {author} {\bibinfo {author} {\bibfnamefont {H.}~\bibnamefont {Singh}}, \bibinfo {author} {\bibfnamefont {R.}~\bibnamefont {Vasseur}}, \bibinfo {author} {\bibfnamefont {A.~C.}\ \bibnamefont {Potter}},\ and\ \bibinfo {author} {\bibfnamefont {S.}~\bibnamefont {Gopalakrishnan}},\ }\bibfield  {title} {\bibinfo {title} {Mixed-state learnability transitions in monitored noisy quantum dynamics},\ }\href {https://doi.org/10.1103/6b9y-scyz} {\bibfield  {journal} {\bibinfo  {journal} {Phys. Rev. B}\ }\textbf {\bibinfo {volume} {113}},\ \bibinfo {pages} {054305} (\bibinfo {year} {2026})}\BibitemShut {NoStop}%
\bibitem [{\citenamefont {Klocke}\ and\ \citenamefont {Buchhold}(2023)}]{KlockeBuchhold2023MajoranaLoops}%
  \BibitemOpen
  \bibfield  {author} {\bibinfo {author} {\bibfnamefont {K.}~\bibnamefont {Klocke}}\ and\ \bibinfo {author} {\bibfnamefont {M.}~\bibnamefont {Buchhold}},\ }\bibfield  {title} {\bibinfo {title} {{Majorana} loop models for measurement-only quantum circuits},\ }\href {https://doi.org/10.1103/PhysRevX.13.041028} {\bibfield  {journal} {\bibinfo  {journal} {Phys. Rev. X}\ }\textbf {\bibinfo {volume} {13}},\ \bibinfo {pages} {041028} (\bibinfo {year} {2023})}\BibitemShut {NoStop}%
\bibitem [{\citenamefont {Klocke}\ \emph {et~al.}(2025)\citenamefont {Klocke}, \citenamefont {Simm}, \citenamefont {Zhu}, \citenamefont {Trebst},\ and\ \citenamefont {Buchhold}}]{KlockeEtAl2025KitaevCircuits}%
  \BibitemOpen
  \bibfield  {author} {\bibinfo {author} {\bibfnamefont {K.}~\bibnamefont {Klocke}}, \bibinfo {author} {\bibfnamefont {D.}~\bibnamefont {Simm}}, \bibinfo {author} {\bibfnamefont {G.-Y.}\ \bibnamefont {Zhu}}, \bibinfo {author} {\bibfnamefont {S.}~\bibnamefont {Trebst}},\ and\ \bibinfo {author} {\bibfnamefont {M.}~\bibnamefont {Buchhold}},\ }\bibfield  {title} {\bibinfo {title} {Entanglement dynamics in monitored {Kitaev} circuits: Loop models, symmetry classification, and quantum {Lifshitz} scaling},\ }\href {https://doi.org/10.1103/PhysRevB.111.224301} {\bibfield  {journal} {\bibinfo  {journal} {Phys. Rev. B}\ }\textbf {\bibinfo {volume} {111}},\ \bibinfo {pages} {224301} (\bibinfo {year} {2025})}\BibitemShut {NoStop}%
\bibitem [{\citenamefont {Buchhold}\ \emph {et~al.}(2021)\citenamefont {Buchhold}, \citenamefont {Minoguchi}, \citenamefont {Altland},\ and\ \citenamefont {Diehl}}]{BuchholdEtAl2021DiracFermions}%
  \BibitemOpen
  \bibfield  {author} {\bibinfo {author} {\bibfnamefont {M.}~\bibnamefont {Buchhold}}, \bibinfo {author} {\bibfnamefont {Y.}~\bibnamefont {Minoguchi}}, \bibinfo {author} {\bibfnamefont {A.}~\bibnamefont {Altland}},\ and\ \bibinfo {author} {\bibfnamefont {S.}~\bibnamefont {Diehl}},\ }\bibfield  {title} {\bibinfo {title} {Effective theory for the measurement-induced phase transition of {Dirac} fermions},\ }\href {https://doi.org/10.1103/PhysRevX.11.041004} {\bibfield  {journal} {\bibinfo  {journal} {Phys. Rev. X}\ }\textbf {\bibinfo {volume} {11}},\ \bibinfo {pages} {041004} (\bibinfo {year} {2021})}\BibitemShut {NoStop}%
\bibitem [{\citenamefont {Ladewig}\ \emph {et~al.}(2022)\citenamefont {Ladewig}, \citenamefont {Diehl},\ and\ \citenamefont {Buchhold}}]{LadewigDiehlBuchhold2022MonitoredOpenFermions}%
  \BibitemOpen
  \bibfield  {author} {\bibinfo {author} {\bibfnamefont {B.}~\bibnamefont {Ladewig}}, \bibinfo {author} {\bibfnamefont {S.}~\bibnamefont {Diehl}},\ and\ \bibinfo {author} {\bibfnamefont {M.}~\bibnamefont {Buchhold}},\ }\bibfield  {title} {\bibinfo {title} {Monitored open fermion dynamics: Exploring the interplay of measurement, decoherence, and free {Hamiltonian} evolution},\ }\href {https://doi.org/10.1103/PhysRevResearch.4.033001} {\bibfield  {journal} {\bibinfo  {journal} {Phys. Rev. Research}\ }\textbf {\bibinfo {volume} {4}},\ \bibinfo {pages} {033001} (\bibinfo {year} {2022})}\BibitemShut {NoStop}%
\bibitem [{\citenamefont {Chahine}\ and\ \citenamefont {Buchhold}(2024)}]{ChahineBuchhold2024MonitoredFermions2D}%
  \BibitemOpen
  \bibfield  {author} {\bibinfo {author} {\bibfnamefont {K.}~\bibnamefont {Chahine}}\ and\ \bibinfo {author} {\bibfnamefont {M.}~\bibnamefont {Buchhold}},\ }\bibfield  {title} {\bibinfo {title} {Entanglement phases, localization, and multifractality of monitored free fermions in two dimensions},\ }\href {https://doi.org/10.1103/PhysRevB.110.054313} {\bibfield  {journal} {\bibinfo  {journal} {Phys. Rev. B}\ }\textbf {\bibinfo {volume} {110}},\ \bibinfo {pages} {054313} (\bibinfo {year} {2024})}\BibitemShut {NoStop}%
\bibitem [{\citenamefont {Poboiko}\ \emph {et~al.}(2024)\citenamefont {Poboiko}, \citenamefont {Gornyi},\ and\ \citenamefont {Mirlin}}]{PoboikoGornyiMirlin2024AboveOneDimension}%
  \BibitemOpen
  \bibfield  {author} {\bibinfo {author} {\bibfnamefont {I.}~\bibnamefont {Poboiko}}, \bibinfo {author} {\bibfnamefont {I.~V.}\ \bibnamefont {Gornyi}},\ and\ \bibinfo {author} {\bibfnamefont {A.~D.}\ \bibnamefont {Mirlin}},\ }\bibfield  {title} {\bibinfo {title} {Measurement-induced phase transition for free fermions above one dimension},\ }\href {https://doi.org/10.1103/PhysRevLett.132.110403} {\bibfield  {journal} {\bibinfo  {journal} {Phys. Rev. Lett.}\ }\textbf {\bibinfo {volume} {132}},\ \bibinfo {pages} {110403} (\bibinfo {year} {2024})}\BibitemShut {NoStop}%
\bibitem [{\citenamefont {Holevo}(1973)}]{Holevo1973InformationBound}%
  \BibitemOpen
  \bibfield  {author} {\bibinfo {author} {\bibfnamefont {A.~S.}\ \bibnamefont {Holevo}},\ }\bibfield  {title} {\bibinfo {title} {Bounds for the quantity of information transmitted by a quantum communication channel},\ }\href {https://www.mathnet.ru/eng/ppi903} {\bibfield  {journal} {\bibinfo  {journal} {Probl. Inf. Transm.}\ }\textbf {\bibinfo {volume} {9}},\ \bibinfo {pages} {177} (\bibinfo {year} {1973})}\BibitemShut {NoStop}%
\bibitem [{\citenamefont {Negari}\ \emph {et~al.}(2024)\citenamefont {Negari}, \citenamefont {Sahu},\ and\ \citenamefont {Hsieh}}]{Negari_2024}%
  \BibitemOpen
  \bibfield  {author} {\bibinfo {author} {\bibfnamefont {A.-R.}\ \bibnamefont {Negari}}, \bibinfo {author} {\bibfnamefont {S.}~\bibnamefont {Sahu}},\ and\ \bibinfo {author} {\bibfnamefont {T.~H.}\ \bibnamefont {Hsieh}},\ }\bibfield  {title} {\bibinfo {title} {Measurement-induced phase transitions in the toric code},\ }\href {https://doi.org/10.1103/PhysRevB.109.125148} {\bibfield  {journal} {\bibinfo  {journal} {Phys. Rev. B}\ }\textbf {\bibinfo {volume} {109}},\ \bibinfo {pages} {125148} (\bibinfo {year} {2024})}\BibitemShut {NoStop}%
\bibitem [{\citenamefont {Jian}\ \emph {et~al.}(2022)\citenamefont {Jian}, \citenamefont {Bauer}, \citenamefont {Keselman},\ and\ \citenamefont {Ludwig}}]{jian2022criticalityentanglementnonunitaryquantum}%
  \BibitemOpen
  \bibfield  {author} {\bibinfo {author} {\bibfnamefont {C.-M.}\ \bibnamefont {Jian}}, \bibinfo {author} {\bibfnamefont {B.}~\bibnamefont {Bauer}}, \bibinfo {author} {\bibfnamefont {A.}~\bibnamefont {Keselman}},\ and\ \bibinfo {author} {\bibfnamefont {A.~W.~W.}\ \bibnamefont {Ludwig}},\ }\bibfield  {title} {\bibinfo {title} {Criticality and entanglement in nonunitary quantum circuits and tensor networks of noninteracting fermions},\ }\href {https://doi.org/10.1103/PhysRevB.106.134206} {\bibfield  {journal} {\bibinfo  {journal} {Phys. Rev. B}\ }\textbf {\bibinfo {volume} {106}},\ \bibinfo {pages} {134206} (\bibinfo {year} {2022})},\ \Eprint {https://arxiv.org/abs/2012.04666} {arXiv:2012.04666 [cond-mat.stat-mech]} \BibitemShut {NoStop}%
\bibitem [{\citenamefont {Jian}\ \emph {et~al.}(2023)\citenamefont {Jian}, \citenamefont {Shapourian}, \citenamefont {Bauer},\ and\ \citenamefont {Ludwig}}]{jian2023Born}%
  \BibitemOpen
  \bibfield  {author} {\bibinfo {author} {\bibfnamefont {C.-M.}\ \bibnamefont {Jian}}, \bibinfo {author} {\bibfnamefont {H.}~\bibnamefont {Shapourian}}, \bibinfo {author} {\bibfnamefont {B.}~\bibnamefont {Bauer}},\ and\ \bibinfo {author} {\bibfnamefont {A.~W.~W.}\ \bibnamefont {Ludwig}},\ }\href {https://arxiv.org/abs/2302.09094} {\bibinfo {title} {Measurement-induced entanglement transitions in quantum circuits of non-interacting fermions: {Born}-rule versus forced measurements}} (\bibinfo {year} {2023}),\ \Eprint {https://arxiv.org/abs/2302.09094} {arXiv:2302.09094 [cond-mat.stat-mech]} \BibitemShut {NoStop}%
\bibitem [{\citenamefont {Oseledets}(1968)}]{Oseledets1968}%
  \BibitemOpen
  \bibfield  {author} {\bibinfo {author} {\bibfnamefont {V.~I.}\ \bibnamefont {Oseledets}},\ }\bibfield  {title} {\bibinfo {title} {A multiplicative ergodic theorem: {Lyapunov} characteristic numbers for dynamical systems},\ }\href {https://www.mathnet.ru/eng/mmo214} {\bibfield  {journal} {\bibinfo  {journal} {Trans. Moscow Math. Soc.}\ }\textbf {\bibinfo {volume} {19}},\ \bibinfo {pages} {197} (\bibinfo {year} {1968})}\BibitemShut {NoStop}%
\bibitem [{\citenamefont {MacKinnon}\ and\ \citenamefont {Kramer}(1983)}]{MacKinnonKramer1983}%
  \BibitemOpen
  \bibfield  {author} {\bibinfo {author} {\bibfnamefont {A.}~\bibnamefont {MacKinnon}}\ and\ \bibinfo {author} {\bibfnamefont {B.}~\bibnamefont {Kramer}},\ }\bibfield  {title} {\bibinfo {title} {The scaling theory of electrons in disordered solids: Additional numerical results},\ }\href {https://doi.org/10.1007/BF01578242} {\bibfield  {journal} {\bibinfo  {journal} {Z. Phys. B: Condens. Matter}\ }\textbf {\bibinfo {volume} {53}},\ \bibinfo {pages} {1} (\bibinfo {year} {1983})}\BibitemShut {NoStop}%
\bibitem [{\citenamefont {Yan}\ \emph {et~al.}(2026{\natexlab{b}})\citenamefont {Yan}, \citenamefont {Bao},\ and\ \citenamefont {Vijay}}]{yan2026algebraicstrongtoweaksymmetrybreaking}%
  \BibitemOpen
  \bibfield  {author} {\bibinfo {author} {\bibfnamefont {S.~W.}\ \bibnamefont {Yan}}, \bibinfo {author} {\bibfnamefont {Y.}~\bibnamefont {Bao}},\ and\ \bibinfo {author} {\bibfnamefont {S.}~\bibnamefont {Vijay}},\ }\href {https://arxiv.org/abs/2610.40567} {\bibinfo {title} {Algebraic strong-to-weak symmetry breaking and partial decodability in the surface code}} (\bibinfo {year} {2026}{\natexlab{b}}),\ \Eprint {https://arxiv.org/abs/2610.40567} {arXiv:2610.40567 [quant-ph]} \BibitemShut {NoStop}%
\bibitem [{\citenamefont {Winter}(2016)}]{Winter_2016_continuity}%
  \BibitemOpen
  \bibfield  {author} {\bibinfo {author} {\bibfnamefont {A.}~\bibnamefont {Winter}},\ }\bibfield  {title} {\bibinfo {title} {Tight uniform continuity bounds for quantum entropies: Conditional entropy, relative entropy distance and energy constraints},\ }\href {https://doi.org/10.1007/s00220-016-2609-8} {\bibfield  {journal} {\bibinfo  {journal} {Commun. Math. Phys.}\ }\textbf {\bibinfo {volume} {347}},\ \bibinfo {pages} {291} (\bibinfo {year} {2016})},\ \Eprint {https://arxiv.org/abs/1507.07775} {arXiv:1507.07775 [quant-ph]} \BibitemShut {NoStop}%
\bibitem [{\citenamefont {Chen}\ and\ \citenamefont {Rouz{\'{e}}}(2025)}]{chen2025quantumgibbsstateslocally}%
  \BibitemOpen
  \bibfield  {author} {\bibinfo {author} {\bibfnamefont {C.-F.}\ \bibnamefont {Chen}}\ and\ \bibinfo {author} {\bibfnamefont {C.}~\bibnamefont {Rouz{\'{e}}}},\ }\href {https://arxiv.org/abs/2504.02208} {\bibinfo {title} {Quantum {Gibbs} states are locally {Markovian}}} (\bibinfo {year} {2025}),\ \Eprint {https://arxiv.org/abs/2504.02208} {arXiv:2504.02208 [quant-ph]} \BibitemShut {NoStop}%
\bibitem [{\citenamefont {Kliesch}\ \emph {et~al.}(2014)\citenamefont {Kliesch}, \citenamefont {Gogolin}, \citenamefont {Kastoryano}, \citenamefont {Riera},\ and\ \citenamefont {Eisert}}]{KlieschEtAl2014LocalityTemperature}%
  \BibitemOpen
  \bibfield  {author} {\bibinfo {author} {\bibfnamefont {M.}~\bibnamefont {Kliesch}}, \bibinfo {author} {\bibfnamefont {C.}~\bibnamefont {Gogolin}}, \bibinfo {author} {\bibfnamefont {M.~J.}\ \bibnamefont {Kastoryano}}, \bibinfo {author} {\bibfnamefont {A.}~\bibnamefont {Riera}},\ and\ \bibinfo {author} {\bibfnamefont {J.}~\bibnamefont {Eisert}},\ }\bibfield  {title} {\bibinfo {title} {Locality of temperature},\ }\href {https://doi.org/10.1103/PhysRevX.4.031019} {\bibfield  {journal} {\bibinfo  {journal} {Phys. Rev. X}\ }\textbf {\bibinfo {volume} {4}},\ \bibinfo {pages} {031019} (\bibinfo {year} {2014})},\ \Eprint {https://arxiv.org/abs/1309.0816} {arXiv:1309.0816 [quant-ph]} \BibitemShut {NoStop}%
\bibitem [{\citenamefont {Hastings}(2007)}]{Hastings2007QuantumBeliefPropagation}%
  \BibitemOpen
  \bibfield  {author} {\bibinfo {author} {\bibfnamefont {M.~B.}\ \bibnamefont {Hastings}},\ }\bibfield  {title} {\bibinfo {title} {Quantum belief propagation: An algorithm for thermal quantum systems},\ }\href {https://doi.org/10.1103/PhysRevB.76.201102} {\bibfield  {journal} {\bibinfo  {journal} {Phys. Rev. B}\ }\textbf {\bibinfo {volume} {76}},\ \bibinfo {pages} {201102} (\bibinfo {year} {2007})},\ \Eprint {https://arxiv.org/abs/0706.4094} {arXiv:0706.4094 [cond-mat.str-el]} \BibitemShut {NoStop}%
\bibitem [{\citenamefont {Schultz}\ \emph {et~al.}(1964)\citenamefont {Schultz}, \citenamefont {Mattis},\ and\ \citenamefont {Lieb}}]{SchultzMattisLieb1964IsingFermions}%
  \BibitemOpen
  \bibfield  {author} {\bibinfo {author} {\bibfnamefont {T.~D.}\ \bibnamefont {Schultz}}, \bibinfo {author} {\bibfnamefont {D.~C.}\ \bibnamefont {Mattis}},\ and\ \bibinfo {author} {\bibfnamefont {E.~H.}\ \bibnamefont {Lieb}},\ }\bibfield  {title} {\bibinfo {title} {Two-dimensional {Ising} model as a soluble problem of many fermions},\ }\href {https://doi.org/10.1103/RevModPhys.36.856} {\bibfield  {journal} {\bibinfo  {journal} {Rev. Mod. Phys.}\ }\textbf {\bibinfo {volume} {36}},\ \bibinfo {pages} {856} (\bibinfo {year} {1964})}\BibitemShut {NoStop}%
\bibitem [{\citenamefont {Kastening}(2001)}]{Kastening2001IsingSpectrum}%
  \BibitemOpen
  \bibfield  {author} {\bibinfo {author} {\bibfnamefont {B.}~\bibnamefont {Kastening}},\ }\bibfield  {title} {\bibinfo {title} {Simplifying {Kaufman}'s solution of the two-dimensional {Ising} model},\ }\href {https://doi.org/10.1103/PhysRevE.64.066106} {\bibfield  {journal} {\bibinfo  {journal} {Phys. Rev. E}\ }\textbf {\bibinfo {volume} {64}},\ \bibinfo {pages} {066106} (\bibinfo {year} {2001})},\ \Eprint {https://arxiv.org/abs/cond-mat/0111380} {arXiv:cond-mat/0111380 [cond-mat.stat-mech]} \BibitemShut {NoStop}%
\bibitem [{\citenamefont {Duminil-Copin}\ \emph {et~al.}(2020)\citenamefont {Duminil-Copin}, \citenamefont {Goswami},\ and\ \citenamefont {Raoufi}}]{DuminilCopinGoswamiRaoufi2020Truncated}%
  \BibitemOpen
  \bibfield  {author} {\bibinfo {author} {\bibfnamefont {H.}~\bibnamefont {Duminil-Copin}}, \bibinfo {author} {\bibfnamefont {S.}~\bibnamefont {Goswami}},\ and\ \bibinfo {author} {\bibfnamefont {A.}~\bibnamefont {Raoufi}},\ }\bibfield  {title} {\bibinfo {title} {Exponential decay of truncated correlations for the {Ising} model in any dimension for all but the critical temperature},\ }\href {https://doi.org/10.1007/s00220-019-03633-y} {\bibfield  {journal} {\bibinfo  {journal} {Commun. Math. Phys.}\ }\textbf {\bibinfo {volume} {374}},\ \bibinfo {pages} {891} (\bibinfo {year} {2020})},\ \Eprint {https://arxiv.org/abs/1808.00439} {arXiv:1808.00439 [math.PR]} \BibitemShut {NoStop}%
\bibitem [{\citenamefont {Izyurov}\ \emph {et~al.}(2024)\citenamefont {Izyurov}, \citenamefont {Kemppainen},\ and\ \citenamefont {Tuisku}}]{IzyurovKemppainenTuisku2024Energy}%
  \BibitemOpen
  \bibfield  {author} {\bibinfo {author} {\bibfnamefont {K.}~\bibnamefont {Izyurov}}, \bibinfo {author} {\bibfnamefont {A.}~\bibnamefont {Kemppainen}},\ and\ \bibinfo {author} {\bibfnamefont {P.}~\bibnamefont {Tuisku}},\ }\bibfield  {title} {\bibinfo {title} {Energy correlations in the critical {Ising} model on a torus},\ }\href {https://doi.org/10.1214/23-AAP1968} {\bibfield  {journal} {\bibinfo  {journal} {Ann. Appl. Probab.}\ }\textbf {\bibinfo {volume} {34}},\ \bibinfo {pages} {1699} (\bibinfo {year} {2024})},\ \Eprint {https://arxiv.org/abs/2104.01084} {arXiv:2104.01084 [math-ph]} \BibitemShut {NoStop}%
\bibitem [{\citenamefont {Bravyi}(2005)}]{Bravyi_2005_FLO}%
  \BibitemOpen
  \bibfield  {author} {\bibinfo {author} {\bibfnamefont {S.}~\bibnamefont {Bravyi}},\ }\bibfield  {title} {\bibinfo {title} {{Lagrangian} representation for fermionic linear optics},\ }\href {https://doi.org/10.26421/QIC5.3-3} {\bibfield  {journal} {\bibinfo  {journal} {Quantum Inf. Comput.}\ }\textbf {\bibinfo {volume} {5}},\ \bibinfo {pages} {216} (\bibinfo {year} {2005})},\ \Eprint {https://arxiv.org/abs/quant-ph/0404180} {arXiv:quant-ph/0404180 [quant-ph]} \BibitemShut {NoStop}%
\bibitem [{\citenamefont {Bravyi}\ and\ \citenamefont {Terhal}(2009)}]{BravyiTerhal2009Cleaning}%
  \BibitemOpen
  \bibfield  {author} {\bibinfo {author} {\bibfnamefont {S.}~\bibnamefont {Bravyi}}\ and\ \bibinfo {author} {\bibfnamefont {B.}~\bibnamefont {Terhal}},\ }\bibfield  {title} {\bibinfo {title} {A no-go theorem for a two-dimensional self-correcting quantum memory based on stabilizer codes},\ }\href {https://doi.org/10.1088/1367-2630/11/4/043029} {\bibfield  {journal} {\bibinfo  {journal} {New J. Phys.}\ }\textbf {\bibinfo {volume} {11}},\ \bibinfo {pages} {043029} (\bibinfo {year} {2009})},\ \Eprint {https://arxiv.org/abs/0810.1983} {arXiv:0810.1983 [quant-ph]} \BibitemShut {NoStop}%
\bibitem [{\citenamefont {Beenakker}(1997)}]{Beenakker1997}%
  \BibitemOpen
  \bibfield  {author} {\bibinfo {author} {\bibfnamefont {C.~W.~J.}\ \bibnamefont {Beenakker}},\ }\bibfield  {title} {\bibinfo {title} {Random-matrix theory of quantum transport},\ }\href {https://doi.org/10.1103/RevModPhys.69.731} {\bibfield  {journal} {\bibinfo  {journal} {Rev. Mod. Phys.}\ }\textbf {\bibinfo {volume} {69}},\ \bibinfo {pages} {731} (\bibinfo {year} {1997})}\BibitemShut {NoStop}%
\bibitem [{\citenamefont {Poboiko}\ \emph {et~al.}(2023)\citenamefont {Poboiko}, \citenamefont {P{\"{o}}pperl}, \citenamefont {Gornyi},\ and\ \citenamefont {Mirlin}}]{PoboikoEtAl2023ProjectiveMeasurements}%
  \BibitemOpen
  \bibfield  {author} {\bibinfo {author} {\bibfnamefont {I.}~\bibnamefont {Poboiko}}, \bibinfo {author} {\bibfnamefont {P.}~\bibnamefont {P{\"{o}}pperl}}, \bibinfo {author} {\bibfnamefont {I.~V.}\ \bibnamefont {Gornyi}},\ and\ \bibinfo {author} {\bibfnamefont {A.~D.}\ \bibnamefont {Mirlin}},\ }\bibfield  {title} {\bibinfo {title} {Theory of free fermions under random projective measurements},\ }\href {https://doi.org/10.1103/PhysRevX.13.041046} {\bibfield  {journal} {\bibinfo  {journal} {Phys. Rev. X}\ }\textbf {\bibinfo {volume} {13}},\ \bibinfo {pages} {041046} (\bibinfo {year} {2023})}\BibitemShut {NoStop}%
\end{thebibliography}
